\documentclass[a4,oneside,11pt]{book}
\usepackage{palatino}   

\usepackage{amsfonts}
\usepackage{latexsym}
\usepackage{amssymb}
\usepackage{amsmath}
\usepackage{amsthm}
\usepackage[mathscr]{eucal}
\usepackage{eufrak}
\usepackage{amssymb}
\usepackage{epsf}
\usepackage{epsfig}

\usepackage{rotating}

\usepackage{sectsty}    

\usepackage{fancyhdr}   

\usepackage{bbold}      

\usepackage{vmargin}    

\chapterfont{\centering\mdseries\sc}
\sectionfont{\centering\mdseries\sc}
\subsectionfont{\mdseries\sc\large}
\subsubsectionfont{\mdseries\sc\large}

\numberwithin{equation}{section}

\setmarginsrb{25mm}{20mm}{25mm}{20mm} %
                {5mm}{5mm}{5mm}{8mm}

\newtheorem{thm}{Theorem}[section]
\newtheorem{lem}[thm]{Lemma}
\newtheorem{prop}[thm]{Proposition}

\def \be  {\begin{equation}}
\def \ee  {\end{equation}}
\def \benn{\begin{equation*}}
\def \eenn{\end{equation*}}
\def \bea {\begin{eqnarray}}
\def \eea {\end{eqnarray}}
\def \beann{\begin{eqnarray*}}
\def \eeann{\end{eqnarray*}}
\def \ba  {\begin{array}}
\def \ea  {\end{array}}
\def \nn  {\nonumber}

\def \btab{\begin{center}\begin{tabular}}
\def \etab{\end{tabular}\end{center}}

\def \la  {\label}
                              
\def \fes {field equations}                             \def \fesb  {field equations }
\def \hs  {heterotic string}                            \def \hsb   {heterotic string }
\def \ks  {Killing spinor}                              
\def \kss {Killing spinors}                             
\def \kse {Killing spinor equation}                     
\def \kses{Killing spinor equations}                    \def \ksesb {Killing spinor equations }
\def \sm  {sigma-model}                                 \def \smb   {sigma-model }
\def \sg  {supergravity}                                \def \sgb   {supergravity }
\def \st  {spacetime}
\def \inv {invariant}

\def \a   {\alpha }                     \def \ab   {\bar{\alpha}}
\def \b   {\beta}                       \def \betab{\bar{\beta}}
\def \g   {\gamma}                      \def \gb   {\bar{\gamma}}
\def \d   {\delta}                      \def \db   {\bar{\delta}}
\def \e   {\epsilon}
\def \ve  {\varepsilon}
\def \z   {\zeta}
\def \th  {\theta}
\def \i   {\iota}
\def \k   {\kappa}
\def \l   {\lambda}
\def \m   {\mu}
\def \s   {\sigma}
\def \t   {\tau}
\def \ph  {\phi}
\def \vp  {\varphi}
\def \ps  {\psi}
\def \o   {\omega}

\def \D   {\Delta}
\def \G   {\Gamma}                      \def \Gh   {\hat{\Gamma}}
\def \L   {\Lambda}
\def \Ph  {\Phi}
\def \O   {\Omega}
\def \S   {\Sigma}
\def \U   {\Upsilon}

\def \bb  {\mathbb}     
\def \bR  {\bb{R}}      
\def \bC  {\bb{C}}      
\def \bO  {\bb{O}}  

\def \bP  {\bb{P}}  
\def \bS  {\bb{S}}  

\def \bK  {\bb{K}}

\def \id  {\bb{1}}      

\def \go  {\mathring{g}}
\def \Ho  {\mathring{H}}
\def \Pho {\mathring{\Ph}}
\def \Ao  {\mathring{A}}
\def \Fo  {\mathring{F}}

\def \Oo  {\mathring{\Omega}}

\def \Vc  {V_{\bC}}     
\def \Uc  {U_{\bC}}     

\def \Dc  {\D^{\bC}}            
\def \Dm  {\D_{32}}     

\def \Dt  {\D^{\bC}}
\def \Dsm {\D^{SM}}

\def \Lf  {\L^2(\bC^5)}

\def \sut  {$SU(2)$}

\def \suf  {$SU(5)$}
\def \sufr {$SU(4)$}
\def \rt   {$\bR^3$}
\def \spf  {$Spin(1,4)$}
\def \spfr {$Spin(4)$}
\def \spt  {$Spin(1,10)$}
\def \sps  {$(Spin(7)\ltimes\bR^8)\times \bR$}

\def \gg    {\mathfrak{g}}
\def \gperp {\mathfrak{g}^{\perp}}
\def \h     {\mathfrak{h}}

\def \son   {\mathfrak{so}(n)}
\def \sonperp   {\mathfrak{so}(n)^{\perp}}
\def \sotm  {\mathfrak{so}(2m)}

\def \so    {\mathfrak{so}}
\def \su    {\mathfrak{su}}

\def \um    {\mathfrak{u}(m)}
\def \umperp    {\mathfrak{u}(m)^{\perp}}
\def \umm   {\mathfrak{u}(m-1)}
\def \ummperp   {\mathfrak{u}(m-1)^{\perp}}

\def \sumperp   {\mathfrak{su}(m)^{\perp}}
\def \summ  {\mathfrak{su}(m-1)}
\def \summperp  {\mathfrak{su}(m-1)^{\perp}}

\def \orbsut  {${\cal O}_{SU(2)}$}
\def \orbrt   {${\cal O}_{\bR^3}$}
\def \orbsufr {${\cal O}_{SU(4)}$}
\def \orbsuf  {${\cal O}_{SU(5)}$}
\def \orbrt   {${\cal O}_{\bR^3}$}
\def \orbsps  {${\cal O}_{Spin(7)}$}

\def \orbsutth{${\cal O}_{SU(2)\times SU(3)}$}

\def \zsut   {\zeta^{SU(2)}}
\def \etasuf {\eta^{SU(5)}}
\def \thsuf  {\theta^{SU(5)}}
\def \etasufr{\eta^{SU(4)}}
\def \thsufr {\theta^{SU(4)}}

\def \cl  {\mathcal{C}l}        

\def \ccl {\bC l}               

\def \we        {\wedge}      
\def \h         {\star}     
\def \lc        {\lrcorner}   
\def \nat       {\natural}    
\def \na        {\nabla}
\def \nao       {\mathring{\nabla}}
\def \naplus    {\nabla^{(+)}}
\def \napluso   {\mathring{\nabla}^{(+)}}

\def \ww  {\mathrm{w}}      
\def \vv  {\mathrm{v}}      

\def \A   {\mathfrak{A}} 
\def \B   {\mathfrak{B}} 

\def \Z   {\mathscr{Z}} 

\def \ord {\mathscr{O}} 

\def \n   {N_s}         
\def \Gten{\G_{\nat}}   

\begin{document}
        \setcounter{page}{1}
        \pagestyle{empty}\renewcommand{\baselinestretch}{1.5} \normalsize
%
%
%
\begin{center}
%
%
%
\resizebox{4cm}{!}{\includegraphics{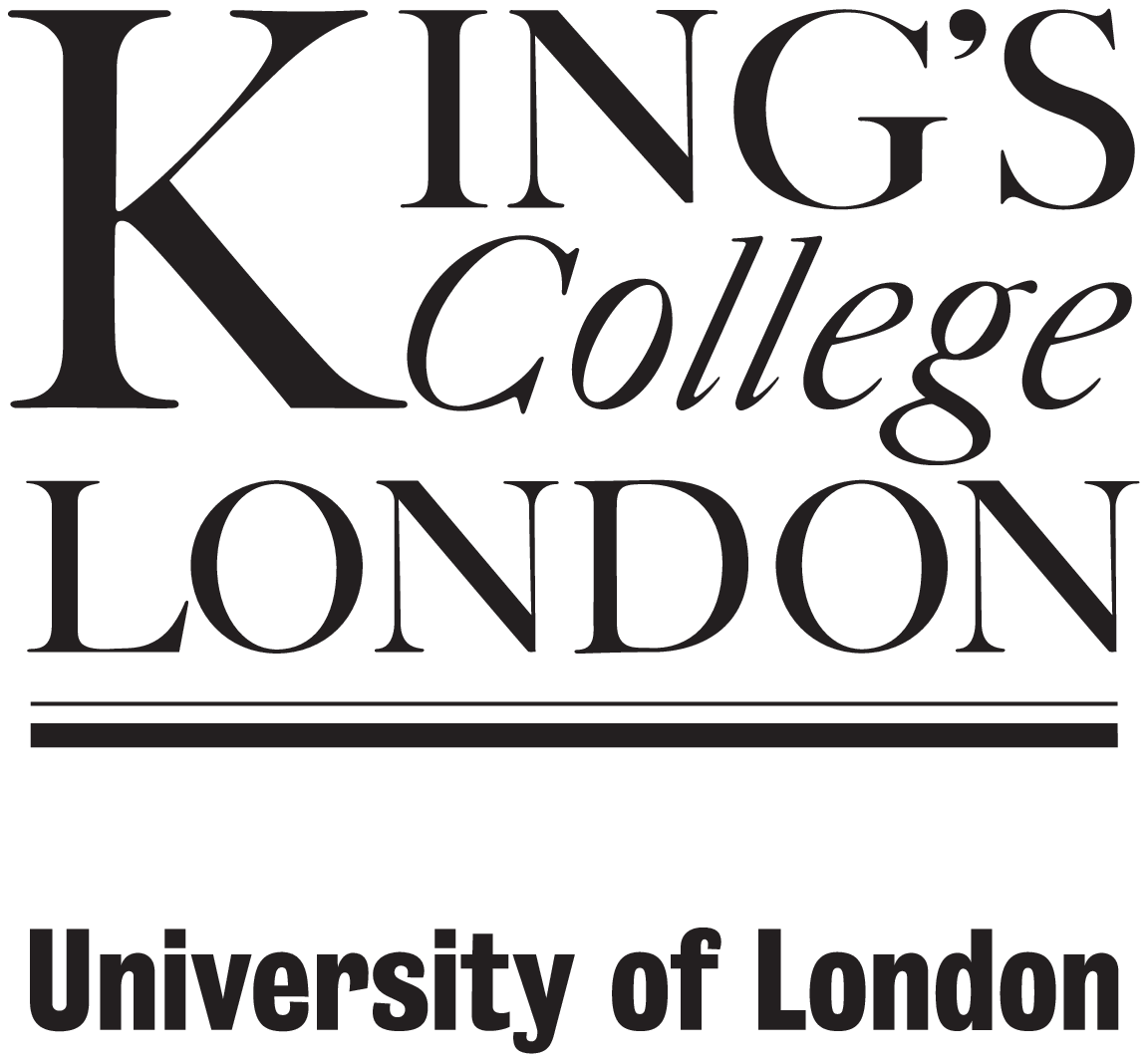}} \\ \ 
\vspace{1cm} \\ \
\Huge{\scshape Spinorial Geometry} \\
\scshape and Supergravity \\
            \vspace{3cm} 
\Large{\scshape Joseph J. Gillard} \\ \medskip
\small{Department of Mathematics,} \\
{King's College, London} \\
            \vspace{3cm}
\large{Ph.D. Thesis} \\
\normalsize{Supervisor: Dr G. Papadopoulos} \\ \ \\
\normalsize{Submitted in September 2005} \\ \ \\
\normalsize{Examined and passed by Prof. P. K. Townsend} \\
\normalsize{and Dr A. Brandhuber in February 2006} \\
\end{center}
        \pagestyle{plain} \renewcommand{\baselinestretch}{1.0} \normalsize
%
%
%
%
\newpage
\begin{center}
\LARGE \scshape{Abstract} \\ \ \\
%
%
\end{center}
\normalsize
\begin{slshape}
Heterotic string backgrounds with $(2,0)$ world-sheet supersymmetry are considered first.
 We investigate the consequences of taking $\a'$-corrections into account in the field equations, in order to remain consistent with anomaly cancellation, while requiring that spacetime supersymmetry is preserved.
 We compute the corrections to the fields, and show that the equations are consistent at order $\a'$, provided that an analogue of the $\partial\bar\partial$-lemma and a suitable gauge-fixing condition are valid on the background.
 These results are applied to $(2,0)$ compactifications and it is shown that after corrections, the Calabi-Yau geometry of the internal space is deformed to be Hermitian.

We then go on to propose a new method for solving the \kses\ of a \sg\ theory, and focus on the $D=11$ case in detail.
 The method is based on an explicit description of spinors in terms of exterior forms and also on the gauge symmetries of the supercovariant connection.
 A comprehensive development of this formalism is provided, which serves to underpin its subsequent application.

We investigate backgrounds with one, two, three and four \kss, provided that one of the spinors is \suf-\inv.
 In many cases, we can utilise our knowledge about the orbits of \spt\ and those of its subgroups on its spinor space to deduce canonical forms for the spinors.
 A derivation of the orbits of \suf\ on the spinor space of \spt\ is provided.
 The spacetime forms associated to a given set of spinors may then be calculated, via a specially constructed \spt-\inv\ bilinear form.
 For each configuration of spinors that is considered, the \kses\ are solved and the resulting constraints are then interpreted to provide illuminating information about the geometry of the spacetime.
\end{slshape}
%
%
%
%
%
\newpage
\begin{center}
\LARGE \scshape{Acknowledgements} \\ \ \\
\end{center}
\normalsize
I would like to thank George Papadopoulos for his supervision and expert guidance throughout the course of my studies.
 Thanks also to Ulf Gran and Dimitrios Tsimpis for their fruitful collaborations and willing helpfulness.
 I am also grateful to McKenzie Wang for his valuable correspondence, and to my examiners Paul Townsend and Andreas Brandhuber for their helpful comments.

My endless thanks are extended to my family for their continual love, support and encouragement.

Thanks to my office-mates past and present, whose combined help, distraction and friendship have made the office such an enjoyable place to be.

Also, special thanks to the many fantastic friends and house-mates who have been there to colour my life with fun, helping me to keep the right perspective and giving me countless laughs and excellent memories.

Finally, I thank God for every blessing, including all of the above.
%
%
%
%
\newpage
\begin{center}
\ \\
\vspace{9cm}
\LARGE \scshape {For Mum and Dad.}
\end{center}
%
%
\newpage
        \pagestyle{fancy}
%
%
%
%
%
\renewcommand{\sectionmark}[1]{\markright{\thesection\ #1}}
{\lhead{\nouppercase{\bfseries{\rightmark}}}              
\chead{}
\rhead{\bfseries{\thepage}}                             
\lfoot{}\cfoot{}\rfoot{}
%
%
%
%
\tableofcontents
%
%
        \pagestyle{fancy}
%
%
%
%
%
\chapter{Introduction}
\section{Motivation}
The aim of this thesis is to gain a further understanding of the \sg\ theories that can arise as low-energy descriptions of string and M-theory. Two separate problems will be investigated, but before describing these, we will briefly review the relevance of various supergravities in the context of modern theoretical high-energy physics.

At the present time, Quantum Field Theory (QFT) provides a
successful way of describing the electromagnetic force as well as
the strong and weak nuclear forces by means of the Standard Model,
which is a QFT with gauge group $SU(3)\times SU(2)\times U(1)$. It
has been repeatedly confirmed to be an accurate model, and one of
its major successes is that it describes these three fundamental
forces by one single QFT. On the other hand, Einstein's
revolutionary theory of General Relativity (GR) provides a
thoroughly affirmed description of the fourth fundamental force,
gravity. The central concept in GR is that space and time are
dynamical and are curved by the presence of matter and energy. It is
this curvature which accounts for the motion of objects in a
gravitational field, among other things. However, as a force gravity
is far weaker than electromagnetism and the nuclear forces, and this
leads to difficulties when one tries to quantise GR. In fact, it is
non-renormalisable and as such, it fails to provide a description of
gravity at the quantum level. Therefore, if there is to be a single
quantum field theory which unifies all four fundamental forces, then
it must incorporate both General Relativity and the Standard Model
and unite them in a consistent way.

Supersymmetric string theory has emerged as the most promising way of achieving a unification so far.
 The arena for this is ten-dimensional spacetime, where the contact with four-dimensional physics that resembles the Standard Model is made by the process of `compactification'.
 This involves treating four of the ten dimensions as large and non-compact, to provide a four-dimensional spacetime, while regarding the remaining six dimensions as compact and too small to be detected with our current investigative technology.
 The requirement of supersymmetry ensures that the number of bosonic and fermionic degrees of freedom are equal.
 It turns out that there are five distinct string theories:
\begin{itemize}
\item \textbf{Type I}: this consists of both open and unoriented closed strings;
\item \textbf{Heterotic $E_8\times E_8$ or $SO(32)$ theories}: these are hybrid theories which consist of closed strings and superstrings, and two distinct theories arise from considering the gauge group to be either $SO(32)$ or $E_8\times E_8$;
\item \textbf{Type IIA}: a theory of closed strings;
\item \textbf{Type IIB}: a theory of closed strings.
\end{itemize}
One might wonder at having five separate consistent theories which are supposed to provide a single unifying theory, and indeed this was the cause of some concern, until in the 1990's a host of duality relationships were found between them (see for example, \cite{schwarz}, \cite{giveon}, \cite{townsendrev} for reviews and extensive references).
 Broadly speaking, these dualities come in three varieties, which can be essentially described as follows.
 Two theories $A$ and $B$ are \textit{S-dual} if $A$ at strong-coupling is equivalent to $B$ at weak-coupling, and vice versa, \cite{sen1}, \cite{schwarz1}.
 They are said to be \textit{T-dual} if compactifying $A$ on a space of large volume yields an equivalent theory to compactifying $B$ on a space of small volume, and vice versa \cite{giveon}.
 Finally, \textit{U-duality} occurs when compactifying $A$ on a space of large (or small) volume is equivalent to $B$ at strong (or weak) coupling, and it also incorporates both $S$ and $T$-dualities \cite{ht3}.

Thus the five apparently distinct theories may be shown to be equivalent in a certain sense.
 Furthermore, it was found that these theories could actually all be unified by a single theory which requires eleven \st\ dimensions \cite{witten1}. This unifying theory was termed `M-theory', and the different string theories arise as perturbative expansions with respect to different limits.

To study the interactions of the massless fields in these string
theories, one must consider the low-energy effective actions. During
the late seventies and early eighties, these actions were
constructed for each of the five string theories. Remarkably, each
was found to be described by a ten-dimensional \sg\ action (plus a
Yang-Mills action in some cases), while the low-energy limit of
M-theory can be described by \sg\ in eleven dimensions. The
equations of motion for the bosonic fields can be derived from the
beta-functions associated to the relevant sigma-model, when one
imposes conformal invariance. The following table shows the
correspondence between each theory and its associated low-energy
effective limit:

\begin{center}
\begin{tabular}{|c||c|} \hline
\textbf{Theory}          & \textbf{Low-energy dynamics}                     \\ \hline\hline
Type I String Theory    & $N=1$, $D=10$ \sg\ / YM with gauge group                      \\
            & $SO(32)$                              \\ \hline
Type IIA String Theory  & Non-chiral $N=2$, $D=10$ \sg                          \\ \hline
Type IIB String Theory  & Chiral $N=2$, $D=10$ \sg                              \\ \hline
Heterotic String Theory & $N=1$, $D=10$ \sg\ / YM with gauge group                      \\
            & either $SO(32)$ or $E_8\times E_8$                    \\ \hline
M-Theory                & $N=1$, $D=11$ \sg                                 \\  \hline
\end{tabular}
\end{center}

So we see that theories of \sg\ are key ingredients for the understanding of the dynamics of the massless fields which occur in string theory. This in itself is motivation enough to pursue a greater understanding of \sg. In addition though, there is much to be said for studying \sg\ theories from a mathematical point of view. For the modelling of \st\ as a curved manifold, as in Einstein's original formulation of GR, brings to light a deep and fascinating interplay between many pure geometric concepts and theoretical physics. When one considers the consequences of adding supersymmetry to the situation, then there is even more structure that warrants investigation.
 For instance, supersymmetric compactification of the ten-dimensional heterotic string to four-dimensions requires that the compact six-dimensional `internal' manifold satisfies strict requirements.
 As we will see in Chapter \ref{het}, when the field strength vanishes this internal space is constrained to be a Calabi-Yau manifold, highlighting a deep connection between string theory and algebraic geometry.

Another example of this interplay is one with which we will be much concerned in this thesis, namely the existence of parallel spinors on a manifold.
 In terms of physics, the number of supersymmetries that a \sg\ solution preserves is equal to the number of linearly independent Killing spinors which exist in the \st.
 This physical motivation requires us to gain a precise mathematical insight into the existence and nature of the spinors.
 For example, one is led to ask questions such as, how can we most conveniently represent these spinors?
 What are the consequences for the manifold if there are many spinors with differing symmetry properties?
 Also, the existence of Killing spinors has a deep connection with the holonomy of the manifold, which in turn opens up an even wider range of mathematical machinery which can be applied to the theory.
 In this way, we see that to understand the physical situations that can occur in a given theory, one must have a handle on the rigorous nature of the underlying mathematical structures.
 It is predominantly this rich and rewarding application of differential geometry and spin geometry to the investigation of \st\ physics, that motivates the author to pursue this work.
\section{Outline of Research}
In regards to \sg\ we will investigate two separate problems.
 The first may be thought of as a question of the \textit{accuracy} of a \sg\ theory, while the second is more concerned with the \textit{completeness} of understanding that one may have about a given theory.
\subsection*{Compactifications of the Heterotic String}
In Chapter \ref{het} the first problem is addressed, by considering whether and how the effective field equations and \kses\ for the heterotic string can be solved when $\a'$-corrections are taken into account.
 Such corrections arise in the following way.
 To derive the effective field equations that govern the low-energy dynamics of the string, the beta-functions for the string sigma-model are calculated as perturbative expansions in $\a'$, where $\a'$ is used as the loop-expansion parameter.
 It is necessary that the model be conformally invariant, which requires that the beta-functions vanish.
 It is exactly the vanishing of the beta-functions for each of the fields that provides the corresponding field equation in the effective theory.
 Thus the field equations arise as perturbative expansions, where the $\a'$-corrections can be thought of as `stringy' corrections to the underlying \sg\ theory.

However, there is a complicating factor which crops up in $N=1$, $D=10$ \sg\ theory, namely the presence of anomalies arising from the graviton and the fermions.
 This presents a conflict, since the heterotic string theory must be anomaly-free.
 To counteract this, the \textit{Green-Schwarz anomaly cancellation mechanism} must be imposed to ensure that the effective theory remains anomaly-free.
 This amounts to adding a correction of order $\a'$ to the two-form field $b$, which in turn modifies its Bianchi identity.
 The two-form is closed to zeroth order, so that $db=0$, but this no longer holds once the anomaly cancellation mechanism has been implemented, since $db$ is now proportional to a term which is of order $\a'$.
 This necessary modification of the $b$-field surely warrants that we must now consider all the equations of the theory in their $\a'$-corrected form.
 To impose anomaly cancellation while still working with the zeroth order uncorrected equations seems an inconsistent way to proceed.
 In particular, one might wonder the circumstances under which a solution to the uncorrected equations will remain a solution to the corrected equations.

Given these considerations, in Chapter \ref{het} the task is undertaken of solving the field and \kses\ when their first-order $\a'$-corrections are taken into account.
 The procedure is fairly intuitive, in that we begin with a set of fields which solve the uncorrected field equations and treat a first-order solution as a small deformation of the uncorrected solution.
 This is implemented by expanding the fields in powers of $\a'$ and working to linear order, from which we find that the $\a'$-corrected equations impose constraints on the deformations of the fields.
 The constraints can then be analysed, providing information about the nature of the allowed deformations.
 This in turn may be interpreted geometrically, as conditions on the \st.

In particular, we consider the consequences for compactifications of the heterotic string to four dimensions, which preserve $(2,0)$ \st\ supersymmetry.
 It was shown by Strominger in \cite{strom} that such compactifications which preserve supersymmetry are necessarily those in which the internal space is an Hermitian (but not K\"{a}hler) manifold equipped with a non-vanishing $(3,0)$-form.
 Then, in \cite{gpsi} it was shown that compactifications with the spin-connection embedded in the gauge connection are completely ruled out. Furthermore, the only possible compactifications to four dimensions with either four or eight remaining supersymmetries, are onto Calabi-Yau $3$-folds with a constant dilaton field.
 In Chapter \ref{het}, we show that when the $\a'$-corrections are taken into account, the internal manifold is deformed from being Calabi-Yau to just being compact and Hermitian.
 Thus, there is a much broader class of solutions once the stringy corrections are taken into account.
\subsection*{Classifying Supersymmetric Solutions}
In Chapter \ref{11d} we turn to the second problem, which deals with the question of how complete an understanding we may have of the \sg\ theories with which we work.
 As has already been mentioned above, effective supergravity theories can be used to describe the low-energy dynamics of the various string theories, as well as M-theory.
 Because of this, there has been a lot of research over the last ten years into the understanding of supersymmetric solutions to \sg\ theories in ten and eleven-dimensions, due to the insights that these provide into the corresponding high energy theory.
 Also, supergravities in fewer dimensions have become active areas of research, for instance in five dimensions, due to new `rotating black ring' solutions which have been found \cite{blackrings}.

It would therefore be ideal to have a systematic classification of the possible solutions that can occur any given \sg\ theory.
 One huge benefit of such a classification is that it would provide a complete set of possible background geometries for the various strings, from the low-energy point of view.
 Until now, a host of exact solutions have been found to the \sg\ field equations by positing ans\"atze based on physical intuition and educated guess-work.
 For instance, one may make certain assumptions about the symmetries of the theory in question, or one may impose some constraints on the field strength, which will lead to some simplification when inserted into the field equations.

Some famous examples in eleven dimensions are the $M2$- and $M5$-brane solutions, first established in \cite{m2} and \cite{m5} respectively, and also the intersecting brane solutions of \cite{intersectingbranes}.
 These and many other solutions have provided some ground-breaking insight into the nature of $M$- and string theory in their low-energy limits.
 However, although these are valid and extremely useful in understanding certain scenarios, the truth is that until we understand the full set of situations that can occur in a given theory, we will be vastly restricted in terms of the insight that we can gain from the low-energy approach to string theory.
 It is inevitable that in determining the general classes of solution that can occur, one will discover solutions which were either missed or simply unobtainable via the method of using ans\"atze, as enlightening as this approach may be in its own right.
 Therefore, to seek a full classification of supersymmetric solutions is to strive for a complete understanding of the low-energy dynamics of string theory and M-theory.

However, clearly such a classification is a formidable task which may be hindered by many things.
 For instance, the \sg\ analogue of the Einstein equation naturally leads to non-trivial Lorentzian manifolds as its solutions. We know that Riemannian manifolds may be classified by their restricted holonomy groups, which is made possible by combining Berger's holonomy theorem \cite{berger} with classical results about reducibility and symmetric spaces (see for example, Ch. 3 of \cite{joyce}).
 However, there is no Berger-type classification of Lorentzian holonomy groups, which are typically more exotic \cite{jose}, \cite{bry}.
 Another problem is that the supercovariant connection which arises naturally in the \kse, is an object which does not have an apparent geometric interpretation.
 For example, in $D=11$ the Levi-Civita connection contains products of two gamma matrices, and so takes its values in the $Spin(1,10)$ subalgebra of the Clifford algebra $\cl_{1,10}$.
 This enables us to relate it to the Levi-Civita connection on the tangent bundle, providing a geometric interpretation of the spin-connection.
 In contrast however, the supercovariant connection contains terms which have products of three and five gamma matrices.
 This means that it takes its values in the Clifford algebra outside of the spin subalgebra. This makes it much more difficult to analyse in comparison to the Levi-Civita spin connection.
 Another complicating factor is that even once one manages to determine all possible classes of solution to a certain \sg, then it may still be a difficult task to construct the explicit solutions occurring within that class.

Having said all this, some extremely positive progress has been made so far in a number of areas, and using a variety of approaches.
 We will mention a few notable examples.
 In \cite{jose}, Figueroa-O'Farrill classified all $D=11$ zero-flux vacua, by employing the machinery of Lorentzian holonomy groups.
 This analysis is made tractable by assuming that the flux vanishes, so that the supercovariant connection reduces to the spin-connection.
 Then, in \cite{josegp1,josegp2}, Figueroa-O'Farrill and Papadopoulos classified all maximally supersymmetric solutions to ten and eleven-dimensional \sg. In this approach, they made use the zero curvature condition which arises from maximal supersymmetry as well as the algebraic \kse\ which arises in the $D=10$ cases.

At another extreme, the authors of \cite{pakis,ggp1} solved the $D=11$ \kses\ for one Killing spinor, applying the theory of $G$-structures as an organisational tool.
 This method has also proved successful in classifying solutions for the minimal $D=5$, minimal gauged $D=5$ and minimal $D=6$ \sg\ theories in \cite{5dclass}, \cite{gauged5dclass} and \cite{6dclass} respectively.
 Moreover, as early as 1983, a classification of the $N=2$, $D=4$ theory was achieved by Tod \cite{tod}.

In this thesis, a new approach to organising the classification of supersymmetric backgrounds is proposed, which may be applied with equal validity to all \sg\ theories.
 This method provides a systematic way to solve the \kses\ of a given \sg, using a formalism which was has previously remained unexploited in this context.
 At the heart of it is an explicit description of the spinors in terms of exterior forms, and the realisation of the spin representation as simple endomorphisms of this particular spinor module.

Using this machinery, we can write the spinors of the theory in a simple and concrete way.
 It is then a fairly straightforward process to determine the subgroup of the spin group which leaves a given spinor invariant, since much is known about the action of groups on the exterior algebra.
 Once this stability subgroup is determined, we can then find a canonical form for the orbit of spinors which possess the same stabiliser.
 This explicit representative of the orbit can be substituted into the \kses, which may then be solved directly, so as to derive expressions for the field strength in terms of the background geometry, as well as additional constraints on the functions which parametrise the spinor.

This calculation is made simpler by an appropriate choice of basis for the spinors, so that the \kses\ reduce to a set of differential and algebraic conditions on the fields, which don't involve the gamma matrices themselves.
 Once this process is performed for a single spinor, the field strength constraints can be substituted back into the \kses.
 Then, we can choose a second spinor with a particular stability subgroup, and solve the \kses\ to obtain further restrictions on the geometry.
 This can be repeated for any number of spinors, depending on how much supersymmetry is being investigated.
 In this way, we can classify the solutions of the theory preserving any given fraction of supersymmetry according to the stability subgroups of the corresponding Killing spinors.
 The foundational concepts of the formalism will be explained in some detail in Chapter \ref{prelim}.
 We note that a different basis of spinors has been employed to solve the \kses\ in seven dimensions \cite{maco1}, and this method has also been extended to eleven-dimensions in \cite{maco2}, \cite{maco3}.

It is worth emphasising at this point that in determining the stabilisers of the spinors, we are looking for subgroups of the gauge group, not of the holonomy group.
 In \sg, the gauge group of the supercovariant connection is a spin group, whereas its holonomy group is a much larger object.
 For instance, in $D=11$ the gauge group is $Spin(1,10)$ whereas it was shown in \cite{gpdt1} that the holonomy group is $SL(32-N,\bR)\ltimes(\oplus^N\bR^{32-N})$, where $N$ is the number of \kss\ which exist in the background.
 The former consists of the gauge transformations $U$ which leave the form of the supercovariant connection invariant, up to a local Lorentz rotation of the frame and the field strength:
\be
\mathcal{D}_A(e,F)\longrightarrow U^{-1}\mathcal{D}_AU=\mathcal{D}_A(e',F')
\ee
 In contrast, a holonomy transformation will typically act non-trivially on the Levi-Civita connection and field strength terms, so that the form of the supercovariant connection changes.
 This distinction is important to make because in finding a canonical form for a spinor, we want to find a representative up to gauge transformations which preserve the supercovariant connection.
 Therefore in $D=11$ we seek a canonical form up to $Spin(1,10)$ gauge transformations rather than up to $SL(32-N,\bR)\ltimes(\oplus^N\bR^{32-N})$ transformations.
 Similarly, an analogous strategy must be employed when investigating any other \sg.

Another key ingredient in our analysis is the choice of a spin-invariant inner product on the space of spinors, which can be used to construct the linearly independent \st\ exterior forms associated to a given set of parallel spinors.
 In \cite{wang}, Wang presents an explicit construction of such inner products for the Riemannian spin groups, which we straightforwardly extend to apply to the Lorentzian spin groups in Chapter \ref{prelim}.
 The resulting spin-invariant forms are useful in characterising the geometry of the background and will be used throughout.

The main results occur in Chapter \ref{11d}, where we apply the formalism to $D=11$ \sg.
 We begin by solving the \kses\ in the presence of one \suf-\inv\ \ks\ in what proves to be an efficient way, to find that our results agree with those of \cite{pakis}.
 We then go on to analyse some of the cases that can arise when there are two \kss.
 In particular we solve the equations for two \suf-\inv\ Killing spinors and for two \sufr-\inv\ \kss, to obtain new results about the geometry of such spacetimes.
 In doing this, we have to look carefully at the orbits of \suf\ on the spinor space of \spt, and determine canonical forms for the spinors in each case.
 We then go on to discuss how the equations may be solved for certain classes of background with more than two \kss, and again obtain previously unknown results, and interpret them in terms of the geometry of the spacetime.

Also, in Appendix \ref{5d}, we apply the formalism to the \kses\ of $D=5$ minimal \sg, as a further illustrative example.
 The case of \sut-\inv\ spinors is investigated, and the geometric consequences derived.
 The possible supersymmetric solutions to this theory have already been classified by Gauntlett \textit{et al.} in \cite{5dclass}, and so it is an instructive process to compare the results gained by our formalism with what has already been found. Indeed, the results of \cite{5dclass} are confirmed, and this example gives another indication of the efficiency of our formalism.

The geometric consequences for the resulting solutions are emphasised throughout the analysis.
%
%
%
%
%
%
\chapter{Foundational Material} \label{prelim}
Throughout this thesis, many geometric and algebraic structures will arise naturally in the course of studying supersymmetric solutions to \sg\ theories.
 The intention is that this chapter may provide a brief, preliminary review of the particular concepts which will form the backbone of much of the work in later chapters.
 It would be a formidable, if not impossible task to provide rigorous introductions to each of these topics, so the aim is to describe and emphasise the aspects which are salient to our analysis.
\section{Elements of KT Geometry} \label{kt}
In our investigation of the supersymmetric backgrounds of heterotic string theory in Chapter \ref{het}, we will be lead to study the properties of K\"ahler with torsion manifolds, which we will define presently.
 We will assume familiarity with the elementary concepts of Hermitian geometry and holonomy.
 Some useful references for these topics are \cite{besse}, \cite{joyce}, \cite{kob1}, \cite{kob2}, \cite{nak}, \cite{salamon}, \cite{yano}.

Let $M$ be an Hermitian manifold of dimension $n=2m$, with a compatible metric $G$ and orthogonal complex structure $J$.
 Define connections with skew-symmetric torsion, $\na^{(\pm)}$, via their components,
\be
\G^{(\pm)}{}_A{}^M{}_N = \G_A{}^M{}_N \pm \frac{1}{2} H^M{}_{AN} \;,
\ee
where $\G_A{}^M{}_N$ are the Christoffel symbols of the Levi-Civita connection on $M$, and $H$ is a three-form.
 Then $(M,G,J)$ is a \textit{K\"ahler with torsion} (KT) manifold if
\be
\nabla^{(+)}J=0 \;.
\ee
In the mathematical literature, the connection with skew-symmetric torsion $\nabla^{(+)}$ is known as the \textit{Bismut connection}, see for example \cite{bismut}, \cite{yano}, \cite{jgsigp}.
 We note that $\naplus$ is also compatible with $G$.

Since $M$ is Hermitian, it possesses a $U(m)$-structure or equivalently, the tensors $(G,J)$ are $U(m)$-\inv.
 Also, both tensors are parallel with respect to $\naplus$, therefore $Hol(\nabla^{(+)})\subseteq U(m)$ \cite{joyce}.
 This in turn implies that the connection is $\mathfrak{u}(m)$-valued, and it is straightforward to see that for $J$ to be $\naplus$-parallel, we must have
\be \label{umhol}
\boxed{
\G^{(+)}_i{}^\a{}_{\betab} = 0 }\;.
\ee
Here, we use Latin indices to denote a $2m$-dimensional real basis of $M$, whereas lower-case barred and unbarred Greek indices denote the antiholomorphic and holomorphic coordinates of the natural unitary frame associated to the $U(m)$-structure, taking values from $1$ to $m$.

Using the $U(m)$-holonomy condition (\ref{umhol}), we see that the independent component of the torsion is
\be \label{torsion}
H_{\a\b\gb}=-\partial_\a G_{\b\gb}+\partial_\b G_{\a\gb} \;.
\ee
Also, by considering the integrability of $J$, i.e. the vanishing of the Nijenhuis tensor, $N$, it can be seen that the $(3,0)$ and $(0,3)$ components of $H$ vanish.
 For instance, calculating the component $N^{\gb}{}_{\a\b}$ leads to the condition $\G^{(+)}{}_{[\a}{}^{\gb}{}_{\b]}=0$, which gives $H_{\a\b\g}=0$.
 Therefore, $H$ is a $(2,1)+(1,2)$ form, which is specified uniquely in terms of the metric and complex structure of $(M, G,J)$.

Another useful tensor is the \textit{Lee form} of the KT geometry, which can be written in components as
\be
\th_i = \frac{1}{2} J^j{}_i H_{jkl} \O^{kl} \;,
\ee
where $\O_{ij}= G_{ik} J^k{}_j$ is the K\"ahler form.
In complex coordinates, we have
\be \label{lee}
\th_\a = \partial_\a{\rm ln}\det(G_{\b\gb}) - G^{\b\gb} \partial_\b G_{\a\gb} \;,
\ee
with $\th_{\ab}= (\theta_\a)^*$.

The connection is $\mathfrak{u}(m)$-valued, but since $\mathfrak{u}(m)=\mathfrak{su}(m)\oplus\mathfrak{u}(1)$, it may be decomposed under $\mathfrak{su}(m)$, by separating it into its trace plus a traceless piece.
 The $\mathfrak{u}(1)$ piece of the connection is thus given by tracing the $\mathfrak{u}(m)$-indices with the complex structure,
\be
\o_i=\G^{(+)}_i{}^j{}_k J^k{}_j \;.
\ee
This is the connection on the canonical bundle $K=\L^{m,0}(M)$, which is induced by $\nabla^{(+)}$ \cite{joyce}.
 In complex coordinates we find
\be \label{trace}
\o_\a = i\G^{(+)}{}_\a{}^\b{}_\b - i\G^{(+)}{}_\a{}^{\betab}{}_{\betab}
    = 2i G^{\b\gb}\partial_\b G_{\a\gb} - i G^{\b\gb}\partial_\a G_{\b\gb} \;,
\ee
with $\o_{\ab} = (\o_\a)^*$.
 Let $\rho=d\omega$ be the curvature of the $U(1)$ connection $\o$.
 Then $Hol(\nabla^{(+)})\subseteq SU(m)$ if and only if $\rho=0$ \cite{joyce}.
 $\rho$ is in fact the Ricci form of $G$.

A K\"ahler with torsion manifold is \textit{conformally balanced} if there exists a function $\phi$ on $M$
such that $\th=2d\phi$, i.e. its Lee form is exact.
 In \cite{gpsib} it is shown that if $(M, J,G)$ is a conformally balanced KT manifold with $Hol(\nabla^{(+)})\subseteq SU(m)$, (so that $\rho=0$), then
\be \label{confbalfe}
R_{ij} + \frac{1}{4} H^k{}_{il} H^l{}_{jk} + 2\nabla_i\partial_j\Phi
= \frac{1}{4} J^k{}_i (dH)_{kjmn} \O^{mn} \;,
\ee
where $\nabla$ is the Levi-Civita connection of the metric $G$.
\section[Complex Spinor Representations]{The Complex Spinor Representations of the Clifford Algebras}
In the investigation of \sg\ \kses\ which we will pursue in Chapter \ref{11d}, we will make use of a formalism for explicitly describing spinors in terms of exterior forms.
 Some excellent references for the topics of Clifford algebras and their spinors are \cite{abs}, \cite{lawson}, \cite{harvey}, \cite{lounesto}.
 For expositions more geared towards physical applications, see \cite{benn}, \cite{west}, \cite{trau}, \cite{vanproy}.

However, more often than not, the literature focuses largely on the \textit{real} Clifford algebras.
 For this reason, a review of the complex spin representations is presented here.
 This is mostly an adaptation of the results for the real case which may be found, for example, in \cite{lawson}, \cite{harvey}.
 Moreover, the explicit formalism presented below is not widely documented and as yet has been unexploited in theoretical physics, although it has appeared in mathematical literature in various guises\footnote{See, for example, the work of Wang in \cite{wang}, on which our formalism is based.
 An older account, which is very informative despite its somewhat obscure terminology, may be found in \cite{rasev}.
 For a very different application to the one explored in this work, see \cite{pressley}.}.
 Therefore, it is hoped that the following pages will provide an apt foundation to the concepts and techniques which prove to be of immense value in our method of solving \kses\ in later chapters.
\subsection{Complex Clifford Algebras} \label{complca}
Let $V$ be an $n$-dimensional vector space over a field $\bK$ of characteristic $\neq2$, equipped with a non-degenerate quadratic form $Q$ of signature $(r,s)$.
 The \textit{Clifford algebra} $\cl(V,Q)$ of $(V,Q)$ is the associative $\bK$-algebra with unit generated by vectors $v,w\in V\subset\cl(V,Q)$, subject to the relations
\be \label{gencarels}
v \cdot w + w \cdot v = Q(v,w)\id \;,
\ee
where $Q(v,w)=\frac{1}{2}(Q(v+w)-Q(w-v))$ is the polarisation of $Q$.
 In the mathematical literature, one will often see a more formal definition of the Clifford algebra as the quotient of the tensor algebra of $V$ by the ideal generated by elements of the form $v\otimes v + Q(V)1$, for $v\in V$.
 However, we choose to work with the definition given above, due to the importance of the explicit relations (\ref{gencarels}) in physical applications.

Before turning to some specific cases, we will first state without proof some important properties of $\cl(V,Q)$.
 Firstly, there is the following isomorphism of vector spaces (see, for example, \cite{harvey}):
\be \label{fundisom} \boxed{
\cl(V,Q)\cong\L^*(V) \;.    }
\ee
That is, the Clifford algebra of $V$ is isomorphic as a vector space (but \textit{not} as an algebra), to the exterior algebra of $V$.
 This allows us to define \textit{Clifford multiplication} $\,\cdot\,$ in a concrete way with reference to the exterior algebra, as
\be \label{cliffmult}
v\cdot\o = v\we\o - v\lc\,\o \;,
\ee
where $v\in V$ and $\o\in\cl(V,Q)$.
 Here, $\lc$ denotes the \textit{interior product}, which is the formal adjoint of the exterior product with respect to the bilinear form $Q(\,,\,)$.

Next, let us define an algebra automorphism in the following way.
 Consider the map $\a:V\longrightarrow V$ such that $\a(v)=-v$, for all $v\in V$.
 By the universal characterisation of Clifford algebras \cite{lawson}, this extends to an algebra automorphism $\a:\cl(V,Q)\longrightarrow\cl(V,Q)$, known as the \textit{canonical automorphism} of $\cl(V,Q)$.
 In fact, $\a$ is an involution since $\a^2=\id$.
 The importance of this map is apparent when we note that since $\a$ is an automorphism, it changes the sign of each individual algebra generator in a product.
 This means that if $\ph\in\cl(V,Q)$ is a product of an even number of generators, then $\a(\ph)=\ph$, whereas if $\ph$ is the product of an odd number of generators, then $\a(\ph)=-\ph$.
 Thus, $\a$ induces a natural splitting
\be
\cl(V,Q)=\cl^0(V,Q)\oplus\cl^1(V,Q) \;,
\ee
into the \textit{even} and \textit{odd} parts of $\cl(V,Q)$ respectively, defined by
\be
\cl^a(V,Q)=\{ \ph\in\cl(V,Q):\a(\ph)=(-1)^a\ph \} \quad \hbox{for}\;\;a=0,1 \;.
\ee
Of particular importance to our later considerations is the even part $\cl^0(V,Q)$, which forms a subalgebra of $\cl(V,Q)$, and is generated by products of even numbers of elements of $V$.
 Under the vector space isomorphism of (\ref{fundisom}), we have the natural identifications
\be \label{evenodd}
\cl^0(V,Q)\cong\L^{Even}(V) \qquad\hbox{and}\qquad \cl^1(V,Q)\cong\L^{Odd}(V) \; .
\ee
Also, we denote by $\cl^\times(V,Q)$ the group of multiplicative units of $\cl(V,Q)$, i.e. the elements which possess a multiplicative inverse.

Now, we come to look at some Clifford algebras which are of relevance in \sg.
 First, let $\bK=\bR$ and $V=\bR^{r,s}$ with orthonormal basis $\{e_M\}_{M=1}^n$, and let $Q$ be the standard pseudo-Riemannian quadratic form.
 Then, the \textit{real Clifford algebra}, denoted by $\cl_{r,s}=\cl(\bR^{r,s},Q)$, is generated by the orthonormal basis of $\bR^{r,s}$, subject to the relations
\be \label{realca}
e_M \cdot e_N + e_N \cdot e_M = 2\eta_{MN}^{r,s} \;,
\ee
where $\eta^{r,s}=diag(-1,\cdots,-1,1,\cdots,1)$ is the diagonal pseudo-Riemannian metric associated to $Q$, with signature $(r,s)$.
 The relations (\ref{realca}) dictate that each basis element squares to $\pm1$ and anticommutes with the other elements.
 From (\ref{fundisom}) we have the vector space isomorphism $\cl_{r,s}=\L^*(\bR^{r,s})$.
 Thus, $dim_{\bR}(\cl_{r,s})=2^{r+s}$, and a natural choice of basis for $\cl_{r,s}$ is the set of elements
\be
\{ 1 \,, \;\, e_M \,, \;\, e_{M_1}e_{M_2} \,, \,\cdots\, ,\;\, e_{M_1}\ldots e_{M_{n-1}} \,, \;\, e_1e_2 \ldots e_n \}\;,
\ee
where Clifford multiplication is understood in the products.

At this point, we define some important groups that occur within $\cl_{r,s}$.
 Firstly, there is the \textit{Clifford group}
\be
\g_{r,s} = \{ \ph\in\cl^\times_{r,s} \;:\; \a(\ph)v\ph^{-1} \in\bR^{r,s} ,\; \forall\, v\in \bR^{r,s} \} \;.
\ee
Then the \textit{Pin} and \textit{Spin} groups associated to $\bR^{r,s}$ are given by
\bea
Pin(r,s) &=& \{ \ph\in\g_{r,s} : Q(\ph)=\pm1 \} \;, \\
Spin(r,s) &=& Pin(r,s)\cap\cl^0_{r,s}  \;.
\eea

Now, let $\bK=\bC$ and $V=\bC^{r,s}$ with orthonormal basis $\{e_M\}_{M=1}^n$, and let $Q$ be a non-degenerate quadratic form of signature $(r,s)$.
 Then, the \textit{complex Clifford algebra} $\ccl_{r,s}=\cl(\bC^{r,s},Q)$, is generated by the orthonormal basis of $\bC^{r,s}$, subject to the relations
\be \label{compca}
e_M \cdot e_N + e_N \cdot e_M = 2 Q(e_M,e_N) \;,
\ee
where $Q(\,,\,)$ is the polarisation of $Q$.
 However, all of the complex inner product spaces $(\bC^{r,s},Q)$ of the same dimension $n=r+s$ are isometric \cite{harvey}.
 This means that there is in effect a unique complex Clifford algebra in each dimension, $\ccl_n$, which satisfies relations (\ref{compca}), where $Q$ is now the standard (complexified) \textit{Euclidean} inner product on $\bC^n$.
 Furthermore, we also have the isomorphism \cite{harvey}, \cite{lawson}
\be \boxed{
\ccl_n \cong \cl_{r,s}\otimes_{\bR}\bC \;, \quad\forall\, r+s=n \;.
}
\ee
In other words, for any of the real Clifford algebras $\cl_{r,s}$ in $n=r+s$ dimensions with quadratic form $Q$, the complex Clifford algebra $\ccl_n$ is simply the complexification, with complexified quadratic form $Q\otimes\bC$.

Also, there is an isomorphism which is invaluable when determining the irreducible representations of the spin groups \cite{harvey}, \cite{lawson}:
\be \boxed{
\cl_{r,s}\cong\cl^0_{r+1,s} \;.
}
\ee
From the above discussion of the complexified Clifford algebras, the following relationship results:
\be
\ccl_n\cong\ccl^0_{n+1} \;.
\ee
This concludes the brief introduction to complex Clifford algebras.
\subsection{Complex Spin Representations of $Spin(r,s)$}
The aim of the following few pages is to gain an understanding of the irreducible complex representations of the spin groups, as these will be of great importance in our later study of \sg\ theories.
 We will state the well-known classification of these representations, and then provide an explicit formalism which will enable us to work with them in a concrete way.
 Let us first clarify some terminology that will be used throughout.

We will refer to the irreducible representations of a complex Clifford algebra $\ccl_n$ as its \textit{pinor representations}, whereas the irreducible representations of its \textit{even part} $\ccl^0_n$ are called the \textit{spin} or \textit{spinor} representations\footnote{Terminology varies within the literature, and some references simply refer to the irreducible representations of the full Clifford algebra as its `spin representations'. However, for the sake of clarity we choose to make the distinction between pinors and spinors.}.
 Their corresponding irreducible modules will be referred to as their pinor and spinor spaces respectively, whose elements are the pinors and spinors associated to the algebra $\ccl_n$.

Since $Pin(r,s)\subset\ccl_n$ and $Spin(r,s)\subset\ccl^0_n$ for $r+s=n$, and the irreducible representations of the algebras restrict to irreducible representations of the groups in both cases \cite{harvey}, we therefore refer to the restrictions to $Pin(r,s)$ and $Spin(r,s)$ as the \textit{complex pinor} and \textit{complex spinor representations}, respectively.

A substantial derivation of the complex spin representations is
provided in Appendix \ref{complreprs}, but here we simply state the
results that are at the heart of much of our later work.
 A particularly simple classification arises:
\paragraph{Even Dimensions: }
For $n=r+s=2m$, the group $Spin(r,s)$ has two inequivalent
irreducible complex representations $\rho^{\pm}_{2m}$, whose spinor
spaces $\D^{\pm}_{2m}$ are complex vector spaces with
$dim_{\bC}(\D^{\pm}_{2m})=2^m$.
\paragraph{Odd Dimensions: }
For $n=r+s=2m+1$, the group $Spin(r,s)$ has a unique irreducible
complex representation $\rho_{2m+1}$, whose spinor space $\D_{2m+1}$
is a complex vector space with $dim_{\bC}(\D_{2m+1})=2^m$.
\section{An Explicit Formalism for the Spin Representations} \label{explreprs}
In this section, we introduce an explicit formalism for describing the complex spinors associated to $Spin(n)$ and $Spin(1,n)$, for $n$ even and odd.
 Groups such as these play a large role in \sg, particularly in regards to the gauge symmetries of the supercovariant connection.

A natural way to describe their irreducible representations is by identifying them with certain complex matrix algebras acting explicitly on a Euclidean space\footnote{See \cite{trau} for a recent review focusing on the matrix classification, as well as the classic references \cite{harvey} and \cite{lawson}.}.
 Alternatively, another useful method is to utilise the isomorphism between Clifford algebras and exterior algebras, so enabling us to describe spinors explicitly in terms of exterior forms.
 The construction that we will follow closely was presented by Wang in \cite{wang}, in order to describe the parallel spinors and bilinear invariants associated to manifolds with special holonomy, although similar methods have been employed elsewhere in various differing contexts (see, for example, \cite{pressley}, \cite{rasev}).

The formalism will be explained in some detail in the coming pages, with an emphasis on the connection between our explicit notions and the general theory of pinor and spinor representations.
 It is hoped that this will provide a thorough foundation to the construction, thereby serving to justify its extensive application to the \kses\ of \sg, which will be explored in later chapters.
\subsection{$Spin(2m)$}
We will begin with the Riemannian case in even dimensions, with $n=2m$, and proceed to build an explicit representation of the complex Clifford algebra $\ccl_{2m}$ on a module which is isomorphic to $\L^*(\bC^m)$, and then restrict it to obtain the complex representations of $Spin(2m)$.

Let us begin with the real vector space $V=\bR^{2m}$, equipped with the standard Euclidean inner product
\be
(u,v) = \d_{ij}u^iv^j \;, \quad u,v\in V \;,
\ee
where summation over repeated indices is understood.
 In this case, $\cl(V)=\cl_{2m}$.

From (\ref{fundisom}) , there is a natural vector space isomorphism
\be
\cl_{2m} \cong \L^*(\bR^{2m}) \; ,
\ee
and also (\ref{evenodd}) tells us that
\bea
&& \cl_{2m}^0 \cong \L^{Even}(\bR^{2m})=\bigoplus_{k=0}^m \L^{2k}(\bR^{2m}) \;, 
\\
&& \cl_{2m}^1 \cong \L^{Odd}(\bR^{2m}) = \bigoplus_{k=0}^{m-1}\L^{2k+1}(\bR^{2m}) \; .
\eea

Let us begin by setting up a suitable basis on $V$.
 It is a result from elementary linear algebra that every even-dimensional real vector space locally admits an almost complex structure, i.e. a type $(1,1)$ tensor that squares to $-\id$ at each point.
 Therefore, we choose one such almost complex structure on $V$, and denote it $J$.
 Then, the presence of this tensor enables us to identify $V$ with $\bC^m$, by defining complex scalar multiplication as
\be
(a+ib) v = av + bJv \;,\quad\hbox{for }\; a,b\in\bR  \;\hbox{ and }\; v\in V \;.
\ee
In this way, if $\{e_1,\cdots,e_m\}$ is an orthonormal basis for $V$ as a complex vector space of complex dimension $m$, then
\be
\{e_1,\,\cdots\,,\,e_m,\,J(e_1),\,\cdots\,,\,J(e_m)\}
\ee
is an orthonormal basis for $V$ as a real vector space of dimension $2m$. (See, for example, \cite{kob2}, for details of these facts about almost complex structures).
 Thus, we have the splitting $V=U\oplus J(U)$, where $U=\bR^m$ is the real subspace spanned by $\{e_1,\cdots,e_m\}$.
 Also, we define $e_{i+m}=J(e_i)$ for $i=1,\cdots,m$, so that $\{e_1,\cdots,e_{2m}\}$ is an orthogonal real basis of $V$.
 This leads to the condition $(J(u),J(v))=(u,v)$ for all $u,v\in V$, so that $J$ is orthogonal with respect to the inner product.
 In other words, $V$ possesses a $U(n)$-structure.

The next step in the construction is to complexify, and so we put $\Vc=\bC\otimes V\cong\bC^{2m}$ and complexify $(\,,\,)$, so that $\cl(V_{\bC})=\ccl_{2m}$.
 We can also extend the complexified inner product to the natural Hermitian one on the complex space $\Vc$,
\be
<\,w,z\,> = <\, w^ie_i \,,\, z^je_j \,> = \d_{ij} (w^i)^* z^j \; , \quad w,z\in \Vc \;,
\ee
where $*$ denotes complex conjugation in $\Vc=\bC^{2m}$.
 This inner product can be extended yet further, to act on any basis $\{b_k\}_{k=1}^{2^{2m}}$ of $\ccl_{2m}\cong\L^*(\bC^{2m})$ in an analogous way, by requiring that
\be
<\,\s,\t\,> = <\,\s^p b_p\,,\,\t^q b_q\,>=\d_{pq}(\s^p)^*\t^q \;,
\ee
for $\s,\,\t\in\cl(\Vc)$.
 In particular, observe that only exterior forms of identical degree yield a non-vanishing inner product.

Now, consider the complex subspace $\Uc=\bC\otimes U\cong\bC^m\subset\Vc$, equipped with the Hermitian inner product induced by restricting that of $\Vc$, and define
\be \label{even:spinorspace}
\boxed {\bP = \cl(\Uc) \cong \L^*(\bC^m) \;. }
\ee
This is a complex vector space with $dim_{\bC}(\bP)=2^m$.
 In fact, it is actually an irreducible module for the algebra $\ccl_{2m}$, which we will now demonstrate by defining the pinor representation of $\ccl_{2m}$ on $\bP$.

We begin by defining a homomorphism from $\Vc$ into the complex endomorphism algebra of $\bP\cong\L^*(\Uc)$.
 Let $\;\G: \Vc\longrightarrow End_{\bC}(\bP)\;$ be given by
\bea \label{even:gamma}
\G_{j}    &:=&  \G(e_j)      =  e_j \we\cdot \, + e_j \lc\cdot      
\nn\\
\G_{j+m}  &:=&  \G(e_{j+m})  =  ie_j\we\cdot \, - ie_j \lc\cdot  
\eea
for $1\leq j \leq m$.
 Here, $\lc$ denotes the interior product as before, i.e. the formal adjoint of $\we$ with respect to the inner product $<\;,\;>$ on $\L^*(\Uc)$.
 We will refer to these maps as \textit{gamma matrices}, and it is straightforward to see that they satisfy the Clifford algebra relations $\{\G_I,\G_J\}=2\d_{IJ}$ for $I,J=1,\cdots,2m$, so that indeed, $\G$ is a homomorphism.
 Note that the endomorphisms defined in (\ref{even:gamma}) are Hermitian with respect to the inner product\footnote{Note that our construction differs from that of \cite{wang}, in that our gamma matrices are Hermitian, rather than skew-Hermitian.}, i.e. $<\G_I\eta\,,\,\z>=<\eta\,,\,\G_I\z>$.

Now, from the \textit{universal characterisation} property of the Clifford algebras \cite{lawson}, $\G$ extends uniquely to a $\bC$-algebra homomorphism $\G: \ccl_{2m}\longrightarrow End_{\bC}(\bP)$.
 In other words, $\G$ provides a representation of $\ccl_{2m}$ on $\bP$.
 Actually, we can go further still, and state that
\be \label{even:endalg}
\boxed{
\G:\ccl_{2m}\cong End_{\bC}(\bP) \;.
}
\ee
To see this, observe first that since $dim_{\bC}(\bP)=2^m$, we have that $dim_{\bC}(End_{\bC}(\bP))=2^m\times2^m = dim_{\bC}(\ccl_{2m})$.
 Secondly, $\G$ bijectively maps each basis element $e_{i_1}e_{i_2}\cdots e_{i_k}$ to the corresponding endomorphism $\G_{i_1}\G_{i_2}\cdots\G_{i_k}$ for $i_1< i_2<\cdots< i_k$, $1\leq k \leq 2m$, so that the latter elements form a basis of $End_{\bC}(\bP)$.
 Given that the gamma matrices also satisfy the Clifford algebra relations, (\ref{even:endalg}) is indeed satisfied.

Thus, in our construction, $\G$ is the unique irreducible pinor representation of $\ccl_{2m}$ on the pinor space $\bP=\L^*(\Uc)$, which is of complex dimension $2^m$.
 However, we wish to go one step further, to obtain the irreducible complex spinor representations of the group $Spin(2m)\subset\ccl^0_{2m}$.
 Using the general theory presented in Appendix \ref{complreprs}, we proceed as follows.

Consider the complex volume element of $\ccl_{2m}$, $\,\o^{\bC}_{2m}=i^m e_1\cdots e_{2m}$.
 In our representation, we have
\be
\G(\o^{\bC}_{2m})=i^m\G_1\G_2\cdots\G_{2m} \;.
\ee
Now, we can define the projections
\be
\pi^\pm=\frac{1}{2} \left( 1\pm\G(\o^{\bC}_{2m}) \right) \;,
\ee
which satisfy $\pi^++\pi^-=1$, $(\pi^\pm)^2=\pi^\pm$ and $\pi^\pm\cdot\pi^\mp=0$.
 From (\ref{even:spinreprs}), we know that the pinor representation $\G$ breaks into two inequivalent irreducible representations of $\ccl^0_{2m}\cong\ccl_{2m-1}$, which we denote $\G_\pm$, with their corresponding irreducible modules specified by $\bS^\pm=\pi^\pm\cdot\bP$.
 These are the complex spinor representations.
 However, since an irreducible representation of $\ccl^0_{2m}$ restricts to an irreducible representation of $Spin(2m)$, we therefore arrive at
\be \label{spin2mreprs}
\boxed{
\G_\pm:Spin(2m)\subset\ccl^0_{2m}\cong End_{\bC}(\bS^\pm) \;.
}
\ee
These are the inequivalent complex spin representations of $Spin(2m)$, and the spinor modules $\bS^\pm$ are each of complex dimension $2^{m-1}$.
 Thus when working with $Spin(2m)$, we have the freedom to make a choice of spinor representation and spinor module $\bS^\pm$.
 Also, if a spinor lies in $\bS^+$ then we say it has \textit{positive chirality} whereas if it lies in $\bS^-$ then it possesses \textit{negative chirality}.
\subsection{$Spin(2m+1)$}
An explicit form of the complex spin representations of $\ccl_{2m+1}$ can be obtained by extending the construction for $\ccl_{2m}$ in a straightforward way.
 Let $V=\bR^{2m}$ as before, and put $V'=V\oplus \bR e_{2m+1}$ where $(e_{2m+1},e_{2m+1})=1$, so that $V'\cong\bR^{2m+1}$.
 Also, let $\G$ be as defined in (\ref{even:gamma}).

Complexifying, we have $V'_{\bC}=\bC\otimes V'\cong\bC^{2m+1}$ and $\cl(V'_{\bC})\cong\ccl_{2m+1}$.
 Next, we can define a projection map by
\bea \label{proj1}
&& \Pi: \ccl_{2m+1}\longrightarrow\ccl_{2m}  \nn \\
\hbox{where}\;\;\; && \Pi(e_{2m+1})= \o^{\bC}_{2m}
\quad\hbox{and}\quad \Pi(e_I)=e_I \;,\; I=1,\cdots,2m \; . \eea
Then, defining $\G_{2m+1}=\G\cdot\Pi(e_{2m+1})$, the $(2m+1)$ gamma
matrices are
\bea \label{odd:gamma}
\G_{j}    &:=&  \G\cdot\Pi(e_j) = e_j \we\cdot \, + e_j \lc\cdot
\nn \\
\G_{j+m}  &:=&  \G\cdot\Pi(e_{j+m})  =  ie_j\we\cdot \, - ie_j \lc\cdot
\nn \\
\G_{2m+1} &:=&  \G\cdot\Pi(e_{2m+1}) =  i^{m}\G_1\G_2\cdots\G_{2m} \;,
\eea
where $1 \leq j \leq m$. 
 Note that $\G_{2m+1}$ is Hermitian, and also that $(\G_{2m+1})^2=\id$ and $\{\G_{2m+1},\G_I\}=0$, $I=1,\cdots,2m$, so that the Clifford algebra relations are satisfied.
 Thus, $\G\cdot\Pi$ provides a representation of $\ccl_{2m+1}$ on the pinor space $\bP$, which was defined in (\ref{even:spinorspace}).

However, from Appendix \ref{complreprs}, we know that there are two inequivalent pinor representations of $\ccl_{2m+1}$ of complex dimension $2^m$, which are distinguished by their action on the complex volume form $\o^{\bC}_{2m+1}=i^{m}e_1e_2\cdots e_{2m+1}$, and are given by
\be
\rho_\pm:\ccl_{2m+1}\cong End_{\bC}(\bP^\pm) \; ,
\ee
where $\rho_\pm(\o^{\bC}_{2m+1})=\pm1$.
 The pinor spaces $\bP^\pm\cong\bP$ are each of complex dimension $2^m$.

Now, with the projection defined as above, so that $\Pi(e_{2m+1})=\o^{\bC}_{2m}$, a short calculation shows that $\G\cdot\Pi(\o^{\bC}_{2m+1})=+1$.
 This means that $\G\cdot\Pi=\rho_+$ and the pinor space is identified as $\bP=\bP^+$ in this case.
 Alternatively, we could have defined $\Pi(e_{2m+1})=-\o^{\bC}_{2m}$.
 Then, $\G\cdot\Pi$ would still have formed a representation of $\ccl_{2m+1}$, but we would instead have $\G\cdot\Pi(\o^{\bC}_{2m+1})=-1$, so that $\G\cdot\Pi=\rho_-$ with pinor module $\bP=\bP^-$.
 However, from the results of Appendix \ref{complreprs}, we know that on restriction from $\ccl_{2m+1}$ to $Spin(2m+1)\subset\ccl^0_{2m+1}\cong\ccl_{2m}$, $\,\rho_+$ and $\rho_-$ are equivalent and irreducible.
 Therefore, by making the identification $\G\cdot\Pi=\rho_+$, equations (\ref{proj1}) and (\ref{odd:gamma}) explicitly provide the $2^m$-dimensional complex spin representation of $Spin(2m+1)$ on the spinor space $\bS=\bP^+\cong\L^*(\Uc)$:
\be
\boxed{
\G\cdot\Pi : Spin(2m+1)\subset\ccl^0_{2m+1} \cong End_{\bC}(\bS) \;.
}
\ee
\subsection{$Spin(1,2m)$}
The construction in this case is the most relevant for the later work on eleven-\linebreak dimensional \sg, and is somewhat analogous to the preceding Riemannian case in odd dimensions.

We begin again with the real vector space $V=\bR^{2m}$ with orthonormal basis \linebreak $\{e_1,\cdots,e_{2m}\}$, standard inner product $(\,,\,)$ and orthogonal complex structure $J$.
 As before, $J$ induces a natural splitting $V=U\oplus J(U)$, where $U$ is the real subspace spanned by the first $m$ basis elements and $J(e_i)=e_{i+m}$ for $i=1,\cdots,m$.
 Again, we let $\G$ be defined as in (\ref{even:gamma}).
 Now, let $V'=V\oplus\bR e_0$ where $(e_0,e_0)=-1$, so that $V'\cong\bR^{1,2m}$, and complexify to obtain $V'_{\bC}=\bC\otimes V'\cong\bC^{1,2m}$.
 Since there is essentially one unique complex Clifford algebra in each dimension regardless of signature \cite{harvey}, we have $\cl(V'_{\bC})=\ccl_{2m+1}$.

In this case, we define a projection map by
\bea \label{proj2}
&& \check\Pi: \ccl_{2m+1}\longrightarrow\ccl_{2m}  \nn \\
\hbox{where}\;\;\; && \check\Pi(e_0)= i\o^{\bC}_{2m}
\quad\hbox{and}\quad \check\Pi(e_I)=e_I \;,\; I=1,\cdots,2m \; ,
\eea
and $\o^{\bC}_{2m}$ is defined as above.

Now, defining $\G_0=\G\cdot\check\Pi(e_0)$, we see that this projection encapsulates the Lorentzian nature of the underlying space $V'$, since $\left(\G_0\right)^2=-1$ so that $\G_0$ is timelike.
 Thus the $(2m+1)$ gamma matrices are
\bea \label{odd:gamma2}
\G_{j}    &:=&  \G\cdot\check\Pi(e_j)      =  e_j \we\cdot \, + e_j \lc\cdot
\nn \\
\G_{j+m}  &:=&  \G\cdot\check\Pi(e_{j+m})  =  ie_j\we\cdot \, - ie_j \lc\cdot
\nn \\
\G_{0} &:=&  \G\cdot\check\Pi(e_0) =  i^{m+1}\G_1\G_2\cdots\G_{2m} \;,
\eea
where $1 \leq j \leq m$.
 Note that $\G_{0}$ is anti-Hermitian, and also that $\{\G_0,\G_I\}=0$, $I=1,\cdots,2m$, so that the Clifford algebra relations in Lorentzian signature are satisfied, $\{\G_A,\G_B\}=2\eta_{AB}\id$, for $A,B=0,1,\cdots,2m$.
 Thus $\G\cdot\check\Pi$ also provides a representation of the algebra $\ccl_{2m+1}$ on the space $\bP=\L^*(\Uc)$.

Now, the complex volume form in this case is $\check\o_{2m+1}^{\bC}=i^me_0e_1\cdots e_{2m}$.
 As above, we distinguish the two inequivalent pinor representations $\rho_\pm$ of $\ccl_{2m+1}$ by their action on the volume form.
 With the gamma matrices defined as in (\ref{odd:gamma2}) we find that $\G\cdot\check\Pi\left(\check\o_{2m+1}^{\bC}\right)=i$, so that we have the identification $\G\cdot\check\Pi=\rho_+$ and the $2^m$-dimensional pinor module is then $\bP^+=\bP=\L^*(\Uc)$.
 Note that the other representation occurs, i.e. $\G\cdot\check\Pi=\rho_-$, if we set $\check\Pi(e_0)=-i\o^{\bC}_{2m}$.

From the results of Appendix \ref{complreprs}, we know that on restriction from $\ccl_{2m+1}$ to \linebreak $Spin_{2m+1}\subset\ccl^0_{2m+1}\cong\ccl_{2m}$, the representations $\rho_+$ and $\rho_-$ are equivalent and irreducible.
 Therefore, by making the identification $\G\cdot\check\Pi=\rho_+$, equations (\ref{proj2}) and (\ref{odd:gamma2}) explicitly provide the $2^m$-dimensional complex spin representation of \linebreak $Spin(1,2m)$ on the spinor space $\bS=\bP^+=\L^*(\Uc)$:
\be
\boxed{
\G\cdot\check\Pi : Spin(1,2m)\subset\ccl^0_{2m+1} \cong End_{\bC}(\bS) \;.
}
\ee
The representation $\rho_-$ can be specified in an analogous way.
\subsection{$Spin(1,2m+1)$}
Here, we will just outline the procedure, since the construction involves a simple combination of the ideas used in the previous two cases.
 An explicit formulation of the two inequivalent spin representations of $Spin(1,2m+1)$ can be obtained by restricting the unique pinor representation of $\ccl_{2m+2}$.
 In this case, we start with the real vector space $V\cong\bR^{2m+1}$ equipped with standard Euclidean inner product $(\,,\,)$, orthonormal basis $\{e_1,\cdots,e_{2m+1}\}$ and orthogonal almost complex structure $J$.
 Then with $V'=V\oplus\bR e_0$ where $(e_0,e_0)=-1$, we have that $\cl(\Vc')\cong\bC\otimes\cl_{1,2m+1}=\ccl_{2m+2}$.

Proceeding as in the previous cases, we can determine the gamma matrices which provide the inequivalent $2^{m+1}$-dimensional irreducible representations of $Spin(1,2m+1)$ on the spinor modules which in this case, are isomorphic to $\L^*(\Uc)$, where $\Uc$ is the complex span of the basis elements $\{e_1,\cdots,e_{m+1}\}$.
 This concludes the outline of the formalism which will be extensively utilised in later chapters.
\section{Spin-invariant Inner Products}\label{forms}
In this section, we will briefly review a method for constructing the spin-invariant bilinear forms associated to a given set of spinors, which can be found in \cite{wang}.
 This will be of great use in later chapters, where we can use this procedure to determine the set of spacetime exterior forms which arise from a set of \kss.
 A great deal of further information regarding spin-\inv\ inner products can also be found in \cite{benn}, for example.

We begin from the representation of $Spin(2m)$ described above in section \ref{explreprs}.
 Define the following endomorphisms of $\bP=\cl(\Uc)=\L^*(\Uc)$:
\benn
\a=\G_1\G_2\cdots\G_m \qquad , \qquad \b=\G_{m+1}\G_{m+2}\cdots\G_{2m}  \;.
\eenn
Now, let $\s,\t\in\bP$. Define $A(\s)=\a(\s^*)$ and $B(\s)=\b(\s^*)$, where $*$ is the complex conjugation operation with respect to $\bP=\cl(\Uc)$.
 Then to each of these operators we associate a bilinear form by
\benn
\A(\s,\t)=<A(\s) , \t > \qquad \hbox{and} \qquad \B(\s,\t)=<B(\s),\t> \; ,
\eenn
Using the explicit representation, it is straightforward to show that $A$ and $B$ possess the following properties\footnote{Note that our results differ from those of Wang in \cite{wang}, due to our differing conventions in defining the gamma matrices.}:
\begin{lem} \
\begin{enumerate}
\item
\beann
        \begin{array}{lcl}
{A^2 = \left\{ \begin{array}{cl} 1 & \; \hbox{if } m \equiv 0,1 \hbox{(mod\,4)} \\
                           -1 & \; \hbox{if } m \equiv 2,3 \hbox{(mod\,4)}
\end{array}
      \right.}
& \qquad &
{B^2 = \left\{ \begin{array}{cl} 1 & \; \hbox{if } m \equiv 0,3 \hbox{(mod\,4)} \\
                           -1 & \; \hbox{if } m \equiv 1,2 \hbox{(mod\,4)}
\end{array}
      \right.}
        \end{array}
\eeann
\item\label{prop1} If $m$ is even, $A$ and $B$ preserve chirality. If $m$ is odd, $A$ and $B$ reverse chirality.
\item $A$ and $B$ are $Spin(2m)$-equivariant, conjugate-linear operators on $\bP$.
\item For odd $m$, $A$ is $Pin(2m)$-equivariant and for even $m$, $B$ is $Pin(2m)$-equivariant.
\end{enumerate}
\end{lem}
As described in section \ref{explreprs}, on restricting from $\ccl_{2m}$ to $Spin(2m)$, the space of pinors splits into two spinor modules of differing chirality as $\bP=\bS^+\oplus\bS^-$.
 In terms of the isomorphism with the exterior algebra of $\Uc$, the spinor modules respect the grading,
\be
\bS^+ \cong \L^{Even}(\Uc) \qquad\hbox{and}\qquad \bS^- \cong \L^{Odd}(\Uc) \;.
\ee
Suppose that $m$ is even, so that $A$ and $B$ preserve chirality as in the second part of the preceding lemma.
 Then, we find that $\bS^+$ and $\bS^-$ are in fact orthogonal with respect to both $\A$ and $\B$.
 This is because the Hermitian inner product between two exterior forms is only non-vanishing when they are of the same degree, and this certainly cannot be the case when the forms are of differing chirality.
 Also, observe that when $m$ is odd, $A$ and $B$ reverse chirality, which implies that they provide conjugate-linear isomorphisms between the equivalent modules $\bS^+$ and $\bS^-$.

Now we come to a result which is crucial for constructing bilinear forms which are spin-invariant:
\begin{prop}\label{spininvprop} \
\begin{enumerate}
\item   \begin{tabular}{|l||c|c|c|c|}  \hline
        $m$ (mod4) & 0 & 1 & 2 & 3  \\ \hline\hline
        $\A$       & Symmetric & Symmetric & Skew & Skew  \\ \hline
        $\B$       & Symmetric & Skew & Skew & Symmetric  \\ \hline
        \end{tabular}
\item $\A$ and $\B$ are $Spin(2m)$-invariant.
\item For even $m$, $\A$ is $Pin(2m)$-invariant; for odd $m$, $\B$ is $Pin(2m)$-invariant.
\end{enumerate}
\end{prop}
Again, these results can be proved using the explicit representation given in equation (\ref{even:gamma}).
 Now, as a consequence of this proposition, we can prove that
\begin{thm} \label{spininvprop1}
For even $m$, $\A$ is $Spin(1,2m)$-invariant; for odd $m$, $\B$ is $Spin(1,2m)$-invariant.
\end{thm}
It is straightforward to see this, when we consider that the Lie algebra of \linebreak $Spin(1,2m)$ is generated by elements of the form $\G_J\G_K$ and $\G_0\G_K$, where $J,K=1,\cdots,2m$.
 The former products generate the group elements of $Spin(2m)$, so from Proposition \ref{spininvprop} we know that both bilinear forms are invariant under such transformations.
 However, products of the form $\G_0\G_K$ contain an odd number of $\ccl_{2m}$ gamma matrices, since $\G_K$ will annihilate itself with the $\G_K$ that occurs in $\G_0$.
 Thus elements of the form $\G_0\G_K$ are generators of the Lie algebra of $Pin(2m)$, so that $\A$ and $\B$ are $Spin(1,2m)$-\inv\ precisely when they are $Pin(2m)$-\inv, whence the result.

Analogous results can easily be obtained for $Spin(1,2m+1)$ by considering the necessary explicit representation, but since we will only have cause to investigate $Spin(1,10)$ and $Spin(1,4)$ in later chapters, we will not pursue the $Spin(1,2m+1)$ case here.
\subsection{Spacetime Forms from Complex Spinors}
Now, suppose that $(M,g)$ is a complete and simply-connected real Lorentzian spin-manifold, of dimension $n$.
 Let $\{e_0,e_1,\cdots,e_{n-1}\}$ be an orthonormal basis for $M$, and construct the subspace $U$ and complex pinor module $\bP\cong\L^*(\Uc)$ according to whether $n$ is even or odd, as in section \ref{explreprs}.
 Then, we have the well-known isomorphism \cite{trau}, \cite{wang}
\be \label{tensprod}
\bP^\star\otimes\bP\cong End_{\bC}(\bP) \;,
\ee
where $\bP^\star$ is the dual representation of $\bP$.
 Since $End_{\bC}(\bP)\cong\ccl_{n}\cong\L^*(\bC^{n})$ according to our formalism, (\ref{tensprod}) implies that
\be
\bP^\star\otimes\bP\cong\L^*(\bC^{n}) \cong \L^*(TM)\otimes\bC \;,
\ee
where the second isomorphism follows from the fact that complexification does not depend on signature \cite{trau}.
 This tells us that every tensor product of pinors can be written as an element of the complexified exterior algebra of the spacetime.
 Since the associated spinor spaces satisfy $\bS\subseteq\bP$ regardless on whether $n$ is even or odd, therefore every tensor product of \textit{spinors} can also be represented by a spacetime exterior form.
 This means that the spin-\inv\ inner products $\A$ and $\B$, which were defined in the previous section, can be used to construct the full set of spacetime exterior forms associated to a given set of spinors.
 Since the product is spin-\inv, consequently so are the exterior forms \cite{wang}.

Let us outline this procedure explicitly.
 Suppose that $M$ is as above with dimension $n$, which may be odd or even, and with spinor module $\bS$.
 Take $\{\s^a\}_{a=1}^{\n}$ to be a set of $\n$ elements of $\bS$.

Now, let $\mathfrak{F}$ denote whichever of $\A$ or $\B$ is $Spin(1,n-1)$-\inv\ (this can be obtained from Theorem \ref{spininvprop1} in the case of $Spin(1,2m)$).
 Then the associated exterior $p$-forms are simply defined by
\be\boxed{
\a^{ab}_{(p)}\,=\,\frac{1}{p!}\,\mathfrak{F}(\,\s^a\,,\,\G_{i_1}\cdots\G_{i_p}\s^b\,)\;e^{i_1}\we\cdots\we e^{i_p} \;,
}
\ee
for $a,b=1,\cdots,\n$ and $p=0,\cdots,n$.
 These are the spacetime forms which are invariant under $Spin(1,n-1)$.

Although initially it looks as if we may have a large number of forms to compute from a given set of spinors, however due to the isomorphism $\L^p\cong\L^{n-p}$ arising from Poincar\'e invariance, we need only compute forms up to degree $\frac{n}{2}$ if $n$ is even, and degree $\frac{n-1}{2}$ if $n$ is odd.
%
%
%
%
%
%
%
\chapter[$\a'$-Corrections to Heterotic String Backgrounds]{$\a'$-Corrections to Heterotic String Backgrounds}\label{het}
The heterotic superstring may be described by a supersymmetric
non-linear \sm\ coupled to $(1,0)$ \sg\
 \cite{gross}, \cite{sen}, \cite{ht2}, \cite{hull}.
 In the \sm\ approach, anomalies arise from the requirements of general coordinate invariance and invariance of the Yang-Mills gauge field, yet these can be cancelled by the implementation of a Green-Schwartz anomaly-cancellation mechanism.
 This amounts to correcting the Bianchi identity for the three-form field strength with a term which is proportional to $\a'$.

On the other hand, the low-energy dynamics of the heterotic string may be described by $N=1$, $D=10$ effective \sg\ coupled to a Yang-Mills gauge field.
 The equations of motion are derived from requiring that the \sm\ be conformally invariant \cite{callan}.
 This is equivalent to the vanishing of the beta-functions, which are calculated perturbatively with $\a'$ as the loop expansion parameter.
 At zeroth order in $\a'$ the vanishing of the beta functions yields the field equations for $N=1$ \sg.
 Higher order terms are associated with `stringy' corrections to the background.

Since anomaly-cancellation requires that a term of order $\a'$ be introduced in the Bianchi identity, in this chapter we investigate the consequences of taking $\a'$-corrections to the field and Killing spinor equations into account.
 We argue that this is a necessary consideration if the background solution is to be consistent with anomaly-cancellation.

In particular, we focus on heterotic backgrounds which are associated to \sm s possessing $(2,0)$ world-sheet supersymmetry.
 This is motivated by the fact that such models are required for the preservation of supersymmetry in compactifications of the form $M\times X$, where $M$ is Minkowski space and $X$ is a six-dimensional compact internal manifold \cite{candelas}, \cite{ht2}.
 We compute the $\a'$-corrections to each field, demonstrate how the field and \kses\ may be satisfied, then apply these results to the case of $(2,0)$ Calabi-Yau compactifications.
\section{Field and Killing Spinor Equations}
The action for a sigma-model possessing $(2,0)$ world-sheet supersymmetry can be written in $(1,0)$ superfields as \cite{hullwitten}, \cite{ht2}
\be
I = -\frac{1}{4\pi\a'}\int d^2xd\th \big\{\left( g_{ij} + b_{ij} \right) i D_+\Ph^i\partial_-\Ph^j
 + \Psi_{-a} \left( D_+\Psi_-^a + A_i{}^a{}_b D_+\Ph^i\Psi^b \right)\big\} \;,
\ee
where $g$ and $b$ depend on the superfields $\Ph^i$, $i,j=0,\cdots,9$, the super-covariant derivative is $D_+={\partial}/{\partial\th_-} + i\th_-\partial_+$ and $\partial_\pm={\partial}/{\partial x^0}\pm{\partial}/{\partial x^1}$.

The low-energy dynamics of the \hsb can be described by $N=1$ effective \sgb in ten dimensions \cite{callan}. The bosonic fields of this theory are the spacetime metric $g$, the Neveu-Schwarz two-form potential $b$ with fieldstrength $H$, the gauge connection $A$ and the dilaton $\Ph$. We will be investigating heterotic backgrounds with non-zero torsion, and so we define connections with torsion by
\benn
\nabla^{(\pm)}_M Y^N=\nabla_MY^N\pm {1\over2} H^N{}_{MR} Y^R \;,
\eenn
where $\nabla$ is the Levi-Civita connection of the metric $g$ and $M,N, R=0,1\dots,9$ are spacetime indices. It is also useful to define the components of the connection with torsion as
\benn
\G^{(\pm)}{}^{N}_{M}{}_{R} = \G_M^N{}_R \pm \frac{1}{2}H^N{}_{MR} \; ,
\eenn
and to denote their associated curvature tensors by $R^{(\pm)}{}_{MN}{}^P{}_Q$.

As has already been mentioned, in $N=1$, $D=10$ \sgb there are
anomalies arising from the graviton and the gauge field. However,
since we require that the \hsb is anomaly-free, the
\textit{Green-Schwarz anomaly cancellation mechanism} must be
imposed to ensure that the \sgb theory provides a consistent
low-energy description of the \hs. This amounts to adding a suitable
finite local counterterm to the Lagrangian as a means of removing
the anomaly, which is achieved by modifying the Yang-Mills gauge
transformation of the two-form $b$ so that the gauge-invariant field
strength is of the form\footnote{Our form conventions are
$\o_{(k)}={1\over k!} \o_{i_1, \dots, i_k} dx^{i_1} \we\cdots \we
dx^{i_k}$.} \cite{hull} \be \label{H} H = db - \frac{\a'}{4} \left(
Q_3(\G^{(-)})-Q_3(A) \right) + \ord(\a'^2) \;, \ee where the $Q_3$
terms are the Chern-Simons three-forms associated to the connections
$\G^{(-)}$ and $A$ respectively. Consequently, we have \be
\label{dH} dH = - \a'P + \mathscr{O}(\a'^2) \;, \quad \hbox{where}
\quad P = \frac{1}{4}\left[{\rm tr} (R^{(-)}\we R^{(-)})-{\rm tr}(
F\we F)\ \right] \;, \ee and the trace on the gauge indices is taken
as \benn {\rm tr} F\wedge F= F^a{}_b\wedge F^b{}_a \;, \qquad
F=dA+A^2 \;. \eenn Similarly for the trace of $R^{(-)}$, which is
the curvature of the connection $\nabla^{(-)}$. Our conventions are
determined by \benn R_{MN}{}^P{}_Q = \partial_M \G^P_{NQ} -
\partial_N \G^P_{MQ} + \G^P_{MR}\G^R_{NQ} - \G^P_{NR} \G^R_{MQ} \; ,
\eenn and the Ricci tensor is $R_{MN}=R_{PM}{}^P{}_N$. We note that
the four-form $P$ is proportional to the difference of the
Pontrjagin forms of the spacetime tangent bundle and the Yang-Mills
bundle of the \hs.

The anomaly cancellation mechanism requires that the field strength receives a correction of order $\a'$, and so for consistency, we will investigate the \fesb to first-order in $\a'$. These are obtained from the vanishing of the two-loop beta functions in the \smb perturbation theory, which is a consequence of conformal invariance \cite{sen}, \cite{callan}. The effective \fesb of the \hs\footnote{We work in the string frame.} are thus
\begin{eqnarray}
R_{MN}+\frac{1}{4} H^R{}_{ML} H^L{}_{NR}+2\nabla_M\partial_N\Ph  \qquad & & \nn \\
+ \a'\frac{1}{4} [ R^{(-)}{}_{MPQR} R^{(-)}{}_N{}^{PQR} - F_{MP} F_N{}^{P} ] + \ord(\a'^2) &=& 0 \nn \\
\nabla_M \big( e^{-2\Ph} H^M{}_{RL}\big)+\ord(\a'^2) &=& 0  \nn \\
\nabla^{(+)}{}^M(e^{-2\Ph} F_{MN}) + \ord(\a'^2) &=& 0 \;,
\label{fes}
\end{eqnarray}
where gauge indices on $F$ have been suppressed.
 We will refer to the field equations at zeroth-order in $\a'$ as the \textit{uncorrected \fes}.
 Note that there is also a field equation for the dilaton $\Ph$.
 However, this is implied by the first two equations above and so we will not consider it explicitly.

Now, let $\{\G^M: \;M=0,\cdots,9\}$ generate a basis of the Clifford algebra $\cl(1,9)$, so that $\G^M\G^N+\G^N\G^M\,=\,2g^{MN}$ and define $\G^{M_{1}\cdots M_{k}}=\G^{[M_{1}}\cdots\G^{M_{k}]}$. The group $Spin(1,9)$ has two inequivalent sixteen-dimensional irreducible spinor representations, with associated spin bundles $S_{\pm}$. Let $\e$ be a section of $S_+$. Then the \ksesb in the string frame are \cite{strom}
\bea \label{gravitino}
\nabla^{(+)} \e + \ord(\a'^2) &=& 0  \\
\label{dilatino}
\big(\G^M \partial_M \Ph - \frac{1}{12} H_{MNR} \G^{MNR}\big)\e + \ord(\a'^2) &=& 0 \\
\label{gaugino} F_{MN}\G^{MN}\e + \ord(\a'^2) &=& 0 \; . \eea These
equations are associated with the supersymmetry transformations of
the gravitino, dilatino and gaugino, respectively. Note that the
first equation is the parallel transport equation for the connection
with torsion $\nabla^{(+)}$.
\section[Heterotic Backgrounds]{Heterotic String Backgrounds} \label{hetbgs}
In the remainder of this chapter, the aim is to take a background solution of the uncorrected \fesb and deform it slightly, by adding perturbations of order $\a'$ to each field.
 Then by substituting these perturbed fields into the two-loop \fes\ (\ref{fes}) and the \kses\ (\ref{gravitino}), (\ref{dilatino}) and (\ref{gaugino}), we can find constraints on the deformations and try to solve them. In achieving this, we therefore find what type of deformations are allowed and investigate the geometric consequences for the spacetime manifold.
\subsection{Uncorrected Backgrounds}
Let us start with a solution to the uncorrected \fes, i.e. a zeroth-order background:
\bea
ds^2 &=& \go_{MN}dX^MdX^N = ds^2(\bR^{10-2m}) + ds^2(X_m) \nn \\
\Ho &=& \frac{1}{3!} \Ho_{ijk}(y)dy^i\we dy^j\we dy^k \nn \\
\Pho &=& \Pho(y) \nn \\
\Ao &=& \Ao_i(y) dy^i \; \label{zeroth}
\eea
where $\{y^i:i=1,\cdots,2m\}$ are coordinates on an internal manifold $X_m$, of complex dimension $m\le 4$. At this stage, $X_m$ may be compact or non-compact.
 Also, we see from (\ref{dH}) that $d\Ho=0$.

The requirement that the background $(\go,\Ho,\Pho,\Ao)$ be
compatible with $(2,0)$ world-sheet supersymmetry constrains $X_m$
to be an Hermitian manifold with complex structure $J$, which is
parallel with respect to the connection $\nabla^{(+)}$
\cite{candelas}.
 These conditions mean that $(X_m,J,\go)$ is a \textit{K\"{a}hler with torsion} (KT) manifold (see section \ref{kt} for more details).

Although $(\go,\Ho,\Pho,\Ao)$ is a solution to the uncorrected \fes, it does not necessarily mean that these fields also solve the \kses. However, in \cite{strom} and \cite{gpsi} it has been shown that the fields will solve the gravitino, dilatino and gaugino equations provided that the following conditions hold, respectively:
\bea
Hol(\nabla^{(+)}) &\subseteq& SU(m) \nn \\
\th &=& 2 d\Pho \nn \\
\Fo_{2,0} = \Fo_{0,2} = 0  \quad&\hbox{and}&\quad \Oo^{ij}\Fo_{ij} = 0   \; , \label{don0}
\eea
where
\beann
&& Hol(\nabla^{(+)}) \;\;\; \hbox{is the holonomy of the connection $\nabla^{(+)}$}  , \\
&& \th_i=\frac{1}{2}J^j{}_i\Ho_{jkl}\O^{kl} \;\;\; \hbox{is the \textit{Lee form} of the KT manifold $(X_m,J,\go)$}, \\
&& \Fo=d\Ao+\Ao\we\Ao \; , \\
&& \Oo_{ij}=\go_{ik}J^k{}_j \;\;\; \hbox{is the K\"{a}hler form on $X_m$}.
\eeann
The conditions on the field strength tell us that $\Fo$ is a traceless $(1,1)$-form with respect to the complex structure, and so it must take values in the Lie algebra $\mathfrak{su}(m)$. Furthermore, the condition $\th=2d\Pho$ says that the Lee form is \textit{exact}.
 KT manifolds with an exact Lee form are called \textit{conformally balanced}, hence the preservation of world-sheet and spacetime supersymmetry implies that $X_m$ is a conformally balanced KT manifold. It can also be shown that backgrounds which satisfy equations (\ref{don0}) preserve $2^{1-m}$ of the spacetime supersymmetry.
\subsection{Background Deformations} \label{bgdefs}
As has been explained, the $D=10$ \sgb anomaly can be removed by the
Green-Schwarz mechanism, provided the three-form field strength $H$
receives a correction of order $\a'$, as in (\ref{H}). So, to find
solutions of the \fesb that are consistent with anomaly
cancellation, we must consider the effective \fesb to order $\a'$,
as in equations (\ref{fes}). These are the zeroth-order equations
plus $\a'$-corrections. Therefore, we treat a solution to these
equations as a small perturbation or deformation of a solution to
the zeroth-order equations. This is implemented by expanding the
background fields in terms of $\a'$:
\bea \label{fieldexps}
g &=& \go + \a' h + \ord(\a'^2) \nn \\
H &=& \Ho + \a' T + \ord(\a'^2) \nn \\
\Ph &=& \Pho + \a'\vp + \ord(\a'^2)\nn \\
A &=& \Ao + \a' C + \ord(\a'^2) \; .
\eea
We assume that $(\go,\Ho,\Pho,\Ao)$ is a solution to the uncorrected effective \fes, and so $(h,T,\vp,C)$ can be thought of as the first-order deformation of the zeroth-order solution.
 Hence we refer to $(g,H,\Ph,A)$ variously as the 'corrected', 'deformed', 'perturbed' or 'first-order' solution to the \fes.

Once the $\a'$-corrections are taken into account, we expect a solution of (\ref{fes}) to be of the form
\bea
ds^2 &=& g_{MN}dX^MdX^N = ds^2(\bR^{10-2m}) + d\tilde{s}^2(X_m) \nn \\
H &=& \frac{1}{3!} H_{ijk}(y)dy^i\we dy^j\we dy^k \nn \\
\Ph &=& \Ph(y) \nn \\
A &=& A_i(y) dy^i \; , \label{first}
\eea
where $d\tilde{s}^2(X_m)= g_{ij}dy^idy^j$.

Now, an important point to consider is the issue of how the geometry of $X_m$ changes as the zeroth-order solution receives $\a'$ corrections. However, this was addressed in \cite{phgp1}, where it was shown that $(2,0)$ world-sheet supersymmetry can be preserved at all orders in \smb perturbation theory. For us, this means that if the manifold $(X_m,J,\go)$ is KT, then since $(2,0)$ world-sheet supersymmetry is preserved at higher loops, the corrected manifold $(X_m,J,g)$ will still be KT, although the torsion $H$ is no longer necessarily closed at first-order in $\a'$, due to (\ref{dH}).

The final preparatory considerations are regarding the number of
spacetime supersymmetries of the deformed solution. Even though
$(\go,\Ho,\Pho,\Ao)$ is assumed to be spacetime supersymmetric,
there is no guarantee that the corrected solution $(g,H,\Ph,A)$ will
be. Since we wish to find supersymmetry-preserving deformations, we
therefore insist that the deformed solution $(g,H,\Ph,A)$ preserves
all $2^{(1-m)}$ of the supersymmetry. Also, the \kses\ do not
receive additional corrections at order $\a'$, and so the form of
the first-order equations with respect to the first-order deformed
fields is identical to the form of the zeroth-order equations with
respect to the undeformed fields\footnote{However, this is not the
case at higher orders, where the equations receive loop corrections,
\cite{bonora}.}. Thus the arguments of \cite{strom}, \cite{gpsi}
hold for the two-loop \ksesb with respect to the corrected fields.
This means that the corrected solution $(X_m,J,g)$ should also
satisfy equations (\ref{don0}), with all quantities expressed in
terms of the deformed metric, gauge field and dilaton. In
particular, it is also a conformally balanced manifold, and its
holonomy (with respect to the deformed metric) is contained in
$SU(m)$.

To summarise the above discussion, we can state \textit{a priori} that the deformations of the geometry of $X_m$ to first-order in $\a'$ satisfy the following two properties:

\begin{itemize}
\item the holonomy of $X_m$ with respect to the deformed metric is contained in $SU(m)$;
\item $X_m$ is a conformally balanced KT manifold.
\end{itemize}
Using these properties and the fact that $(\go,\Ho,\Pho,\Ao)$ solves the field and \ksesb at zeroth-order in $\a'$, we can substitute the deformed fields $(g,H,\Ph,A)$ into equations (\ref{fes}), (\ref{gravitino}), (\ref{dilatino}) and (\ref{gaugino}) and find the resulting constraints on the deformations $(h,T,\vp,C)$.
\section{Solution of the Killing Spinor Equations} \label{defkses}
We begin by turning our attention to the solution of the gravitino and dilatino \kses.
 To solve these, we must specify the deformations
\be \label{fielddefs}
(g,H,\Ph,A)= (\, \go+\a'h \,,\; \Ho+\a'T \,,\; \Pho+\a'\vp \,,\; \Ao+\a'C \,)
\ee
which preserve the properties that $Hol(\nabla^{(+)})\subseteq SU(m)$ and $(X_m, J, g)$ is a conformally balanced KT manifold, see equations (\ref{don0}).
 In this analysis, we consider  deformations which preserve the hermiticity of the metric with respect to the complex structure $J$.
 In other words, the deformations of the metric must satisfy $h_{\alpha\beta}=0$, where $\a,\b=1,\cdots,m$ label the holomorphic coordinates on $X_m$.

Let us first consider the gravitino equation (\ref{gravitino}), which is solved automatically if $Hol(\nabla^{(+)})\subseteq SU(m)$.
 Recall from section \ref{kt} that the requirement of $Hol(\nabla^{(+)})\subseteq U(m)$ leads to the condition $\G^{(+)}_i{}^\a{}_{\betab}=0$.
 This condition then enables us to derive the independent component of the torsion, see equation (\ref{torsion}).
 Now, substituting the field expansions for $(g,H)$ into this equation and neglecting terms of second order in $\a'$, it follows that $Hol(\nabla^{(+)})\subseteq U(m)$ provided that the deformation for the torsion is
\be \label{tordef}
\boxed{
T_{\a\b\gb} = -\nao_\a h_{\b\gb} + \nao_\b h_{\a\gb} } \;,
\ee
where $\nao$ is the Levi-Civita connection of the metric $\go$ and $T_{\bar\alpha\bar\beta\gamma}= (T_{\alpha\beta\bar\gamma})^*$.
 The rest of the components of $T$ vanish, since the torsion of a KT geometry is a $(2,1)+(1,2)$-form.

Next, the deformation of the $U(1)$ part of the connection, i.e. the connection of the canonical bundle of $X_m$ induced by $\naplus=\naplus(g,H)$, can be found from (\ref{trace}) to be
\bea \label{cancon}
\o_\a = \o(\go)_\a + \a' \left\{ 2i \nao_\b h_\a{}^\b - i \nao_\a h^\b{}_\b + i \Ho_{\d\b}{}^\b h_{\a}{}^\d - i\Ho_{\a\b\gb} h^{\b\gb} \right\} + \ord(\a'^2) \;,
\eea
where $\o(\go)$ is the connection of the canonical bundle induced by $\napluso=\nabla^{(+)}(\go,\Ho)$, and we have $\o_{\ab}=(\o_{\a})^*$.
 Indices have been raised with the zeroth-order metric $\go$.

As we have seen in section \ref{kt}, a necessary and sufficient condition for $Hol(\nabla^{(+)})\subseteq SU(m)$ is that the curvature of the canonical bundle vanishes, $d\o(g)=0$.
 For the connection (\ref{cancon}), since the zeroth-order manifold is KT, we have $d\o(\go)=0$, so that the zeroth-order part of the curvature vanishes.
 Therefore, a sufficient condition for the the curvature of $\o(g)$ to vanish, and hence for $Hol(\naplus)$ to remain in $SU(m)$ after $\a'$-corrections, is that
\be \label{ricciflat}
\boxed{
2 \nao_\b h_\a{}^\b - \nao_\a h^\b{}_\b + \Ho_{\d\b}{}^\b h_{\a}{}^\d - \Ho_{\a\b\gb} h^{\b\gb} = 0 }\;.
\ee

Observe that (\ref{ricciflat}) and its complex conjugate provide $2m$ constraints on the possible deformations of the metric.
 This is the same as the number of degrees of freedom that we have for infinitesimal diffeomorphisms of the manifold $X_m$.
 Furthermore, there is some redundancy in specifying the deformation $h$ up to an infinitesimal diffeomorphism generated by a vector field $v$, i.e.
\be \label{gah} h'_{\a\betab} = h_{\a\betab} + \nao_\a v_{\betab} +
\nao_{\betab} v_{\a} \;. \ee
On these grounds, we conjecture that (\ref{ricciflat}) is a
gauge-fixing condition for allowable diffeomorphisms of $X_m$.
 Later on, we will see further indications that this is a justifiable gauge choice.

Now, we turn to the solution of the dilatino \kse\ (\ref{dilatino}).
 From (\ref{don0}), this equation can be re-cast in the form $\th(\go)=2d\Pho$, from which we see that $(X_m,\go,J)$ being conformally balanced is equivalent to the zeroth-order dilatino equation being satisfied.
 Therefore, to show that (\ref{dilatino}) is satisfied to first-order in $\a'$, we must demonstrate that $X_m$ remains conformally balanced after the field deformations are taken into account.

Using the formula for the Lee form given in equation (\ref{lee}), we can compute its first-order deformation to find
\be \label{leedef}
\th_\a = \th(\go)_\a + \a' \left\{ \nao_\a ( \go^{\b\gb} h_{\b\gb} ) - \nao_\b h_{\a\gb} \go^{\b\gb}
+ \frac{1}{2} \Ho_{\a\b\gb} h^{\b\gb} - \frac{1}{2} \Ho^{\db}{}_{\b\gb} \go^{\b\gb} h_{\a\db} \right\} \;,
\ee
where $\th(\go)$ is the Lee form of the uncorrected geometry.
 Substituting (\ref{ricciflat}) into (\ref{leedef}) yields
\be \label{leedef2}
\th_\a = \th(\go)_\a + \frac{\a'}{2} \nao_\a ( \go^{\b\gb} h_{\b\gb} ) \;.
\ee
Now, setting
\be \label{dilksedef}
\vp = \frac{1}{4} \go^{\b\gb} h_{\b\gb} = \frac{1}{2}\go^{ij}h_{ij} \;,
\ee
we obtain
\be
\th = 2d\Ph \;,
\ee
so that the corrected Lee form remains exact, as required.
 This means that $X_m$ remains conformally balanced to first-order in $\a'$, and hence the dilatino \kse\ is indeed solved to order $\a'$.
 Moreover, from equation (\ref{dilksedef}), we now see that the deformation of the dilaton is also determined in terms of that of the metric:
\be
\boxed{
\Ph = \Pho + \frac{\a'}{4} \go^{\b\gb} h_{\b\gb} + \ord(\a'^2) } \;.
\ee
Therefore, the condition given in equation (\ref{ricciflat}) is sufficient to to ensure that both the gravitino and dilatino \kses\ are satisfied to first-order in $\a'$.

So far, we have determined the deformation of the torsion and the dilaton in terms of the metric correction.
 It remains to examine the gaugino \kse\ and the conditions on the gauge connection.
 However, we postpone this until after the investigation of the field equations of the metric and the two-form gauge field strength.
\section{Solution of the Field Equations} \label{deffes}
Having derived the conditions for the deformations to satisfy the gravitino and dilatino \kses, we  now focus on the solutions of the field equations for the metric and the NS two-form potential\footnote{As was previously mentioned, these equations imply the field equation for the dilaton.}.
 In particular, we show that at first-order in $\a'$ both of these field equations are satisfied provided that the heterotic anomaly-cancellation condition holds.
 We begin by assuming that the background $(\go,\Ho,\Pho,A)$ satisfies the field equations (\ref{fes}) at zeroth order in $\a'$.
\subsection{The Einstein Equation}
The preliminary step is to substitute the field expansions (\ref{fieldexps}) into the field equation for the metric, (\ref{fes}), and collect the terms linear in $\a'$.
 However, let us first define the \textit{Lichnerowicz operator} $\D_L$, which arises naturally in this calculation.
 For any Riemannian manifold $(M,\go)$ with associated Levi-Civita connection $\nao$, we have
\be
R_{ij}(\go+\e h) = R_{ij}+\e \D_L h_{ij} + \mathcal{O}(\e^2)~.
\ee
In other words, $\D_L$ is the first-order deformation of the Ricci tensor under a small perturbation of the metric.
 One can show that
\bea
\D_L h_{ij} = &-& {1\over2}\nao^2 h_{ij} - \mathring{R}_{ik jl} h^{kl} + {1\over2}\nao_i\nao^kh_{kj} +
{1\over2}\nao_j\nao^kh_{ki}
\nn \\
&-&{1\over2}\nao_i\nao_j h^k{}_k + {1\over2} \mathring{R}_{ki} h^k{}_j + {1\over2} \mathring{R}_{kj} h^k{}_i \;.
\label{lop}
\eea
 Now, substituting the field expansions into the Einstein equation (\ref{fes}), we find the term linear in $\a'$ to be
\bea \nn
\D_L h_{ij} &-& \frac{1}{4}\Ho_{imn} T_j{}^{mn} - \frac{1}{4} \Ho_{jmn} T_i{}^{mn}
+ \frac{1}{2} h^{mn} \go^{kl} \Ho_{imk} \Ho_{jnl}
\\
&+& 2\nao_i\partial_j \vp - \go^{kl} (\nao_i h_{jk} + \nao_j h_{ik} - \nao_k h_{ij}) \partial_l\Pho + S_{ij}=0 \;,
\label{licheq}
\eea
where we have defined
\be
S_{ij} = \frac{1}{4} \left\{ R^{(-)}{}_{iklm} R^{(-)}{}_j{}^{klm} - \Fo_{ik ab} \Fo_j{}^{kab} \right\} \;,
\ee
the two-loop contribution to the beta function.
 The curvature $R^{(-)}$ is with respect to $(\go,\Ho)$, and  $\Fo=F(\Ao)$.

To show that the Einstein equation can be solved for the deformed metric $g$, we must solve equation (\ref{licheq}) in terms of the deformation $h$, which clearly will be a non-trivial process.
 However, as was mentioned in section \ref{bgdefs}, if we assume that spacetime supersymmetry is preserved despite the deformation of the fields, then this means that $X_m$ remains conformally balanced and possesses $SU(m)$ holonomy at order $\a'$.
 Therefore, we are entitled to use condition (\ref{confbalfe}) to simplify the field equation for the metric deformation in the following way.
 We proceed by substituting the field expansions (\ref{fielddefs}) into (\ref{confbalfe}).
 Neglecting terms of order $\a'^2$, the terms which are linear in $\a'$ are related by
\bea \nn
\D_L h_{ij} &-& \frac{1}{2}\Ho_{mn(i} T_{j)}{}^{mn} + \frac{1}{2} h^{mn} \go^{kl} \Ho_{imk} \Ho_{jnl}
+ 2\nao_i\partial_j\vp \\
&-& \go^{kl} \left(\nao_i h_{jk} + \nao_j h_{ik} - \nao_kh_{ij} \right) \partial_l\Pho
= \frac{1}{4} J^k{}_i dT_{kjmn} \Oo^{mn} \;, \label{defconfbalfe}
\eea
where $\Oo$ is the K\"ahler form with respect to $\go$.
 Note that there is no explicit contribution from the metric deformation to the right-hand side.
 This is because the torsion is closed at zeroth order, $d\Ho=0$.

Now, we can simplify the first-order part of the Einstein equation by substituting (\ref{defconfbalfe}) into (\ref{licheq}), to give
\be \label{metricfe}
\frac{1}{4} J^k{}_i dT_{kjmn} \Oo^{mn} + S_{ij}=0 \;.
\ee
In fact, we will now see that the anomaly-cancellation condition implies the field equation of the metric.
 To first order in $\a'$, (\ref{dH}) gives
\be \label{anomcanc}
dT = - P = -\frac{1}{4}\left[ R^{(-)k}{}_l\wedge R^{(-)l}{}_k - F^a{}_b\wedge F^b{}_a \right] \;,
\ee
where $P$ depends on the uncorrected fields $(\go, \Ho, \Ao)$.
 Using the facts that $Hol(\naplus)\subseteq SU(m)$ and that $R^{(-)}_{ij,kl}=R^{(+)}_{kl,ij}$ provided $d\Ho=0$ \cite{hull}, we can deduce some useful identities for the curvature:
\be \label{Rconds}
R^{(-)}_{mn}{}^i{}_j J^m{}_k J^n{}_l = R^{(-)}_{kl}{}^i{}_j \;, \qquad \Oo^{mn}R^{(-)}_{mn}{}^i{}_j = 0 \;.
\ee
Also, as a consequence of assuming that the zeroth-order background is supersymmetric, the gaugino equation provides similar conditions for the field strength $F$ \cite{strom}:
\bea \label{Fconds}
F_{mn}{}^a{}_bJ^m{}_k J^n{}_l = F_{kl}{}^a{}_b \;, \qquad \Oo^{mn}F_{mn}{}^a{}_b = 0 \;.
\eea
Now, contracting the anomaly-cancellation condition (\ref{anomcanc}) with the zeroth-order \linebreak K\"ahler form $\Oo$ and  using conditions (\ref{Rconds}) and (\ref{Fconds}), we obtain the field equation for the deformation of the metric (\ref{metricfe}) exactly.
 Therefore, in order to solve this equation, it is sufficient to show that the anomaly-cancellation condition (\ref{anomcanc}) can be satisfied, which we now proceed to do.

Substituting (\ref{tordef}) into (\ref{anomcanc}), we find that
\be \label{ddbar}
P= -2i \partial \bar\partial \U \;,
\ee
where we have defined the $(1,1)$-form $\U_{ij}= h_{ik} J^k{}_j$, which can be thought of as the first-order deformation of the K\"ahler form.
 The global anomaly-cancellation condition requires that $P$ be exact.
 It is therefore an exact, real $(2,2)$-form.
 On a compact K\"ahler manifold, the global $\partial\bar\partial$-lemma says that there exists a globally defined $(1,1)$-form such that (\ref{ddbar}) is satisfied.
 However, $X_m$ is Hermitian but at this stage need not necessarily be either compact or K\"ahler. Therefore the $\partial\bar\partial$-lemma does not apply in this form in general\footnote{See \cite{jgsigp} for a more general formulation of the lemma on Hermitian manifolds.}.
 This means that under these circumstances, (\ref{ddbar}) cannot be solved globally in general.
 Nevertheless, we can use this constraint to say that a \textit{sufficient} condition for (\ref{ddbar}) to be solved globally, is that the $\partial\bar\partial$-lemma \textit{is} valid, in a manner analogous to the K\"ahler case.
 In other words, a sufficient condition for a spacetime supersymmetric background with non-zero torsion to satisfy the metric field equation to first-order in $\a'$, is that there exists a globally defined $(1,1)$-form $\U$ such that (\ref{ddbar}) is satisfied.
 Assuming the validity of this generalised lemma, we see that the field equation for the deformation of the metric, (\ref{licheq}), is now solved via the anomaly-cancellation condition (\ref{anomcanc}).

Further work needs to be done to obtain a necessary condition for the solution to the Einstein equation.
 Such a condition would constitute a concrete starting point from which we may investigate deformations.
 Nevertheless, given that the $\partial\bar\partial$-lemma can be generalised locally to Hermitian manifolds, the sufficient condition which we have obtained above is a reasonable assumption to ensure that the metric of $X_m$ solves its field equation.
 In this way, we have determined a particular class of heterotic backgrounds which solve the field and Killing spinor equations to first-order in $\a'$ and also satisfy the anomaly-cancellation condition.
 A different approach would be required to obtain a full classification of such solutions.

Observe that the solution to the anomaly-cancellation condition (\ref{ddbar}) is not \linebreak unique.
 Indeed, if $\U$ is a solution, then
\be \label{gaU}
\U' = \U + \partial \bar w+\bar\partial w
\ee
for some $(1,0)$-form $w$, is also a solution.
 However, setting $v=-iw$, this equation tells us that
\be
\U'_{\a\betab} = -i\left( h_{\a\betab} + \nao_\a v_{\betab} + \nao_{\betab} v_\a \right) \;,
\ee
so that the gauge freedom in specifying $\U$ is equivalent to the freedom we have in specifying the metric deformation $h$ up to an infinitesimal diffeomorphism generated by the vector $v$, as in (\ref{gah}).
 Therefore having determined $\U$ from (\ref{ddbar}), we still have the gauge freedom to solve the supersymmetry condition (\ref{ricciflat}).
\subsection{The NS Two-form Field Equation}
It remains to show that the field equation for the NS two-form gauge potential can also be satisfied by the deformed fields (\ref{fielddefs}).
 We demonstrate this by utilising an identity proved in \cite{gpsib} (Corollary 3.2), which
can be stated as follows: Let $(X_m,g,J)$ be a conformally balanced KT manifold with torsion $H$, such that $dH\neq0$, and $Hol(\naplus)\subseteq SU(m)$.
 Then,
\be
\na^iH_{ijk} = \th_i\, H^i{}_{jk} \;,
\label{feqktb}
\ee
where $\na$ is the Levi-Civita connection on $X_m$.

We know that both KT manifolds $(X_m,\go,J)$ and $(X_m, g,J)$ are conformally balanced and have $Hol(\naplus)\subseteq SU(m)$, because of supersymmetry requirements.
 Therefore, (\ref{feqktb}) is valid for both the uncorrected and corrected KT structures, and so the torsion and Lee forms for each structure satisfy (\ref{feqktb}).

For the $\a'$-corrected  background $(g, H,\Ph,A)$, the field equation for the NS two-form is
\be \label{ns}
-2\partial_i \Ph H^i{}_{jk} + \na^iH_{ijk} + \ord(\a'^2) = 0 \;.
\ee
Using (\ref{feqktb}), we can write this as
\be
(\th_i - 2\partial_i \Ph ) H^i{}_{jk} + \ord(\a'^2) = 0 \;,
\ee
which vanishes identically, since  $(X_m, J, g)$ is conformally balanced with $\th=2d\Ph$ (see section \ref{defkses}).

By assumption, the NS field equation holds for the uncorrected background \linebreak $(\go, \Ho, \Pho, \Ao)$, and we have shown here that it also holds for the deformed fields \linebreak $(g, H, \Ph, A)$.
 Consequently, the part of (\ref{ns}) which is linear in $\a'$ must be satisfied automatically.
 Therefore the field equation for the NS two-form gauge potential is solved without any additional conditions on the metric deformation $h$.
\section{The Gaugino and Gauge Field Equations} \label{gfe}
In this section we show that the gaugino \kse\ (\ref{gaugino}) implies the field equation of the gauge field after the $\a'$-corrections are taken into account.
 First, we will prove a result which gives a relationship between these two equations on a conformally balanced KT manifold.
\subsection{On K\"ahler Manifolds}
We begin by describing a well-known relation between the gaugino \kse\ and the field equation of a gauge connection on a \textit{K\"ahler} manifold.
 Let $E$ be a vector bundle over a K\"ahler manifold $(M,G,J)$, which is equipped with a connection $A$ and corresponding curvature $F$.
 The following conditions, known as the \text{Donaldson equations}, constitute a reformulation of the gaugino \kse\ (see, for example, \cite{strom}, \cite{gsw2}):
\be \label{don}
F_{2,0} = F_{0,2} = 0 \;, \qquad\qquad \O^{ij} F_{ij} = 0 \;.
\ee
It is known that if the gauge field $A$ satisfies these equations, then it also solves the field equation
\be \label{gfeq}
\na^i F_{ij} = 0 \;,
\ee
where $\na$ is the Levi-Civita connection on the K\"ahler manifold.
 This can be seen as follows.

With $F_{\a\b}=F_{\ab\betab}=0$, the independent component of the field equation is
\be
G^{\g\betab}\na_\g F_{\betab\a} = 0 \;.
\ee
Since $M$ is K\"ahler, we have $Hol(\na)\subseteq U(m)$, which means that $\G_i^{\a}{}_{\betab}=0$ for the components of the Levi-Civita connection.
 Consequently, the Bianchi identity $\na_{[i}F_{jk]}=0$ yields
\be
\na_\g F_{\betab\a}=-\na_\a F_{\g\betab}
\ee
on a K\"ahler manifold.
 The field equation becomes
\be
\na_\a \left(G^{\g\betab} F_{\g\betab}\right) = 0 \;,
\ee
which is automatically satisfied if the condition $\O^{ij}F_{ij}=0$ holds.
 Thus, Donaldson's equations (\ref{don}) imply the field equation for the gauge field $A$.
 Furthermore, it has been shown by Donaldson \cite{don} and Uhlenbeck and Yau \cite{uhyau} that if $E$ is a \textit{stable} bundle over a complex surface, then there is a unique connection $A$ which solves (\ref{don}).
\subsection{On Conformally Balanced KT Manifolds}
Next, we modify the previous result to include the more general class of manifolds with which we are presently concerned.
 Let $E$ be a vector bundle over a \textit{non-K\"ahler}, conformally-balanced KT manifold $(M,G,J)$, equipped with a connection $A$ and corresponding curvature $F$.
 Donaldson's equations (\ref{don}) can be straightforwardly generalised to KT manifolds by allowing $\O$ to be the K\"ahler form of the Hermitian metric $G$.
 We will now show that these equations imply the field equations
\be \label{ktgfeq}
\na^{(+)}{}^i(e^{-2\Ph} F(A)_{ij})=0 \;,
\ee
where $\naplus$ is the connection of the KT structure with torsion $H$ and Lee form $\th=2d\Phi$.

Let us choose complex coordinates with respect to the complex structure $J$ and and expand (\ref{ktgfeq}) to give
\bea \label{ktgfeq1}
0 = -2 G^{\g\betab} \left(\partial_\g \Ph \right) F_{\betab\a} + G^{\g\betab} \na_\g F_{\betab\a}
- \frac{1}{2} G^{\g\betab}H^{\db}{}_{\g\betab} F_{\db\a}
- \frac{1}{2} G^{\g\betab} H^\d{}_{\g\a} F_{\betab\d} \;,
\eea
where $\na$ is the Levi-Civita connection of the metric $G$.
 Now, the Bianchi identity implies that
\be
\na_\g F_{\betab\a } = - \na_\a F_{\g\betab }- \na_{\betab} F_{\a\g} \;.
\ee
We know that $Hol(\naplus)\subseteq SU(m)$, so that $\G^{(+)}_i{}^\a{}_{\betab}=0$ and consequently $\G_i^\a{}_{\betab}=\frac{1}{2}H^\a{}_{i\betab}$.
 Hence,
\be \label{bi1}
\na_\g F_{\betab\a } = - \na_\a F_{\g\betab} - \frac{1}{2} H^{\db}{}_{\betab\a} F_{\db\g}
- \frac{1}{2} H^{\db}{}_{\betab\g} F_{\a\db} \;.
\ee
Assuming that $\O^{ij}F_{ij}=0$ holds and using that the Lee form (\ref{lee}) can be written in complex coordinates as
\be
\th_\a = - H_{\a\b\gb}G^{\b\gb} \;,
\ee
we can contract (\ref{bi1}) with $G^{\g\betab}$ to obtain
\be
G^{\g\betab}\na_\g F_{\betab\a} = -\frac{1}{2}H^{\db}{}_{\betab\a} G^{\betab\g} F_{\db\g}
- \frac{1}{2} \th_\g G^{\g\betab} F_{\a\betab} \;.
\ee
Substituting this into (\ref{ktgfeq1}), we find that the torsion terms cancel and the gauge field equation reduces to
\be
(\th_\g - 2\partial_\g\Ph) G^{\g\betab} F_{\betab\a} = 0 \;,
\ee
which vanishes identically since $(M, J,G)$ is conformally balanced with $\th=2d\Ph$.
 Therefore Donaldson's equations (\ref{don}) and the Bianchi identity imply the field equations (\ref{ktgfeq}) for a conformally balanced KT manifold.
\subsection{Solution to the Gauge Field Equation}
We shall use the result of the previous section to show that the gauge field equation (\ref{ktgfeq}) can be satisfied when we implement the $\a'$ field deformations (\ref{fielddefs}).
 As discussed previously, we assume that both the uncorrected background $(\go,\Ho,\Pho,\Ao)$ and the $\a'$-corrected background $(g,H,\Ph,A)$ preserve spacetime supersymmetry, so that their corresponding sets of \kses\ are satisfied, or equivalently, conditions (\ref{don0}) hold in each case.
 This means that the corresponding KT structures $(X_m,\go,J)$ and $(X_m,g,J)$ are conformally balanced, and the holonomies of their $\naplus$ connections are contained in $SU(m)$.
 In addition, both the uncorrected gauge connection $\Ao$ and its deformation $A$ satisfy conditions (\ref{don}), the latter up to linear order in $\a'$.

Applying the result which was proved in the previous section, the Donaldson equations for each background $(\go,\Ho,\Pho,\Ao)$ and $(g,H,\Ph, A)$ ensure that their corresponding field equations (\ref{ktgfeq}) are satisfied.
 This means that since (\ref{ktgfeq}) is satisfied at both zeroth and first-order in $\a'$, on expanding in terms of $\a'$ for the corrected background, the complicated term which is linear in $\a'$ will vanish identically and therefore this term does not provide additional constraints on $h$.
 In other words, we have demonstrated that the field equation for the gauge potential is satisfied automatically if supersymmetry is preserved, without any additional conditions on the deformations.

Using equations (\ref{don}) at both zeroth and first-order in $\a'$, the following conditions on the gauge field deformation $C$ result from the terms which are linear in $\a'$:
\bea
\na_\a C_\b - \na_\b C_\a = \na_{\ab} C_{\betab} - \na_{\betab} C_{\ab} &=& 0
\nn \\
h^{\a\betab} F(\Ao)_{\a\betab} + g^{\a\betab} \left( \na_\a C_{\betab} - \na_{\betab} C_\a \right) &=& 0 \;,
\label{gaugedef}
\eea
where $\na$ is the covariant derivative with respect to the gauge connection $\Ao$.
 The solution of these equations in terms of $C$ is a complicated procedure and will be pursued no further here.
 Rather, we observe that conditions (\ref{gaugedef}) are satisfied automatically if the \kses\ are satisfied.
\section{$(2,0)$ Compactifications of the Heterotic String}
We now turn our attention to spacetime supersymmetric compactifications of the heterotic string which also preserve $(2,0)$ world-sheet supersymmetry.
 The ans\"atze for such solutions is given in (\ref{zeroth}), with the additional requirement that the internal space $X_m$ is compact.
 This situation was considered in \cite{strom}, where it was shown that no warp factor is allowed for the non-compact part of the metric.
 From the considerations of the previous sections, we know that such backgrounds are expected to receive $\a'$-corrections if they are to remain consistent with anomaly-cancellation.
 Below, we shall investigate the deformations of these backgrounds that arise on taking $\a'$-corrections into account.
\subsection{The Zeroth-order Solution}
From section \ref{hetbgs}, if a compactification is to preserve $(2,0)$ world-sheet supersymmetry and $2^{1-m}$ of spacetime supersymmetry in $(10-2m)$-dimensions, then it is required that the internal manifold $X_m$ be conformally balanced, KT with $Hol(\naplus)\subseteq SU(m)$, and also in this case, compact\footnote{This assumption is sufficient for the spectrum in $(10-2m)$ dimensions to be discrete \cite{candelas}.}.
 In this section we will consider compactifications to four dimensions of the form $M\times X_3$, where $X_3$ is compact with holonomy contained in $SU(3)$.
 Furthermore we assume that the fields are smooth on $X_3$ and that the torsion is closed at zeroth-order, $d\Ho=0$.
 These assumptions place strong restrictions on the geometry of $X_3$ and in \cite{gpsi}, it was shown that under these conditions, $X_3$ is Calabi-Yau , $\Ho=0$ and the dilaton $\Pho$ is constant.
 In addition, at zeroth-order in $\a'$ the gauge connection $\Ao$ satisfies equations (\ref{don}) on $X_3$.
 These facts form the basis of our $\a'$ expansion.

Let $(\go, \Ho=0, \Pho, \Ao)$ be the uncorrected Calabi-Yau background.
 In general, the deformed background $(g,H,\Ph,A)$ has non-vanishing torsion $H$, and the anomaly-cancellation
mechanism requires that it is not closed, $dH\neq0$.
 Since the zeroth-order term in $H$ vanishes, $H$ is purely first-order in $\a'$, so that
\be
H = \a' T + \ord(\a'^2) \;.
\ee
Let us now examine the deformations of this background.
\subsection{The First-order Solution}
In section \ref{hetbgs}, we saw that the deformations of the torsion and dilaton $(T,\vp)$ are given by equations (\ref{tordef}) and (\ref{dilksedef}) and furthermore, both of these are directly determined by the metric deformation $h$.
 Also, since $X_3$ is Calabi-Yau, the $\partial\bar\partial$-lemma is applicable, which is sufficient to ensure that the Einstein equation may be solved.
 We have also seen that the remaining field and Killing spinor equations are satisfied without further constraints on the corrected fields, except for the correction to the gauge field $C$, which automatically satisfies equations (\ref{gaugedef}) if the background is supersymmetric.
 So it remains for us to show that condition (\ref{ricciflat}) may be solved on $X_3$.

Since the torsion vanishes at zeroth-order in $\a'$, we can rewrite (\ref{ricciflat}) as
\be \label{CYgauge}
\nao^{\betab} h_{\a\betab} - \frac{1}{2} \nao_\a (g^{\g\betab} h_{\g\betab}) = 0 \;.
\ee
In real coordinates, this equation becomes
\be \label{CYgauge1}
\nao^j h_{ji} - \frac{1}{4} \nao_i h^j{}_j = 0 \;.
\ee
This can be thought of as a gauge-fixing condition for the deformations associated with infinitesimal diffeomorphisms of $X_3$ and furthermore, we will now see that this condition can always be attained.

First, suppose that $h$ does not solve (\ref{CYgauge1}), but that $h'_{ij}=h_{ij}+\nao_i v_j+\nao_j v_i$ does, for some vector $v$.
 We can show that such a vector exists in the following way.
 If $h'$ satisfies (\ref{CYgauge}), then $v$ is determined by
\be
\nao^k \nao_k v_i + {1\over2}\nao_i \nao^k v_k = \nao^j h_{ji} - {1\over4} \nao_i h^j{}_j~.
\label{ddf}
\ee
The left-hand side is an elliptic operator acting on $v$, and can be inverted if its kernel is orthogonal to the right-hand side of the equation \cite{lawson}, \cite{joyce}.
 Now, suppose that $X_3$ is an \textit{irreducible} Calabi-Yau manifold and that $w$ is in the kernel of the operator.
 Then, we have
\be
0 = \int_{X_3}  w^i \left( \nao^k \nao_k w_i + {1\over2} \nao_i \nao^k w_k \right) d{\rm vol}
  = - \int_{X_3} \left(\nao_k w_i \nao^k w^i + {1\over2} (\nao^k w_k)^2 \right) d{\rm vol} \;.
\ee
The integral is equated with zero since $w$ is in the kernel, but the integrand itself can be written as a sum of squares, as in the second equality.
 Therefore, each term must vanish separately, which implies that $w$ is parallel with respect to the Levi-Civita connection.
 However, since $X_3$ is irreducible, there are no parallel one-forms on $X_3$, so that $w$ is zero and therefore the kernel vanishes.
 This means that the right-hand side of equation (\ref{ddf}), which we have assumed is non-zero, is indeed orthogonal to the kernel of the elliptic operator.
 Therefore the equation can be inverted and solved for $v$, so that the gauge condition (\ref{CYgauge1}) can always be attained.

Since this gauge can be satisfied for $(2,0)$ compactifications, it follows that the torsion and dilaton deformations for such backgrounds are determined by $h$ as in equations (\ref{tordef}) and (\ref{dilksedef}), so that the field equations are satisfied to first-order in $\a'$.
 Thus the Calabi-Yau metric $\go$ is deformed to a Hermitian one $g=\go+\a' h\,$ at first-order in $\a'$.
\section{Conclusion}
The salient feature of the foregoing investigation is that the three key ingredients which govern heterotic backgrounds associated to $(2,0)$ sigma models, namely the field equations, \kses\ and the anomaly-cancellation condition, are indeed consistent with one another at order $\a'$.
 Sufficient conditions have been provided, which consist of a gauge-fixing condition for possible diffeomorphisms of the underlying manifold, equation (\ref{ricciflat}), as well as the applicability of a generalised $\partial\bar\partial$-lemma.

In other words, these conditions determine a class of solutions which are guaranteed to solve the first-order equations.
 Presumably there will be other classes of solution which do not satisfy these conditions yet still solve the first-order equations, so we have by no means attempted to provide a classification.
 Further work is required to identify all the possible classes of solution to the zeroth-order field and Killing spinor equations which also solve the corrected equations and anomaly-cancellation condition.
 Thus we have found a particular class of solution that remains consistent to first-order, although this is only a subset of all possible solutions.
 Note that in the case of $(2,0)$ Calabi-Yau compactifications, the internal geometry is deformed to be Hermitian.

We may also draw some conclusions about the space of allowed deformations of the metric.
 Consider the expression given in (\ref{ddbar}), where the field equation for the metric deformation is reduced to
\be
P= -2i \partial \bar\partial \U \;,
\ee
where $\U_{ij}= h_{ik} J^k{}_j$.
 It is straightforward to see that $\U$ is a solution up to gauge transformations
\be
\U' = \U + \partial \bar w+\bar\partial w
\ee
for some $(1,0)$-form $w$.
 Thus the classes of independent solutions are parametrised by the $i\partial\bar\partial$-cohomology class
\be
V^{1,1}(X_m) = \frac{Ker\left( i\partial\bar\partial : \L^{1,1}(X_m) \rightarrow \L^{2,2}(X_m)  \right)}
        {\partial\L^{0,1}(X_m) + \bar\partial\L^{1,0}(X_m)}  \;.
\ee
These classes are otherwise known as elements of the \textit{Aeppli group}, see \cite{jggpdt}, \cite{jgsigp}.

In the case of Calabi-Yau manifolds, the $\partial\bar\partial$-lemma applies, and so we have the isomorphism
\be
V^{1,1}(X_3)\cong H^{1,1}(X_3) \;,
\ee
so that the dimension of the space of metric deformations is the Hodge number $h^{1,1}$.
 However, not all of these deformations will be supersymmetric, since to be so they must also satisfy condition (\ref{ricciflat}).

The machinery developed in this chapter can also be applied to known compactification solutions for the heterotic string, to determine the allowable $\a'$ corrections to the backgrounds.
 For instance, in \cite{jggpdt} the deformations were computed for the conifold and the $U(n)$-\inv\ Calabi-Yau metric.
%
%
%
%
%
%
\chapter[Spinorial Geometry in $D=11$ Supergravity]{Spinorial Geometry in $D=11$ Supergravity} \label{11d}
In this chapter, we describe a systematic method for solving the \kses\ of \sg, which was first proposed in \cite{jguggp}.
 The formalism is developed and applied extensively in the context of eleven-dimensional \sg.

The procedure for solving the \kses\ will be organised by means of the stability subgroups of the \kss, which will require some knowledge of the orbits of \spt\ and its subgroups on its space of spinors, $\Dc$.
 Utilising the explicit representation of \spt\ which may be derived from the construction in section \ref{explreprs}, it is a straightforward process to determine a canonical form for a spinor with a certain stabiliser.
 For the case of one \ks, $\n=1$, once a canonical form has been determined, it can then be substituted into the \kses, whose solution is greatly facilitated by the simplicity of the spinor.

In the $\n=2$ case, we may use the stability subgroup $G\subset Spin(1,10)$ of the first spinor to obtain a canonical form for the second spinor up to $G$ transformations, and so on for $\n>2$.
 Although there will be a large number of different cases to consider, even for $\n=2$, this formulation has the advantage of allowing a systematic study of solutions.

The second key ingredient in simplifying the problem of solving these equations is a judicious choice of basis for the gamma matrices.
 The \kses\ can then be expanded in terms of this basis, and the coefficients in this expansion will provide a set of constraints relating the field strength $F$ to the connection and to the spacetime functions which parametrise the spinor.
 They also give constraints on the connection itself, which can very often be interpreted as important information about the geometry of the background.

We will solve the \ksesb for the case of one \ks\ with stability \suf, confirming the results of \cite{pakis}.
 Then, we will go on to consider some different situations that can arise with two \kss.
 Solutions will be presented for certain configurations of $SU(5)$- and $SU(4)$-invariant spinors, and this will require some investigation into the orbits of \suf\ on $\Dc$.
 We will then investigate the cases for $\n=3$ and $4$ in which all of the spinors have their stability subgroups contained in \suf.
 In each case we use the constraints arising from the \kses\ to draw conclusions about the geometry of the background.
\section{$D=11$ Supergravity} \label{11d:sugra}
In $D=11$ \sg\ \cite{cjs}, \cite{nahm}, the bosonic fields consist of a metric $g$, and a three-form potential $A$ with four-form field strength $F=dA$.
 The bosonic part of the action involves the sum of an Einstein-Hilbert term, a generalised Maxwell term and a Chern-Simons term,
\be
I = \frac{1}{2\k^2} \int_{M} \mathcal{R} \; d{\rm vol} - \frac{1}{2} F \we\h F - \frac{1}{6} A \we F \we F \;,
\ee
where $\mathcal{R}$ is the scalar curvature of $g$ and $d{\rm vol}=+\sqrt{|g|}dx^0\we\cdots\we dx^{10}$ is the oriented volume element.
 We will neglect the higher-order correction term, since this will not concern us in our analysis.

The resulting equations of motion are  
\bea
0 &=& R_{MN}-\frac{1}{2}g_{MN}\mathcal{R} - \frac{1}{12}\left( F_{MQRS}F_N{}^{QRS} - \frac{1}{8}g_{MN}F_{PQRS}F^{PQRS} \right) \\
0 &=& d\h F +\frac{1}{2} F\we F \;.
\eea
Spacetime indices are labelled by $M,N,P,Q,\cdots=0,1,\cdots,9,10$.

In considering bosonic solutions, the gravitino is set to zero so that the supersymmetry variations of $g$ and $A$ vanish automatically.
 Thus the condition for supersymmetry to be preserved is simply that the supersymmetry variation of the gravitino should remain zero,
\be
\d_\e\psi = \mathcal{D} \e = 0 \;,
\ee
where $\e$ is a Majorana spinor.
 The supercovariant derivative of eleven-dimensional \sg\ is given by
\be \label{11d:sucovder}
\mathcal{D}_M = \na_M + \S_M \;,
\ee
where
\be
\na_M = \partial_M + \frac{1}{4}\O_{M,AB}\G^{AB}
\ee
is the spin-connection induced by the Levi-Civita connection, and
\be
\S_M = - \frac{1}{288}\left( \G_M{}^{PQRS}F_{PQRS} - 8F_{MQRS}\G^{QRS} \right)
\ee
is the $F$-dependent part of the connection.
 Therefore, a \sg\ background preserves supersymmetry if there exists at least one spinor $\e$ which is parallel with respect to $\mathcal{D}$,
\be \label{11d:susy}
\mathcal{D}_M\e = 0 \; .
\ee
These supercovariantly constant spinors are known as the \textit{Killing spinors} of the background.
 The number of supersymmetries preserved by a solution, $\n$, is the maximum number of linearly independent solutions to (\ref{11d:susy}), i.e. the number of independent \kss.
\section{An Explicit Representation of \spt}\label{11d:spinrepr}
\subsection{Spinors and Gamma Matrices}
First of all, we will summarise the construction of the spinor space and gamma matrices for \spt, following the prescription set down in section \ref{explreprs}.
 That is, we first build a representation of $Spin(10)$ and then extend it to a representation of $Spin(1,10)$.
 It is worth stating our index conventions at this point.
 Denoting the tenth spatial direction by the symbol $\nat$, we have:
\begin{itemize}
\item Indices $i,j,\cdots$ run from $1$ to $5$ ,
\item Indices $I,J,\cdots$ run from $1$ to $\nat$ ,
\item Indices $A,B,\cdots$ run from $0$ to $\nat$ .
\end{itemize}

Now, let us begin with the real vector space $V=\bR^{10}$ equipped with Euclidean inner product $(\,,\,)$.
 Locally, $V$ admits an almost complex structure $J$ which is orthogonal with respect to $(\,,\,)$, and also a set of vectors $\{e_1,\cdots,e_5\}$ such that
\be \label{11d:basis}
\{e_1,\cdots,e_5,J(e_1),\cdots,J(e_5)\}
\ee
is an orthonormal basis for $V$ (see section \ref{explreprs}).
 Thus $J$ induces an orthogonal splitting $V=U\oplus J(U)$, where $U$ is the real subspace spanned by $\{e_1,\cdots,e_5\}$.
 We make the identifications $e_{j+5}=J(e_j)$, for $j=1,\cdots,5$.
 Observe that since $V$ is even-dimensional with a Riemannian metric $(\,,\,)$ and orthogonal almost complex structure $J$, it possesses a $U(5)$-structure.

Now from section \ref{explreprs}, the space of complex pinors associated to $V$ is obtained as follows.
 First, we complexify the vector space, $\Vc=V\otimes\bC$, and form the natural Hermitian inner product $<\,,\,>$ which is induced by $(\,,\,)$.
 This can then be extended to act on the exterior algebra $\L^*(\Vc)\cong\cl(\Vc)\cong\ccl_{10}$ in the natural manner, with the inner product of two basis elements of degree $r$ given by
\be
<\, e_{J_1\cdots J_r} \,,\, e_{K_1\cdots K_r} \,>  =   r!\, \d_{[J_1}^{K_1}\cdots\d_{J_r]}^{K_r} \;,
\ee
and the inner product between elements of differing degrees is zero.
 Clifford multiplication is understood in the above, with $e_{IJ}=e_{I}\cdot e_{J}$.
 With respect to the vector space isomorphism between the Clifford and exterior algebras, we have $e_{J_1\cdots J_r}=e_{J_1}\we\cdots\we e_{J_r}$.

Then, the pinor module of $\ccl_{10}$ is given by the $2^5$-dimensional complex vector space
\be
\bP= \L^*(\Uc) \;.
\ee
Also, the unique pinor representation which maps $\ccl_{10}$ into $End_{\bC}(\bP)$ is provided by the following gamma matrices:
\bea \label{spin10repr}
\G&:&\ccl_{10}\cong End_{\bC}(\bP) \nn\\
\G_j      &=& \G(e_j) = e_j \we \cdot + e_j \lc \cdot \\
\G_{j+5}  &=& \G(e_{j+5}) = ie_j \we \cdot - ie_j \lc \cdot  \;\;,
\eea
where $\lc$ is the adjoint of $\we$ with respect to $<\,,\,>$ and $j=1,\cdots,5$.
 It is straightforward to show that $\G_j$ and $\G_{j+5}$ are Hermitian with respect to $<\,,\,>$, and that $\{\G_i,\G_j\} = \{\G_{i+5},\G_{j+5}\} = 2\delta_{ij} $ and $\{\G_i,\G_{j+5}\} = 0$, as required.

Now, we know from (\ref{spin2mreprs}) that $\bP$ decomposes into the sum of the two irreducible complex spinor modules of $Spin(10)$, each of dimension $2^4=16$.
 However, we wish to investigate the group $Spin(1,10)$, so we continue by introducing the time-component gamma matrix as in (\ref{odd:gamma2})\footnote{In fact, we have the choice $\G_0=\pm\G_1\G_2\cdots\G_9\Gten$, corresponding to the two inequivalent pinor representations of $\ccl_{11}$, but we choose the plus sign as our convention.},
\be \label{gamma10}
\G_0=\G_1\G_2\cdots\G_9\Gten \; .
\ee
It is elementary to show that $(\G_0)^2=-\id$, $\G_0$ anticommutes with $\G_I$, and together they satisfy the algebra of $\cl_{1,10}\otimes\bC$, namely $\{\G_A,\G_B\}=2\eta_{AB}$, where $\eta=diag(-1,1,\cdots,1)$.

Following the details of section \ref{explreprs}, on restricting to $Spin(1,10)\subset\ccl_{11}$, the representation specified by (\ref{spin10repr}) and (\ref{gamma10}) provides the unique spin representation of $Spin(1,10)$ on the spinor module
\be \label{11dimspinors}
\boxed{
\Dc=\bP=\cl(\Uc)\cong\L^*(\bC^5) \;,
}
\ee
where we have changed notation from $\bS$ to $\Dc$ to conform with the literature relating to spinorial geometry.
 This is also known as the space of \textit{Dirac spinors}.
 Thus every spinor of \spt\ can be written as an element of the exterior algebra of $\bC^5$ in this representation.
\subsection{The \spt-invariant Inner Product and Spacetime Forms}\label{11d:forms}
Following the theory of section \ref{forms}, which is based on the work of Wang in \cite{wang}, one can write down a \spt-invariant inner product, from which the \spt-invariant spacetime forms can be constructed.
 From the choice of $\A$ or $\B$, we choose to work with $\B$, since it is the $Pin(10)$-invariant one.
 This is required so that the inner product extends to being \spt-\inv\ (see section \ref{forms} for details).

Defining the map $\b=\G_6\cdots\Gten$, a \spt-invariant bilinear form is given by
\bea
\B(\s,\t)&=& <\,\b(\s^*)\,,\,\t\,> \\
         &=& <\, \G_6\cdots\G_\nat(\s^*)\;,\;\t\;> \;.
\eea
Note that $\B$ is a skew-symmetric inner product on $\Dc$.

Now, suppose we have $\n$ Killing spinors $\{\s^a\}_{a=1}^{\n}$.
 Then from the construction of section \ref{forms}, the \spt-\inv\ \st\ $p$-forms associated to each pair of spinors $\s^a,\s^b\in\Dc$ are defined by
\be \label{11d:stforms}
\boxed {\a_{(p)}^{ab} = \frac{1}{p!}\, \B(\s^a,\G_{A_1\ldots A_p}\s^b) \, e^{A_1}\we\cdots\we e^{A_p} } \qquad
p=0,\cdots,\nat  \; .
\ee
At first sight there are potentially a lot of forms to calculate for a given set of spinors, but in fact this number is reduced considerably for the following two reasons. Firstly, the symmetry properties of the gamma matrices and the skewness of $\B$ imply that $\a^{ab}=\a^{ba}$ for $p=1,2,5$ and $\a^{ab}=-\a^{ba}$ for $p=0,3,4$. Hence it is sufficient to compute only the forms with $a\leq b$. Secondly, one need only compute the forms up to degree $p=5$, as the rest can be found using Poincar\'{e} duality.
\subsection{The Majorana Condition}\label{11d:majcond}
In eleven-dimensional \sg, we work with $32$-dimensional Majorana spinors, and so we must impose a reality condition on the space of complex Dirac spinors to pick out the required subspace, which we denote $\Dm$.
 This may be achieved by setting
\be \label{11d:maj}
\boxed{
\eta^*=\G_0\b(\eta)=\G_{1\cdots5}(\eta) \;.
}
\ee
In other words, the Majorana spinors of eleven-dimensional \sg, $\Dm$, are those elements of $\Dc$ which satisfy (\ref{11d:maj}).
 Also, the bilinear form $\B$ restricts to a \spt-\inv\ skew-symmetric inner product on $\Dm$.
\subsection{Orbits of $Spin(1,10)$ in $\Dm$} \label{11d:orbits}
In \cite{bry}, Bryant has shown that there are precisely two types of non-trivial orbit of $Spin(1,10)$ on the space of spinors $\Dm$.
 His method involves representing \spt\ on the space $\bO^4\cong\Dm$, where $\bO$ denotes the ring of octonians, and then considering the level sets of a certain non-negative \spt-\inv\ polynomial $\mathbf{p}$ defined on $\bO^4$.
 Although the procedure is transparent in this formalism, we will not present explicit details of the derivation, but rather highlight the salient points of the analysis.
 For full details, see \cite{bry} and \cite{bry1}.

The first important thing that was shown is that there are two types of non-trivial orbit of \spt\ on its spinor space.
 These correspond to the positive level sets of $\mathbf{p}$, and the non-zero elements of the zero level set.
 These level sets are of dimension $31$ and $25$ respectively.

Furthermore, a homomorphism can be defined from \spt\ to $SO(1,10)^\uparrow$, the identity component of $SO(1,10)$, which maps the orbits of the former into those of the latter.
 It turns out that the image of the \spt\ orbits in $\bR^{1,10}$ is the union of the origin, the forward null-cone, and the future-directed timelike vectors.
 From this, it can be seen that the spinors which lie in the orbit of the positive level sets of $\mathbf{p}$, are mapped to timelike vectors in $\bR^{1,10}$ under the homomorphism.
 The stability subgroup of such a spinor can then be found to be isomorphic to $SU(5)\subset Spin(10)\subset Spin(1,10)$, therefore we denote this orbit by \orbsuf.
 As an elementary check, note that since
\be
\mathcal{O}_{SU(5)}\cong\frac{Spin(1,10)}{SU(5)} \;,
\ee
we have that $dim($\orbsuf$)=55-24=31$, as required.

On the other hand, non-trivial spinors which lie in the zero orbits of $\mathbf{p}$ are mapped to null vectors in $\bR^{1,10}$, and their stabilisers can be found to be isomorphic to \sps$\subset Spin(1,10)$.
 This orbit we denote by \orbsps.
 Here, observe that
\be
\mathcal{O}_{Spin(7)}\cong\frac{Spin(1,10)}{(Spin(7)\ltimes\bR^8)\times\bR} \;,
\ee
so that the dimension of this orbit is $dim(\mathcal{O}_{Spin(7)})=55-30=25$, as required according to the corresponding level set of $\mathbf{p}$.
\subsection{An Antiholomorphic Basis of Spinors}\label{11d:hermbasis}
For many computations, solving the \ksesb can be aided by using a new, Hermitian basis of gamma matrices.
 We define these as
\bea
\Gh_{\a} &=& \frac{1}{\sqrt2} \left( \G_{\a} - i \G_{\a +5}  \right)  \\
\Gh_{\betab} &=& \frac{1}{\sqrt2} \left( \G_{\b} + i \G_{\b +5}  \right) \quad\quad \a, \b = 1,\cdots,5 \; ,
\eea
where the hats differentiate the Hermitian gamma matrices from the original basis, which was defined in
 (\ref{spin10repr}).

From these definitions, we find that
\bea
\Gh_{\a}&=&\sqrt2e_\a\we  \\
\Gh_{\ab}&=&\sqrt2e_\a\lc \quad\quad\quad \a = 1,\cdots,5 \; ,
\eea
where $e_\a$ is one of the basis elements of the real vector space $V$.
 Therefore, the Hermitian gamma matrices correspond to creation and annihilation operators respectively, on the space of spinors $\Dc$.

Their holomorphic and antiholomorphic indices are raised and lowered using the standard Hermitian metric $g_{\a \betab} = \delta_{\a \betab}$ on $\bC^5$ and the Clifford algebra relations in this basis become $\{\Gh_{\a},\Gh_{\betab}\}=2g_{\a \betab}$ and $\{\Gh_{\a},\Gh_{\b} \} = \{\Gh_{\ab},\Gh_{\betab}\} = 0$.

It is also useful to note that we can write the 5-form basis element as
\be
e_{12345} = \frac{1}{8.5!}\e_{\ab_1\cdots\ab_5}\Gh^{\ab_1\cdots\ab_5}(1) \; ,
\ee
where $\e_{\bar1\bar2\bar3\bar4\bar5}=\sqrt2$ is the antiholomorphic volume-form, normalised for later convenience.

Now, starting from the \textit{Clifford vacuum} $1\in\Dc$, and acting with the gamma matrices which have antiholomorphic upper indices, i.e. the creation operators, we obtain a new basis for $\D^{\bC}\cong\L^*(\bC^5)$:
\be \label{11d:hermbasis1}
\{ 1, \; \Gh^{\ab}(1), \; \Gh^{\ab_1\ab_2}(1), \; \Gh^{\ab_1\ab_2\ab_3}(1), \;
\Gh^{\ab_1\ab_2\ab_3\ab_4}(1), \; \Gh^{\ab_1\ab_2\ab_3\ab_4\ab_5}(1)  \}  \; .
\ee
This basis will greatly simplify the calculations involved in solving the \kses\ for certain spinor configurations.
%
%
%
\section{$\n=1$ Backgrounds}\label{11d:n=1}
In this section, backgrounds with one \ks\ will be investigated with the major aim of solving the \kses, and the strategy will be as follows.
 As we have already seen in section \ref{11d:orbits}, there are two non-trivial orbits of \spt\ in its spinor space $\Dc$, which are characterised by their stability subgroups: one of the orbits is precisely the set of \suf-invariant spinors, and the other is precisely the set of \sps-invariant spinors.
 They are denoted by \orbsuf\ and \orbsps, respectively.
 This means we may classify $\n=1$ backgrounds according to the subgroup of \spt\ which leaves its Killing spinor invariant.
 However, we will focus on the case in which there is one \suf-\inv\ Killing spinor $\eta$.

A key concept in our procedure is the gauge group of the supercovariant connection $\mathcal{D}$.
 This is the group of transformations which leaves the form of $\mathcal{D}$ invariant, up to a local Lorentz transformation of the frame and field strength.
 In eleven dimensions this is \spt, and must not be confused with the \textit{holonomy group} of the supercovariant connection, which in this case is $SL(31,\bR)\ltimes\bR^{31}$, since there is one \ks\ \cite{gpdt1}.
 The latter will in general alter the form of $\mathcal{D}$, and so is inappropriate for the approach we wish to take.
 Instead, we will utilise the \spt\ gauge-invariance, by transforming $\eta\in$\orbsuf\ into a simple representative of its orbit, via \spt\ transformations.
 We will refer to this representative as a \textit{canonical form} for its \spt\ orbit.
 In other words, we may simplify the spinor without making the \kses\ any more complicated.
 The \spt-invariance of $\mathcal{D}$ means that any two spinors lying in the same orbit of \spt\ yield identical solutions to the \kses, up to local Lorentz transformations of the fields.
 This means that we can use the canonical form for the orbit to make the solution of the \kses\ relatively straightforward.

Another key ingredient in our method is the Hermitian basis of gamma matrices described in section \ref{11d:hermbasis}.
 We may expand the \kses\ in terms of this basis and after some straightforward computation, the equations reduce to a set of algebraic and differential constraints on the fields of the theory.
 We will see that these constraints provide important information about the geometry of the \st.

The solution to the $D=11$, $\n=1$ \kses\ for an \suf-\inv\ \ks\ was performed in \cite{pakis} using a different method which relied on the use of Fierz identities, and we find that our results agree with those of that paper.
\subsection{The \suf\ Orbit}\label{11d:orbsuf}
The first step in the method is to determine a canonical form for a spinor which has stability subgroup $SU(5)\subset Spin(10)\subset Spin(1,10)$.
 We can begin by using our knowledge of the representations of the complex Clifford algebras to decompose the pinor representation $\Dc$ under $Spin(10)$ as
\benn
\Dc = \D^+ \oplus \D^- \; ,
\eenn
where $\D^{\pm}$ are the inequivalent spinor representations of $Spin(10)$, of complex dimension $16$ (cf. (\ref{spin2mreprs})).
 The Majorana condition selects a subspace of spinors $\Dm$ which intersects both of $\D^\pm$.

Now, from the isomorphism between the Clifford and exterior algebras, we have that
\bea
\D^+ &\cong& \L^{Even}(\Uc) = \L^{0}_{\mathbf{1}}(\bC^5)
            \oplus\L^{2}_{\mathbf{10}}(\bC^5) \oplus \L^{4}_{\bar{\mathbf{5}}}(\bC^5) \\
\D^- &\cong& \L^{Odd}(\Uc)  = \L^{1}_{\mathbf{5}}(\bC^5)
            \oplus\L^{3}_{\bar{\mathbf{10}}}(\bC^5) \oplus \L^{5}_{\bar{\mathbf{1}}}(\bC^5) \;,
\eea
where the decomposition is into irreducible representations of \suf, and the bold subscripts denote the complex dimension of each representation.
 This decomposition tells us that the one-dimensional representations are carried by the spinors\footnote{Our convention will be to write spinors with downstairs multi-indices, where Clifford multiplication is implicit, whereas the spacetime exterior forms will be written with indices upstairs and in terms of the wedge product.} $1$ and $e_{12345}$.
 In other words, these are the spinors which transform trivially under \suf\ and so they span the orbit \orbsuf, so that a general \suf-invariant spinor is of the form
\be
\eta = a 1 + b e_{12345} \; , \qquad a,b\in\bC \;.
\ee
Here, $a$ and $b$ are complex numbers, but when we look at the \kses, we must treat the parameters of the spinor as arbitrary spacetime functions.

Now, on imposing the Majorana condition (\ref{11d:maj}), we find that $b=a^*$.
 Therefore, a general \suf-invariant Majorana spinor of \orbsuf\ can be written as
\be
\eta = a 1 + a^*e_{12345} \; , \qquad a\in\bC \;.
\ee
Thus, a basis for the \suf-invariant Majorana spinors is given by:
\bea
\etasuf &=& \frac{1}{\sqrt2}(1+e_{12345}) \label{suf1} \\
\thsuf  &=& \frac{i}{\sqrt2}(1-e_{12345}) \label{suf2}
\eea
They have been normalised so as to have length $1$, with respect to the Hermitian inner product on $\L^*(\Uc)$.

Observe that $\etasuf$ and $\thsuf$ are related by a $Spin(10)\subset Spin(1,10)$ transformation,
\be
\thsuf=\G_0\etasuf=\G_{1\cdots\nat}\etasuf \; ,
\ee
which serves as a confirmation that they indeed represent the same orbit of \spt\ in $\Dc$.
 Since they are equivalent in this way, we can choose either of them as a canonical representative of \orbsuf.
 When we analyse the \kses\ in later sections, we will make the choice of canonical form for an \suf-invariant spinor to be
\be \label{11d:1stsuf}
\boxed{
\eta_1=f(x)\etasuf } \;,
\ee
where $f$ is an arbitrary \textit{real} function on the \st.
\subsection{Spacetime Forms Associated to $\etasuf$}\label{11d:su5forms}
Next, we compute the spacetime forms associated to the spinor $\etasuf$, using formula (\ref{11d:stforms}) and the elementary results that
\bea
 \B(1,1) \; = &0& = \; \B(e_{12345},e_{12345}) \; = \; 0 \\
 \B(1,e_{12345}) \; = &-i& = \; - \B(e_{12345},1) \;,
\eea
noting that $\B$ skew-symmetric.
 Using these facts, we find that the non-vanishing forms associated to $\etasuf$ are as follows.
\begin{itemize}
\item \textbf{One-form}:
\bea
\boxed{
\k^{SU(5)}=\B(\eta^{SU(5)}, \G_0\eta^{SU(5)}) e^0=- e^0
}
\eea
\item \textbf{Two-form}:
\bea
\boxed{
\o^{SU(5)}= -e^1\we e^6 - e^2\we e^7 - e^3\we e^8 - e^4\we e^9 - e^5\we e^{\nat}
}
\label{kahlersuf}
\eea
\item \textbf{Five-form}:
\bea
\boxed{
\t^{SU(5)} = {\rm Im}\left\{ (e^1+i e^6)\we\cdots\we(e^5+ie^\nat)\right\} + \frac{1}{2}e^0\we\o^{SU(5)}\we\o^{SU(5)}
}
\eea
\end{itemize}
These are the spacetime forms associated to the orbit \orbsuf.

Each of these exterior forms are \suf-\inv\ since the associated spinor is \suf-\inv.
 Indeed, we will now show that $\o^{SU(5)}$ is the K\"ahler form of the Hermitian space $(V,J)$.
 In our conventions, the K\"ahler form is defined to be
\be
\hat\o = \frac{1}{2}g_{IK}J^K{}_J\,e^I\we e^J \;.
\ee
Now, we define the natural Hermitian basis of exterior forms by
\bea
\hat{e}^\a  &=&  \frac{1}{\sqrt2}\left( e^\a + ie^{\a+5} \right) \nn \\
\hat{e}^{\ab}   &=&  \frac{1}{\sqrt2}\left( e^\a - ie^{\a+5} \right)  \; , \label{11d:hermforms}
\eea
where $\a=1,\cdots,5$ and $e^\a$ is one of the original basis elements from (\ref{11d:basis}).
 Now, in this new basis of forms we have $g_{\a\betab}=\d_{\a\betab}$ and $J^\a{}_{\b}=i\d^\a_{\b}$, as in section \ref{11d:hermbasis}.
 Therefore, we see that
\bea
\hat\o \,=\, -i\d_{\a\betab}\,\hat{e}^\a\we\hat{e}^{\betab} \,=\, -\left( e^1\we e^6 + \cdots e^5 \we e^\nat \right)
\,=\, \o^{SU(5)} \;.
\eea
Thus, $\o^{SU(5)}$ is the K\"ahler form in these conventions.

Also, consider the first part of $\t^{SU(5)}$, which we denote
\be
\hat{\t} = {\rm Im}\left\{(e^1+i e^6)\we\cdots\we(e^5+ie^\nat)\right\} \;.
\ee
Note that this is proportional to the imaginary part of the holomorphic volume form\footnote{We have the factor of $\sqrt2$ in $\e$ as a result of the earlier definition of $\e_{\bar1\bar2\bar3\bar4\bar5}=\sqrt2$.}
\be
\e  = {\sqrt2}\;\hat{e}^{1}\we\cdots\we\hat{e}^{5} = \frac{1}{4}\,(e^1+i e^6)\we\cdots\we(e^5+ie^\nat) \;.
\ee
In fact, we can obtain $\e$ from $\hat\t$ using the almost complex structure, in the following way.
 Define a map $I:\L^p(\bC^5)\longrightarrow\L^p(\bC^5)$ by (cf. \cite{cabrera})
\be
I\cdot\a_{(p)}=\frac{1}{p!}J^k{}_{i_1}\a_{ki_2\cdots i_p}e^{i_1}\we\cdots\we e^{i_p} \;,
\ee
where $\a_{(p)}$ is an arbitrary $p$-form.
 Then, we have
\be
I\cdot\e = i\e \qquad\hbox{and}\qquad I\cdot\bar\e = -i\bar\e \;,
\ee
where $\bar\e=(\e)^*$ is the antiholomorphic volume form, with $\e_{\bar1\bar2\bar3\bar4\bar5}=\sqrt2$.

Now, we observe that
\be
\hat\t = 4{\rm Im}(\e) = 2i\left( \bar\e - \e \right) \;,
\ee
which implies that
\be
I.\hat\t = 2\left( \bar\e + \e \right) = 4{\rm Re}(\e) \;.
\ee
Therefore, we have
\be
\e = \frac{1}{4} \left( I\cdot\hat\t + i\hat\t \right) \;,
\ee
so that $\hat\t$ gives rise to the holomorphic volume form on the Hermitian space \linebreak spanned by the basis vectors $\{e_1,e_2,\cdots,e_\nat\}$.
 This means that the space possesses an $SU(5)$-structure determined by $(g,J,\e)$, or equivalently by $(\hat\o,\hat\t)$.
 In other words, the space is in fact a \textit{special Hermitian manifold}, \cite{cabrera}.

Furthermore, in the context of \sg\ backgrounds, we shall see that $\k^{SU(5)}$ gives rise to a timelike Killing vector, which enables us to express the spacetime metric in terms of coordinates which are adapted to this vector.

It is also noteworthy that having found these explicit expressions for the forms associated to \orbsuf, it is straightforward to establish their algebraic relationships, e.g. $\i_{\k}\t=\frac{1}{2}\o\we\o$.
 Many of these identities were calculated in \cite{pakis}, using different conventions and a method involving the use of Fierz rearrangements.
\subsection{Analysis of the $\n=1$ Killing Spinor Equations}\label{11d:n=1kses}
\subsubsection{The Killing Spinor Equations}
We begin by introducing a real orthonormal frame $\{e^{A}\}_{A=0}^{\nat}$ for the background $M$, and writing the spacetime metric as
\be
ds^2= -(e^0)^2+(e^1)^2+\cdots+(e^\nat)^2 \;.        
\ee
Next, we expand the four-form field strength $F$ into electric and magnetic parts as
\be
F = \frac{1}{3!} e^0\we G_{ijk} e^i\we e^j\we e^k + \frac{1}{4!} F_{ijkl} e^i\we e^j\we e^k\we e^l \; .
\ee
The Levi-Civita spin-connection $\O_{A,MN}$ lives in the space $T^*(M)\otimes\so(1,10)$ (see Appendix \ref{GrayHerv}).
 Singling out the time direction, the independent non-vanishing components are
\be
\O_{0, ij}~,~~~~\O_{0, 0j}~,~~~~\O_{i,0j}~,~~~~\O_{i,jk}~.
\ee
Now, we can decompose the \kse\ (\ref{11d:susy}) into time and spatial components, and simultaneously expand out the field-strength and connection to give
\bea \label{11d:susy1}
0 &=& \partial_0 \z + {1\over 4} \O_{0, ij}\G^{ij}\z - {1\over2}\O_{0,0i} \G_0\G^{i}\z
- {1\over 288} \bigl( \G_0 \G^{ijkl} F_{ijkl} - 8 G_{ijk} \G^{ijk} \bigr) \z \; , \nn \\
0 &=& \partial_i \z + {1\over4} \O_{i,jk} \G^{jk}\z - {1\over2} \O_{i,0j} \G_0 \G^j\z
- {1\over 288} \bigl( \G_i{}^{jklm} F_{jklm}\cr
&& +4 \G_0 \G_i{}^{jkl} G_{jkl} - 24 \G_0 G_{ijk} \G^{jk}- 8 F_{ijkl} \G^{jkl} \bigr) \z \;,
\eea
where $\z$ is a spinor in $\Dm$.

Suppose now that $\z$ is in the orbit of \suf-\inv\ Majorana spinors.
 We can capitalise on the \spt\ gauge-invariance of equations (\ref{11d:susy1}), by bringing $\z$ into the canonical form
\be
\boxed{
\eta_1 = f(x) \etasuf  } \;,
\ee
via a \spt\ transformation, where $f$ is a real function on the spacetime which parametrises the spinor.
 The form of the supercovariant derivative will not change under this transformation, yet we now have a simple and explicit expression for the spinor, which can be substituted into the equations.

Inserting the canonical form $\eta_1$ into the equations gives
\bea
0 &=& \partial_0 \log(f) \eta^{SU(5)} + {1\over4}\O_{0, ij}\G^{ij}\eta^{SU(5)}-{1\over2}\O_{0,0i} \G_0\G^{i}\eta^{SU(5)} \nn \\
&-& {1\over 288} \bigl( \G_0 \G^{ijkl} F_{ijkl} - 8 G_{ijk}\G^{ijk} \bigr)\eta^{SU(5)} \;, \nn\\
0 &=& \partial_i\log(f) \eta^{SU(5)} + {1\over4} \O_{i,jk} \G^{jk}\eta^{SU(5)} - {1\over2} \O_{i,0j} \G_0 \G^j\eta^{SU(5)}    \nn \\
&-& \!\!\!\!{1\over 288} \bigl( \G_i{}^{jklm} F_{jklm} + 4 \G_0 \G_i{}^{jkl} G_{jkl} - 24 \G_0 G_{ijk} \G^{jk} - \! 8 F_{ijkl} \G^{jkl} \bigr) \eta^{SU(5)} .
\label{11d:kses}
\eea
The next step is to use the properties
\be
\G_0 1=i 1 \qquad\hbox{and}\qquad \G_0 e_{12345}=-i e_{12345}
\ee
to eliminate $\G_0$ from (\ref{11d:kses}) and thus express the equations entirely in terms of the ten spatial gamma matrices.

Then, we expand each gamma matrix in terms of the Hermitian basis constructed in section \ref{11d:hermbasis}.
 This requires that we also expand the metric, spin-connection and field strength in terms of the associated Hermitian basis of tensors.
 Using $\Gh^\a 1=0$ and $\Gh^{\ab} e_{12345}=0$ we can commute gamma matrices through to hit these spinors, so that certain terms are annihilated.
 The final step is to collect together the terms proportional to each Hermitian basis element, so that the \kses\ become expansions in the Hermitian basis.
 Schematically, the resulting expression for each \kse\ is of the form
\bea \label{11d:hermexp}
0 &=& C_0(1) \,+\, (C_1)_{\ab}\Gh^{\ab}(1) \,+\, (C_2)_{\ab_1\ab_2}\Gh^{\ab_1\ab_2}(1)
\,+\, (C_3)_{\ab_1\ab_2\ab_3}\Gh^{\ab_1\ab_2\ab_3}(1)
\nn \\
&& +\, (C_4)_{\ab_1\ab_2\ab_3\ab_4}\Gh^{\ab_1\ab_2\ab_3\ab_4}(1)
\,+\, (C_5)\Gh^{\bar1\bar2\bar3\bar4\bar5}(1) \;,
\eea
where each coefficient is a linear combination of connection and fieldstrength components.
 Now, since this expansion is equated with zero, we argue that the coefficient of each basis element must vanish separately.
 Thus each \kse\ reduces to a set of at most six constraints of the form $C_0=\cdots=C_5=0$, one for each component of the Hermitian basis of (\ref{11d:hermbasis1}).
 However, these are not all independent, as we shall see shortly.
 (See \cite{systematics} for general formulae describing the relationships between these constraints.)
 The independent equations are the conditions that we seek to solve and interpret.

Let us first consider the time component $\mathcal{D}_0\eta_1=0$.
 The independent constraints which result from the above procedure are:
\bea
0 &=& \partial_0 \log f + {1\over2} \O_{0,{\a\bar\b}} g^{\a\bar\b} - {i\over24} F_{\a}{}^{\a}{}_{\b}{}^{\b}
\label{11d:n=1t1} \\
0 &=& i \O_{0,0 \ab} + {1\over3} G_{\ab\b}{}^{\b}{} + {i\over 72} F_{\b_1\b_2\b_3\b_4} \e^{\b_1\b_2\b_3\b_4}{}_{\ab}  \label{11d:n=1t2} \\
0 &=& \O_{0,\ab\bar\b}-{i\over6} F_{\ab\bar\b\g}{}^\g{} - {1\over 18} G_{\g_1\g_2\g_3}
 \e^{\g_1\g_2\g_3}{}_{\ab\bar\b} \label{11d:n=1t3}
\eea
Also, there are three more conditions, but they are simply the complex conjugate equations to the ones above, and therefore provide no additional independent constraints.

The spatial component of the \kse\ decomposes into two parts, according to whether the derivative is along the holomorphic or anti-holomorphic spatial frame directions, i.e. as to whether $i=\a$ or $i=\ab$.
 However, we find that the $D=11$ supercovariant derivative is real, in the sense that the holomorphic and antiholomorphic components are related by complex conjugation and dualisation with respect to the antiholomorphic spinor basis.
 Thus they give an identical set of constraints, and we need only consider either the holomorphic or antiholomorphic component.

We find that the independent constraints which arise from $\mathcal{D}_{\ab}\eta_1=0$ are:
\bea
0 &=& \partial_{\ab} \log f + {1\over2} \O_{\ab, \b\bar\g} g^{\b\bar\g} + {i\over12} G_{\ab\g}{}^\g
- {1\over 72} \e_{\ab}{}^{\b_1\b_2\b_3\b_4} F_{\b_1\b_2\b_3\b_4}
\label{11d:n=1a1} \\
0 &=& \partial_{\ab} \log f - {1\over2} \O_{\ab, \b\bar\g} g^{\b\bar\g} + {i\over4} G_{\ab\g}{}^\g
\label{11d:n=1a2} \\
0 &=& i \O_{\ab,0\bar\b} + {1\over6} F_{\ab\bar\b\g}{}^\g - {i\over 18} \e_{\ab\bar\b}{}^{\g_1\g_2\g_3}
G_{\g_1\g_2\g_3} \label{11d:n=1a3} \\
0 &=& i\O_{\ab,0\b} + {1\over12} g_{\ab\b} F_\g{}^\g{}_\d{}^\d + {1\over2} F_{\ab\b\g}{}^\g
\label{11d:n=1a4} \\
0 &=& \O_{\ab, \bar\b\bar\g} + {i\over 6} G_{\ab\bar\b\bar\g} - {1\over 12} \e_{\ab\bar\b\bar\g}{}^{\g_1\g_2}
F_{\g_1\g_2\d}{}^\d - {1\over 12} F_{\ab\g_1\g_2\g_3} \e^{\g_1\g_2\g_3}{}_{\bar\b\bar\g}
\label{11d:n=1a5} \\
0 &=& \O_{\ab,\b\g} - {i\over 2} G_{\ab\b\g} - {i\over3} g_{\ab[\b} G_{\g] \d}{}^\d
- {1\over 36} F_{\ab\bar\g_1\bar\g_2\bar\g_3} \e^{\bar\g_1\bar\g_2\bar\g_3}{}_{\b\g}
\label{11d:n=1a6}
\eea
Let us now proceed to solve these equations, to express the field-strength in terms of the geometry of the spacetime and to find a set of constraints for the connection and the function $f$.
\subsubsection{Solving the Killing Spinor Equations}
The constraints form a relatively simple set of equations which we will now analyse, in order to derive relationships between the fields and the geometry of the background.

Two immediate consequences arise from (\ref{11d:n=1t1}) and its complex conjugate:
\bea
\partial_0\log f &=& 0 \label{11d:df=0} \\
F_{\a}{}^{\a}{}_{\b}{}^{\b} &=& -12i\O_{0,{\a\bar\b}} g^{\a\bar\b}  \; . \label{11d:tr1}
\eea
Also, equation (\ref{11d:df=0}) implies that $\partial_0 f=0$, so that $f$ does not depend on the frame time direction.
 Constraint (\ref{11d:tr1}) tells us that the trace of the magnetic part of $F$ is determined by the spin-connection.

Next, we subtract (\ref{11d:n=1a2}) from (\ref{11d:n=1a1}), to find that
\be
\O_{\ab, \b}{}^\b - {i\over6} G_{\ab \b}{}^\b - {1\over72}\e_{\ab}{}^{\b_1\dots\b_4}F_{\b_1\dots\b_4} = 0\;.
\label{11d:n=1a1a}
\ee
Together with (\ref{11d:n=1t2}) this equation gives two conditions:
\bea
F_{\b_1\dots\b_4} &=& {1\over2} (-\O_{0,0\ab} + 2 \O_{\ab, \b}{}^\b) \e^{\ab}{}_{\b_1\dots\b_4}
\label{11d:F1} \\
G_{\ab \b}{}^\b &=& -2i \O_{\ab, \b}{}^\b-2i \O_{0,0\ab}  \; .
\label{11d:G1}
\eea
Now, substituting (\ref{11d:F1}) and (\ref{11d:G1}) back into (\ref{11d:n=1a1}), we find that
\be
\partial_{\ab} \log f + \frac{1}{2} \O_{0,0\ab} = 0 \;.
\label{11d:daf}
\ee

Equation (\ref{11d:n=1a4}) and its complex conjugate imply
\be
\O_{\ab,0\b} + \O_{\b,0\ab} = 0 \; ,
\label{11d:conn1}
\ee
and in the next section, it will be shown that this is a geometric condition which can always be satisfied by an appropriate choice of frame.

Next we take the trace of (\ref{11d:n=1a4}) to obtain
\be
F_{\a}{}^\a{}_\b{}^\b = 12 i \O_{\ab,0\b} g^{\ab\b} \; .
\label{11d:tr2}
\ee
Together with (\ref{11d:tr1}), this constraint requires that
\be
- \O_{0,\ab}{}^{\ab} + \O_{\ab,0}{}^{\ab} = 0 \;.
\label{11d:conn4}
\ee
Again, we will see later that this condition is satisfied in a certain choice of frame. Substituting (\ref{11d:tr2}) back into (\ref{11d:n=1a4}), we have
\be
F_{\b\ab\g}{}^\g = 2i \O_{\ab,0\b} + 2i g_{\ab\b} \O_{\bar\g, 0 \d} g^{\bar\g\d} \; .
\label{11d:F2}
\ee

Now, tracing (\ref{11d:n=1a6}) and using (\ref{11d:F1}) and (\ref{11d:G1}) gives the condition
\be
\O_{0,0\g} = \O_{\ab, \b\g} g^{\ab\b} - \O_{\g,\b}{}^\b \; .
\label{11d:conn2}
\ee
This is another constraint on the geometry of spacetime which will be investigated shortly.
 Substituting (\ref{11d:conn2}) back into (\ref{11d:n=1a6}), we have
\be
G_{\ab\b\g} = -2i \O_{\ab,\b\g} + 2i g_{\ab[\b|} \O_{0,0|\g]} \; .
\label{11d:G2}
\ee
Also, equations (\ref{11d:n=1t3}) and (\ref{11d:n=1a3}) imply that
\be
\O_{0, \ab\bar\b}=\O_{\ab, 0\bar\b} \; .
\label{11d:conn3}
\ee
This is a geometric constraint which comes from the torsion-free condition of the Levi-Civita connection, which will be considered in the next subsection.

We now turn to equations (\ref{11d:n=1a3}) and (\ref{11d:n=1a5}). Together they imply the following two constraints:
\bea
G_{\ab_1\ab_2\ab_3} &=& 6i \O_{[\ab_1,\ab_2\ab_3]}
\label{11d:G(0,3)} \\
F_{\ab\b_1\b_2\b_3} &=& {1\over2} \left(\O_{\ab,\bar\g_1\bar\g_2} + 3\O_{[\ab,\gb_1\gb_2]}  \right)\e^{\bar\g_1\bar\g_2}{}_{\b_1\b_2\b_3}  + 6 i \O_{[\b_1|,0|\b_2} g_{\b_3]\ab} \; .
\label{11d:F(1,3)}
\eea

Each equation has now been analysed, and so it remains to summarise the results and interpret them accordingly.
\subsubsection{Summary of Results for the $\n=1$ Equations}  \label{11d:n=1summary}
Having considered each equation, we summarise the results for the field strength in a table, for easy reference:
\begin{center} \ \\
\begin{tabular}{|c|c|} \hline
Component       & Solution                                                              \\ \hline\hline
$G^{3,0}$       & $G_{\b_1\b_2\b_2}=-i\O_{[\b_1,\b_2\b_3]}$                                     \\ \hline
$G^{2,1}$       & $G_{\ab\b_1\b_2}=-2i\O_{\ab,\b_1\b_2} - 2i\O_{0,0[\b_1}g_{\b_2]\ab}$          \\ \hline
$F^{4,0}$       & $F_{\b_1\b_2\b_3\b_4}
          = \frac{1}{2}\left(-\O_{0,0\ab}+2\O_{\ab,\g}{}^{\g}\right)\e^{\ab}{}_{\b_1\b_2\b_3\b_4}$\\ \hline
$F^{3,1}$       & $F_{\ab\b_1\b_2\b_3} = {1\over2} \left(\O_{\ab,\bar\g_1\bar\g_2} + 3\O_{[\ab,\gb_1\gb_2]}  \right)\e^{\bar\g_1\bar\g_2}{}_{\b_1\b_2\b_3}  + 6 i \O_{[\b_1|,0|\b_2} g_{\b_3]\ab}$
\\ \hline
trace$(F^{2,2})$& $F_{\b}{}^\b{}_\g{}^\g = 12 i \O_{\ab,0\b} g^{\ab\b}$                         \\ \hline
$F^{2,2}_0$       & Undetermined by \kses         \\ \hline
\end{tabular}
\end{center}
~ \\
Also, we have the following constraints on the connection and \st\ function $f$:
\begin{center} \ \\
\begin{tabular}{|c|c|c|} \hline
$\O_{\ab,0\b} = - \O_{\b,0\ab}$  &
$\O_{0, \ab\bar\b}=\O_{\ab, 0\bar\b}$  &
$\O_{0,\ab}{}^{\ab} = \O_{\ab,0}{}^{\ab}$ \\ \hline
$\O_{0,0\g} = \O_{\ab, \b\g} g^{\ab\b} - \O_{\g,\b}{}^\b$  &
$\partial_{\ab} \log f = - \frac{1}{2} \O_{0,0\ab}$ & $\partial_0 f=0$ \\ \hline 
\end{tabular}
\end{center}
~ \\
These conditions may be interpreted as relationships between the various intrinsic torsion modules for an $\su(5)$-structure in an $\so(1,10)$ manifold, as described in Appendix \ref{GrayHerv}.

To make contact with representation theory, it is useful to see that splitting the four-form field strength $F$ into the above components corresponds to its decomposition into representations of \suf.
 This may be seen most easily by counting the degrees of freedom of the indices.
 We begin with $F_{ABCD}$, which is a four-form transforming under $SO(1,10)$, and so it has $\frac{1}{4!}11.10.9.8=330$ degrees of freedom.
 The first splitting is into representations of $SO(10)$, and this is achieved by separating out the time index to give $G_{JKL}=F_{0JKL}$ and $F_{IJKL}$.
 This corresponds to the decomposition $\mathbf{330}\longrightarrow\mathbf{120}\oplus\mathbf{210}$ under $SO(10)$.

Now, $SO(10)$ acts irreducibly on each exterior power \cite{joyce}, \cite{salamon}.
 However, irreducible representations of \suf\ are carried by \textit{traceless} $(p,q)$-forms, for $p+q\leq5$, relative to the complexification of the exterior algebra.
 Therefore to decompose the $SO(10)$ representations into \suf\ irreps, we must decompose the spaces of exterior forms accordingly, and remove their traces.
 Firstly, we have (see \cite{salamon}, for example)
\be
\L^3(\bR^{10}) \cong \L^{(3,0)+(0,3)}\oplus\L^{(2,1)+(1,2)}_{0}\oplus\L^{(1,0)+(0,1)} \;,
\ee
where the spaces on the right-hand side are exterior powers of $\bC^5$, and the subscript $0$ denotes that the trace has been removed.
 For instance, the trace of a $(2,1)$-form is a $(1,0)$-form.
 By counting dimensions, this can be seen to correspond to the decomposition
\be
\mathbf{120} \longrightarrow \mathbf{10}\oplus\bar{\mathbf{10}} \oplus \mathbf{45}\oplus\bar{\mathbf{45}}
\oplus\mathbf{5}+\bar{\mathbf{5}} \;.
\ee
Similarly, the component $F_{IJKL}$ respects the decomposition
\be
\L^4(\bR^{10}) \cong \L^{(4,0)+(0,4)}\oplus\L^{(3,1)+(1,3)}_{0} \oplus\L^{(2,2)}_0
\oplus\L^{(2,0)+(0,2)} \oplus \L^{(1,1)}_0 \oplus \bR \;.
\ee
This makes explicit the branching\footnote{See \cite{slansky} for comprehensive tabulations of the representations, tensor products and branching rules for many commonly occurring Lie groups, such as $SO(10)$ and $SU(5)$.} of the $\mathbf{210}$ of $SO(10)$ under \suf, as
\be
\mathbf{210} \longrightarrow \mathbf{5}\oplus\bar{\mathbf{5}}\oplus\mathbf{40}\oplus\bar{\mathbf{40}}
\oplus\mathbf{75}\oplus\mathbf{10}\oplus\bar{\mathbf{10}}\oplus\mathbf{24}\oplus\mathbf{1} \;.
\ee

From the table, we see that the $\n=1$ \kses\ determine all components of the field strength except for the traceless part of the $(2,2)$-component of the magnetic flux, denoted $F^{2,2}_0$.
 This corresponds to the $\mathbf{75}$ of \suf.
 Note also that the traceless part of the connection, $\O_{i,\b\gb}-\frac{1}{5}\O_{i,\d}{}^{\d}g_{\b\gb}$, is not involved in the solution either.
\subsubsection{The Geometry of the Spacetime} \label{11d:n=1geom}
We will now turn to the analysis of the results found in the
previous section, and use these to help draw some important
geometric conclusions about the nature of a spacetime which admits
an \suf-\inv\ \ks.
 Since $\eta_1=f\etasuf$, we have the associated
one-form $\k^f=-f^{2}\k^{SU(5)}=f^{2}e^0$.
 It is straightforward to see that the Killing vector condition $\nabla_{(A}\k^f_{B)}=0$ is satisfied, using the constraints from the last section.
 Indeed, the $(A,B)=(0,0)$ component is satisfied vacuously.
 The $(A,B)=(0,\ab)$ component leads to the equation
\be
\partial_{\ab} f^2 +  \O_{0, 0\ab} f^2 = 0 \; ,
\ee
which is satisfied using the condition $\partial_{\ab} \log f = - \frac{1}{2} \O_{0,0\ab}$. In a similar way, the $(A,B)=(\a,\b)$ and $(A,B)=(\a,\betab)$ components are satisfied using the other connection conditions from the table.
 Hence $\k^f$ is associated to a Killing vector field.
 Also, since $(\k^f)^2=-f^4$, we see that $\k^f$ is in fact a \textit{timelike Killing vector field} in the spacetime.

Since we have a timelike Killing vector field, we can always choose an adapted coordinate system in which the metric may be written as
\be
ds^2= - f^4 (dt + \a)^2+ ds^2_{10} \; ,
\ee
where $ds^2_{10}$ is a metric on the ten-dimensional base-space $B$ which is transverse to the orbits of $\k^f$, and both $f$ and $\a$ are independent of the time coordinate $t$.

Now, a natural choice of non-coordinate frame is to take $e^0=f^2(dt + \a)$, and $\{e^J\}_{J=1}^{\nat}$ to be an orthonormal frame on $B$, so we have
\be \label{11d:n=1frame}
ds^2 = -(e^0)^2 + \sum^{\nat}_{J=1} (e^J)^2 \; .
\ee
Denoting the Levi-Civita connection of this frame by $\O$, we consider the torsion-free condition
\be
de^A + \O^A{}_B \we e^B = 0 \; .
\ee
Since the frame $\{e^J\}$ does not depend on the time coordinate $t$, the torsion-free condition implies that
\be
\O_{I,0J}=\O_{0,IJ} \; , \quad I,J=1,\cdots,\nat \; .
\ee
Therefore, this frame is consistent with the constraints (\ref{11d:conn1}), (\ref{11d:conn4}) and (\ref{11d:conn3}).

It remains to interpret conditions (\ref{11d:daf}) and ({\ref{11d:conn2}).
 Firstly, we use  (\ref{sumit}) and (\ref{sonit}) to relate a component of the intrinsic torsion of the $\so(10)$-structure to those of the $\su(5)$-structure,
\be \label{11d:n=1so10}
(\rm{y}_4)_{\g} = (\rm{w_4})_\g + (\rm{w_5})_\g \;.
\ee
We can write this in terms of a Gray-Hervella intrinsic torsion module for comparison with \cite{pakis} as follows.
 In our conventions, we set $(W_5)_i=\frac{1}{80}\e^{j_1\cdots j_5}\na_{[i}\e_{j_1\cdots j_5]}$, where $\e$ is the antiholomorphic volume form\footnote{In normalising $W_5$ we have altered the numerical factor from the standard definition to compensate for $\e_{\bar1\bar2\bar3\bar4\bar5}=\sqrt{2}$ in our conventions.}.
 This gives
\be \label{11d:W5}
(W_5)_\g = \frac{1}{2} \left( \O_{\betab,}{}^{\betab}{}_\g - \O_{\g,\b}{}^\b \right) \;,
\ee
so that
\be
2(W_5)_\g = (\rm{w_4})_\g + (\rm{w_5})_\g \;.
\ee

Now, we combine equations (\ref{11d:daf}) and (\ref{11d:conn2}) to give
\be
W_5 = - d f \; ,
\label{11d:base1}
\ee
where $d$ is the exterior derivative on $B$, so that $W_5$ is exact.
 Since this equation involves no time component, it is purely a restriction on the geometry of the ten-dimensional almost Hermitian manifold $B$.
 Thus the \kses\ lead to an explicit constraint on $B$.

Our results agree with those of \cite{pakis}, up to numerical factors arising from differing conventions and normalisations.

By way of conclusion, let us here summarise the conditions on the spacetime that arise from the \kses\ for one \suf-\inv\ \ks:
\begin{itemize}
\item All components of $F$ are determined by the connection, as specified the table, except for the traceless $(2,2)$-piece, which is undetermined by the \kses.
 This corresponds to the $\mathbf{75}$ irreducible representation of \suf\ arising from the decomposition of $F$, as described in the previous section.
\item There is a timelike Killing vector field associated to the \st, which enables us to choose adapted coordinates in which the frame may be written in the form (\ref{11d:n=1frame}).
\item The spatial manifold $B$ is special almost Hermitian, i.e. it possesses an \suf-structure, which may be determined as in section \ref{11d:su5forms}.
\item An important constraint on the geometry is that its Gray-Hervella module $W_5$ is exact.
\end{itemize}
\subsection{The \sps\ Orbit}
From section \ref{11d:orbits} we know that aside from \suf-\inv\ spinors, the other non-trivial orbit of \spt\ in $\Dm$ is \orbsps, the subspace of \sps-\inv\ spinors.
 Although we will not consider this case in detail here, we will provide a canonical form to represent \orbsps.
 As a guiding principle, we use the fact that such a spinor must be associated to a null vector \cite{bry}.
 Therefore, consider the following spinor,
\be
\eta = a e_1+b e_{2345}~,~~~~~~~a,b\in \bC~.
\ee
The Majorana condition implies that
\be
a=b^*.
\ee
This means that $\eta$ is a linear combination of the two Majorana spinors
\bea
&&e_1+e_{2345}
\\
&&i(e_1-e_{2345})~.
\eea
Now, we set
\bea
\eta^{Spin(7)}&=&{1\over  2} (i(1-e_{12345})+e_1+e_{2345})~.
\eea
Computing the associated vector we find the non-vanishing components
\bea
\k_0(\eta^{Spin(7)}, \eta^{Spin(7)})&=& \B(\eta^{Spin(7)}, \G_0 \eta^{Spin(7)})=-1
\\
\k_1(\eta^{Spin(7)}, \eta^{Spin(7)})&=& \B(\eta^{Spin(7)}, \G_1 \eta^{Spin(7)})=1 \;,
\eea
so that
\be
\kappa^{Spin(7)}=-e^0+e^1 \;.
\ee
This vector is null, and therefore we may take $\eta^{Spin(7)}$ as a representative of the orbit \orbsps.

We will not pursue the analysis of the \kses\ for this spinor here, but rather continue to consider the consequences of having two or more \kss\ in the background.
 However, the analysis has been performed for a $Spin(7)\ltimes\bR^8$-\inv\ spinor in the context of IIB \sg\ in \cite{IIB}, although the gauge group in that case is $Spin(1,9)$ and so a different spinor representation is used.
 As a matter of comparison with the $D=11$ case, it is interesting to note that in IIB \sg\ there are three distinct orbits for a single spinor of $Spin(1,9)$.
 These are the orbits of the $Spin(7)\ltimes\bR^8$, $SU(4)\ltimes\bR^8$ and $G_2$-\inv\ spinors \cite{IIB}.
%
%
%
\section{$\n=2$ Backgrounds
}\label{11d:n=2su5}
In this section, we investigate $D=11$ \sg\ backgrounds with two \kss\ $\eta_1,\eta_2\in\Dc$.
 Here, the situation is more complicated and there are many more cases that can occur than for just one spinor.
 These result from the different combinations of stability subgroups that two spinors can have.
 We know that the only non-trivial orbits of \spt\ in $\Dt$ are \orbsuf\ and \orbsps, so these are the only broad classes of solution to the \kses\ in the $\n=1$ case.
 However, the second spinor may have trivial stabiliser, or it may lie in either orbit, and furthermore it could have a subgroup of \suf\ or \sps\ as its stabiliser.
 In this way, there is a large number of possible combinations to consider, depending on the combination of stability subgroups that the spinors possess.
 In this section, we will focus on some specific cases that can occur when $\eta_1$ is in \orbsuf.
\subsection{Orbits of $SU(5)$ in $\Dm$}
Let $\eta_1\in\,$\orbsuf.
 Then, as discussed in the $\n=1$ case, we can use \spt\ gauge transformations to bring it into canonical form, so that $\eta_1=a\eta^{SU(5)}$, $a\in\bC$.
 Now consider $\eta_2\in\Dm$.
 Firstly, note that this spinor must be linearly independent of $\eta_1$ at every \st\ point, because if they are linearly dependent at a single point, then they are dependent everywhere.
 Now generically, $\eta_2$ will be a linear combination of exterior forms of each degree, and will have the identity as its stability subgroup.
 However, there will be special cases, in which some of its parameters vanish or are restricted in some way, so that the stabiliser is non-trivial.
 These are the cases that we will focus on in this section, with the aim of determining canonical forms for some different possibilities that may occur.

However, the \spt\ gauge-invariance has been used to transform $\eta_1$ into canonical form.
 This means that in transforming $\eta_2$ into a canonical form, we only have freedom to use \suf\ transformations, since these leave the canonical form of $\eta_1$ fixed.
 Any \spt\ transformation which lies outside of $SU(5)\subset Spin(1,10)$ will destroy the simple form of $\eta_1$.
 Therefore, we are required to gain some understanding of the orbits of $SU(5)$ in the spinor space $\Dm$, to enable us to derive canonical forms for each case.
 Whereas there are only two non-trivial orbits of \spt, since \suf\ is a much smaller group, there will be considerably more orbits.
 To the investigation of these we now turn.

Recall the decomposition of the complex spinor module of \spt\ under
$Spin(10)$.
 We have
\be
\Dc\longrightarrow\D^+_{\mathbf{16}}\oplus\D^-_{\mathbf{16}} \;.
\ee
Now, from section \ref{11d:orbsuf}, the isomorphism between the Clifford and exterior algebras makes it straightforward to write down the decomposition of the $Spin(10)$ modules into irreducible representations of \suf:
\bea \label{11d:n=2decomp}
\D^+_{\mathbf{16}} &=& \L^0_\mathbf{1}(\bC^5) \oplus\L^2_{\mathbf{10}}(\bC^5) \oplus\L^4_{\mathbf{\bar5}}(\bC^5) \\
\D^-_{\mathbf{16}} &=& \L^1_\mathbf{5}(\bC^5) \oplus\L^3_{\mathbf{\bar{10}}}(\bC^5) \oplus\L^5_{\mathbf{\bar1}}(\bC^5)
\eea
As in the $\n=1$ case, the superscript denotes the degree of the exterior forms, while the bold subscript labels the dimension of the representation.

Since we are looking for Majorana spinors, we can simplify the next steps in the analysis by focusing on the half of $\eta_2$ which lies in $\D^+$, since the components which lie in $\D^-$ are precisely determined by the Majorana condition (\ref{11d:maj}).
 In other words, a generic spinor of $\Dm$ can be written as
\bea
\eta_2 = b 1 + \chi + \hbox{m.c.} \; , \quad b\in\bC \; ,
\label{11d:n=2gen}
\eea
where $\chi\in\L^2_{\mathbf{10}}(\bC^5) \oplus\L^4_{\mathbf{\bar5}}(\bC^5)$, and the abbreviation `m.c.' indicates the Majorana conjugates of the components $b1$ and $\chi$.

Now, using decomposition (\ref{11d:n=2decomp}), the four possible cases for $\chi$ are as follows:

\begin{itemize}
\item \textsc{Case} 1:~~~~$\chi=0$
\item \textsc{Case} 2:~~~~$\chi\in \L_{\mathbf{\bar 5}}^4(\bC^5)$
\item \textsc{Case} 3:~~~~$\chi\in \L_{\mathbf{{10}}}^2(\bC^5)$
\item \textsc{Case} 4:~~~~$\chi \in\L_{\mathbf{{10}}}^2(\bC^5)\oplus\L_{\mathbf{{\bar 5}}}^4(\bC^5)$
\end{itemize}

We will examine the action of \suf\ on $\chi$ in each case, determine the subgroup of \suf\ which leaves it invariant, and then provide a canonical form for each possibility.
\\ \ \begin{center} \underline{\underline{\textsc{Case} 1:~~~$\eta_1=a\etasuf$ ~~and~~ $\chi=0$\;}}\end{center}
Here, we have $\;\eta_2=b1+b^*e_{12345}\;$ so that in fact, $\eta_2\in\,$\orbsuf.
 From the discussion of the $\n=1$ case and equations (\ref{suf1}) and (\ref{suf2}), we know that $\etasuf$ and $\thsuf$ form a basis for \orbsuf, and so the second \ks\ can be written as
\be \label{11d:2ndsuf}
\eta_2 = b_1\etasuf + b_2\thsuf \; , \qquad b_1, b_2 \in \bR \; .
\ee
Thus, $\eta_2$ is the most general \suf-\inv\ spinor.

In the analysis of the \kses, the constants $a,b_1,b_2\in\bR$ will be treated as \st\ functions, so in general the \kses\ will involve three unknown real functions and constrain them accordingly. Also, we require that the spinors are non-vanishing and linearly independent, hence both $a$ and $b_2$ must be non-zero. However, a special case can arise when $b_1=0$, in which case $\eta_1=a\etasuf$ and $\eta_2=b_2\thsuf$.
\\ \ \begin{center} \underline{\underline{\textsc{Case} 2:~~~$\eta_1=a\etasuf$ ~~and~~ $\chi\in\L^{4}_{\mathbf{\bar{5}}}(\bC^5)$\;:}}
\end{center}
First, observe that $\L^{4}_{\mathbf{\bar{5}}}(\bC^5)$ is the dual representation to the standard vector representation of \suf\ on $\L^{1}_{\mathbf{{5}}}(\bC^5)$.
 Since dualisation is an isomorphism via the \suf-\inv\ antiholomorphic volume form, the orbit structure of these two representations must be identical.
 Therefore, since the vector representation $\L^{1}_{\mathbf{{5}}}(\bC^5)$ has one non-zero orbit, namely the orbit of elements which are preserved by $SU(4)$, this is also true of $\L^{4}_{\mathbf{\bar{5}}}(\bC^5)$.
 Thus $\chi$ has $SU(4)$ as its stabiliser, and we will denote its orbit by \orbsufr.

To find a simple orbit representative, we begin by decomposing $\L^4(\bC^5)$ under \sufr, as
\be
\L^4_{\mathbf{{5}}}(\bC^5) \cong \L^4_{\mathbf{{1}}}(\bC^4) \oplus \L^3_{\mathbf{{4}}}(\bC^4) \;.
\ee
If we consider the embedding of \sufr\ in \suf\ whereby \sufr\ acts on the space spanned by $\{e_1,e_2,e_3,e_4\}$, then this decomposition corresponds to the splitting of a four-form $\l\in\L^4(\bC^5)$ as
\be
\frac{1}{4!} \l^{ijkl}e_{ijkl} = \l^{1234}e_{1234} + \frac{1}{3!}\l^{abc5}e_{abc5} \;,
\ee
where $i,j,k,l=1,\cdots,5$ and $a,b,c=1,\cdots,4$.
 We see that the one-dimensional representation is carried by the first term in the splitting, and so the \sufr-\inv\ piece is spanned by the spinor
\benn e_{1234} \; . \eenn

Thus, imposing condition (\ref{11d:maj}), a general \sufr-\inv\ Majorana spinor is of the form
\be
\chi = re_5 + se_{1234} \; , \qquad  r,s\in\bC \; .
\ee
As in the \suf\ case, the Majorana condition tells us that $s=r^*$, so that there are two linearly independent Majorana spinors which span the orbit \orbsufr:
\bea
\etasufr &=& \frac{1}{\sqrt2} \left( e_5 + e_{1234} \right) \label{sufr1} \\
\thsufr &=& \frac{i}{\sqrt2} \left( e_5 - e_{1234} \right) \label{sufr2} \;.
\eea
The normalisation is such that each spinor has unit length with respect to the Hermitian inner product on $\Dt$.

Observe the following relationships between the spinors:
\bea
\etasufr &=& \G_5\etasuf \\
\thsufr &=& -\G_5\thsuf = \Gten\etasuf \\
\thsufr &=& -\G_0\etasufr \;.
\eea
Since the canonical forms (\ref{sufr1}) or (\ref{sufr2}) both represent the orbit \orbsufr, they are equally suitable to use as the canonical form for $\chi\in\L^{4}_{\mathbf{\bar{5}}}(\bC^5)$.
 We will choose to use $\etasufr$, so that $\chi=b_3\etasufr$.
 Therefore, the most general \sufr-\inv\ spinor $\eta_2$ can be written as
\be  \label{11d:gensufr}
\eta_2 = b_1\etasuf + b_2\thsuf + b_3\etasufr   \; , \qquad b_1,b_2,b_3\in\bR \; .
\ee
Note that $SU(4)\cong Spin(6)$ also occurs as a subgroup of $(Spin(7)\ltimes\bR^8)\times\bR$, so there exist \sufr-\inv\ spinors in the orbit \orbsps.
 Therefore, in (\ref{11d:gensufr}) we have in fact obtained the most general \sufr-\inv\ spinor up to \suf\ transformations.

In the analysis of the $\n=2$ \kses\ for this case, with $\eta_1=a\etasuf$, there will be in general four unknown real \st\ functions to deal with: $a,b_1,b_2,b_3$.
 It is required that $b_3\neq 0$, because otherwise this would reduce to the \suf-\inv\ case examined previously.
 However, a special case that can arise is when $\eta_1=a\etasuf$ and $\eta_2=b_3\etasufr$.
 Then, $\eta_1$ and $\eta_2$ lie on different orbits of \spt\ in $\Dm$, as can be seen from the relationship
\benn
\etasufr = \G_5 \etasuf \;.
\eenn
In other words, the spinors are related by the $Pin(10)$ transformation $\G_5$ and not a $Spin(1,10)$ transformation, which means that they do not lie on the same orbit of $Spin(1,10)$.
 Since $\eta_1\in$\,\orbsuf\ and the only non-trivial orbits of \spt\ are \orbsuf\ and \orbsps, in this special case we must have $\eta_2\in$\,\orbsps.
 Therefore, its stabiliser is actually a subgroup of \sps\ \,rather than of \suf, arising from the isomorphism
\benn
SU(4)\cong Spin(6)\subset Spin(7) \;.
\eenn
\
\
\begin{center}\underline{\underline{\textsc{Case} 3:~~~$\eta_1=a\etasuf$ ~~and~~ $\chi\in\L^{2}_{\mathbf{10}}(\bC^5)$\;:}} \end{center}
To examine this case we must investigate the orbit structure of \suf\ acting on the vector space $\L^{2}(\bC^5)$.
 We seek to find how many distinct orbits there are, to determine their stability subgroups and to provide a canonical form for each case.
\\ \ \\
- \underline{\textbf{Orbits of \suf\ in $\L^{2}(\bC^5)$\;:}}
\\
An \suf\ element acts on an arbitrary two-form by the transformation
\benn
\o \longrightarrow \o'=U^T\o U \; .
\eenn
We will begin by considering a generic two-form $\rho\in\L^2(\bC^5)$.
 This is a vector space of complex dimension ten, so that $\rho$ depends on twenty parameters in the most general case.
 We begin by determining a canonical form for $\rho$, and then we will describe the special cases which can arise from this general solution.

\begin{lem}
A canonical form for the most general orbit of \suf\ in $\L^{2}(\bC^5)$ is
\be
\boxed{
\s = \l e^1\we e^2 + \m e^3\we e^4 \;, \qquad \hbox{where } \l \neq\m\;,\;\l,\m\in\bR \;.
}
\ee
\end{lem}
Indeed, we can show that any general element such as $\rho\in\Lf$ can be obtained by applying an \suf\ transformation to $\s$.
 A convenient way to approach this is to write $U=e^{i\a H}$, where $H$ is a traceless Hermitian matrix and $\a$ is the infinitesimal parameter. Then, to first order in $\a$,
\benn
\s \longrightarrow \s + i\a \{ H^T\s + \s H \}
\eenn
Now, let
\benn
H =
\begin{pmatrix}
   x_1   &   y_1   &   y_2   &   y_3   &   y_4    \\
  y_1^*  &   x_2   &   z_1   &   z_2   &   z_3    \\
  y_2^*  &  z_1^*  &   x_3   &   w_1   &   w_2    \\
  y_3^*  &  z_2^*  &   w_1^* &   x_4   &    v     \\
  y_4^*  &  z_3^*  &   w_2^* &   v^*   &   x_5
\end{pmatrix} \; ,\;\;
\hbox{where }\; v,w_i,y_i,z_i\in\bC,\; x_i\in\bR \;.
\eenn
We also have the constraint $x_1+x_2+x_3+x_4+x_5=0$, since $H$ is traceless. Let us assume that the complex upper triangular entries in $H$ are independent from one another, i.e. they are not linearly related or conjugate to each other. Therefore in general, $H$ depends on $24$ real parameters, as required for an $\su(5)$ matrix.

Next, we calculate the two-form
\beann
\rho &=& \s + i\a \{ H^T\s + \s H \}  \\
&=&
\begin{pmatrix}
0 & \l\{ 1+i\a(x_1+x_2) \}      & i\a(\l z_1 -\m y_3^*) & i\a(\l z_2 + \m y_2^*)        & i\a\l z_3     \\
  &            0                & -i\a(\l y_2+\m z_2^*) & i\a(-\l y_3 + \m z_1^*)       & -i\a\l y_4    \\
  &                             &           0           & \m\{ 1+i\a(x_3+x_4) \}        & i\a\m v       \\
  &                             &                       &          0                    & -i\a\m w_2    \\
  &                             &                       &                               &      0
\end{pmatrix}
\; ,
\eeann
where only the upper-triangular entries have been written, for convenience.

We now argue that $\rho$ is a generic two-form, so that its entries are independent from one another, in the following way.
 By assumption, $\l$ and $\m$ are independent, and also the entries of $H$ are independent.
 Now, from these assumptions, it is straightforward to see that the entries $\rho_{13},\;\rho_{14},\;\rho_{15},\;\rho_{23},\;\rho_{24},\;\rho_{25},\;\rho_{35},\;\rho_{45}$ are independent.
 However, $\rho_{12}$ and $\rho_{34}$ may be related, until we note that since $x_1+x_2+x_3+x_4+x_5=0$, the real numbers $(x_1+x_2)$ and $(x_3+x_4)$ are independent.
 Therefore, since $\l$ and $\m$ are unrelated, we see that $\rho_{12}$ and $\rho_{34}$ are indeed independent.

Hence $\rho$ is a generic two-form with arbitrary entries.
 This affirms that we can reach a general element of $\Lf$ by applying an \suf\ transformation to $\s$, so that $\s$ is a canonical form for the \suf-orbit of generic two-forms in $\Lf$, as required.
 We will now investigate the stability subgroup of $\s$ along with some special cases that can arise.
\\ \
\\ \ \\
- \underline{\textbf{Stabiliser of $\s$\;:}}
\\
To explicitly find the stability subgroup of $\s=\l e^1\we e^2 + \m e^3\we e^4$, one must determine the form of the \suf\ matrices which leave it invariant,
\be
\textrm{Stab}(\s)=\{ U\in SU(5):\; U^T\s U = \s \} \;.
\ee
A more practical way to determine these matrices is to solve the equivalent condition $\s U=U^*\s$.
 This condition restricts the number of parameters in $U$ and constrains the matrices to lie in a certain subgroup of \suf.
 In the most general case, where $\l\neq\m$, it is straightforward to show that the stabiliser of $\s$ is $SU(2)\times SU(2)$.
 Thus, the orbit of the most generic element of $\Lf$ is given by
\be
\mathcal{O}_{SU(2)\times SU(2)} \cong \frac{SU(5)}{SU(2)\times SU(2)} \; .
\ee
As a useful confirmation, we can count dimensions for the orbit of the most generic two-form.
 Observe that
\bea
&& \text{dim}_{\bR}\left(\mathcal{O}_{SU(2)\times SU(2)}\right)=24-6=18 \;\;\text{and} \\
&& \text{dim}_{\bR}\left(\L^2(\bC^5)\right) = 5\cdot4 = 20 \; ,
\eea
and hence $\text{codim}_{\bR}\left(\mathcal{O}_{SU(2)\times SU(2)}\right)=2$, which is equal to the number of independent parameters $\l$ and $\m$, as expected.

There are also some special cases that arise for certain values of $\l$, $\m$, and these cases give rise to different stabiliser groups. The results are summarised in the table below.
\begin{center}
\begin{tabular}{|c|c|} \hline
$\l,\,\m\in\bR$                 &  $\textrm{Stab}(\s)$  \\ \hline\hline
$\l,\m\neq0$, $\l\neq\m$        &  $SU(2)\times SU(2)$  \\ \hline
$\l=\m\neq0$                    &  $Sp(2)$              \\ \hline
$\l=0$ or $\m=0$                &  $SU(2)\times SU(3)$  \\ \hline
$\l=\m=0$                       &  $SU(5)$              \\ \hline
\end{tabular}
\end{center} \
\\
- \underline{\textbf{Majorana Spinor Representatives\;:}}
\\
Having found that there are \textit{three} non-trivial orbits of \suf\ in $\Lf$, we wish to find canonical Majorana representatives in each case.

We begin by considering the most general orbit, with stability subgroup $SU(2)\times SU(2)$ and spinor representative
\be \label{11d:suttcanform}
 \l e_{12} + \m e_{34} \; , \qquad \l\neq\m\neq 0 \;, \; \l,\m \in\bR \;.
\ee
Applying the Majorana condition shows that this is a linear combination of two independent Majorana spinors
\bea
\eta^{SU(2)\times SU(2)} &=& \frac{1}{\sqrt2}\left(\l e_{12} + \m e_{34} - \l e_{345} - \m e_{125}\right)
\nn \\
\th^{SU(2)\times SU(2)}  &=& \frac{i}{\sqrt2}\left(\l e_{12} + \m e_{34} + \l e_{345} + \m e_{125}\right) \;,
\eea
subject to the condition $\l^2+\m^2=1$, which arises from normalising with respect to the Hermitian inner product.

The second orbit has stability subgroup $Sp(2)$, denoted $\mathcal{O}_{Sp(2)}$, and arises as a special case of the generic orbit when\footnote{Strictly speaking, we may have $\l=\pm\m$, which corresponds to different embeddings of $Sp(2)$ in \suf. However, we choose to work with the plus sign.} $\l=\m$.
 Thus, it has representative
\benn
 e_{12} + e_{34} \; ,
\eenn
and again we find a basis of two Majorana spinors for this orbit, given by
\bea
\eta^{Sp(2)} &=& \frac{1}{2} \left( e_{12} + e_{34} - e_{345} - e_{125} \right) \; ,
\nn \\
\th^{Sp(2)} &=& \frac{i}{2} \left( e_{12} + e_{34} + e_{345} + e_{125} \right) \; .
\eea

Also, in this case there is another $Sp(2)$-\inv\ spinor which lies in the space $\L^4(\bC^5)$, and is obtained by wedging $e_{12} + e_{34}$ with itself, to give a spinor proportional to $e_{1234}$.
 As has already been mentioned, a general element of $\L^4(\bC^5)$ has stability subgroup \sufr, and so this extra spinor corresponds to the embedding $Sp(2)\subset SU(4)$.
 The associated Majorana spinors are
\bea
\zeta^{Sp(2)}   &=& \frac{1}{\sqrt2} \left( e_{5} + e_{1234} \right) \; ,
\nn \\
\ph^{Sp(2)}     &=& \frac{i}{\sqrt2} \left( e_{5} - e_{1234} \right) \; .
\eea

Finally, we have the orbit \orbsutth\ which arises when either $\l=0$ or $\m=0$, with stability subgroup $SU(2)\times SU(3)$.
 A representative is given by
\benn
e_{12} \; ,
\eenn
and the associated Majorana spinors which span \orbsutth\ are
\bea
\eta^{SU(2)\times SU(3)} &=& \frac{1}{\sqrt2} \left( e_{12} - e_{345} \right) \; ,
\nn \\
\th^{SU(2)\times SU(3)}  &=& \frac{i}{\sqrt2} \left( e_{12} + e_{345} \right) \; .
\eea
This concludes the discussion of the orbits of \suf\ in $\Lf$.
\\ \ \begin{center} \underline{\underline{\textsc{Case} 4:~~~$\eta_1=a\etasuf$ ~~and~~ $\chi\in\L^{2}_{\mathbf{10}}(\bC^5)\oplus\L^{4}_{\mathbf{\bar{5}}}(\bC^5)$\;:}} \end{center}
The final possibility for $\chi$ is that it has both a two-form and a four-form component.
 In this case, we must investigate the orbits of \suf\ acting non-trivially on both subspaces of $\L^{2}_{\mathbf{10}}(\bC^5)\oplus\L^{4}_{\mathbf{\bar{5}}}(\bC^5)$, otherwise it will reduce to one of the cases considered previously.

As before, we use \spt\ transformations to bring $\eta_1$ into canonical form \linebreak $a\etasuf$.
 Now, we are restricted to using \suf\ transformations to bring $\chi$ to some sort of canonical form.
 Let us begin by bringing the two-form component into the form of equation (\ref{11d:suttcanform}) by an \suf\ gauge transformation.
 Now, we may only use $SU(2)\times SU(2)$ transformations to simplify the four-form component, since this is the stabiliser of (\ref{11d:suttcanform}).
 Let us proceed in the following way.

Consider the dual space $\L^1(\bC^5)$.
 If we take the standard embedding of $SU(2)\times SU(2)$ in \suf, with the two copies of \sut\ acting on the spaces spanned by $\{e_1,e_2\}$ and $\{e_3,e_4\}$ respectively, then a one-form decomposes under $SU(2)\times SU(2)$ as
\be
\l_ie^i = \l^te_t + \l^ve_v + \l^5e_5   \;,
\ee
where $i=1,\cdots,5$, $t=1,2$ and $v=3,4$.
 Now, using $SU(2)\times SU(2)$ we can transform the $(\mathbf{2},\mathbf{1})$ and $(\mathbf{1},\mathbf{2})$ representations $e_t$ and $e_v$ into their respective \sut\ canonical forms $e_1$ and $e_3$, so that a canonical form for $\l$ up to $SU(2)\times SU(2)$ transformations is given by
\be
c_1 e_1 + c_2 e_3 + c_3 e_5 \;,
\ee
for some complex constants $c_1$, $c_2$, $c_3$.
 Dualising, this translates to a canonical form for $\L^4(\bC^5)$ up to $SU(2)\times SU(2)$ transformations given by
\be
c_1 e_{2345} + c_2 e_{1245} + c_3 e_{1234} \;.
\ee
Consequently, a representative of the most general orbit of \suf\ in $\L^{2}_{\mathbf{10}}(\bC^5)\oplus\L^{4}_{\mathbf{\bar{5}}}(\bC^5)$ is
\be  \label{11d:generic}
c_1 e_{2345} + c_2 e_{1245} + c_3 e_{1234} + b \left( \l_1 e_{12} + \l_2 e_{34} \right) \;,
\ee
where $\l_1,\l_2\in\bR$, $c_1,c_2,c_3,b \in\bC$, and $\l_1\not=\l_2\not=0$ and $c_1\neq c_2\neq c_3\neq0$.

The maximal common stability subgroup of $\eta_1=a\eta^{SU(5)}$ and (\ref{11d:generic}) is  $\{\id\}$, so that this case has the least residual symmetry.
 The  complex spinor in (\ref{11d:generic}) will give rise to Majorana spinors which may be used as Killing spinors for $\n=2$ backgrounds.
 However, we will now describe some special cases that arise when certain parameters are fixed.
 This gives rise to spinors with non-trivial stability subgroups and therefore more symmetry.
 Since we are interested in the case where $\chi\in \L_{\mathbf{10}}^2(\bC^5)\oplus\L_{\bar{\mathbf{5}}}^4(\bC^5)$, we assume that at least one of $c_1, c_2, c_3$ and at least one of $\l_1, \l_2$ do not vanish.
 Some examples will now be described briefly.

First suppose that $\l_1\not=\l_2\not=0$.
 Then, if either $c_1$ or $c_2$ vanishes, the stability subgroup is enlarged from $\{\id\}$ to $SU(2)$.
 Again, this is most easily seen by looking at the one-form dual to the four-form piece, as above.
 However, if both $c_1=c_2=0$, then the stability subgroup is $SU(2)\times SU(2)$.
 If $c_1, c_2\not=0$ and $c_3=0$, the stability subgroup is $\{\id\}$.
 On the other hand if $\l_1=\l_2$ and either $c_1$ or $c_2$ vanishes, then
the stability subgroup is $Sp(1)\cong SU(2)$.
 This also holds when in addition $c_3=0$.
 Also, if $c_1=c_2=0$, then the stabiliser of (\ref{11d:generic}) is $Sp(2)$.

Next, suppose that $\l_2=0$.
 If $c_1, c_3\not=0$ and $c_2=0$, then the stability subgroup is $SU(2)$.
 If $c_1=c_2=0$ and $c_3\not=0$, then (\ref{11d:generic}) has stabiliser $SU(2)\times SU(2)$.
 Also, if $c_2=c_3=0$ and $c_1\not=0$, then the stability subgroup is $SU(3)$.
 Analogous results hold for $\l_1=0$.

In the following section, we summarise the foregoing results as well as presenting some more special cases arising from (\ref{11d:generic}).
\subsubsection{Summary of the Most General Case} \label{11d:n=2generalcase}
From the preceding discussion, the most general choice of spinors for $\n=2$ backgrounds, provided one of them represents the orbit \orbsuf, is
\bea
\eta_1 &=& a \eta^{SU(5)}
\\ \label{11d:generic1}
\eta_2 &=& b_1 1 + c_1 e_{2345} + c_2 e_{1245} + c_3 e_{1234} + b_2 (\l_1 e_{12} + \l_2 e_{34}) + {\rm m.c}
\;,
\eea
where the parameters are as in (\ref{11d:generic}), and in addition $b_1\in \bC$ and $b_2=b$.

For the generic case the maximal common stability subgroup of $\eta_1$ and $\eta_2$ is $\{\id\}$.
 However the table below provides many special cases in which the stability subgroup is non-trivial.
\begin{equation}
\begin{array}{|c|c|}\hline
\hbox{Component of $\eta_2$ in $\D^+$} & \hbox{Stability subgroup}  \\  \hline
b_1         &  SU(5)        \\
b_1 1 + c_1 e_{2345}        &  SU(4)        \\
b_1 1 + b_2  e_{12}     &  SU(2)\times SU(3)    \\
b_1 1 + b_2 (\l_1 e_{12} + \l_2 e_{34})         &  SU(2)\times SU(2)    \\
b_1 1 + b_2 (e_{12} + e_{34})   &  Sp(2)        \\
b_1 1 + c_3 e_{1234} + b_2 (\l_1 e_{12} + \l_2 e_{34})      &  SU(2)\times SU(2)    \\
b_1 1 + c_3 e_{1234} + b_2 ( e_{12} +  e_{34})      &  Sp(2)        \\
b_1 1 + c_3 e_{1234} + b_2  e_{12}      &  SU(2)\times SU(2)    \\
b_1 1 + c_1 e_{2345} + b_2  e_{12}      &  SU(3)        \\
b_1 1 + c_1 e_{2345} + c_3 e_{1234} + b_2 (\l_1 e_{12} + \l_2 e_{34})   &  SU(2)        \\
b_1 1 + c_1 e_{2345} + c_2 e_{1245} + b_2 (\l_1 e_{12} + \l_2 e_{34})       &  \{\id\}      \\
\hbox{Most general case, equation (\ref{11d:generic1})} & \{\id\}\\ \hline
\end{array}
\end{equation}
In the table, only the part of $\eta_2$ lying in $\D^+$ has been presented, and as usual we must apply the Majorana condition to find the corresponding piece in $\D^-$.

Note that in some cases, further simplification of $\eta_2$ may occur.
 For example, there may be some scenarios analogous to the $SU(4)$-invariant case, in which we were able to exclude the presence of either $\etasufr$ or $\thsufr$ in $\eta_2$.
\subsection{Spacetime Forms Associated to the \sufr-\inv\ Spinors} \label{11d:n=2forms}
Shortly, we will solve the \kses\ for two spinors with stability subgroups \suf\ and \sufr.
 However, before we do so, we present the \st\ forms which are associated to them.
 These forms will prove useful when we come to describe the \st\ geometry of the solutions.

Since we wish to investigate \sufr-\inv\ spinors, we take
\be
\eta_1=a\eta^{SU(5)} \qquad \hbox{and} \qquad \eta_2=b1+\chi +\text{m.c.} \; ,
\ee
where $\chi\in\L^4_{\mathbf{\bar 5}(\bC^5)}$, so that $\eta_2\in$\orbsufr\ (see last section).
 Also, the inner product is linear, therefore to find the \st\ forms associated to $\eta_1$ and $\eta_2$ it is sufficient to compute the forms for the following linearly independent \sufr-\inv\ spinors:
\bea
&& \etasuf = \frac{1}{\sqrt2} \left( 1 + e_{12345} \right) \;, \qquad
\thsuf  = \frac{i}{\sqrt2} \left( 1 - e_{12345} \right) \;, \\
&& \etasufr = \frac{1}{\sqrt2} \left( e_5 + e_{1234} \right) \;, \qquad
\thsufr = \frac{i}{\sqrt2} \left( e_5 - e_{1234} \right) \;.
\eea
The forms associated to $\eta_1$ and $\eta_2$ can be found as linear combinations of those associated to the above spinors.

Following some straightforward computation using the theory from section \ref{11d:forms}, the forms associated to these spinors are given in the table below.
\begin{center}
\begin{tabular}{|c||c|}\hline
Spinors & Associated Spacetime Forms \\ \hline\hline
$\etasuf$, $\etasuf$    & $\k=-e^0$ \\
\                       & $\o=-e^1\we e^6 - e^2\we e^7 - e^3\we e^8 - e^4\we e^9 - e^5\we e^{\nat}$ \\
\                       & $\t={\rm Im}[(e^1+i e^6)\we\cdots\we(e^5+ie^\nat)]+\frac{1}{2}e^0\we\o\we\o$ \\
\hline
$\thsuf$, $\thsuf$      & $\k=- e^0$ \\
\                       & $\o=-e^1\we e^{6}-e^2\we e^{7}-e^3\we e^{8}-e^4\we e^{9}-e^5\we e^{\nat}$ \\
\                       & $\t=-{\rm Im}[(e^1+i e^6)\we\cdots\we(e^5+ie^\nat)]
                                +\frac{1}{2}e^{0}\we \o\we \o$ \\ \hline
$\etasuf$, $\thsuf$     & $\a=-1$ \\
\                       & $\z=\frac{1}{2}\o^{SU(5)}\we\o^{SU(5)}$ \\
\                       & $\t={\rm Re}[(e^1+i e^6)\we\cdots\we(e^5+ie^\nat)]$ \\
\                       & where \; $\o^{SU(5)}=\o(\etasuf,\etasuf)=\o(\thsuf,\thsuf)$ \\ \hline
$\etasufr$, $\etasufr$  & $\k=-e^0$ \\
\                       & $\o=e^1\we e^6+e^2\we e^7+e^3\we e^8+e^4\we e^9-e^5\we e^\nat$ \\
\                       & $\t={\rm Im}[(e^1+i e^6)\we\ldots\we(e^5+ie^\nat)]
                                +\frac{1}{2}e^{0}\we \o\we \o$ \\ \hline
$\etasuf$, $\etasufr$   & $\k=e^\nat$ \\
\                       & $\o=-e^0\we e^5~$ \\
\                       & $\xi=-\o^{SU(4)}\wedge e^5~$ \\
\                       & $\z={\rm Im}[(e^1+ie^6)\we\cdots\we(e^4+ie^9)]-e^0\we \o^{SU(4)}\we e^\nat$ \\
\                       & $\t=-e^0\we {\rm Re} [(e^1+ie^6)\we\cdots\we (e^4+ie^9)]
                                -{1\over2}\o^{SU(4)}\we\o^{SU(4)}\we e^\nat $ \\
\                       & where \; $\o^{SU(4)}=e^1\we e^{6}+e^2\we e^{7}+e^3\we e^{8}+e^4\we e^{9}$ \\ \hline
$\thsuf$, $\thsufr$     & $\k= e^5$ \\
\                       & $\o=e^0\we e^\nat$ \\
\                       & $\xi=\o^{SU(4)}\we e^\nat$ \\
\                       & $\z={\rm Re}[(e^1+i e^6)\we\cdots\we (e^4+i e^9)]-e^0 \we \o^{SU(4)}\we e^5$ \\
\                       & $\t=e^0\we {\rm Im} [(e^1+ie^6)\we\dots\we (e^4+ie^9)]
                                -{1\over2}\o^{SU(4)}\we\o^{SU(4)}\we e^5 $ \\ \hline
\end{tabular}
\end{center}

In particular, the table shows that
\beann
\k(\etasuf,\etasuf)&=&\k(\thsuf,\thsuf)=\k^{SU(5)} \quad\hbox{and} \\
\o(\etasuf,\etasuf)&=&\o(\thsuf,\thsuf)=\o^{SU(5)} \; ,
\eeann
whereas $\,\t(\etasuf,\etasuf)=\t^{SU(5)}\,$ is linearly independent from $\,\tau(\thsuf,\thsuf)$. Also, note that the inner product of $\etasuf$ and $\thsuf$, given by the zero-form $\a$, is non-degenerate.

In the $\n=1$ case, we know that the single spinor is associated to either a timelike or null vector. However, the forms $\k(\etasuf,\etasufr)$ and $\k(\thsuf,\etasufr)$ show that in the $\n\geq2$ case, the spinors can give rise to spacelike vectors as well.

As a final observation, we record the algebraic identity
\be \label{11d:algid1}
\o^{SU(4)} \,=\, \k(\etasuf,\etasufr) \,\lc\; \xi(\thsuf,\thsufr) \,=\, - \k(\thsuf,\thsufr) \,\lc\; \xi(\etasuf,\etasufr) \;.
\ee
This will prove useful in examining the geometry of $D=11$ \sg\ backgrounds with more than two \sufr-\inv\ \kss.
\subsection{Two \suf-\inv\ Killing Spinors} \label{11d:2su5}
\subsubsection{The Killing Spinor Equations}
We now turn to the solution of the \kses\ for the most general \suf-\inv\ spinors (see equations (\ref{11d:1stsuf}) and (\ref{11d:2ndsuf})),
\bea \label{11d:n=2gensufr1}
&&\boxed{\eta_1 = f(x) (1+e_{12345}) }\\
&&\boxed{\eta_2 = g_1(x) (1+e_{12345}) + i g_2(x) (1-e_{12345}) \label{11d:n=2gensufr2} } \;,
\eea
where the parameters $f, g_1$ and $g_2$ in the spinors are now treated as real functions of the spacetime.
 We assume that $g_2\not=0$ because otherwise $\eta_2$ would be linearly dependent on $\eta_1$.

Denoting the supercovariant derivative by $\mathcal{D}$, the requirement for an $\n=2$ background to be supersymmetric is that $\,\mathcal{D}_M\eta_1=0=\mathcal{D}_M\eta_2$. The \kses\ were solved for the first spinor in section \ref{11d:n=1kses}. Observe that
\benn
0=\mathcal{D}_M\eta_1=\mathcal{D}_M[f(1+e_{12345})] = (\partial_Mf)(1+e_{12345})+f\mathcal{D}_M(1+e_{12345}) \; ,
\eenn
and therefore
\be
\mathcal{D}_M(1+e_{12345})=-\left(\partial_M\log f\right)(1+e_{12345}) \; .
\ee
This expression helps us to rewrite the \kse\ for $\eta_2$ in a more convenient way.
 We have
\beann
0 &=& \mathcal{D}_M[g_1(1+e_{12345}) + ig_2(1-e_{12345})]  \\
  &=& \!\!\!(\partial_M g_1)(1+e_{12345}) \!+ i (\partial_M g_2)(1-e_{12345})
        \!+ g_1\mathcal{D}_M(1+e_{12345}) \!+ ig_2\mathcal{D}_M(1-e_{12345}) \;.
\eeann
Multiplying this equation by $g_2^{-1}$, the second \kse\ can now be expressed as
\bea\nn
0 &=& g_2^{-1}\left[ \partial_M(g_1+ig_2)-g_1\partial_M\log f \right]\,1
+ g_2^{-1} \left[ \partial_M(g_1-ig_2)-g_1\partial_M\log f \right]\,e_{12345} \\
&& + i\mathcal{D}_M\left(1-e_{12345}\right) \label{11d:n=2kse} \; .
\eea
To analyse this equation, we use a similar procedure to the $\n=1$ case. First, we separate the free index into time and spatial components, $M=(0,I)$. Next, we commute all $\G_0$ matrices through the supercovariant connection so that they hit the spinors $1$ and $e_{12345}$. Using $\G_0(1)=i$ and $\G_0(e_{12345})=-ie_{12345}$, we can remove $\G_0$ from the equations, so that they are expressed in terms of the ten-dimensional spatial gamma matrices.
 Then, we expand each gamma matrix term with respect to the Hermitian basis described in section \ref{11d:hermbasis}, simplify the products of gamma matrices and collect terms together.
 We are left with a basis expansion equated with zero, as in equation (\ref{11d:hermexp}).
 Therefore, the coefficient of each Hermitian basis element must vanish independently, from which we obtain the constraints on the field strength, the connection and the functions $f,g_1,g_2$.

The constraints coming from the equation for $\eta_2$ are very
similar to those from the equations for $\eta_1$, with differences
arising due to the factor of $i$ and the sign difference for
$e_{12345}$ in the last term of (\ref{11d:n=2kse}). We proceed to
analyse the constraints from both sets of equations together.

First, we find the following independent constraints which arise
from the time-components of the \kses\ for $\eta_1$ and $\eta_2$:
\bea
0 &=& \partial_0 f = \partial_0 (g_1+i g_2) \label{11d:n=2t1} \\
0 &=& \O_{0,\a\bar\b} g^{\a\bar\b}-{i\over12} F_\a{}^\a{}_\b{}^\b \label{11d:n=2t2} \\
0 &=& G_{\a_1\a_2\a_3} = F_{\a_1\a_2\a_3\a_4} \label{11d:n=2t3} \\
0 &=& i\O_{0,0\ab}+{1\over 3} G_{\ab \b}{}^\b \label{11d:n=2t4} \\
0 &=& \O_{0,\ab\bar\b}-{i\over6} F_{\ab\bar\b\g}{}^\g=0 \label{11d:n=2t5}
\eea

At this stage, we can make a simplification regarding the \st\ functions $f,g_1,g_2$. Comparing the spatial components of the \kses\ for $\eta_1$ and $\eta_2$, we see that
\be
\partial_{\ab} \log \left(\frac{g_1}{f}\right) = \partial_{\ab} \log\left(\frac{g_2}{f}\right) = 0 \; .
\ee
In particular, this and its complex conjugate implies
\be
g_1 = c_1(t)f  \quad\hbox{and}\quad  g_2 = c_2(t)f \;,
\ee
where $c_1,c_2$ are arbitrary real functions of the time coordinate.
 However, differentiating this with respect to time and using (\ref{11d:n=2t1}), it is clear that $c_1$ and $c_2$ must in fact be real constants.
 Therefore $g_1=c_1f$ and $g_2=c_2f$.

Furthermore, \kss\ are determined only up to a constant scale, so we may scale $\eta_2$ arbitrarily while retaining the same solution to the \kses.
 So without loss of generality we may normalise with respect to the Hermitian inner product so that $c_1^2+c_2^2=1$, which enables us to introduce an angle $\vp\in[0,2\pi)$ such that $g_1=\cos\vp\,f$ and $g_2=\sin\vp\,f$.

Hence we can write the second \ks\ as
\be
\eta_2 = f \left[ \cos\vp (1+e_{12345}) + i \sin\vp (1-e_{12345}) \right] \; ,
\ee
where $\vp\neq0,\pi$, so that $\eta_1$ and $\eta_2$ remain linearly independent.
 Other than these disallowed values, $\vp$ is undetermined by the \kses.

Thus both \kss\ depend on the same time-independent \st\ function, $f$. Using this fact, the conditions arising from the spatial components of the \kses\ are:
\bea
0 &=& \partial_{\ab} \log f+{1\over2} \O_{\ab, \b\bar\g} g^{\b\bar\g}+{i\over12} G_{\ab\g}{}^\g \label{11d:n=2s1} \\
0 &=& \partial_{\ab} \log f-{1\over2} \O_{\ab, \b\bar\g} g^{\b\bar\g}+{i\over4} G_{\ab\g}{}^\g \label{11d:n=2s2}  \\
0 &=& i \O_{\ab,0\bar\b}+{1\over6} F_{\ab\bar\b\g}{}^\g  \label{11d:n=2s3}  \\
0 &=& i\O_{\ab,0\b}+{1\over12} g_{\ab\b} F_\g{}^\g{}_\d{}^\d+{1\over2} F_{\ab\b\g}{}^\g  \label{11d:n=2s4} \\
0 &=& \O_{\ab, \bar\b\bar\g}  \label{11d:n=2s5}  \\
0 &=& \e_{\ab\bar\b\bar\g}{}^{\g_1\g_2} F_{\g_1\g_2\d}{}^\d + F_{\ab\g_1\g_2\g_3}
\e^{\g_1\g_2\g_3}{}_{\bar\b\bar\g}  \label{11d:n=2s6}  \\
0 &=& \O_{\ab,\b\g}-{i\over 2} G_{\ab\b\g}-{i\over3} g_{\ab[\b} G_{\g] \d}{}^\d  \label{11d:n=2s7}
\eea
As in the $\n=1$ case, these equations can be solved to determine components of the field strength in terms of the connection and to reveal some important information about the geometry of the spacetime.
\subsubsection{Solving the Killing Spinor Equations}
An immediate consequence for the field strength, arising from (\ref{11d:n=2t3}), is that the components $F^{(4,0)+(0,4)}$ and $G^{(3,0)+(0,3)}$ vanish. Also, (\ref{11d:n=2t2}), (\ref{11d:n=2t4}) and (\ref{11d:n=2t5}) imply respectively that
\be
F_{\a}{}^{\a}{}_{\b}{}^{\b}=-12i\O_{0,\b}{}^{\b} \; , \quad G_{\ab\b}{}^{\b}=-3i\O_{0,0\ab} \; , \quad
F_{\ab\betab\g}{}^{\g}=-6i\O_{0,\ab\betab} \; . \label{11d:n=2FGF}
\ee
Next, subtracting (\ref{11d:n=2s2}) from (\ref{11d:n=2s1}) gives
\be
G_{\ab\b}{}^{\b}=-6i\O_{\ab,\b}{}^{\b} \;,
\ee
which in conjunction with the above expression for $G_{\ab\b}{}^{\b}$, tells us that
\be
2\O_{\ab,\b}{}^{\b}-\O_{0,0\ab}=0 \;. \label{11d:n=2conn1}
\ee
Hence (\ref{11d:n=2s1}) now implies that
\be \label{11d:s12}
\O_{\ab,\b}{}^{\b}=\frac{1}{2}\O_{0,0\ab}=-\partial_{\ab}\log f \;.
\ee
Next, (\ref{11d:n=2s3}) gives
\be
F_{\ab\betab\g}{}^{\g}=-6i\O_{\ab,0\betab} \;, \label{11d:n=2F1}
\ee
which, together with (\ref{11d:n=2FGF}) provides the geometric constraint
\be
\O_{\ab,0\betab}=\O_{0,\ab\betab} \;. \label{11d:n=2conn2}
\ee
Also, (\ref{11d:n=2s4}) and its complex conjugate imply that
\be
\O_{\ab,0\b}=-\O_{\b,0\ab} \;. \label{11d:n=2conn3}
\ee
Shortly, we will show that this condition as well as (\ref{11d:n=2conn1}) and (\ref{11d:n=2conn2}) are satisfied in an appropriate choice of frame.

Now, tracing (\ref{11d:n=2s4}) gives
\be
F_{\a}{}^{\a}{}_{\b}{}^{\b}=12i\O^{\b}{}_{,0\b} \;,
\ee
which is compatible with the similar expression from (\ref{11d:n=2FGF}) in a suitable frame. Equation (\ref{11d:n=2s4}) now implies that
\be
F_{\b\ab\g}{}^{\g} = 2i\O_{\ab,0\b} - 2i\O_{0,\g}{}^{\g}g_{\ab\b} \; .
\ee
Next, using (\ref{11d:n=2F1}), the solution to (\ref{11d:n=2s6}) is given by
\be
F_{\ab\b_1\b_2\b_3} = 6i\O_{0,[\b_1\b_2}g_{\b_3]\ab} \;.
\ee
The final equation to be investigated is (\ref{11d:n=2s7}). Tracing the $(\ab,\b)$ indices and using our expression for $G_{\ab\b}{}^{\b}$, we find that
\be \label{11d:s62}
\O^{\b}{}_{,\b\g} + \O_{\g,\b}{}^{\b} = 0 \; .
\ee
Substituting this back into (\ref{11d:n=2s7}) gives
\be
G_{\ab\b\g} = -2i\O_{\ab,\b\g} + 2i\O_{0,0[\g}g_{\b]\ab} \; .
\ee
This concludes the solution of the \kses\ for two \suf-\inv\ spinors, and it remains to summarise the results and interpret them geometrically.
\subsubsection{Summary of Results for Two \suf-\inv\ Killing Spinors}
To summarise in a clear way, we present the solutions in the following table:
\begin{center}
\begin{tabular}{|c|c|} \hline
Component       & Solution      \\ \hline\hline
$G^{3,0}$       & $G_{\a_1\a_2\a_3}=0$           \\ \hline
$G^{2,1}$       & $G_{\ab\b_1\b_2}=-2i\O_{\ab,\b_1\b_2} - 2i\O_{0,0[\b_1}g_{\b_2]\ab}$    \\ \hline
$F^{4,0}$       & $F_{\b_1\b_2\b_3\b_4}=0$  \\ \hline
$F^{3,1}$       & $F_{\ab\b_1\b_2\b_3} =6i\O_{0,[\b_1\b_2}g_{\b_3]\ab}$  \\ \hline
trace$(F^{2,2})$& $F_{\b}{}^\b{}_\g{}^\g = 12 i \O_{\ab,0\b} g^{\ab\b}$                         \\ \hline
$F^{2,2}_0$       & Undetermined by \kses         \\ \hline
\end{tabular}
\end{center}

The components $G^{(2,1)+(1,2)}$, $F^{(3,1)+(1,3)}$ and the trace of the $(2,2)$-component are thus determined by the geometry. The traceless $(2,2)$ piece, $F^{2,2}_0$, is undetermined by the \kses, as in the $\n=1$ case.

Also, we have the following constraints on the connection and \st\ function $f=g_1=g_2$:
\begin{center}
\begin{tabular}{|c|ccc} \hline
\multicolumn{4}{|c|}{$\O_{\ab,0\betab}=\O_{0,\ab\betab}$ \vline~
$\O_{\ab,0\b}=-\O_{\b,0\ab}$  \vline~
$\O_{\ab,\b}{}^{\b}=\frac{1}{2}\O_{0,0\ab}=-\partial_{\ab}\log f$ \vline~
$\O_{\betab,}{}^{\beta}{}_{\g} + \O_{\g,\b}{}^{\b} = 0$ }   \\
\hline
\multicolumn{1}{|c|}{$\O_{\ab,\betab\gb}=0$} &  &  & \\
\cline{1-1}
\end{tabular}
\end{center}
%
%
\subsubsection{The Geometry of the Spacetime}
We now proceed to extract important geometric information about the spacetime from the constraints found above.

As in the $\n=1$ case, $\eta_1$ gives rise to the vector $\k^f=-f^2 e^0$.
 Note also that the vector associated to $\eta_2$ is proportional to $\k^f$, since $g_1=c_1f$ and $g_2=c_2f$, where $c_1,c_2\in\bR$.
 So there is essentially one independent vector associated to these spinors.
 We can see that $\k^f$ is in fact a Killing vector, by an analogous calculation to that of section \ref{11d:n=1geom}, so we will not repeat it here.
 This allows us to choose adapted coordinates in which to write the spacetime metric as
\be
ds^2 = - f^4 (dt+\a)^2+ ds_{10}^2 \;.
\ee
As in the $\n=1$ case, we can now choose the frame in which $e^0=f^2(dt+\a)$ and $\{e^I\}_{I=1}^{\nat}$ is an orthonormal frame on the ten-dimensional manifold $B$, which is transverse to the orbits of $\k^f$.
 Then, the metric is of the form
\be \label{11d:n=2frame}
ds^2 = -(e^0)^2 + \sum^{\nat}_{I=1} (e^I)^2 \; .
\ee
The torsion-free condition of the Levi-Civita connection in this frame implies that $\O_{I,0J}=\O_{0,IJ}\,$ as before, which ensures that the geometric conditions (\ref{11d:n=2conn2}) and (\ref{11d:n=2conn3}) are satisfied.

The remaining geometric conditions are (\ref{11d:n=2conn1}), (\ref{11d:s12}), (\ref{11d:s62}) and (\ref{11d:n=2s5}).
 The latter condition implies that the almost Hermitian manifold $B$ is in fact complex.
 This can be seen using the torsion free condition as follows.
 In particular, equation (\ref{11d:n=2s5}) implies that
\be
d\hat{e}^{\ab} = -\O_\b{}^{\ab}{}_{\bar\g} e^\b\wedge e^{\bar\g} - \O_{\betab}{}^{\ab}{}_\g  e^{\betab}\wedge e^\g
- \O_{\betab}{}^{\ab}{}_{\bar \g} e^{\bar\b}\wedge e^{\bar\g} \;,
\ee
and
\be
d\hat{e}^\a = -\O_\b{}^{\a}{}_{\g} e^\b\we e^{\g} - \O_{\betab}{}^{\a}{}_\g  e^{\betab}\we e^\g
- \O_{\b}{}^{\a}{}_{\gb} e^{\b}\we e^{\gb} \;.
\ee
We see that that the $(2,0)$ part of $d\hat{e}^{\ab}$ and the $(0,2)$ part of $d\hat{e}^{\a}$ vanishes.
 This implies that the almost complex structure is integrable (see, for example, \cite{kob2}), so that $B$ is complex and therefore a \textit{Hermitian} manifold.

Alternatively, in terms of the classification of almost Hermitian
manifolds\footnote{See Appendix \ref{GrayHerv} for details.},
condition (\ref{11d:n=2s5}) implies that
\be
\ww_1=\ww_2=0 \;,
\ee
which again implies that $B$ is Hermitian \cite{grayherv}.

Equation (\ref{11d:n=2conn1}) provides a relationship between intrinsic torsion modules of the $\su(5)$ and $\so(10)$ structures on the $\so(1,10)$ manifold $M$,
\be \label{11d:n=2so10}
(\rm y_4)_{\ab} = 2 (\rm w_5)_{\ab} \;.
\ee

It remains to interpret conditions (\ref{11d:s12}) and (\ref{11d:s62}).
 In terms of the intrinsic torsion, (\ref{11d:s62}) implies that
\be
(\rm w_4)_\g = (\rm w_5)_\g \;,
\ee
which explains the difference between equations (\ref{11d:n=2so10}) and (\ref{11d:n=1so10}).
 Furthermore, using (\ref{11d:W5}), we see that in this case
\be
(W_5)_\g = (\rm w_5)_\g \;.
\ee

Now, we can express (\ref{11d:s12}) as
\be
W_5 = - df \;,
\ee
which is identical to the condition on the geometry which arises for $\n=1$ backgrounds.
 In other words, $W_5$ is exact.

Let us now summarise these important consequences for $D=11$ \sg\ backgrounds possessing two \suf-\inv\ \kss:
\begin{itemize}
\item All components of $F$ are determined by the spin-connection, as in the table, except for the traceless $(2,2)$ piece, $F^{2,2}_0$.
 Again, this corresponds to the $\mathbf{75}$ irrep of \suf.
 Furthermore, $G^{3,0}$ and $F^{4,0}$ vanish.
\item There is a timelike Killing vector field associated to the spacetime, according to which we may choose an adapted frame to express the spacetime metric as in (\ref{11d:n=2frame}).
\item The ten-dimensional spatial manifold $B$ transverse to the orbits of the Killing vector is complex, hence Hermitian.
\item There is an \suf-structure on $B$, specified explicitly by the Levi-Civita connection satisfying the constraints imposed by the \kses, as detailed in the preceding pages. Equivalently, it may be specified by the \suf-\inv\ tensors associated to $\eta_1$ and $\eta_2$.
 Thus $B$ is in fact a special Hermitian manifold
\item In terms of Gray-Hervella classes by which we may classify the manifold, we have $\rm w_1=\rm w_2=0$ and $\rm w_4 = \rm w_5$$=W_5=-df$.
\end{itemize}
\subsection{Two \sufr-\inv\ Killing Spinors} \label{11d:2su4}
\subsubsection{The Killing Spinor Equations}
We now consider a background in which we have two \sufr-\inv\ spinors.
 Take $\eta_1$ as before, and let $\eta_2$ be given as in (\ref{11d:gensufr}).
 The most general \sufr-\inv\ spinors are then represented by
\bea
\eta_1 &=& f(x) (1+e_{12345}) \\
\eta_2 &=& g_1(x) (1+e_{12345})+ i g_2(x) (1-e_{12345}) + \sqrt2g_3(e_5+e_{1234}) \;,
\eea
where $f, g_1, g_2$ and $g_3$ are real functions of the spacetime, which will become constrained by the \kses.
 In this section, we will investigate a special case in which $g_1=g_2=0$ and $g=g_3\neq0$, so that the spinors are of the form
\bea
&&\boxed{ \eta_1 = f(x) (1+e_{12345}) } \\
&&\boxed{ \eta_2 = \sqrt2 g(x) (e_5+e_{1234}) } \;.
\eea
The most general case has been analysed in \cite{IIB}, in the context of supersymmetric IIB backgrounds.

Again, the \kse\ for $\eta_1$ is as in the $\n=1$ case.
 For the second spinor, multiplying the equation by $(\sqrt2 g)^{-1}$, we obtain the following:
\be
0 = \left(\partial_M\log g\right) (e_5 + e_{1234}) + \mathcal{D}_M(e_5 + e_{1234}) \;.
\ee
To solve this equation, we repeat the usual process.
 That is, we commute each $\G_0$ through to hit the spinors, and using $\G_0(e_5)=-ie_5$ and $\G_0(e_{1234})=ie_{1234}$, we write the equations in terms of ten-dimensional gamma matrices.
 We then expand it in the Hermitian basis described in section \ref{11d:hermbasis}.
 However, in this case, because the \sufr\ representative $\eta_2$ has been chosen such that the $e_5$ spatial direction is singled out, it is convenient to split the holomorphic spatial index as $\a=(\a,5)$, and from now on, indices $\a,\b,\g,\d,\cdots=1,2,3,4$.
 This has the effect of decomposing the connection and field strength into representations of \sufr.

Having done this, again we collect terms together to express the \kses\ as an Hermitian basis expansion of the form
\bea
0 &=& C_0(1) \,+\, (C_1)_{\ab}\Gh^{\ab}(1) \,+\, (C_2)_{\bar5}\Gh^{\bar5}(1)
\,+\, (C_3)_{\ab_1\ab_2}\Gh^{\ab_1\ab_2}(1) \,+\, (C_4)_{\ab\bar5}\Gh^{\ab\bar5}(1) \nn \\
&+& (C_5)_{\ab_1\ab_2\ab_3}\Gh^{\ab_1\ab_2\ab_3}(1) \,+\, (C_6)_{\ab_1\ab_2\bar5}\Gh^{\ab_1\ab_2\bar5}(1)
\,+\, (C_7)_{\ab_1\ab_2\ab_3\ab_4}\Gh^{\ab_1\ab_2\ab_3\ab_4}(1) \nn \\
&+& (C_8)_{\ab_1\ab_2\ab_3\bar5}\Gh^{\ab_1\ab_2\ab_3\bar5}(1)
\,+\, (C_9)\Gh^{\bar1\bar2\bar3\bar4\bar5}(1) \;.
\eea
Thus the constraints on the field strength, connection and functions $f$, $g$ can be read off from the vanishing of the basis element coefficients in this expansion.

We begin by considering the time component of the \kses, \linebreak $\mathcal{D}_0\eta_2=0$.
 Putting $\,\e_{\ab_1\ab_2\ab_3\ab_4}=\e_{\ab_1\ab_2\ab_3\ab_4\bar5}\,$, the following independent conditions arise:
\bea
0 &=& -i\O_{0,05}+{1\over 3} G_{5\a}{}^\a -{i\over 72} F_{\a_1\a_2\a_3\a_4} \e^{\a_1\a_2\a_3\a_4}
\label{11d:n=22t1} \\
0 &=& \O_{0,\betab 5}+{i\over 6} F_{\betab 5 \g}{}^\g -{1\over18} G_{\g_1\g_2\g_3} \e^{\g_1\g_2\g_3}{}_{\betab}
\label{11d:n=22t2} \\
0 &=& \partial_0\log g + {1\over 2} \O_{0,\b}{}^\b-{1\over 2} \O_{0,5\bar5}
+{i\over 24} F_{\a}{}^\a{}_\b{}^\b -{i\over 12} F_\a{}^\a{}_{5\bar 5}  \label{11d:n=22t3}  \\
0 &=& {1\over6} G_{\betab_1\betab_2 5}+ [-{1\over8} \O_{0,\g_1\g_2} -{i\over 48} F_{\g_1\g_2\d}{}^\d
+{i\over 48} F_{\g_1\g_2 5\bar5}] \e^{\g_1\g_2}{}_{\betab_1\betab_2}  \label{11d:n=22t4}  \\
0 &=& {i\over36} F_{\g_1\g_2\g_3 \bar 5} \e^{\g_1\g_2\g_3}{}_{\betab} - {i\over2}
\O_{0,0\betab}+{1\over6} G_{\betab\g}{}^\g-{1\over6} G_{\betab5\bar5}   \label{11d:n=22t5}
\eea

Having split the spatial direction into $\a=(\a,5)$, $\a=1,2,3,4$ as described above, the constraints arising from the component of the \kse\ for $\eta_2$ with derivative pointing along the spatial $\ab$ directions, $\mathcal{D}_{\ab}\eta_2=0$, are:
\bea
0 &=& -i\O_{\ab,05}+\frac{1}{6}F_{\ab5\g}{}^\g-\frac{i}{18}g_{\ab\g_1}G_{\g_2\g_3\g_4}\e^{\g_1\g_2\g_3\g_4}
\label{11d:n=22a1}  \\
0 &=& \O_{\ab,\bar\b 5}-\frac{i}{6}G_{\ab\bar\b 5}
- \!\frac{1}{12}\left( F_{\ab\g_1\g_2\g_3}
+ g_{\ab\g_1} ( F_{\g_2\g_3\d}{}^\d  - F_{\g_2\g_3 5\bar 5} ) \right)
\e^{\g_1\g_2\g_3}{}_{\bar\b}
\label{11d:n=22a2}  \\
0 &=& \partial_{\ab} \log g + \frac{1}{2}\O_{\ab,\g}{}^\g-\frac{1}{2}\O_{\ab,5\bar 5} \nn \\
   && \qquad -\frac{i}{12}G_{\ab\g}{}^\g+\frac{i}{12}G_{\ab 5\bar 5}
-\frac{1}{18}g_{\ab\g_1}F_{\g_2\g_3\g_4\bar 5}\e^{\g_1\g_2\g_3\g_4}  \label{11d:n=22a3} \\
0 &=& \!\!F_{\ab\bar\b_1\bar\b_25}
- \!\!\left( \frac{3}{2}\O_{\ab,\g_1\g_2} + \frac{3i}{4}G_{\ab\g_1\g_2}
+ \frac{i}{2}g_{\ab\g_1} (G_{\g_2\d}{}^\d  - G_{\g_2 5\bar 5} ) \right)\!
\e^{\g_1\g_2}{}_{\bar\b_1\bar\b_2}  \label{11d:n=22a4} \\
0 &=& - i \O_{\ab,0\bar\b}  + \frac{1}{6}F_{\ab\bar\b\g}{}^\g - \frac{1}{6}F_{\ab\bar\b 5\bar 5}
+ \frac{i}{6}\e_{\ab\bar\b}{}^{\g_1\g_2}G_{\g_1\g_2\bar 5} \label{11d:n=22a5} \\
0 &=& \!\!\left( i\O_{\ab,0\g} - \frac{1}{2}F_{\ab\g\d}{}^\d + \frac{1}{2}F_{\ab\g 5\bar 5}
+ \frac{1}{12}g_{\ab\g}F_\d{}^\d{}_\s{}^\s + \frac{1}{6}g_{\ab\g}F_{5\bar5\d}{}^\d \right)
\e^{\g}{}_{\bar\b_1\bar\b_2\bar\b_3}  \label{11d:n=22a6}  \\
0 &=& \O_{\ab,\bar\b_1\bar\b_2} - \frac{i}{6}G_{\ab\bar\b_1\bar\b_2}
-  \left( \frac{1}{4}F_{\ab \bar 5\g_1\g_2} + \frac{1}{6}g_{\ab\g_1}F_{\g_2\bar 5\d}{}^\d \right)
\e^{\g_1\g_2}{}_{\bar\b_1\bar\b_2} \label{11d:n=22a7}  \\
0 &=& \partial_{\ab} \log g-\frac{1}{2}\O_{\ab,\g}{}^\g+\frac{1}{2}\O_{\ab,5\bar 5}
    -\frac{i}{4}G_{\ab\g}{}^\g+\frac{i}{4}G_{\ab 5\bar 5}  \label{11d:n=22a8}  \\
0 &=& \frac{1}{3}F_{\ab \bar\b_1\bar\b_2\bar\b_3}
+ \left( \O_{\ab,\g\bar 5} + \frac{i}{2}G_{\ab\g\bar 5} +\frac{i}{6}g_{\ab\g}G_{\bar 5\d}{}^\d \right)
\e^\g{}_{\bar\b_1\bar\b_2\bar\b_3}
\label{11d:n=22a9} \\
0 &=& i\O_{\ab,0\bar 5} - \frac{1}{2}F_{\ab\bar 5\g}{}^\g   \label{11d:n=22a10}
\eea

Finally, the constraints which arise from the \kse\ for $\eta_2$ with derivatives along the spatial $\bar5$ direction, $\mathcal{D}_{\bar5}\eta_2=0$, are the following:
\bea
0 &=& i\O_{\bar5,05}-\frac{1}{12}F_\g{}^\g{}_\d{}^\d+\frac{1}{3}F_{5\bar 5\g}{}^\g  \label{11d:n=2251} \\
0 &=& \O_{\bar 5,\bar \b 5}-\frac{i}{6}G_{\bar \b\g}{}^\g-\frac{i}{3}G_{\bar \b 5\bar 5}
-{1\over36} F_{\bar5\g_1\g_2\g_3} \e^{\g_1\g_2\g_3}{}_{\bar\b}  \label{11d:n=2252} \\
0 &=& \partial_{\bar 5} \log g+\frac{1}{2}\O_{\bar 5,\g}{}^\g -\frac{1}{2}\O_{\bar 5,5\bar 5}
-\frac{i}{4}G_{\bar 5\g}{}^\g  \label{11d:n=2253} \\
0 &=& \frac{1}{2}F_{\bar\b_1\bar\b_2\g}{}^\g + F_{\bar\b_1\bar\b_2 5\bar 5}
+ \left( \frac{3}{4}\O_{\bar 5,\g_1\g_2} + \frac{i}{8}G_{\bar 5\g_1\g_2} \right)
\e^{\g_1\g_2}{}_{\bar \b_1\bar \b_2}  \label{11d:n=2254} \\
0 &=& -i\O_{\bar 5,0\bar\b} - \frac{1}{2}F_{\bar\b\bar 5\g}{}^\g  \label{11d:n=2255} \\
0 &=& -\frac{2i}{3}G_{\bar\b_1\bar\b_2\bar\b_3}
+ \left( i\O_{\bar 5,0\g} - \frac{1}{6}F_{\bar 5\g\d}{}^\d  \right)
\e^{\g}{}_{\bar \b_1\bar \b_2\bar \b_3}  \label{11d:n=2256} \\
0 &=& \O_{\bar 5,\bar \b_1\bar \b_2}-\frac{i}{2}G_{\bar 5\bar \b_1\bar \b_2}  \label{11d:n=2257} \\
0 &=& \!\!\!- \frac{2}{3}F_{\bar\b_1\bar\b_2\bar\b_3\bar\b_4}
+ \! \left( \partial_{\bar 5} \log g - \frac{1}{2}\O_{\bar 5,\g}{}^\g+\frac{1}{2}\O_{\bar 5,5\bar 5}
    -\frac{i}{12}G_{\bar 5\g}{}^\g \right)\e_{\bar \b_1\bar \b_2\bar \b_3\bar \b_4}  \label{11d:n=2258} \\
0 &=& -F_{\bar\b_1\bar\b_2\bar\b_3\bar 5}+\O_{\bar 5,\g\bar 5} \e^\g{}_{\bar\b_1\bar\b_2\bar\b_3}
\label{11d:n=2259}
\eea
We will now turn to analyse these constraints, and then seek to interpret them in terms of the geometry of the spacetime background.
\subsubsection{Solving the Killing Spinor Equations} \label{11d:2sufrsoltn}
Now that the spatial $\bar5$ index is split from the other spatial directions $\ab=\bar1,\bar2,\bar3,\bar4$, there is a greater number of equations to solve.
 However, we can make use of all the expressions for the field strength which arise from the equation for $\eta_1$ (see section \ref{11d:n=1summary}).
 Substituting these into equations (\ref{11d:n=22t1})-(\ref{11d:n=2259}), the constraints associated with the \kse\ for $\eta_2$ provide further conditions on the geometry of the \st.
 This calculation is fairly long and is detailed in Appendix \ref{app:sufr}.
 Here, we simply summarise the results:
\\ \ 
\\
- \underline{\textbf{Geometric constraints}:}
\\
The resulting conditions on the spacetime functions are
\be
g=f~,~~~\partial_0f=(\partial_{5}-\partial_{\bar 5})f=0~.
\label{summary00}
\ee
In fact $g$ is proportional to $f$, but since Killing spinors are determined up to a constant scale, we may equate them in this case.

The conditions on the $\O_{0,0i}$ components are
\bea
&& \O_{0,05}=\O_{0,0\bar 5}=-2\partial_5 \log f=-2 \partial_{\bar5} \log f~,~~~~ \nn \\
&& \O_{0,0\a}=-2\partial_{\a} \log f~.
\label{summary1}
\eea

Next, components of the form $\O_{0,ij}$ satisfy
\bea
&& \O_{0, 5\ab}=\O_{0, 5\a}= \O_{0, 5\bar5}=\O_{0,\b}{}^\b=0~,~~~~~
\nn \\
&& \O_{0, \b_1\b_2}={i\over4} (\O_{5,\bar \g_1\bar\g_2}-
\O_{\bar 5,\bar \g_1\bar\g_2})\e^{\bar\g_1\bar\g_2}{}_{\b_1\b_2} \;.
\label{summary2}
\eea
We find that the traceless part of $\O_{0, \a\bar\b}$ is not determined by the \kses.

The conditions which arise for components of the form $\O_{\ab, ij}$ are
\bea
&& \O_{[\bar\b_1,\bar\b_2\bar\b_3]}=0~,~~~\quad
\O_{\ab, \b_1\b_2}=-\O_{0,0[\b_1} g_{\b_2]\ab}~,~~~
\nn \\
&& \O_{ \b,\ab}{}^\b={3\over2} (\O_{\bar5, \ab \bar 5}-\O_{\bar5, \ab  5})
=-{3\over 2} \O_{0,0\ab}~,
\nn \\
&& \O_{\a,\b}{}^\b = -{1\over2}(\O_{\bar5, \a \bar 5}+\O_{\bar5, \a  5})
=-{1\over2} (\O_{0,0\a}+2\O_{5,\a5})~.
\label{summary3}
\eea

Also, we have
\bea
&&\O_{[\bar\b_1, \bar\b_2]\bar 5}=-\O_{\bar 5,\bar\b_1 \bar\b_2}~,  \qquad
\O_{[\bar\b_1, \bar\b_2] 5}=-\O_{5,\bar\b_1 \bar\b_2}~, \qquad
\O_{(\bar\b_1, \bar\b_2) 5}=\O_{(\bar\b_1, \bar\b_2) \bar 5}~,
\nn \\
&&\O_{(\ab, \b) \bar5}=\O_{(\ab, \b) 5}={1\over2} g_{\ab\b} \O_{0,0\bar 5}~,
\qquad  \O_{\ab, 5\bar5}=0~.
\label{summary4}
\eea

Finally, the conditions on the  $\O_{\bar5, ij}$ components are
\bea
&&\O_{5,\b}{}^\b=\O_{\bar5,\b}{}^\b~, \qquad
 \O_{ 5, \ab 5}=\O_{\bar 5, \ab \bar5}~, \qquad
\O_{ 5, \ab \bar5}=\O_{\bar 5, \ab 5}~,
\nn \\
&&\O_{\bar5, \ab \bar 5}-\O_{\bar5, \ab  5}= -\O_{0,0\ab}~, \qquad
\O_{\bar5, 5\bar 5}=-\O_{5,5\bar 5}=-\O_{0,0\bar 5}~.
\label{summary5}
\eea
The above equations (\ref{summary00})-(\ref{summary5}) along with their complex conjugates, give the full set of conditions on the connection that are required for a background to admit $\n=2$ supersymmetry with Killing spinors given by $\eta_1=f \eta^{SU(5)}$ and $\eta_2=g \eta^{SU(4)}$.
 These may be converted into information about the intrinsic torsion modules corresponding to $\su(4)$-structures on $\so(1,10)$ manifolds, using Appendix \ref{GrayHerv}.
 Note that the traceless part of $\O_{\a, \b\bar\g}$ is not determined by the Killing spinor equations.
\\ \ 
\\
- \underline{\textbf{Field strength constraints}:}
\\
The conditions that have been derived on the spin-connection in Appendix \ref{app:sufr} and summarised above, in turn restrict the form of the field strength.
 We summarise these results in the following table:
\begin{center}
\begin{tabular}{|c|c|} \hline
Component       & Solution      \\ \hline\hline
$G^{3,0}$       & $G_{\a\b\g}=0$          \\
        & $G_{5\b\g}= 2i\O_{5,\b\g}$    \\ \hline
$G^{2,1}$       & $G_{\ab\b\g}=0$                       \\
        & $G_{\ab5\g}=-2i\O_{\ab,5\g}-ig_{\ab\g}\O_{0,05}$  \\
        & $G_{\bar5\b\g}=-2i\O_{\bar5,\b\g}$            \\
        & $G_{\bar55\a}=-2i\O_{\bar5,5\a} + i\O_{0,0\a}$    \\ \hline
$F^{4,0}$       & $F_{\a_1\a_2\a_3\a_4} = \frac{1}{2}\left( -3\O_{0,05}+2\O_{5,\b}{}^\b \right)
        \e_{\a_1\a_2\a_3\a_4}$                  \\
        & $F_{5\a_1\a_2\a_3}=\frac{1}{2}\left( \O_{0,0\betab} - 2\O_{\betab,\g}{}^\g \right)
        \e^{\betab}{}_{\a_1\a_2\a_3}$ \\ \hline

$F^{3,1}$       & $F_{\ab\b_1\b_2\b_3}={1\over2} \left( 2 \O_{\ab, \bar5\bar\g}
        \e^{\bar\g}{}_{\b_1\b_2\b_3}-3 \O_{5,\bar\g_1\bar\g_2}
        \e^{\bar\g_1\bar\g_2}{}_{[\b_1\b_2} g_{\b_3]\ab} \right)$   \\
        & $F_{\bar 5\b_1\b_2\b_3}=-\O_{\bar5,\bar\g\bar5} \e^{\bar
        \g}{}_{\b_1\b_2\b_3}$                       \\
        & $F_{\ab 5\b_1\b_2}={1\over2} \O_{\ab, \bar\g_1\bar \g_2}
        \e^{\bar\g_1\bar \g_2}{}_{\b_1\b_2}$                \\
        & $F_{\a\b5\bar5}={1\over2} \left(\O_{5,\bar\g_1\bar\g_2}-\O_{\bar 5,\bar\g_1\bar\g_2}\right)
        \e^{\bar\g_1\bar \g_2}{}_{\a\b}$                \\ \hline

$F^{2,2}$   & $F_{\ab\bar5 \b_1\b_2}={1\over2} \O_{\ab,\bar\g_1\bar\g_2}
         \e^{\bar\g_1\bar\g_2}{}_{\b_1\b_2}$                \\
        & $F_{\a\bar\b 5\bar5}=-2i\O_{0, \a\bar\b}$         \\
        & $F_{5\bar5\a}{}^\a=0$                     \\
        & $F_{\a\bar\b\g}{}^\g=0$                   \\ \hline
\end{tabular}
\end{center}

The $(2,2)$ component of the field strength $F_{\a_1\a_2\betab_1\betab_2}$ is not determined by the Killing spinor equations.
\subsubsection{The Geometry of the Spacetime}
We will now investigate some consequences for the spacetime geometry which arise from the existence of two \sufr-\inv\ \kss, using equations (\ref{summary00})-(\ref{summary5}).

Firstly, we observe that the spacetime admits a timelike Killing vector field $\k^f$ associated to $\eta_1$, which is inherited from the $\n=1$ case (see section \ref{11d:n=1geom}).
 In this case, we also have exterior forms associated to the spinors $\eta^{SU(5)}$ and $\eta^{SU(4)}$, and these can be found using the results of the table in section \ref{11d:n=2forms}.
 The two-form $\o^{SU(4)}$ and the first part of the four-form $\z$ give rise to an \sufr-\inv\ K\"ahler form and antiholomorphic volume form, and these determine an \sufr-structure on $B$, the space which is transverse to the orbits of $\k^f$.

Another difference between this case and the case in which both spinors are \suf-\inv\ is that here $\O_{\ab,\betab\gb}\neq0$, so that the spatial manifold $B$ is \textit{not} complex, hence not Hermitian.
 To check this explicitly, we need only compute the $(2,0)$ part of $d\hat{e}^{\bar5}$, for example, to see that it is not required to vanish, since $\O_{5,5\a}$ is not constrained to vanish by the \kses.
 Therefore neither of the complex structures associated with $\o(\eta^{SU(5)},\eta^{SU(5)})$ or $\o(\etasufr,\etasufr)$ are integrable, so that $B$ is a non-complex almost Hermitian manifold.

From section \ref{11d:n=2forms}, we also see that there is a one-form which may be constructed from the spinors $\eta^{SU(5)}$ and $\eta^{SU(4)}$, namely $\,\k=e^\nat$.
 Using this, we define a vector field by
\be
\tilde\k^f = \sqrt2f^2\partial_\nat = if^2 (\partial_{5}-\partial_{\bar5})~.
\ee
It is evident that $\tilde\k^f$ is a \textit{spacelike} vector.

It is straightforward to show that $\tilde\k^f$ solves the Killing vector equation $\nabla_A\tilde\k^f_B+\nabla_B\tilde\k^f_A=0$, hence it is a Killing vector field on the spacetime.
 Note that the non-vanishing components of the one-form associated to the
vector field are \linebreak $(\tilde \k^f)_5=- (\tilde \k^f)_{\bar5}=-if^{-2}$.
 For example, let us consider the $(A,B)=(0,0)$ component of the Killing equation.
 Using that $f$ is time-independent (\ref{summary00}), we get
\be
if^{-2} \left( \O_{0,}{}^5{}_0  - \O_{0,}{}^{\bar 5}{}_0 \right) =0 \;,
\ee
which is satisfied due to (\ref{summary1}).
 In a similar way, the rest of the connection conditions ensure that the other components of the Killing vector equation are satisfied, so that $\tilde \k^f$ is a spacelike Killing vector on the spacetime.

In fact, we can also demonstrate that $\tilde \k^f$ preserves the almost complex structure on $B$ given in complex coordinates by
\be
\tilde{J}=\hbox{diag}(\, i\d^\a_\b \,,\, -i \,,\, -i\d^{\ab}_{\betab} \,,\, i \,) \;,
\ee
where $\a,\b=1,2,3,4$.
 The K\"ahler form associated to $\tilde{J}$ is then $\tilde\o=-i\o$, where
\be
\o:=\o(\etasufr,\etasufr)=e^1\we e^6+e^2\we e^7+e^3\we e^8+e^4\we e^9-e^5\we e^\nat \;.
\ee
Therefore to say that $\tilde\k^f$ preserves $\tilde{J}$ is equivalent to saying that it preserves the associated K\"ahler form $\tilde\o$.
 In turn, this is equivalent to the requirement that
\be
{\mathcal L}_{\tilde \k^f}\o=0 \;,
\ee
since $\tilde\o=-i\o$.

To see that this equation holds, we first express the Lie derivative in terms of the spin connection as
\be
({\mathcal L}_{\tilde \k^f}\o)_{AB} \,=
\,-\, 2\partial_{[A}(\tilde\k^f)^D \o_{B]D}
\,+\, 2(\tilde\k^f)^D\O_{[A|}{}^C{}_D\o_{|B]C}
\,-\, 2(\tilde\k^f)^D\O_D{}^C{}_{[A}\o_{B]C} \;.
\ee
Now, using conditions (\ref{summary00})-(\ref{summary5}), we can proceed to show that $\tilde\k^f$ indeed does preserve $\o$.
 For instance, the $(A,B)=(5,\bar5)$ component yields
\be
( \mathcal{L}_{\tilde\k^f}\o)_{5\bar5} = 2f^2\left( - \O_{\bar5,5\bar5}
+ (\partial_5 + \partial_{\bar5})\log f \right) \;,
\ee
which vanishes as a consequence of (\ref{summary1}) and (\ref{summary5}).
 The other components can be shown to hold in a similar way.
 It can similarly be shown that $\k^f$ preserves the almost complex structure of $B$ which is associated to $\o(\etasuf,\etasuf)$.

In addition, we also find that $[\k^f,\tilde \k^f]=0$.
 This results from the vanishing of the connection components of the form $\O_{0,5i}=\O_{5,0i}$ and also condition (\ref{summary00}).
 Since we have a pair of commuting Killing vector fields, we may introduce coordinates $u^a$ adapted to both vectors, and write the metric as \cite{jguggp}
\be \label{11d:n=2metric1}
ds^2=U_{ab} (du^a+\a^a) (du^b+\a^b)+ \g_{IJ} dx^I dx^J~,
\ee
where $\a$ and $\g$ depend on the remaining coordinates $x^I$.

To summarise, we have found the following consequences of having two \sufr-\inv\ \kss\ in a $D=11$ \sg\ background:
\begin{itemize}
\item The field strength    is determined by the spin-connection according to the table in section \ref{11d:2sufrsoltn}, except for the component $F_{\a_1\a_2\betab_1\betab_2}$.
 Since its trace is found to vanish, $F_{\a\betab\g}{}^\g=0$, the undetermined component $F_{\a_1\a_2\betab_1\betab_2}$ corresponds to a traceless $(2,2)$-form.
 In other words, it corresponds to the irreducible representation $\mathbf{20}'$ of \sufr, which arises in the decompostion of the \suf\ traceless $(2,2)$-piece into \sufr\ irreps, given by
\be
\L^{(2,2)}_0(\bC^5)\cong\L^{(2,2)}_0(\bC^4)\oplus\L^{(2,1)+(1,2)}_0(\bC^4)
\oplus\L^{(1,1)}_0(\bC^4) \;,
\ee
or alternatively,
\be
\mathbf{75}\longrightarrow\mathbf{20}'\oplus\mathbf{20}\oplus\bar{\mathbf{20}}\oplus\mathbf{15} \;.
\ee
\item The spacetime admits two commuting Killing vector fields $\k^f$ and $\tilde \k^f$, the former of which is timelike whereas the latter is spacelike.
 We may adapt coordinates according to these vectors to express the metric in the form (\ref{11d:n=2metric1}).
\item The space $B$ which is transverse to the orbits of $\k^f$, is an almost Hermitian manifold but is not complex.
\item $B$ possesses an \sufr-structure, and the Killing vector field $\tilde\k^f$ preserves the almost complex structure associated to $\o(\etasufr,\etasufr)$.
 The \sufr-structure is determined by conditions (\ref{summary1})-(\ref{summary5}) and the results of Appendix \ref{GrayHerv}.
\end{itemize}
%
%
%
%
%
%
\section{$\n>2$ Backgrounds} \label{11d:n>2}
In this section, we will demonstrate how the procedure for solving the \kses\ outlined above may be extended to backgrounds possessing more than two supersymmetries.
 Solutions will be given for certain configurations of three and four \sufr-\inv\ \kss\footnote{These were first presented in \cite{jguggp}.}, and the resulting constraints will be used to draw important conclusions about the geometry of the background.
 The strategy for systematic classification may be extended to $\n>2$ as follows.

As before, we take $\eta_1\in$\orbsuf, and $\eta_2$ to be of the
form (\ref{11d:generic1}).
 In general, both $\eta_1$ and $\eta_2$ have the identity as their common stability subgroup.
 However, if certain parameters in the generic spinor $\eta_2$ vanish, then the spinors are left invariant by some subgroup $H\subseteq SU(5)$.
 As we have seen in section \ref{11d:n=2canforms}, typically $H$ is an $SU(q)$ group or a product of such groups.
 We then decompose the $Spin(10)$ module $\D^+_{\mathbf{16}}$ under the action of $H$ and investigate the possible orbits of $H$ in each component arising in this decomposition.
 It is sufficient to investigate this module, since the corresponding spinors in $\D^-_{\mathbf{16}}$ are determined precisely by Majorana conjugation.
 In other words, it would be entirely equivalent to start by decomposing the module $\D^-_{\mathbf{16}}$ under $H$.

Once the orbits of $H$ in the spinor space have been determined, we can then find a canonical representative for each one, up to $H$ transformations.
 The third \ks\ $\eta_3$ can now be chosen as any linear combination of these representatives, provided it is linearly independent of $\eta_1$ and $\eta_2$.

This procedure may be iterated to find representatives for any number of \kss.
 Its effectiveness is apparent in the cases in which the common stability subgroup of the spinors is large, because then it can be used to restrict the choice of the next spinor.
 This in turn facilitates the solution of the \kses.
 However, in the cases where the stability subgroup of $\eta_1$, $\eta_2$ and $\eta_3$ is small, further progress in solving the \kses\ may still be difficult.

For example, suppose that $\eta_1$ and $\eta_2$ are chosen such that their common stability subgroup is $\{\id\}$.
 In this case, the third spinor may be chosen to be any other spinor which is linearly independent of $\eta_1$ and $\eta_2$.
 Although our formalism can be used, there is no apparent simplification in either the computation of the Killing spinor equations or of the exterior forms associated with $\eta_3$.
 Consequently, the conditions which arise on the geometry will be rather involved, and difficult to interpret.

Therefore, for the sake of illumination, we shall focus on some $\n>2$ backgrounds admitting spinors which possess large symmetry groups.
 In particular, we will now describe the scenario that occurs when the spinors are invariant under $SU$ groups of progressively decreasing dimension.
 This will serve to further illustrate the general procedure of constructing canonical forms for various orbits within $\Dm$.
 Then we will proceed to solve the \kses\ for configurations of three and four \sufr-\inv\ \kss.
\subsection{The $SU$ Series}
Let us consider the situation in which we have multiple spinors whose stability subgroups are consecutively \suf, \sufr, $SU(3)$, \sut\ and $\{\id\}$.
 We refer to this as the \textit{$SU$ series}\footnote{This set-up can also be thought of as a \textit{Calabi-Yau series}, since the Killing spinors given in this section are those expected in M-theory Calabi-Yau compactifications with fluxes \cite{jguggp}. However, we will not pursue this here, but rather focus on the construction of canonical forms for the associated spinors.}.
 We will consider each stabiliser $H\subset SU(5)$ in turn, and determine the maximal number of linearly independent spinors which are \inv\ under $H$.

Recall that the spinor space of \spt, splits as $\Dc\cong\D^+\oplus\D^-$ under \linebreak $Spin(10)\subset Spin(1,10)$, where $\D^\pm$ are inequivalent $Spin(10)$ modules of complex dimension $16$.
 Under $SU(5)\subset Spin(10)$ these decompose as
\bea \label{su5decomp}
&& \D^+_{\mathbf{16}} \cong \L^0_{\mathbf1}(\bC^5) \oplus \L^2_{\mathbf{10}}(\bC^5) \oplus \L^4_{\bar{\mathbf5}}(\bC^5)
\nn \\
&& \D^-_{\mathbf{16}} \cong \L^1_{\mathbf5}(\bC^5) \oplus \L^3_{\bar{\mathbf{10}}}(\bC^5) \oplus \L^5_{\bar{\mathbf1}}(\bC^5) \;,
\eea
where the bold subscripts denote the complex dimension of each irreducible representation.
 We can observe that each space is indeed irreducible under \suf, since via the oscillator basis of gamma matrices (\ref{11d:hermbasis1}) we have the isomorphism
\be
\L^p(\bC^5)\cong\L^{(0,p)}(\bC^5) \;,\quad p\leq5 \;.
\ee
Since $(0,p)$-forms of the complexified exterior algebra are irreducible under the action of \suf\ (see, for example, \cite{joyce}), we conclude that the $p$-form modules in this decomposition are also irreducible \suf\ modules.

In what follows, we wish to determine the Majorana spinors of $\Dm$ that are invariant under $SU(2)\subset SU(3) \subset SU(4) \subset SU(5)$.
 Therefore we need only consider one of the $Spin(10)$ spinor modules, since the Majorana condition precisely determines the corresponding spinor components that arise in the other module.

So, let us consider $\D^-_{\mathbf{16}}$.
 From (\ref{su5decomp}), a general element can be written as the sum of three irreducible components as
\be
\eta = \l^i e_i + \frac{1}{3!}\m^{ijk}e_{ijk} + \nu e_{12345} \;,
\ee
where $\l^i$, $\m^{ijk}$, $\nu\in\bC$ for $i,j,k=1,\cdots,5$.
 Our method will be to decompose each component progressively under the $SU$ series.
 The way we choose to embed the groups in each other is to specify that $SU(q)$ acts on the space $\bC^q$ which is spanned by the vectors $\{e_1,\cdots,e_q\}$, for $q\leq5$.
 Note that we immediately have the \suf-\inv\ spinor $\nu e_{12345}$, which decomposes no further.

We begin by setting
\be
\z = \l^i e_i + \frac{1}{3!}\m^{ijk}e_{ijk} \;,
\ee
and decomposing under \sufr.
 According to the decompositions
\bea
&& \L^1_{\mathbf5}(\bC^5) \cong \L^0_{\mathbf1}(\bC^4)\oplus \L_{\mathbf4}^1(\bC^4)
\\
&& \L^3_{\bar{\mathbf{10}}}(\bC^5) \cong \L_{\mathbf6}^2(\bC^4)\oplus \L_{\mathbf4}^3(\bC^4) \;,
\eea
the general spinor decomposes under \sufr\ as
\be
\z = \l^ae_a + \l^5e_5 + \frac{1}{2}\m^{ab5}e_{ab5} + \frac{1}{3!}\m^{abc}e_{abc} \;,
\ee
where $a,b,c=1,\cdots,4$.
 We see that this yields the one-dimensional \sufr\ representation $\l^5e_5$, so that the spinors in $\D^-$ which possess \sufr\ as their maximal stability subgroups are spanned by
\be \label{11d:su4inv}
e_{12345} \;,\qquad e_5 \;.
\ee

Next, we seek a basis of spinors which are invariant under $SU(3)$.
 Decomposing the non-trivial \sufr\ modules under $SU(3)$ gives
\bea
&&\L^1_{\mathbf4}(\bC^4) \cong \L_{\mathbf1}^0(\bC^3)\oplus \L_{\mathbf3}^1(\bC^3)
\\
&&\L^2_{\mathbf6}(\bC^4) \cong \L_{\mathbf3}^1(\bC^3)\oplus\L_{\mathbf3}^2(\bC^3)
\\
&&\L^3_{\mathbf4}(\bC^4) \cong \L^2_{\mathbf3}(\bC^3)\oplus \L_{\mathbf1}^3(\bC^3) \;.
\eea
In terms of $\z$, we find that
\be
\z = \l^me_m + \l^4e_4 +\l^5e_5 + \m^{m45}e_{m45} + \frac{1}{2}\m^{mn5}e_{mn5} + \frac{1}{2} \m^{mn4}e_{mn4}
+ \frac{1}{3!}\m^{mnp}e_{mnp} \;,
\ee
where now, $m,n,p=1,2,3$.
 Here we have obtained an extra $SU(3)$ singlet $\l^4e_4$, which corresponds to the summand $\L_{\mathbf1}^0(\bC^3)$ in the above decomposition.
 Thus the $SU(3)$-\inv\ spinors of $\D^-$ are spanned by
\be
e_{12345} \;, \quad e_5 \;,\quad e_4 \;.
\ee

To obtain the \sut-\inv\ spinors we decompose the non-trivial $SU(3)$ modules under \sut\ to obtain
\bea
&& \L_{\mathbf3}^1(\bC^3) \cong \L_{\mathbf1}^0(\bC^2)\oplus \L_{\mathbf2}^1(\bC^2)
\\
&& \L_{\mathbf3}^2(\bC^3) \cong \L_{\mathbf2}^1(\bC^2)\oplus\L_{\mathbf1}^2(\bC^2) \;.
\eea
This corresponds to the splitting
\bea
\z &=& \l^te_t + \l^3e_3 + \l^4e_4 + \l^5e_5 + \m^{t45}e_{t45} + \m^{345}e_{345} + \m^{t35}e_{t35}
+ \m^{125}e_{125}
\nn \\ && + \m^{t34}e_{t34} + \m^{124}e_{124} + \m^{123}e_{123} \;,
\eea
where $t=1,2$.
 The terms with a $t$ index are still reducible under $SU(1)\subset SU(2)$, but the remaining terms have no free indices hence provide the trivial representations of \sut.
 In summary then, the \sut-\inv\ spinors in $\D^-$ are spanned by the eight components
\be
e_{12345},\quad e_5,\quad e_4,\quad e_3,\quad e_{345},\quad e_{125},\quad e_{124},\quad e_{123} \;.
\ee
However, we are ultimately concerned with the Majorana spinors of $\Dm$ which may be found in $D=11$ \sg\ backgrounds.
 Therefore, applying the Majorana condition (\ref{11d:maj}) and normalising with respect to the Hermitian inner product, we find that a basis for the \sut-\inv\ Majorana spinors of $\Dm$ is given by the sixteen spinors:
\bea
&& \etasuf = \frac{1}{\sqrt2}(1+e_{12345}) \;,\qquad \thsuf = \frac{i}{\sqrt2}(1-e_{12345})             \;,\\
&& \etasufr = \frac{1}{\sqrt2}(e_5+e_{1234}) \;,\qquad \thsufr = \frac{i}{\sqrt2}(e_5-e_{1234})         \;,\\
&& \eta_1^{SU(3)} = \frac{1}{\sqrt2}(e_4-e_{1235}) \;,\qquad \th_1^{SU(3)} = \frac{i}{\sqrt2}(e_4+e_{1235})     \;,\\
&& \eta_2^{SU(3)} = \frac{1}{\sqrt2}(e_{45}-e_{123}) \;,\qquad \th_2^{SU(3)} = \frac{i}{\sqrt2}(e_{45}+e_{123}) \;,\\
&& \eta_1^{SU(2)} = \frac{1}{\sqrt2}(e_3+e_{1245}) \;,\qquad \th_1^{SU(2)} = \frac{i}{\sqrt2}(e_3-e_{1245})     \;,\\
&& \eta_2^{SU(2)} = \frac{1}{\sqrt2}(e_{12}-e_{345}) \;,\qquad \th_2^{SU(2)} = \frac{i}{\sqrt2}(e_{12}+e_{345}) \;,\\
&& \eta_3^{SU(2)} = \frac{1}{\sqrt2}(e_{35}+e_{124}) \;,\qquad \th_3^{SU(2)} = \frac{i}{\sqrt2}(e_{35}-e_{124}) \;,\\
&& \eta_4^{SU(2)} = \frac{1}{\sqrt2}(e_{34}-e_{125}) \;,\qquad \th_4^{SU(2)} = \frac{i}{\sqrt2}(e_{34}+e_{125}) \;.
\eea
Thus, there is a maximum of sixteen independent \sut-\inv\ \kss\ available for a $D=11$ \sg\ background.
 This is the same number as is expected in $M$-theory compactifications on $K_3$ \cite{jguggp}.
 However, we will not investigate such matters here, but instead we will turn to solving the \kses\ for two particular configurations involving more than two \kss.
\subsection{$\n=3$ Backgrounds with $SU(4)$-invariant Killing Spinors} \label{11d:3sufrsoltn}
Let us consider the class of $\n=3$ backgrounds which possess \kss\ that are all invariant under \sufr.
 In this case it suffices to combine the conditions for $\n=2$ backgrounds with $SU(5)$ and $SU(4)$-invariant spinors, which have been derived in sections \ref{11d:2su5} and \ref{11d:2su4}.
 We will restrict the parameters so that the background has Killing spinors of the form
\be
\boxed{ \eta_1=f_1 \eta^{SU(5)} } \;,\quad
\boxed{ \eta_2=f_2 \theta^{SU(5)} } \;,\quad
\boxed{ \eta_3=f_3 \eta^{SU(4)} } \;.
\ee
Recall that in the case of \sufr-\inv\ spinors, indices $\a ,\b ,\cdots$ run from $1$ to $4$.
\\ \ 
\\
- \underline{\textbf{Geometric constraints}:}
\\
Combining the conditions of the two classes of $\n=2$ backgrounds, we find firstly that
\be
f_1=f_2=f_3=f~,~~~\partial_0f=\partial_5f=\partial_{\bar 5}f=0~.
\label{ssummary00}
\ee
Next, we use the condition $\O_{\ab,\betab\gb}=0$ together with the other conditions summarised in section \ref{11d:2sufrsoltn}, to find the following conditions on the $\O_{0,0i}$ components:
\be
\O_{0,05}=\O_{0,0\bar 5}=0~,~~~~
\O_{0,0\alpha}=-2\partial_{\a} \log f~.
\label{ssummary1}
\ee
The conditions which arise for the $\O_{0,ij}$ components  are
\be
\O_{0, 5\bar\a}=\O_{0, 5\a}= \O_{0, 5\bar5}=\O_{0,\b}{}^\b=0~,~~~~~
\O_{0, \b_1\b_2}={i\over4} \O_{5,\bar \g_1\bar\g_2} \epsilon^{\bar\g_1\bar\g_2}{}_{\b_1\b_2} \;,
\label{ssummary2}
\ee
and we find that the traceless part of $\O_{0, \a\bar\b}$ is not determined by the \kses.

The conditions on the components of the form $\O_{\bar\a, ij}$ are
\bea
\O_{\bar\b_1,\bar\b_2\bar\b_3}&=&0~,~~~
\O_{\bar\a, \b_1\b_2}=-\O_{0,0[\b_1} g_{\b_2]\bar\a}~,~~~
\O_{ \b,\bar\a}{}^\b=-{3\over2} \O_{\bar5, \bar\a  5}
=-{3\over 2} \O_{0,0\bar\a}~,
\cr
\O_{\a,\b}{}^\b&=&-{1\over2}\O_{\bar5, \a  5}
=-{1\over2} \O_{0,0\a}~.
\label{ssummary3}
\eea
In addition, for the components of the type $\O_{i,j5}$, we have
\bea
&&\O_{\bar\b_1, \bar\b_2\bar 5}=\O_{\bar5,\bar\b_1\bar\b_2}=0~,~~~
\O_{[\bar\b_1, \bar\b_2] 5}=-\O_{5,\bar\b_1 \bar\b_2}~,~~~
\O_{(\bar\b_1, \bar\b_2) 5}=0~,
\cr
&&\O_{(\bar\a, \b) \bar5}=\O_{(\bar\a, \b) 5}=0~,~~~
\O_{\bar\a, 5\bar5}=0~.
\label{ssummary4}
\eea
Furthermore, the traceless part of $\O_{\a,\b\bar\g}$ remains undetermined.

Finally, the conditions on the $\O_{\bar5, ij}$ components are
\bea
&&\O_{5,\b}{}^\b=\O_{\bar5,\b}{}^\b=0~,~~~
 \O_{ 5, \bar\a 5}=\O_{\bar 5, \bar\a \bar5}=0~,~~~
\O_{ 5, \bar\a \bar5}=\O_{\bar 5, \bar\a 5}~,
\cr
&&\O_{\bar5, \bar\a  5}= \O_{0,0\bar\a}~,~~~
\O_{\bar5, 5\bar 5}=\O_{5,5\bar 5}=0~.
\label{ssummary5}
\eea

The above constraints together with their complex conjugates, provide the full set of geometric conditions that are required for a background to admit $\n=3$ supersymmetry determined by the \sufr-\inv\ \kss\ $\eta_1$, $\eta_2$ and $\eta_3$.
\\ \ 
\\
- \underline{\textbf{Field strength constraints}:}
\\
We can now substitute the geometric conditions given above into the expressions for the components of the field strength which were derived in sections \ref{11d:2su5} and \ref{11d:2su4}.
 This is a straightforward procedure, and the results are displayed in the following table:
\begin{center}
\begin{tabular}{|c|c|} \hline
Component       & Solution                                          \\ \hline\hline
$G^{3,0}$       & $G_{\a\b\g}=0$                                    \\
        & $G_{5\b\g}=0$                                     \\ \hline
$G^{2,1}$       & $G_{\bar\a \b\g}=0$                                   \\
        & $G_{\bar5 \b\g}=-2i \O_{\bar5,\b\g}$                          \\
        & $G_{\bar\a 5\g}=-2i \O_{\bar\a,5\g}$                          \\
        & $G_{\bar 5 5\a}=-2i \O_{\bar5,5\a}+i\O_{0,0\a}$                   \\ \hline
$F^{4,0}$       & $F_{\a_1\a_2\a_3\a_4}=0$                                  \\
        & $F_{5\a_1\a_2\a_2}=0$                                 \\ \hline
$F^{3,1}$       & $F_{\bar\a\b_1\b_2\b_3} = - {3\over2} \O_{5,\bar\g_1\bar\g_2}
        \e^{\bar\g_1\bar\g_2}{}_{[\b_1\b_2} g_{\b_3]\bar\a}$                    \\
        & $F_{\bar 5\b_1\b_2\b_3}=0$                                \\
        & $F_{\bar\a 5\b_1\b_2}=0$                              \\
        & $F_{\a\b5\bar5}={1\over2} \O_{5,\bar\g_1\bar\g_2}\e^{\bar\g_1\bar \g_2}{}_{\a\b}$ \\ \hline
$F^{2,2}$   & $F_{\bar \a\bar5 \b_1\b_2}=0$                             \\
        & $F_{\a\bar\b 5\bar5}=-2i \O_{0, \a\bar\b}$                        \\
        & $F_{\a\bar\b\g}{}^\g=0$                               \\
        & $F_{5\bar5 \a}{}^\a=0$                                \\ \hline
\end{tabular}
\end{center}

We have used that the (3,0)+(0,3) part of the connection $\O_{i,jk}$ vanishes throughout this derivation.
 We have also used the conditions for $\n=2$ supersymmetry which were derived in Appendix \ref{app:sufr}.
 The field strength components that do not appear in the table are undetermined by the Killing spinor equations.
\subsubsection{The Geometry of the Spacetime}
We shall now investigate some aspects of the spacetime geometry of a background admitting three \sufr-\inv\ \kss, arising from conditions (\ref{ssummary00})-(\ref{ssummary5}).
 We find that the geometry of such $\n=3$ backgrounds combines aspects of the geometries of the $\n=2$ backgrounds which possess $SU(5)$ and $SU(4)$-invariant Killing spinors, that have been investigated in sections \ref{11d:2su5} and \ref{11d:2su4}.

As in all previous cases, the spacetime admits a timelike Killing vector field $\k^f=f^2e^0$, which is inherited from the spinor $\eta_1$.
 From the exterior forms associated with the  spinors $\eta^{SU(5)}$, $\theta^{SU(5)}$ and $\eta^{SU(4)}$, the spacetime admits an $SU(4)$-structure on the space $B$ which is transverse to the orbits of $\k^f$, as in section \ref{11d:2su4}.
 However, unlike the $\n=2$ backgrounds with $SU(4)$-invariant spinors, in this case the space $B$ \textit{is} complex.
 This arises because the (3,0)+(0,3) parts of the connection $\O_{A,BC}$ vanish, making the almost complex structure on $B$ integrable.
 This is similar to the case of $\n=2$ backgrounds with $SU(5)$-invariant spinors.
 Therefore, $B$ is in fact a ten-dimensional Hermitian manifold with an \sufr-structure.

There are also \textit{two} spacelike Killing vector fields
\bea
&& \tilde\k^f = if^2(\partial_{5}-\partial_{\bar5}) \\
&& \hat\k^f   = f^2 (\partial_{5}+\partial_{\bar5}) \;,
\eea
which are associated with the one-forms $\k\left( \etasuf,\etasufr \right)=e^\nat$ and $\k\left( \thsuf,\etasufr \right)=e^5$, respectively.
 Furthermore, the Killing vector fields $\kappa^f$, $\tilde\kappa^f$ and $\hat\kappa^f$ mutually commute.
 This can be shown in a similar way to the case of section \ref{11d:2su4}.

We also find that the vectors $\tilde \kappa^f$ and $\hat \kappa^f$ preserve the  complex structure of $B$,
\be
{\mathcal L}_{\tilde \k^f}\omega(\eta^{SU(5)},\eta^{SU(5)}) = 0 \;.
\ee
Again, the computation is analogous to that which was described for the $\n=2$ background with $SU(4)$-invariant Killing spinors, so details will be omitted here.

The orbits of $\k^f$, $\tilde\k^f$ and $\hat\k^f$ are along the $e^0$, $e^\nat$ and $e^5$ directions respectively.
 Recalling the identity given in (\ref{11d:algid1}), we see that there is an \sufr-\inv\ two-form
\be \label{hatBkahler}
\o^{SU(4)} = e^1\we e^{6}+e^2\we e^{7}+e^3\we e^{8}+e^4\we e^{9}
\ee
which exists on the submanifold $\hat B\subset B$ spanned by $\{e_1,e_2,e_3,e_4,e_6,e_7,e_8,e_9\}$.
 Clearly $\o^{SU(4)}$ is a K\"ahler form for $\hat B$ which is associated to a complex structure.
 Since $\hat B$ is complex as a submanifold of $B$, possesses an Hermitian metric given by restricting that of $B$, and is equipped with the complex structure associated to $\o^{SU(4)}$, therefore $\hat B$ is itself an Hermitian manifold.

Next, using the exterior forms from the table in section \ref{11d:n=2forms}, we can define
\bea \label{su4vol}
\t^{SU(4)} &=& e^0 \lc\, \left(  \t(\etasuf,\etasufr) - i \t(\thsuf,\thsufr)  \right) \nn\\
       &=& \hat{e}^{1}\we\hat{e}^{2}\we\hat{e}^{3}\we\hat{e}^{4} 
\;,
\eea
where $\{\hat{e}^{\a}\}_{\a=1}^{4}$ is a holomorphic basis of forms on $\hat B$, as specified in equation (\ref{11d:hermforms}).
 In other words, by contracting the five-forms associated to the \sufr-\inv\ spinors with $e^0$, we can obtain an \sufr-\inv\ holomorphic volume form $\t^{SU(4)}$ on $\hat B$.
 This means that there is an \sufr-structure defined on $\hat B$ and specified by the tensors $(g|_{\hat B}\,,\,\o^{SU(4)}\,,\,\t^{SU(4)})$, so that $\hat B$ is in fact a \textit{special Hermitian manifold} \cite{cabrera}.

Now, we can adapt coordinates to the three vector fields given above, and write the metric as
\be
ds^2=U_{ab} (du^a+\b^a) (du^b+\b^b)+\gamma_{IJ} dx^I dx^J \;,
\label{11d:n=3metric}
\ee
where  $a,b=0,1,2$, $\,I,J=1,\dots, 8$ and $U$, $\b$ and $\gamma$ depend only on the $x^I$ coordinates.

To summarise, we have found the following consequences of having three \sufr-\inv\ \kss\ in a $D=11$ \sg\ background:
\begin{itemize}
\item The field strength    is determined by the spin-connection according to the table in section \ref{11d:3sufrsoltn}. \item The spacetime admits three commuting Killing vector fields $\k^f$, $\tilde \k^f$ and $\hat\k^f$, the former of which is timelike whereas the latter two are spacelike.
 We may adapt coordinates according to these vectors to express the metric in the form (\ref{11d:n=3metric}).
\item The space $B$ which is transverse to the orbits of $\k^f$ is a complex hence Hermitian manifold.
\item $B$ possesses an \sufr-structure, and the Killing vector field $\tilde\k^f$ preserves the almost complex structure associated to $\o(\etasuf,\etasuf)$.
 This \sufr-structure is determined explicitly using conditions (\ref{ssummary1})-(\ref{ssummary5}) and the results of Appendix \ref{GrayHerv}, which describe the intrinsic torsion of an $\so(1,10)$ manifold possessing an $\su(4)$-structure.
\item The eight-dimensional submanifold $\hat B$ of $B$ which is transverse to the orbits of all three Killing vector fields is also Hermitian, and admits an \sufr-structure which may be specified by the tensors $(g|_{\hat B}\,,\,\o^{SU(4)}\,,\,\t^{SU(4)})$, or equivalently via conditions (\ref{ssummary1})-(\ref{ssummary5}).
\end{itemize}
\subsection{$\n=4$ Backgrounds with $SU(4)$-invariant Killing Spinors} \label{11d:4sufrsoltn}
As a final illustration of our method, we will now investigate a particular class of $\n=4$ backgrounds, namely those which admit four Killing spinors which are all invariant under \sufr,
\be
\boxed{ \eta_1=f_1 \eta^{SU(5)} } \;,\quad
\boxed{ \eta_2=f_2 \theta^{SU(5)} } \;, \quad
\boxed{ \eta_3=f_3\eta^{SU(4)} } \;,\quad
\boxed{ \eta_4=f_4 \theta^{SU(4)} } \;,
\ee
where $f_1,\,f_2,\,f_3$ and $f_4$ are real functions of the spacetime.
 The conditions arising from the three first Killing spinors have already been derived in the previous section.
 Moreover, the constraints coming from $\mathcal{D}_A\eta_4=0$ can be obtained from the formulae in \ref{11d:2su4} by changing the sign of each term which contains the epsilon tensor.
 This is required since the sign of $e_{1234}$ is the only difference between $\eta^{SU(4)}$ and $\theta^{SU(4)}$.
\\ \ 
\\
- \underline{\textbf{Geometric constraints}:}
\\
The constraints that the Killing spinor equations place on the spacetime functions are
\be
f_1=f_2=f_3=f_4=f~,~~~\partial_0f=\partial_{5}f=\partial_{\bar
5}f=0~.
\label{sum0}
\ee
The conditions arising for the $\O_{0,0i}$ components are found to be
\be
\O_{0,05}=\O_{0,0\bar 5}=-2\partial_5 \log f=-2
\partial_{\bar5} \log f=0~,
\qquad \O_{0,0\alpha}=-2\partial_{\a}
\log f~.
\label{sum1}
\ee
Next, the $\O_{0,ij}$ components are required to satisfy
\be
\O_{0, 5\bar\a}=\O_{0, 5\a}= \O_{0,
5\bar5}=\O_{0,\b}{}^\b=\O_{0, \b_1\b_2}=0~,
\label{sum2}
\ee
and we find that the traceless part of $\O_{0, \a\bar\b}$ is undetermined.

Now, the constraints on components of the form $\O_{\bar\a, ij}$ are given by
\bea
\O_{\bar\b_1,\bar\b_2\bar\b_3}&=&0~,\qquad
\O_{\bar\a, \b_1\b_2}=-\O_{0,0[\b_1} g_{\b_2]\bar\a}~,\qquad
\O_{ \b,\bar\a}{}^\b=-{3\over2} \O_{\bar5, \bar\a  5}
=-{3\over 2} \O_{0,0\bar\a}~,
\cr
\O_{\a,\b}{}^\b&=&-{1\over2}\O_{\bar5, \a  5}
=-{1\over2} \O_{0,0\a}~,
\label{sum3}
\eea
and the traceless part of $\O_{\a,\b\bar\g}$ remains undetermined.

In addition, components of the type $\O_{i,j5}$ must obey
\bea
&&\O_{\bar\b_1, \bar\b_2\bar 5}=\O_{\bar\b_1, \bar\b_2
5}=0~,\qquad \O_{(\bar\a, \b) \bar5}=\O_{(\bar\a, \b) 5}=0~,\qquad
\O_{\bar\a, 5\bar5}=0~.
\label{sum4}
\eea
And finally, the conditions on the  $\O_{\bar5, ij}$ components are
\bea
&&\O_{5,\b}{}^\b=\O_{\bar5,\b}{}^\b=0~,\qquad
 \O_{ 5, \bar\a 5}=\O_{\bar 5, \bar\a \bar5}=0~,\qquad
\O_{ 5, \bar\a \bar5}=\O_{\bar 5, \bar\a 5}~,
\nn \\
&&\O_{\bar5, \bar\a  5}= \O_{0,0\bar\a}~,\qquad \O_{\bar5,
5\bar 5}=-\O_{5,5\bar
5}=0~,\qquad\O_{5,\bar\a_1\bar\a_2}=\O_{\bar 5,\bar\a_1\bar\a_2}=0~.
\label{sum5}
\eea
The condition that remains to be examined is (\ref{11d:s62}).
 In this case, we see that
\be
\O_{\bar\a,5}{}^{\bar\a}=0~.
\label{sum6}
\ee
The above constraints, together with their complex conjugates, provide the full set of geometric conditions that are required for a background to admit $\n=4$ supersymmetry determined by the \sufr-\inv\ \kss\ $\eta_1$, $\eta_2$, $\eta_3$ and $\eta_4$.
\\ \ 
\\
- \underline{\textbf{Field strength Constraints}:}
\\
Using the above conditions on the spin-connection, the field strength components can be expressed in terms of the geometry.
 As in the previous cases, we substitute conditions (\ref{sum1})-(\ref{sum6}) into the expressions for the field strength which arise in the $\n=2$ cases.
 The results are summarised below:
\begin{center}
\begin{tabular}{|c|c|} \hline
Component       & Solution                                          \\ \hline\hline
$G^{3,0}$       & $G_{\a\b\g}=0$                                    \\
        & $G_{5\b\g}=0$                                     \\ \hline
$G^{2,1}$       & $G_{\bar\a \b\g}=0$                                   \\
        & $G_{\bar5 \b\g}=0$                            \\
        & $G_{\bar\a 5\g}=-2i \O_{\bar\a,5\g}$                          \\
        & $G_{\bar 5 5\a}=-i\O_{0,0\a}$                 \\ \hline
$F^{4,0}$       & $F_{\a_1\a_2\a_3\a_4}=0$                                  \\
        & $F_{5\a_1\a_2\a_2}=0$                                 \\ \hline
$F^{3,1}$       & $F_{\bar\a\b_1\b_2\b_3}=0$                    \\
        & $F_{\bar 5\b_1\b_2\b_3}=0$                                \\
        & $F_{\bar\a 5\b_1\b_2}=0$                              \\
        & $F_{\a\b5\bar5}=0$    \\ \hline
$F^{2,2}$   & $F_{\bar \a\bar5 \b_1\b_2}=0$                             \\
        & $F_{\a\bar\b 5\bar5}=-2i \O_{0, \a\bar\b}$                        \\
        & $F_{\a\bar\b\g}{}^\g=0$                               \\
        & $F_{5\bar5 \a}{}^\a=0$                                \\ \hline
\end{tabular}
\end{center}
~\\
In comparison to the $\n=3$ and $\n=2$ cases, we can observe how much more constrained the field strength becomes as we add \kss\ with large stability subgroups to the background.
\subsubsection{The Geometry of the Spacetime}
Let us now investigate some aspects of the spacetime geometry for backgrounds admitting four \sufr-\inv\ \kss, that
arise from conditions (\ref{sum1})-(\ref{sum6}).
 We shall not elaborate on the details here, since the geometric properties of the spacetime in this case are similar to those we have seen for $\n=2$ and $\n=3$ backgrounds.
 Rather, we shall present a summary without any preceding discussion:
\begin{itemize}
\item The field strength is determined by the spin-connection according to the table in section \ref{11d:4sufrsoltn}.
\item The spacetime admits three commuting Killing vector fields $\k^f$, $\tilde \k^f$ and $\hat\k^f$, the former of which is timelike whereas the latter two are spacelike.
 We may adapt coordinates according to these vectors to express the metric in the form (\ref{11d:n=3metric}), as in the $\n=3$ case.
\item The space $B$ which is transverse to the orbits of $\k^f$, is an Hermitian manifold with respect to the complex structure associated to the two-form $\o^{SU(5)}$, see (\ref{kahlersuf}).
\item $B$ admits an \sufr-structure, which is explicitly specified by conditions (\ref{sum1})-(\ref{sum5}) and the results of Appendix \ref{GrayHerv}, or equivalently, via the exterior forms associated to the spinors $\eta^{SU(5)}$, $\theta^{SU(5)}$, $\eta^{SU(4)}$
and $\theta^{SU(4)}$.
\item The eight-dimensional submanifold $\hat B$ of $B$ which is transverse to the orbits of all three Killing vector fields is also Hermitian, and admits an \sufr-structure which may be specified by the tensors $(g|_{\hat B}\,,\,\o^{SU(4)}\,,\,\t^{SU(4)})$, or equivalently via conditions (\ref{sum1})-(\ref{sum5}).
\end{itemize}
Note that the explicit components of the $SU(4)$-structures in the $\n=3$ and $\n=4$ backgrounds are different, since some components of the spin-connection vanish in the latter which do not vanish in the former.

This concludes the \kse\ analysis for backgrounds admitting \sufr-\inv\ spinors.
%
%
%
%
%
%
\setcounter{footnote}{1}
\section{Conclusions}
\subsection*{Summary}
We have presented a new and efficient method for solving the \kses\ of \sg.
 The two key ingredients of this method are an explicit description of spinors in terms of exterior forms, and a knowledge of the orbits of the gauge group of the supercovariant connection on the space of spinors.
 This formalism reduces the \kses\ to a set of linear differential and algebraic conditions, which provide concrete relationships between the fieldstrength, the connection and the functions which parametrise the spinors.
 It is advantageous that all dependence on gamma matrices is removed when it comes to analyse the constraints.
 Also, we can use our knowledge of how groups act on the exterior algebra to enable us to explicitly write down the form of a spinor with a particular stability subgroup.
 This leads to the possibility of classifying supersymmetric solutions according to the stability subgroups of their Killing spinors, although obviously the field equations would need to be investigated in each case for a full classification.
 It also provides valuable simplification particularly in the $\n=1$ case, when we can use the gauge group to bring the spinor into a canonical form, so that the $\n=1$ \kses\ can then be solved in a straightforward way.

In this thesis the emphasis of the spinorial geometry is on $D=11$ \sg, which is important to applications in $M$-theory.
 In the case of one \suf-\inv\ spinor, the $\n=1$ equations give us expressions for all the field strength components in terms of the geometry, except for the traceless $(2,2)$-piece, by using the canonical form for the spinor.
 These results independently verify the work of \cite{pakis}.

 Our formalism also serves to make the analysis of the $\n=2$ case tractable, and the \kses\ were solved for some backgrounds possessing two \kss, namely the class in which both spinors are \suf-\inv\ and also the class in which both are \sufr-\inv.
 We have the benefit of being able to substitute all the information derived from the $\n=1$ case back into the \kses\ so that it is automatically incorporated into the $\n=2$ solution.
 It is clear that this method can be continued for $\n>2$, as we have demonstrated in the cases of $\n=3$ and $\n=4$ backgrounds with \sufr-\inv\ \kss.

The method is inherently systematic, and it is this characteristic which makes it a promising ingredient in the classification of supersymmetric solutions to the \sg\ theories which are of relevance today.
 However, that is not to say that the solution of the \kses\ will now be easy for cases in which there are many \kss.
 In the $D=11$ case with two \kss\ there are already a number of different cases to consider, depending on the combination of stability subgroups which the spinors possess.
 For example, the spinors may both have stability subgroup \suf\ or $(Spin(7)\ltimes\bR^8)\times\bR$, or indeed any combination from these groups or their subgroups.
 Each case must be considered separately and will give rise to a distinct geometry, so one can see how the number of possible cases to be considered is very high, even for $\n=2$.
 However, the method is certainly efficient enough so that the calculations are both feasible and illuminating for $\n>2$ when the common stabiliser of the spinors is non-trivial, since many constraints on the geometry occur in $\n=1$ and $\n=2$, which simplify the equations for $\n>2$.
 More constraints will emerge for each extra \ks\ that is introduced, constraining the solutions further and further.
 In this way, we can see how enlightening it will be to systematically study the different cases for all $\n$, as it is certain that a multitude of new classes of solution will arise which as yet are unknown.
\subsection*{Outlook}
The method presented in this thesis has already met with considerable success when applied to some other important cases.
 In \cite{IIB} and \cite{G2} the $\n=1$ \kses\ of IIB \sg\ are solved for each of the three classes of spinor which can occur, namely for spinors with stability subgroup $Spin(7)\ltimes\bR^8$, $SU(4)\ltimes\bR^8$ or $G_2$.
 Some interesting cases with two and four \kss\ are also investigated.
 In \cite{systematics}, a manual for systematically constructing all supersymmetric solutions of $D=11$ \sg\ is presented, and the classification of such solutions is reduced to evaluating the supercovariant derivative and an integrability condition on a certain set of spinors.
 General expressions are presented for each constraint, and some examples are also constructed.
 Then in \cite{IIBsystematics} a similar systematic analysis is performed for supersymmetric solutions in type IIB \sg, and some new illustrative solutions are constructed.
 It would also be extremely worthwhile to perform analogous investigations for the supergravities relevant to type IIA string theory and the heterotic string\footnote{During the interim between the submission and examination of this thesis, these important cases were successfully analysed, see \cite{phil1} and \cite{phil2}.}.

Also, our method may be used as an independent check of the results of work such as \cite{5dclass}, \cite{tod}, \cite{gauged5dclass}, \cite{6dclass} and others, as well as to attempt to classify solutions of other lower-dimensional supergravities, perhaps even those with extended supersymmetry.
 The spinor spaces of such theories are of sufficiently low dimension that the number of orbits of the gauge spin group is reduced, as compared to the higher-dimensional theories.
 This should make the analysis significantly more straightforward.

Leaving the goal of classification aside for a moment, it would also be of great interest to express the \kss\ of well-known backgrounds in terms of exterior forms, using our formalism.
 This may lead to further insight into the properties of ans\"atze-derived solutions such as $M$-branes \cite{m2}, \cite{m5}, for instance.
 In \cite{gencalibs} it was observed that $M2$- and $M5$-brane backgrounds admit generalised calibrations, and it would be interesting to see how these are related to the explicit spacetime forms associated to the \kss\ of these backgrounds\footnote{Since the examination of this thesis, such matters were investigated in \cite{pete}.}.

In the analysis of the $D=11$ \kses, we indicated how the Gray-Hervella classification of almost-Hermitian manifolds may be employed to describe the geometry which arises when the spinors have certain stability subgroups.
 Two examples from the $\n=1$ and $\n=2$ cases are manifolds with \suf- and \sufr-structures.
 $G$-structures such as these have been investigated already (see, for example, \cite{pakis}, \cite{chiossi}, \cite{cabrera}), and some details on $SU(n-1)$-structures in $2n$-dimensional manifolds have been given in Appendix \ref{GrayHerv}.
 However, in the $\n=2$ case, some other structures arise, namely $SU(2)\times SU(2)$, $Sp(2)$ and $SU(2)\times SU(3)$-structures.
 In other cases too, more exotic structures such as these will emerge naturally as different combinations of stability subgroup are considered, therefore such $G$-structures require some further study.
%
%
%
%
%
%
%
\appendix
%
%
%
%
%
\chapter{The Complex Spin Representations} \label{complreprs}
In this appendix, we present a comprehensive derivation of the complex spin representations of the groups $Spin(r,s)$, and demonstrate the simple classification that results.
 Some invaluable references for this topic are \cite{lawson} and \cite{harvey}, although both focus largely on the classification of the real representations.
 The complex case is entirely analogous and simpler in some respects, yet is provided here to serve as a foundation for the explicit formalism presented in section \ref{explreprs}, as well as for the sake of completeness.
 It is hoped that this will provide a useful supplement to those who approach the subject of spinorial geometry with little prior knowledge of the workings of the complex Clifford algebras.

Throughout this section, all notation and definitions are as in section \ref{complca}.
 The fundamental relationships between the Clifford and exterior algebras are also taken for granted here, as well as isomorphisms such as $\ccl_{n}^0\cong\ccl_{n-1}$.
 In addition, the complex volume form of $\ccl_n$ is defined to be
\be
\o_{n}^{\bC} = i^{\left[\frac{n}{2}\right]}e_1e_2\cdots e_n \;,
\ee
where $\left[\frac{n}{2}\right]$ denotes the integer part of $n$, which we note is equal to $m$ both when $n=2m$ and when $n=2m+1$.
 This convention means that $\left( \o_{n}^{\bC} \right)^2=1$, for all $n$.

In the following we will be examining the Clifford algebras in dimensions $n=2m$ and $n=2m+1$ for $m\geq1$, and so for completeness we note here that $\ccl_1\cong\bC\oplus\bC$.
\section{Irreducible Representations of $\ccl_n$}\label{irreps}
Let us begin by summarising some well-known results about the structure of the complex Clifford algebras.
 All of these can be found in \cite{lawson}, for example.

Since $\left(\o_n^{\bC}\right)^2=1$, we can always define projection operators $P^\pm=\frac{1}{2}\left(\id\pm\o_{n}^{\bC} \right)$.
 Using these, we can state the following result.
\begin{prop}\label{castructure} \
\begin{enumerate}
\item $\ccl_{2m}$ is a simple algebra;
\item $\ccl_{2m+1}$ is not simple, but is the direct sum of the two isomorphic subalgebras $\ccl^\pm_{2m+1}=P^\pm\cdot\ccl_{2m+1}$, which are simple.
\end{enumerate}
\end{prop}
It is well-known that a simple algebra possesses only one non-trivial irreducible representation up to equivalence.
 Therefore Proposition \ref{castructure} tells us that $\ccl_{2m}$ has a unique irreducible representation whereas $\ccl_{2m+1}$ has two inequivalent irreducible representations.
 We will refer to the irreducible representations of a complex Clifford algebra $\ccl_n$ as its \textit{pinor representations}, and to those of its even part $\ccl^0_n$ as its \textit{spin} or \textit{spinor representations}.

This leads to the following fundamental classification of the complex Clifford algebras in terms of the endomorphism algebras of certain vector spaces \cite{rasev}.
\begin{prop}\label{endclass}
Let $\bP$, $\bP^+$ and $\bP^-$ denote distinct copies of $2^m$-dimensional complex affine space. Then, the irreducible representations of the complex Clifford algebras can be described as follows:
\begin{enumerate}
\item Denoting the unique pinor representation of $\ccl_{2m}$ by $\rho$, we have
\be
\rho:\ccl_{2m}\cong End_{\bC}(\bP) \; .
\ee
\item Denoting the inequivalent pinor representations of $\ccl_{2m+1}$ by $\rho_\pm$, we have
\be
\rho_+\oplus\rho_-:\ccl_{2m+1}\cong End_{\bC}(\bP^+)\oplus End_{\bC}(\bP^-) \;,
\ee
\end{enumerate}
\end{prop}
From this result, we observe that the pinor representations of $\ccl_{2m}$ and $\ccl_{2m+1}$ each have complex dimension $2^m$, and the irreducible modules $\bP$, $\bP^\pm$ are their \textit{pinor spaces}.

Denoting by $\mathcal{M}_r(\bC)$ the algebra of $r\times r$ matrices with complex entries, an immediate consequence of Proposition \ref{endclass} is the following:
\be
\ccl_{2m}\cong\mathcal{M}_{2^m}(\bC) \qquad\hbox{and}\qquad
\ccl_{2m+1}\cong\mathcal{M}_{2^m}(\bC)\oplus\mathcal{M}_{2^m}(\bC).
\ee
In other words, a natural way to describe the irreducible representations of $\ccl_{2m}$ and $\ccl_{2m+1}$ is by matrices acting on the pinor spaces $\bP=\bP^\pm\cong\bC^{2^m}$.
 However, this is just one particular manifestation of the pinor representations, which are unique only up to equivalence.
 In the formalism presented in section \ref{explreprs}, we choose to represent the algebras $\ccl_n$ on pinor spaces which are isomorphic to $\L^*(\bC^{m})$ for $n=2m$ and $n=2m+1$.

Now that we have gained some insight into the general structure of the complex Clifford algebras, we turn to a pair of lemmas which will prove useful when we restrict the pinor representations to the groups $Spin(r,s)\subset\ccl_n$, for $r+s=n$.
\begin{prop}\label{oddreprs}
Let $\rho:\ccl_n\longrightarrow End_{\bC}(\bP)$ be a pinor
representation on the space $\bP$, for odd $n$. Then exactly one of
the following holds:
\be
\rho(\o_n^{\bC})=\id \qquad\hbox{or}\qquad
\rho(\o_n^{\bC})=-\id \;.
\ee
These possibilities distinguish the
two inequivalent pinor representations in odd dimensions, which we
denote $\rho_{\pm}$ according to $\rho_\pm(\o_n^{\bC})=\pm\id$.
Their corresponding pinor spaces are $\bP^\pm\cong\bP$.
\end{prop}
This is the complex version of a result found in \cite{lawson}, for example, and so the proof is omitted, since it is entirely analogous to the real case.

Next, we have
\begin{prop}\label{evenreprs}
Let $\rho:\ccl_n\longrightarrow End_{\bC}(\bP)$ be the unique pinor representation on the space $\bP$, for even $n$, and define the projections $\pi^\pm=\frac{1}{2}\left(\id\pm\rho(\o_n^{\bC}) \right)$.
 Consider the splitting
\be
\bP=\bS^+\oplus\bS^-
\ee
where $\bS^\pm=\pi^\pm\cdot\bP$.
 Then the subspaces $\bS^\pm$ are invariant under the even subalgebra $\ccl^0_n\cong\ccl_{n-1}$ and correspond to the two inequivalent irreducible representations of $\ccl_{n-1}$.
\end{prop}
\proof
Let $\phi\in\ccl^0_{2m}$. First note that $\phi\cdot\o_{2m}^{\bC}=\o_{2m}^{\bC}\cdot\phi$, so that the volume element commutes with the even part of $\ccl_{2m}$. Therefore,
\bea
\rho(\phi)\cdot \bS^\pm &=& \rho(\phi)\cdot\pi^\pm\cdot \bP \\
                &=& \pi^\pm\cdot\rho(\phi)\cdot \bP \\
                &=& \pi^\pm\cdot \bP \;, \quad\quad\hbox{since } \rho(\phi)\in End_{\bC}(\bP)\,, \\
                &=& \bS^\pm \;,
\eea
and we see that $\ccl^0_{2m}$ leaves $\bS^\pm$ invariant.

To prove the second part of the statement, let us explicitly set up the correspondence $\ccl_{2m-1}\cong\ccl^0_{2m}$ using the algebra isomorphism \cite{lawson} $f:\ccl_{2m-1}\longrightarrow\ccl^0_{2m}$, such that for each basis element, $f(e_a)=e_{2m}\cdot e_a \;,\quad a=1,\cdots,2m-1$.

Under this isomorphism, we see that $\o^{\bC}_{2m-1}\cong\o_{2m}^{\bC}$, where $\o^{\bC}_{2m-1}$ is the volume element for $\ccl_{2m-1}$. Therefore, since $(2m-1)$ is odd, we have from Proposition \ref{oddreprs} that either $\rho(\o^{\bC}_{2m-1})=\id$ or $\rho(\o^{\bC}_{2m-1})=-\id$.
 Moreover, that result tells us that the corresponding representations which are characterised by $\rho_{\pm}(\o^{\bC}_{2m-1})=\pm\id\,$ are indeed the two inequivalent irreducible representations of $\ccl_{2m-1}\cong\ccl^0_{2m}$.
\qed

We ultimately wish to determine the irreducible complex representations of the groups $Spin(r,s)\subset\ccl^0_n\subset\ccl_n$ for $n=r+s$.
 To do so, we must restrict the irreducible representations of $\ccl_n$ to its even subalgebra.
 In other words we must restrict the pinor representations to the spinor representations, and this is where the above two results are of great importance.

Proposition \ref{oddreprs} will prove to be useful in determining the spinor representations, since for odd $n$ we must consider the consequences of restricting each inequivalent pinor representation.
 Therefore we need such a means of distinguishing which pinor representation we are working with.

Proposition \ref{evenreprs} will play an important part in the case of $n=2m$, since it tells us that the pinor representation restricts to two inequivalent spinor representations.
 Furthermore, the space of pinors decomposes as $\bP=\bS^+\oplus\bS^-$, where $\bS^\pm$ are the $2^{m-1}$-dimensional complex \textit{spinor spaces} for $\ccl^0_{2m}$.
\section{Irreducible Representations of $Spin(r,s)$}
In this section, we focus on the groups $Spin(r,s)$ for $n=r+s$.
 Since their complex irreducible representations are found from restricting those of the even subalgebra $\ccl_n^0$, we will refer to them as the \textit{complex spinor representations} of $Spin(r,s)$.
\subsection{Even Dimension}
For $n=2m$, we must restrict the pinor representations of $\ccl_{2m}$ subject to the following inclusions:
\be
Spin(r,s)\subset\cl^0_{r,s}\subset\ccl^0_{2m}\cong\ccl_{2m-1}\subset\ccl_{2m}
\ee
From section \ref{irreps}, we know that in even dimensions $\ccl_{2m}$ has a unique complex pinor representation,
\be
\rho: \ccl_{2m}\cong End_{\bC}(\bP) \;,
\ee
where $dim_{\bC}(\bP)=2^m$.
 We must first consider the restriction of $\rho$ to the even subalgebra $\ccl^0_{2m}\cong\ccl_{2m-1}\subset\ccl_{2m}$.

We proceed by setting $\pi^\pm=\frac{1}{2}(\id\pm\rho(\o_{2m}^{\bC}))$ as before, where $\o_{2m}^{\bC}=i^m e_1\cdots e_{2m}$ is the complex volume element.
 Now, we have the decomposition
\be
\bP=\bS^+\oplus \bS^- \;,\qquad \hbox{where } \bS^\pm=\pi^\pm\cdot \bP \;,
\ee
and Proposition \ref{evenreprs} tells us that the complex subspaces $\bS^\pm$ are invariant under $\ccl_{2m}^0$.
 Furthermore, this result says that on restricting from $\ccl_{2m}$ to $\ccl^0_{2m}$, $\bS^\pm$ correspond to the two inequivalent irreducible representations of $\ccl_{2m}^0$,
\be
\rho=\rho_+\oplus\rho_- :\ccl^0_{2m}\cong End_{\bC}(\bS^+)\oplus End(\bS^-) \;.
\ee
These are the spin representations of $\ccl_{2m}^0$.

Now, an irreducible representation of $\ccl^0_{2m}$ remains irreducible when restricted to $\cl^0_{2m}\subset\ccl^0_{2m}$.
 Indeed, noting that $\ccl^0_{2m}\cong\cl^0_{2m}\otimes\bC$, we can see this as follows.
 First, suppose that the irreducible representation $\rho_+$ of $\ccl^0_{2m}$ is actually reducible on restriction to $\cl^0_{2m}$.
 Then, let $S\subset\bS^+$ be one of the supposed non-trivial complex subspaces which are invariant under the action of $\cl^0_{2m}$.
 Then $\rho_+(\ph_i)\cdot S\subseteq S$ for any two elements $\ph_i\in\cl^0_{2m}$, $i=1,2$.
 However, if we consider the element $\ph_1+i\ph_2\in\ccl^0_{2m}$, then since $\rho_+$ is a complex endomorphism, we find that
\be
\rho_+(\ph_1+i\ph_2)\cdot \s = \rho_+(\ph_1)\cdot \s + i \rho_+(\ph_2) \cdot \s \subseteq S
\ee
for every $\s\in S$.
 This implies that $S$ is an invariant subspace of $\bS^+$ under the action of $\ccl^0_{2m}$, so that $\rho_+:\ccl^0_{2m} \longrightarrow End_{\bC}(\bS^+)$ is actually reducible, which is a contradiction.
 We can argue similarly for $\rho_-$.
 Therefore, the inequivalent representations of $\ccl^0_{2m}\,$ must remain irreducible when restricted to $\cl^0_{2m}\subset\ccl^0_{2m}$.

There is one final restriction which is necessary to obtain the complex spin representations of the spin groups in even dimensions, namely to $Spin(r,s)\subset\cl^0_{2m}$.
 However, we note that every irreducible representation of $\cl^0_{2m}$ restricts to an irreducible representation of $Spin(r,s)$ for $r+s=2m$ \cite{lawson}, \cite{harvey}.
 To summarise, we have:
\begin{center}
\fbox{
\begin{minipage}{14cm}
The \textbf{complex spinor representations} of $Spin(r,s)$ for $r+s=2m$, are the two inequivalent irreducible representations
\be \label{even:spinreprs}
\rho_\pm:Spin(r,s)\subset\ccl^0_{2m}\cong End_{\bC}(\bS^\pm)\;,
\ee
where the spinor spaces $\bS^\pm$ are complex vector spaces, each of dimension $2^m$.
\end{minipage}
}
\end{center}
\subsection{Odd Dimension}
To find the complex spinor representations of $Spin(r,s)$ for $r+s=2m+1$, we must restrict the irreducible representations of $\ccl_{2m+1}$ as follows:
\be
Spin(r,s)\subset\cl^0_{r,s}\subset\ccl^0_{2m+1}\cong\ccl_{2m}\subset\ccl_{2m+1}
\ee
So the first important restriction is that of the two inequivalent pinor representations of $\ccl_{2m+1}$ to the even subalgebra.
 Recall that in odd dimensions there are two inequivalent pinor representations of $\ccl_{2m+1}$,
\be
\rho_\pm:\ccl_{2m+1}\cong End_{\bC}(\bS^\pm)\; ,
\ee
distinguished one from the other by their action on the complex volume element: $\rho_{\pm}(\o_{2m+1}^{\bC})=\pm\id$.
 Their pinor spaces $\bP^\pm$ each have complex dimension $2^m$.

To begin with, let us obtain a nice way of describing the even subalgebra of $\ccl_{2m+1}$, closely following the method of \cite{lawson}.
 Since $\left(\o_{2m+1}^{\bC}\right)^2=1$, we can decompose $\ccl_{2m+1}$ into isomorphic subalgebras as in Proposition \ref{castructure}:
\be \label{decomp}
\ccl_{2m+1}=\ccl^+\oplus\ccl^- \; ,
\ee
where $P^\pm=\frac{1}{2}\left(\id\pm\o_{2m+1}^{\bC} \right)$ and $\ccl^{\pm}=P^\pm\cdot\ccl_{2m+1}$.

Now, since $n$ is odd, the volume form satisfies $\a(\o^{\bC}_{2m+1})=-\o^{\bC}_{2m+1}$, so that
\be
\a\cdot P^\pm=P^\mp\cdot\a \;,
\ee
(see section \ref{complca} for the definition of the canonical automorphism $\a$).
 Consequently, the canonical automorphism acts as in isomorphism of subalgebras, giving
\be
\a(\ccl^{\pm})\cong\ccl^{\mp} \; .
\ee
On the other hand, the even part $\ccl^0_{2m+1}$ is characterised by the property $\a(\ph)=\ph$, for all $\ph\in\ccl^0_{2m+1}$.
 This leads us to the following lemma, which says that the even part lies diagonally within the decomposition (\ref{decomp}):
\begin{lem}
$\quad\ccl^0_{2m+1}=\left\{ \phi\in\ccl_{2m+1}:\,\phi=\vp+\a(\vp) \,,\, \vp\in\ccl^+ \right\}$.
\end{lem}
\proof
Let $\phi\in\ccl^0_{2m+1}$. Then $\phi=\vp^++\vp^-$ for $\vp^{\pm}\in\ccl^{\pm}$. Since $\phi$ is even, we have $\a(\phi)=\phi$, so that $\a(\vp^{\pm})=\vp^{\mp}$. Hence $\phi=\vp^++\a(\vp^+)$.
\qed

Now, we can investigate the consequences of restricting the representations $\rho_{\pm}$ to $\ccl^0_{2m+1}\subset\ccl_{2m+1}$.
 Since they are distinguished from each other by their action on the volume element, $\rho_\pm(\o^{\bC}_{2m+1})=\pm\id$, we see that
\be
\rho_\pm(\a(\o_{2m+1}^{\bC}))=\rho_\pm(-\o_{2m+1}^{\bC})=\mp\id \;.
\ee
In other words,
\be
\rho_\pm\cdot\a=\rho_\mp \; .
\ee

Now, we must investigate the restrictions of both $\rho_\pm$ to the even subalgebra.
 Let us first choose to restrict $\rho_+$.
 Taking $\phi=\vp+\a(\vp)\in\ccl^0_{2m+1}$ to be an arbitrary element, where $\vp\in\ccl^+$, we have
\bea
\rho_+(\phi)\,=\,\rho_+(\vp+\a(\vp))
        \,=\,\rho_+(\vp)+\rho_+(\a(\vp))
        \,=\,\rho_+(\vp)+\rho_-(\vp) \; .
\eea
Similarly, on considering $\rho_-$, we see that $\rho_-(\phi)=\rho_-(\vp)+\rho_+(\vp)$.
 Therefore, when acting on the even subalgebra, the representations coincide,
\be
\rho_+(\phi)=\rho_-(\phi) \;,\quad \forall \phi\in\ccl^0_{2m+1} \;,
\ee
so that $\rho_+$ and $\rho_-$ are \textit{equivalent} irreducible representations of $\ccl^0_{2m+1}$.

Thus, for $n=2m+1$, on restricting from $\ccl_{2m+1}$ to $\ccl^0_{2m+1}$, there is one inequivalent irreducible representation
\be
\rho :\ccl^0_{2m+1}\cong End_{\bC}(\bS) \;,
\ee
where $\rho=\rho_\pm|_{Even}$ and $\bS\cong \bP^\pm$, respectively.

Finally, since every irreducible representation of $\ccl^0_{2m+1}$ restricts to an irreducible representation of $Spin(r,s)$, we have:
\begin{center}
\fbox{
\begin{minipage}{14cm}
The \textbf{complex spinor representation} of $Spin(r,s)$ for $r+s=2m+1$, is the irreducible complex representation
\be \label{odd:spinreprs}
\rho:Spin(r,s)\subset\ccl^0_{2m+1}\cong End_{\bC}(\bS) \;,
\ee
where the spinor space $\bS$ is a complex vector space of dimension $2^m$.
\end{minipage}
}
\end{center}
This concludes our study of the classification of the complex spin representations.
%
%
%
%
%
%
%
\chapter{Classifying Almost Hermitian Manifolds}\label{GrayHerv}
\def \sum{\mathfrak{su}(m)}
A \sg\ solution consists of a Lorentzian manifold on which the fields of the theory exist, subject to the constraints arising from the governing equations.
 This means that a crucial element in attempting to classify the solutions to such a theory, is to be able to classify the possible manifolds which can arise as spacetimes.
 In this section, an extension of the Gray-Hervella classification of almost Hermitian manifolds is summarised, as this will be vital in helping to describe the geometry of spacetimes admitting timelike Killing spinors.
 Some examples of these are provided in Chapter \ref{11d}, where the \kses\ are solved in eleven dimensions.

This brief account is intended to provide a convenient and applicable way of implementing such a classification, and is not intended to be a pedagogical introduction to the notions of $G$-structures and intrinsic torsion.
 For fuller and more rigourous explanations, see for example \cite{joyce}, \cite{salamon}.
\section{The Intrinsic Torsion of Almost Hermitian Manifolds}
Let $(M,g)$ be an $n$-dimensional oriented Riemannian manifold.
 Then its frame bundle $F(M)$ is a principal $SO(n)$-bundle over $M$.
 A \textit{$G$-structure} on $M$ is a reduction of $F(M)$ to a principle $G$-bundle $R(M)$, where $G\subseteq SO(n)$ is a closed, connected subgroup \cite{joyce}, \cite{salamon}.
 A $G$-structure is equivalent to a set of $G$-invariant tensors defined on $M$, and this is often a more convenient approach to take.
 If $\gg\subseteq\son$ is the Lie algebra of $G$, then we may alternatively use the terminology \textit{$\gg$-structure} when working explicitly in Lie algebra terms.

Now, suppose that we can decompose $\son=\gg\oplus\gperp$, with respect to a $\gg$-\inv\ inner product, $<\;,\;>$.
Denoting the Levi-Civita connection on $M$ by $\nabla$, and a
compatible $\gg$-connection by $\nabla^{\gg}$, we define the
\textit{intrinsic torsion} $K$ to be the difference, \be
K=\nabla-\nabla^{\gg} \;. \ee In the cases which we will consider,
there exists a unique connection $\nabla^{\gg}$, such that $K$ takes
values in $T^*(M)\otimes\gperp$ \cite{cabrera}, where
$T^*(M)\cong\bR^n$ is treated as the vector representation of
$\so(n)$.

The independent components of a $\gg$-structure on $M$ correspond to the irreducible representations of $\gg$ in $T^*(M)\otimes \gperp$.
 Equivalently, the possible reductions of the $\son$-structure to a $\gg$-structure can be characterised by the vanishing of one or more irreducible components in the decomposition of $T^*(M)\otimes\gperp$ under $\gg$.
 If  $T^*(M)\otimes \gperp$ decomposes into $r$ $\gg$-irreducible representations, then there are $2^r$ inequivalent compatible reductions of the $\son$-structure to a $\gg$-structure.
 Then we may also say that the manifold admits a $G$-structure.

A $G$-structure is equivalent to a set of $G$-invariant tensors, denoted collectively by $\a$, and the intrinsic torsion can be represented by the $\gg$-irreducible components of $\nabla\a$.
 Below, we show that it can be equivalently described by irreducible components of the Levi-Civita spin-connection on $M$, which is denoted by $\O$.
 This description will prove extremely useful in the work on spinorial geometry, because in solving the \kses\ the field strength components are found to be largely determined by the components of $\O$, for some suitable choice of frame.
 Therefore a description of the intrinsic torsion in terms of the connection $\O$ will allow us to easily translate our results for the field strength into geometric information about the solutions.
\section{$\um$-structures on an $\sotm$-manifold}
A $2m$-dimensional real manifold $M$ equipped with a $\um$-structure is known as an \textit{almost Hermitian manifold}, and we will consider these first (see, for example, \cite{cabrera}).
 The $\um$-structure on $M$ means that there exists a Riemannian metric $g$ and an almost complex structure $J$, which is orthogonal with respect to $g$, and $U(m)$ is the subgroup of $SO(2m)$ which fixes these tensors.
 The \textit{K\"{a}hler form} associated to $(g,J)$ can be defined in the usual way, as $\o(X,Y)=g(X,JY)$.

The classification of almost Hermitian manifolds was performed by Gray and Herv\-ella in \cite{grayherv}.
 It was shown that $T^*(M)\otimes\umperp$ decomposes into four irreducible representations under $\um$, so that there are $2^4=16$ classes of almost Hermitian manifold compatible with an $\sotm$-structure.
 In situations where we have a real frame which is adapted to the $\um$-structure, as in the case of the supersymmetric backgrounds with a timelike Killing vector, the intrinsic torsion can be represented by certain components of the $\sotm$ spin-connection $\O_{A,MN}=\O_{A,[MN]}\,$ \cite{cabrera}, \cite{jguggp}.

The decomposition $\sotm=\um\oplus\umperp$ is essentially equivalent to decomposing the adjoint representation, the space of two-forms on $\bR^{2m}$, as
\be
\L^2 = \L^{(1,1)}\oplus\L^{(2,0)+(0,2)} \; ,
\ee
with respect to the adapted Hermitian frame (see for example, (\ref{11d:hermforms})).
 Or, in \linebreak representation-theoretic terms, this may be expressed as
\be
\mathbf{\frac{1}{2}(2m)(2m-1)} \longrightarrow
\mathbf{m^2}\oplus\mathbf{\frac{1}{2}m(m-1)}\oplus\mathbf{\overline{\frac{1}{2}m(m-1)}} \;.
\ee
 Expanding the $\sotm$-indices according to this decomposition, the independent components of the spin-connection are
\be
\O_{A, \b\gb}\;, \quad \O_{A, \b\g}\;, \quad \O_{A,\betab\gb}\;.
,
\ee
where lower-case barred and unbarred Greek indices denote antiholomorphic and holomorphic indices with respect to the natural Hermitian basis on $M$, respectively.

In this splitting the connection components $\O_{A,\b\gb}$ span the space $T^*(M)\otimes\um$, while the other two components span the complementary space $T^*(M)\otimes\umperp$.
 Splitting the $T^*(M)$-index, the intrinsic torsion is therefore represented by the components
\be \label{it1}
\O_{\a,\b\g}\;, \quad \O_{\a,\betab\gb} \;, \quad \O_{\ab, \b\g}\;, \quad \O_{\ab, \betab\gb}\;.
\ee
These can be related to the standard definition of the intrinsic torsion of a $\um$-structure as $\nabla \o$, by observing that
\bea
\nabla_{\ab}\o_{\b\g}   &=& 2i \O_{\ab,\b\g}
\nn \\
\nabla_{\ab}\o_{\betab\gb} &=&  -2i \O_{\ab,\betab\gb} \;. \label{it2}
\eea
To classify almost Hermitian manifolds, one must decompose the components given in (\ref{it1}) into the four
 irreducible $\um$-representations of $T^*(M)\otimes \umperp$.
 The appearance or vanishing of any of these four components determines the sixteen classes of almost Hermitian manifold to which $M$ may belong.

The first component of the intrinsic torsion can be decomposed under
$\um$ into a trace and a traceless part, as \be (\ww_3)_{\ab\b\g}
=\O_{\ab, \b\g}-\frac{2}{m-1}\O_{\db,}{}^{\db}{}_{[\g} g_{\b]\ab}
\;, \qquad (\ww_4)_{\g}=\O_{\betab,}{}^{\betab}{}_\g
\ee
and the second component can be decomposed as
\be
(\ww_1)_{\ab \betab\gb}=\O_{[\ab, \betab\gb]} \;, \qquad
(\ww_2)_{\ab \betab\gb} =\O_{\ab,\betab\gb} - \O_{[\ab,\betab\gb]} =
\frac{2}{3}\O_{\ab,\betab\gb}-\frac{1}{3}\O_{\gb, \ab\betab}
-\frac{1}{3} \O_{\betab,\gb\ab} \; ,
\ee
so that
$\O_{\ab,\betab\gb}=(\ww_1)_{\ab \betab\gb}+(\ww_2)_{\ab
\betab\gb}$.

The above irreducible components explicitly illustrate the decomposition of \linebreak $T^*(M)\otimes \umperp$ under $\um$.
 For instance, consider the case $m=5$, which will be of some significance in the analysis of eleven-dimensional \sg\ backgrounds with a timelike \ks.
 In this case, $\O_{A,MN}$ lies in the space $T^*(M)\otimes\mathfrak{so}(10)$, and so it carries the $\mathbf{10}\otimes\mathbf{45}$ representation of $\mathfrak{so}(10)$.
 Now, decomposing the indices according to the $\mathfrak{u}(5)$-adapted Hermitian basis shows that the connection carries the $(\mathbf{5}\oplus\mathbf{\bar{5}})\otimes(\mathbf{10}\oplus\mathbf{\bar{10}}\oplus\mathbf{25})$ $\mathfrak{u}(5)$-representation.
 Next, we separate out and disregard the components $\O_{\a,\b\gb}$ and $\O_{\ab,\b\gb}$, since they carry the $\mathbf{25}$ of $\mathfrak{u}(5)$ on their $(\b\gb)$-indices, which means that they correspond to the $T^*(M)\otimes\mathfrak{u}(5)$ piece of the space.
 Therefore, the intrinsic torsion which lies in the complementary space $T^*(M)\otimes\mathfrak{u}(5)^{\perp}$ must correspond to the representation $\left(\mathbf{5}\oplus\mathbf{\bar{5}}\right)\otimes\left(\mathbf{10}\oplus\mathbf{\bar{10}}\right)$.
 The decomposition of this product into irreducible representations of $\su(5)$ is
\be
\left(\mathbf{5}\oplus\mathbf{\bar{5}}\right)\otimes\left(\mathbf{10}\oplus\mathbf{\bar{10}}\right)
\longrightarrow \left(\mathbf{5}\oplus\mathbf{\bar{5}}\right)
\oplus\left(\mathbf{10}\oplus\mathbf{\bar{10}}\right) \oplus\left(\mathbf{40}\oplus\mathbf{\bar{40}}\right)
\oplus\left(\mathbf{45}\oplus\mathbf{\bar{45}}\right) \;,
\ee
where
\bea
\ww_1  \leftrightarrow  \mathbf{10}\oplus\mathbf{\bar{10}}\;, \quad
\ww_2  \leftrightarrow  \mathbf{40}\oplus\mathbf{\bar{40}}\;, \quad
\ww_3  \leftrightarrow  \mathbf{45}\oplus\mathbf{\bar{45}}\;, \quad
\ww_4  \leftrightarrow  \mathbf{5}\oplus\mathbf{\bar{5}} \;.
\eea

More generally, we use (\ref{it2}) to directly relate the classes $\ww_i$ to the classes of the Gray-Hervella classification, which are denoted by $\mathcal{W}_i$ \cite{grayherv}.
 We can then verify the result of that reference which provides the dimensionalities of the intrinsic torsion modules:
\bea
&& dim(\mathcal{W}_1)   = \frac{1}{3}m(m-1)(m-2) \;, \quad
dim(\mathcal{W}_2)  = \frac{2}{3}m(m-1)(m+1) \;, \\
&& dim(\mathcal{W}_3)   = m(m+1)(m-2) \;, \quad
dim(\mathcal{W}_4)  = 2m \;, \\
&& dim(\mathcal{W}) = 2dim(\mathcal{W}_1\oplus\mathcal{W}_2)= 2dim(\mathcal{W}_3\oplus\mathcal{W}_4)
= 2m^2(m-1) \;,
\eea
In our formalism, we need only count the degrees of freedom of each component $\ww_i$ (plus its complex conjugate).

As an illustration of the Gray-Hervella classification of almost Hermitian manifolds, we describe two important classes which arise. Firstly, when the only non-vanishing components are $\ww_3$ and $\ww_4$, then the intrinsic torsion lies in the class $\mathcal{W}_3\oplus\mathcal{W}_4$ and according to the classification, $M$ is Hermitian.
 This proves to be the case when an eleven-dimensional background possesses two \suf-\inv\ \kss, as in section \ref{11d:2su5}.
 Secondly, when all the components vanish, i.e. each module $\mathcal{W}_i$ vanishes, then $M$ is K\"{a}hler.
 Note also that although the above analysis holds for all values of $m$, some classes vanish identically for $m=1,2$.
\section{$\sum$-structures on an $\sotm$-manifold}
It is straightforward to extend the classification of $\um$-structures to the that of $\sum$-structures, for $n\geq3$, \cite{salamon}, \cite{cabrera}.
 If $M$ admits an $\sum$-structure then it is known as a \textit{special almost Hermitian manifold}.
 As well as possessing a Riemannian metric $g$ and almost complex structure $J$, $M$ is also equipped with an $\sum$-\inv\ complex volume form $\Psi$, of unit length with respect to the standard Hermitian inner product which is induced by $g$.

In this case, we have the decomposition $\sotm=\sum\oplus\bR\oplus\umperp$, which can also be expressed in terms of the dimensionality of representations as
\be
\mathbf{\frac{1}{2}(2m)(2m-1)} \longrightarrow
\mathbf{(m^2-1)}\oplus\,\mathbf{1}\,\oplus\mathbf{\frac{1}{2}m(m-1)}\oplus\mathbf{\overline{\frac{1}{2}m(m-1)}} \;.
\ee
Therefore, we have $\sumperp=\bR\oplus\umperp$.
 The intrinsic torsion now takes values in
\bea \label{sumperp}
T^*(M)\otimes\sumperp &=& T^*(M)\otimes\left(\bR\oplus\umperp\right) \nn \\
              &=& T^*(M)\oplus\left(T^*(M)\otimes\umperp\right) \;.
\eea

In terms of the Levi-Civita spin-connection, the space $T^*(M)\otimes\sum$ is spanned by the component which is traceless and $(1,1)$ with respect to its $\sum\subset\sotm$-indices,
\be
\mbox{i.e. }\;\O_{A,\b\gb}^0=\O_{A,\b\gb}-\frac{1}{m}\O_{A,\d}{}^{\d}g_{\beta\gb} \; .
\ee
With this in mind, the space $T^*(M)\otimes\left(\bR\oplus\umperp\right)$ as given in (\ref{sumperp}) is spanned by the four components arising in the $\um$-structure case, as well as an additional component to account for the removal of the trace from the $\O_{A,\b\gb}$.
 Therefore the independent components of the intrinsic torsion are represented by \cite{cabrera}
\be \label{it3}
\O_{\ab, \b\g}\;, \qquad\O_{\ab, \betab\gb}\;, \qquad \O_{\ab,\b}{}^\b \;,
\ee
and their complex conjugates.
 Now, decomposing the intrinsic torsion under $\sum$ gives the five irreducible components
\bea \label{sumit}
(\ww_1)_{\ab \betab\gb} &=& \O_{[\ab, \betab\gb]} \;, \\
(\ww_2)_{\ab \betab\gb} &=& \frac{2}{3}\O_{\ab,\betab\gb}-\frac{1}{3}\O_{\gb, \ab\betab}
                -\frac{1}{3}\O_{\betab,\gb\ab} \;, \\
(\ww_3)_{\ab\b\g}   &=& \O_{\ab,\b\g}-\frac{2}{m-1}\O_{\db,}{}^{\db}{}_{[\g} g_{\b]\ab} \;, \\
(\ww_4)_\a      &=& \O_{\betab,}{}^{\betab}{}_\a \;, \\
(\ww_5)_{\ab}       &=& \O_{\ab,\b}{}^\b \;.
\eea
The vanishing of one or more of the above components characterises the $2^5=32$ classes of special almost Hermitian manifolds.

Let us return to the $m=5$ example.
 The intrinsic torsion now lies in the space $T^*(M)\otimes\mathfrak{su}(5)^{\perp}$.
 The connection components $\O_{A,MN}$, which carry the $\mathbf{10}\otimes\mathbf{45}$ of $\mathfrak{so}(10)$, decompose under $\mathfrak{su}(5)$ in the following way:
\be
\mathbf{10}\otimes\mathbf{45} \longrightarrow
\left(\mathbf{5}\oplus\mathbf{\bar{5}}\right)
\otimes\left(\mathbf{1}\oplus\mathbf{24}\oplus\mathbf{10}\oplus\mathbf{\bar{10}}\right) \;.
\ee
However, the $\mathbf{24}$ corresponds to the traceless piece $\O^0_{A,\b\gb}$ which spans the space \linebreak $T^*(M)\otimes\mathfrak{su}(5)$, and so we disregard this term.
 The intrinsic torsion lies in the space $T^*(M)\otimes\mathfrak{su}(5)^{\perp}$, which decomposes into irreducible $\mathfrak{su}(5)$-representations as \cite{pakis}
\benn
\left(\mathbf{5}\oplus\mathbf{\bar{5}}\right)
\otimes\left(\mathbf{1}\oplus\mathbf{10}\oplus\mathbf{\bar{10}}\right)
\longrightarrow
\left(\mathbf{5}\oplus\mathbf{\bar{5}}\right)\oplus\left(\mathbf{5'}\oplus\mathbf{\bar{5}'}\right)
\oplus\left(\mathbf{10}\oplus\mathbf{\bar{10}}\right) \oplus\left(\mathbf{40}\oplus\mathbf{\bar{40}}\right)
\oplus\left(\mathbf{45}\oplus\mathbf{\bar{45}}\right) \;.
\eenn
The components $\ww_1,\ww_2,\ww_3,\ww_4$ correspond to the same pairs of irreducible representations as in the $\mathfrak{u}(5)$-structure case.
 In addition, in this case we have the correspondence $\ww_5\rightarrow\mathbf{5'}\oplus\mathbf{\bar{5}'}$, arising from the extra trace piece.

In passing, we note that all the components of the intrinsic torsion in the $\um$-structure case can be derived from $\nabla\o$.
 However, $\ww_5$ does not arise from this derivative, therefore some extra information is needed to specify this extra component.
 In fact, this information is given by the complex volume form $\Psi$, and in \cite{cabrera} it is shown how the intrinsic torsion of an $\sum$-structure may be derived completely from either $\nabla\mathcal{R}e(\Psi)\,$ or $\,\nabla\mathcal{I}m(\Psi)$.
\section{$\umm$-structures on an $\sotm$-manifold}
Now we turn to investigate $\umm$-structures on an $\sotm$ manifold.
 There are a number of equivalent ways to embed $\umm$ into $\um$, but for our purposes we will achieve this by requiring that $m$-direction of the adapted basis be left invariant under $\umm$.
 We can express this by splitting all tensor indices according to $\alpha=(i,m)$, where $i=1,\cdots,m-1$, and the $i$-indices transform under $\umm$.
 The following analysis works for all values of $m$, but note that for $m\leq 4$ some of the classes vanish identically.

Here, we have $\sotm=\umm\oplus\ummperp$ and the intrinsic torsion lies in the space $T^*(M)\otimes\ummperp$.
 We see that
\be
dim(\ummperp)=\frac{1}{2}(2m)(2m-1)-(m-1)^2=m^2+m-1 \; ,
\ee
so that the intrinsic torsion carries the $\mathbf{2m}\otimes(\mathbf{m^2+m-1})$ representation of $\sotm$.
 Again, we use the spin-connection to represent the space $T^*(M)\otimes\ummperp$.
 Decomposing the indices of $\O_{A,MN}$ with respect to the Hermitian basis, splitting off the $m$-index and disregarding the components $\O_{A,i\bar j}$ which carry the adjoint of $\umm$ on the $(i\bar j)$ indices, we obtain the following independent components:
\bea \label{it4}
&&      \O_{\bar i, \bar j\bar k} \;,   \quad \O_{\bar i, j k} \;,
    \quad \O_{\bar i, \bar j \bar m} \;,    \quad \O_{\bar i, j m}  \;,
    \quad \O_{\bar i, \bar j m} \;,     \quad \O_{\bar i, j \bar m} \;,
\\
&&      \O_{\bar m, jk} \;,     \quad  \O_{\bar m, \bar j\bar k} \;,
    \quad \O_{\bar i, m\bar m} \;,      \quad \O_{\bar m, m\bar m} \;,
    \quad \O_{\bar m, mj} \;,       \quad \O_{\bar m, \bar m j} \;,
    \quad \O_{\bar m, m \bar j} \;,     \quad \O_{\bar m, \bar m \bar j} \;.
\nn
\eea
However, the components in the first row are further reducible under $\umm$.
 We decompose these into irreducible representations of $\umm$ by symmetrisation, antisymmetrisation and by removing traces where appropriate, to give the independent components
\begin{center}
\begin{tabular}{ll} \label{ummit1}
$(\ww_1)_{\bar i\bar j\bar k} = \O_{[\bar i, \bar j\bar k]}\;,$
& $(\ww_2)_{\bar i\bar j\bar k} = \frac{2}{3}\O_{\bar i, \bar j\bar k} - \frac{1}{3}\O_{\bar k, \bar i\bar j} - \frac{1}{3}\O_{\bar j, \bar k \bar i} \;,$ \\
$(\ww_3)_{\bar i j k} = \O_{\bar i, j k} - \frac{2}{(m-2)}\O_{\bar m}{}^{\bar m}{}_{[k} g_{j]\bar i} \;,$
& $(\ww_4)_{k} = \O_{\bar j,}{}^{\bar j}{}_k  \;,$
\\
$(\vv_1)_{\bar i \bar j\bar m}  = \O_{[\bar i,\bar j]\bar m} \;, \quad$
$(\vv_2)_{\bar i \bar m\bar m}  = \O_{(\bar i,\bar j)\bar m} \;,$
& $(\vv_5)_{\bar i \bar j m}  = \O_{[\bar i,\bar j] m} \;, \quad$
$(\vv_6)_{\bar i \bar j m}  = \O_{(\bar i,\bar j) m} \;,$
\\
$(\vv_3)_{\bar i j m}  = \O_{\bar i, j m} - \frac{1}{m-1}g_{\bar i j}\O_{\bar k,}{}^{\bar k}{}_{m} \;,$
& $(\vv_4)_{m}  = \O_{\bar k,}{}^{\bar k}{}_{m} \;,$
\\
$(\vv_7)_{\bar i j \bar m}  = \O_{\bar i, j \bar m} - \frac{1}{m-1}g_{\bar i j}\O_{\bar k,}{}^{\bar k}{}_{\bar m} \;,$
& $(\vv_8)_{\bar i j \bar m}  = \O_{\bar k,}{}^{\bar k}{}_{\bar m} \; .$
\end{tabular}
\end{center}
The remaining components in the bottom row of (\ref{it4}) are already irreducible representations of $\umm$, and so we can use them to define the remaining independent components,
\begin{center}
\begin{tabular}{llll} \label{ummit2}
$(\ww_5)_{jk} = \O_{\bar m, jk} \;,$        &  $(\ww_6)_{\bar j\bar k} = \O_{\bar m, \bar j\bar k} \;,$
& $(\ww_7)_{\bar i} = \O_{\bar i, m\bar m} \;,$ &  $\ww_8=\O_{\bar m, m\bar m}\;,$ \\
$(\vv_9)_{j} = \O_{\bar m, m j} \;,$        &  $(\vv_{10})_{\bar j} = \O_{\bar m, \bar m\bar j} \;,$
& $(\vv_{11})_{j} = \O_{\bar m, \bar m j} \;,$  &  $(\vv_{12})_{\bar j} = \O_{\bar m, m\bar j} \;.$
\end{tabular}
\end{center}
We see that the space $T^*(M)\otimes\ummperp$ decomposes into twenty irreducible modules under $\umm$, so that the existence or vanishing of any combination of these modules characterises the $2^{20}$ inequivalent classes of $\sotm$-manifold possessing a $\umm$-structure.
\section{$\summ$-structures on an $\sotm$-manifold}
To investigate the $\summ$-structures that may arise on an $\so(2m)$-manifold, we begin be observing that $\umm=\summ\oplus\bR$, so that $\sotm=\summ\oplus\bR\oplus\ummperp$.
 This means that the intrinsic torsion must now lie in the space
\bea \label{summperp}
T^*(M)\otimes\summperp  &=& T^*(M)\otimes\left( \bR\oplus\ummperp \right) \nn \\
            &=& T^*(M)\oplus\left( T^*(M)\otimes\ummperp\right) \;.
\eea
Also, we can see that
\be
dim(\summperp)=\frac{1}{2}(2m)(2m-1)-((m-1)^2-1)=m(m+1) \; ,
\ee
so that the intrinsic torsion carries the $\mathbf{2m}\otimes(\mathbf{m(m+1)})$ representation of $\sotm$.

In terms of the Levi-Civita spin-connection, the space $T^*(M)\otimes\summ$ is spanned by the component which is traceless and $(1,1)$ with respect to its $\summ\subset\sotm$-indices,
\be
\mbox{i.e. }\;\O_{A,j \bar k}^0 = \O_{A,j \bar k} - \frac{1}{m-1}\O_{A,k}{}^{k}g_{j\bar k} \; .
\ee
With this in mind, the space $T^*(M)\otimes\left(\bR\oplus\umperp\right)$ as given in (\ref{summperp}) is spanned by the twenty components arising in the $\umm$-structure case, see equation (\ref{it4}), as well as additional components to represent the trace of $\O_{A,j\bar k}$.
 We can therefore define the additional $\summ$-irreducible components by \cite{jguggp}
\be \label{it5}
(\ww_9)_{\bar i} = \O_{\bar i,j}{}^{j} \;, \qquad (\ww_{10}) = \O_{\bar m, j }{}^{j} \;,
\ee
and their complex conjugates.
 The twenty classes $\ww_1,\cdots,\ww_8$ and $\vv_1,\cdots,\vv_{12}$ are as in the $\umm$-structure case.
 Thus the space $T^*(M)\otimes\summperp$ decomposes into twenty-two irreducible modules under $\summ$, so that the existence or vanishing of any combination of these modules characterises the $2^{22}$ inequivalent classes of $\sotm$-manifold possessing an $\summ$-structure.

We see examples of $\su(4)$ structures in $\so(10)$ manifolds in eleven-dimensional \sg\ backgrounds possessing three or four \sufr-\inv\ \kss.
 The existence of a timelike Killing vector in the spacetime effectively leads to a ten-dimensional almost Hermitian spatial manifold $B$, transverse to the orbits of this vector.
 These configurations of \kss\ also lead to an eight-dimensional submanifold $\hat{B}\subset B$ possessing an $\su(4)$-structure.
 The non-vanishing intrinsic torsion modules are described in section \ref{11d:n>2}.
\section{Lorentzian Signature}
So far we have investigated $\gg$-structures for manifolds of Euclidean signature.
 In fact, the analysis can be extended to Lorentzian signature manifolds in a straightforward way.
 We shall not examine this in detail but rather we shall describe the $\son$-structures which can arise in an $\mathfrak{so}(n,1)$ manifold, as this is the case which is of relevance in the context of supersymmetric spacetime backgrounds.

In this case, $\mathfrak{so}(n,1)=\son\oplus\sonperp$, and the intrinsic torsion lies in the space $T^*(M)\otimes\sonperp$, where $T^*(M)\cong\bR^{n,1}\,$ is treated as the vector representation of $\so(n,1)$.
 We see that $dim(\sonperp)=\frac{1}{2}(n+1)n-\frac{1}{2}n(n-1)=n$, so that the intrinsic torsion carries the representation $(\mathbf{n+1})\otimes\mathbf{n}$ of $\mathfrak{so}(n,1)$.
 In terms of the spin-connection $\O_{A,MN}$, which lies in the space $T^*(M)\otimes\mathfrak{so}(n,1)$, we can split the frame index as $A=(0,i)$, where $i=1,\cdots,n$.
 Then the independent components of the connection are
\be
\O_{0,0i} \;,\qquad \O_{i,j0} \;, \qquad \O_{0,jk} \;, \qquad \O_{i,jk} \;.
\ee
However, the latter two carry the adjoint representation of $T^*(M)\otimes\son$ on their $(jk)$ indices, and so we may disregard them.
 The independent components which represent the intrinsic torsion in $T^*(M)\otimes\sonperp$ are therefore
\be
\O_{0,0i} \;, \qquad \O_{i,j0} \;.
\ee
Indeed, together these components possess $n(n+1)$ degrees of freedom, as required.

Now, we can decompose these components under $\son$ into four irreducible representations,
\bea \label{sonit}
(\rm{y}_1)_{ij} = \O_{[i,j]0} \;, ~~~ (\rm{y}_2)_{ij} = \O_{(i,j)0}-\frac{1}{n}g_{ij}\O_{k,}{}^k{}_0 \;,
~~~ (\rm{y}_3) = \O_{k,}{}^k{}_0 \;, ~~~ (\rm{y}_4)_i = \O_{0,0 i} \;.
\eea
The vanishing of one or more of the above components of the intrinsic torsion characterises the $2^4=16$ inequivalent $\son$-structures of an $\mathfrak{so}(n,1)$ manifold.

We can combine the results of this section with those we have presented for the case of Euclidean signature manifolds.
 In particular, we can determine the $\um$-, $\sum$-, $\umm$- and $\summ$-structures which can arise on an $\mathfrak{so}(2m,1)$ manifold.
 For example, for each of the $16$ $\sotm$-structures on an $\mathfrak{so}(2m,1)$ manifold, there are $16$ inequivalent $\um$-structures.
 Thus there are $256$ possible $\um$-structures in an $\mathfrak{so}(2m,1)$ manifold.

The conditions for $\n=1$, $2$, $3$ and $4$ supersymmetry in eleven-dimensional \sg, which are derived in Chapter \ref{11d}, can be viewed as particular $\mathfrak{su}(5)$- and $\mathfrak{su}(4)$- structures in an $\mathfrak{so}(10,1)$ manifold.
 In our method, the conditions which arise from the \kses\ are naturally derived in terms of spacetime Levi-Civita spin-connection, which is the reason for the analysis of this appendix.
%
%
%
%
%
%
%
\chapter{Spinorial Geometry in $D=5$ Supergravity}\label{5d}
In this section, as a simple application of the formalism outlined in Chapters \ref{prelim} and \ref{11d} for solving the \kses, we will briefly consider the case of minimal $D=5$, $N=1$ \sg.
 The solutions to this theory have already been classified in \cite{5dclass}, but we investigate it here as a further illustrative example of our method.

In five dimensions, there are two distinct non-trivial orbits of $Spin(1,4)$ on its spinor space, namely the orbits of \sut\ and \rt-\inv\ spinors \cite{bry}.
 In this appendix, we will restrict our attention to the former orbit.
\section{Minimal Supergravity in $5$ Dimensions}
In minimal $D=5$ \sg\ \cite{crem} the bosonic fields are the metric $g$ and a one-form gauge potential $A$ with corresponding two-form field-strength $F$.
 The bosonic part of the action is
\be
I = \frac{1}{4\pi G} \int_{M} -\frac{1}{4}\mathcal{R} d{\rm vol} - \frac{1}{2} F \we\h F - \frac{2}{2\sqrt3} A \we F \we F \;,
\ee
where $\mathcal{R}$ is the scalar curvature of $g$ and $d{\rm vol}=+\sqrt{|g|}dx^0\we\cdots\we dx^{4}$ is the oriented volume element.

The resulting bosonic field equations are
\bea
0 &=& R_{MN} + 2\left( F_{MP}F_{N}{}^{P} - \frac{1}{6} g_{MN}F^2 \right)
 \\
0 &=& d \h F + \frac{2}{\sqrt3} F\we F \;,
\eea
where spacetime indices are labelled by $M,N,P=0,1,2,3,4$.

In considering bosonic solutions, the gravitino is set to zero so that the supersymmetry variations of $g$ and $A$ vanish.
 Then, the condition for supersymmetry to be preserved is that the supersymmetry variation of the gravitino itself should remain zero.
 In other words, there exists a non-zero spinor $\e$ satisfying \cite{5dblackholes}
\bea \label{5d:kse}
0 = \d_{\e}\ps_M = \mathcal{D}_M\e = \left\{ \partial_M + \frac{1}{4}\O_{M,NP}\G^{NP} +
\frac{i}{4\sqrt3}\left\{ \G_M{}^{NP} - 4\d_M^N\G^P \right)F_{NP} \right\}\e  \;,
\eea
where $\O$ is the Levi-Civita spin-connection\footnote{The factor of $i$ in the supercovariant derivative is a consequence of using the `mostly plus' metric.}.

In $D=5$ \sg, there are no Majorana spinors but instead we have \textit{symplectic Majorana} (SM) spinors \cite{crem}, \cite{trau}, \cite{west}.
 Denoting the complex representation of \spf\ by $\Dc$, a symplectic Majorana spinor consists of a doublet
\be
\z = \begin{pmatrix} \z_1 \\ \z_2 \end{pmatrix} \in \Dc\oplus\Dc \;,
\ee
which transforms under $SU(2)\cong Sp(1)$ and satisfies a generalised Majorana condition which relates components of each copy of $\Dc$.

To be more precise, the complex spin representation of \spf\ has complex dimension four \cite{lawson}, hence it has $8$ real degrees of freedom.
 However the requirement of \sut\ internal symmetry demands that our spinors live the space $\Dc\oplus\Dc$, which has $16$ real dimensions.
 Therefore, we must impose a reality condition to relate the components of the two copies of $\Dc$, in order to reduce the number of degrees of freedom to $8$.
 The appropriate constraint is provided by the \textit{symplectic Majorana condition} \cite{crem}
\be \nn
\z^{\dagger}_{k} \G_0 = (\z^k)^{T} B \;,
\ee
from which we derive
\be
\z^* = - \left( \G_0 B \otimes \e \right) \z \; , \quad \mbox{ie.} \; \begin{pmatrix} \z^{1*} \\ \z^{2*} \end{pmatrix} = \begin{pmatrix} -\G_0 B \z^2 \\ \G_0 B \z^1 \end{pmatrix} \; .
\ee
\section{An Explicit Representation of $Spin(1,4)$}
We will now apply the formalism presented in section \ref{explreprs} to the five-dimensional case.
 We begin by constructing the \spfr\ module $\Dc$, and then extend the representation to one of \spf.
\subsection{Spinors and Gamma Matrices}
Let $V=\bR^{4}\cong\bC^2$ be equipped with Euclidean inner product $(,)$.
 Locally, $V$ admits an almost complex structure $J$ which is orthogonal with respect to $(\,,\,)$, and also a set of vectors $\{e_1,e_2\}$ such that
\be
\{e_1,e_2,J(e_1),J(e_2)\}
\ee
is an orthonormal basis for $V$ (see section \ref{explreprs} and references therein).
 Thus $J$ induces an orthogonal splitting $V=U\oplus J(U)$ where $U$ is the real subspace spanned by $\{e_1,e_2\}$.
 We make the identifications $e_{j+2}=J(e_j)$, for $j=1,2$.
 We note in passing that $V$ possesses a $U(2)$-structure specified by the tensors $(g,J)$.

Now from section \ref{explreprs}, the space of complex pinors associated to $V$ is obtained as follows.
 First, we complexify the vector space, $\Vc=V\otimes\bC$, and form the natural Hermitian inner product $<\,,\,>$ which is induced by $(\,,\,)$.
 This Hermitian inner product can then be extended to act on the exterior algebra $\L^*(\Vc)\cong\cl(\Vc)\cong\ccl_{4}$ in the natural manner.
 Then, the pinor module of $\ccl_{4}$ is given by the $2^2$-dimensional complex vector space $\bP= \L^*(\Uc)$.
 Also, the unique pinor representation which maps $\ccl_{4}$ into $End_{\bC}(\bP)$ is provided by the following gamma matrices:
\bea \label{spin4repr}
\G_j      &=& e_j \we \cdot + e_j \lc \cdot \\
\G_{j+2}  &=& ie_j \we \cdot - ie_j \lc \cdot  \;\;,
\eea
where $\lc$ is the adjoint of $\we$ with respect to $<\,,\,>$ and $j=1,2$.
 It is straightforward to show that $\G_j$ and $\G_{j+2}$ are Hermitian with respect to $<,>$, and that $\{\G_i,\G_j\} = \{\G_{i+2},\G_{j+2}\} = 2\delta_{ij} $ and $\{\G_i,\G_{j+2}\} = 0$, as required.

Now, we wish to investigate the group $Spin(1,4)$, so we continue by introducing the time-component gamma matrix as in (\ref{odd:gamma2})\footnote{In fact, we have the choice $\G_0=\pm i\G_1\G_2\G_3\G_4$, corresponding to the two inequivalent pinor representations of $\ccl_{5}$, but we choose the plus sign as our convention.},
\be \label{gamma4}
\G_0=i\G_1\G_2\G_3\G_4 \; .
\ee
It is elementary to show that $(\G_0)^2=-\id$, $\G_0$ anticommutes with $\G_I$, for $I=1,2,3,4$, and together they satisfy the algebra of $\cl_{1,4}\otimes\bC$, namely $\{\G_A,\G_B\}=2\eta_{AB}$, where $\eta_{AB}=diag(-1,1,1,1,1)$.

Following the details of section \ref{explreprs}, on restricting to $Spin(1,4)\subset\ccl_{5}$, the representation specified by (\ref{spin4repr}) and (\ref{gamma4}) provides the unique spin representation of $Spin(1,4)$ on the spinor module
\be \label{5dimspinors}
\boxed{
\Dc=\bP=\cl(\Uc)\cong\L^*(\bC^2) \;.
}
\ee
This is also known as the space of \textit{Dirac spinors}.
 Thus, every spinor of \spf\ can be written as an element of the exterior algebra of $\bC^2$ in this representation.
\subsection{A Basis of Symplectic Majorana Spinors}
Recall that the spinors of five-dimensional \sg\ are symplectic Majorana (SM).
 In other words, a spinor $\z$ is of the form
\be \label{5d:SMcond}
\begin{pmatrix} \z^{1*} \\ \z^{2*} \end{pmatrix} = \begin{pmatrix} -\G_0 B \z^2 \\ \G_0 B \z^1 \end{pmatrix}  \in \Dc\oplus\Dc\; .
\ee
We denote the space of SM spinors, i.e. the subspace of $\Dc\oplus\Dc$ satisfying (\ref{5d:SMcond}), by $\Dsm$.
 Let us now derive a basis for this space.

Consider an arbitrary spinor in $\Dc\oplus\Dc$ with first component $\z^1=a1+se_1+te_2+be_{12}$, for $a,b,s,t\in\bC$.
 Applying condition (\ref{5d:SMcond}), we obtain
\be
\z = \begin{pmatrix} a \\ -ib^* \end{pmatrix} 1   + \begin{pmatrix} s \\ it^* \end{pmatrix} e_1
   + \begin{pmatrix} t \\ -is^* \end{pmatrix} e_2 + \begin{pmatrix} b \\ ia^* \end{pmatrix} e_{12} \;.
\ee
Therefore, this is the general form of a spinor in $\Dsm$.
 We see that $\z$ is parametrised by the $8$ real parameters given by the real and imaginary parts of $a,b,s,t\in\bC$, and it is clear that the second component of the spinor, $\z^2$, is completely determined by the first.
 Thus, the space of SM spinors, $\Dsm$, is precisely the subspace of $\Dc\oplus\Dc$ for which the second component is related to the first in the above way.

Expanding the complex numbers into real and imaginary parts, we see that $\z$ is a linear combination of the following $8$ independent SM spinors:
\begin{center}
\begin{tabular}{ccc}
    $\begin{pmatrix} 1 \\  0 \end{pmatrix} 1 + \begin{pmatrix} 0 \\ i \end{pmatrix} e_{12}$ \;,\qquad
&       $\begin{pmatrix} i \\  0 \end{pmatrix} 1 + \begin{pmatrix} 0 \\ 1 \end{pmatrix} e_{12}$ \;,\qquad
&       $\begin{pmatrix} 0 \\ -i \end{pmatrix} 1 + \begin{pmatrix} 1 \\ 0 \end{pmatrix} e_{12}$ \;,\nn  \\
    $\begin{pmatrix} 0 \\ -1 \end{pmatrix} 1 + \begin{pmatrix} i \\ 0 \end{pmatrix} e_{12}$ \;,\qquad
&       $\begin{pmatrix} 1 \\ 0 \end{pmatrix} e_1 + \begin{pmatrix} 0 \\ -i \end{pmatrix} e_2$  \;,\qquad
&       $\begin{pmatrix} i \\ 0 \end{pmatrix} e_1 + \begin{pmatrix} 0 \\ -1 \end{pmatrix} e_2$  \;,\nn \\
    $\begin{pmatrix} 0 \\ i \end{pmatrix} e_1 + \begin{pmatrix} 1 \\  0 \end{pmatrix} e_2$  \;,\qquad
&   $\begin{pmatrix} 0 \\ 1 \end{pmatrix} e_1 + \begin{pmatrix} i \\  0 \end{pmatrix} e_2$  \;.
&   ~
\end{tabular}
\end{center}
This set forms a basis for the SM spinors of $N=1$, $D=5$ \sg.
\subsection{A $Spin(1,4)$-invariant Inner Product and Spacetime Forms}\label{sutip}
Following the theory of \ref{forms}, one can write down a \spf-invariant inner product, from which the \spf-invariant spacetime forms associated to the spinors of $\Dsm$ can be constructed.
 From the choice of $\A$ or $\B$, we choose to work with $\B$, since it is the $Pin(4)$-invariant one.
 This is required so that the inner product extends to being \spf-\inv\ (see section \ref{forms} for details).

Defining the map $\b=\G_3\G_4$, a $Spin(1,4)$-invariant inner product on $\Dsm$ is given by
\bea
\mathcal{B}(\zeta,\eta) &=& \frac{1}{2} < \, \b(\zeta^{i})^*\,,\,\e_{ij}\,\eta^j \, >  \;,
\eea
where $*$ denotes standard complex conjugation in $\Lambda^{*}(U_{\bC})$.
 Note that we must incorporate the $SU(2)$-\inv\ symplectic form $\e_{ij}$ to contract the symplectic indices, so that the output of the inner product is purely a complex number, as required.

Using this bilinear form, we can construct every \spf-\inv\ exterior form associated to a given pair of spinors. For $\eta^k,\xi^l\in\D^{\bC}$, the associated forms are given by
\be \nn
\a_{A_1\cdots A_p}(\eta^k,\xi^l) =  B(\eta^k,\G_{A_1\cdots A_p}\xi^l) \;, \quad\hbox{where $A_i=0,\cdots,4$ and $\;i=0,\cdots,p$.}
\ee
Note that for this definition of the spacetime forms, the symplectic indices remain free, i.e. they are not contracted with the symplectic form. Therefore, we can form three distinct $p$-forms from a pair of spinors $\z,\xi\in\Dsm$ with components $\eta^k,\z^l$, corresponding to $(k,l)=(1,1),(1,2)$ or $(2,2)$.

Also, we need only compute forms up to degree $2$, since the forms of higher degrees are related by Poincar\'{e} duality.
\section{Orbits of \spf\ in $\Dc$}
We saw in the eleven-dimensional case that there are two non-trivial orbits of \linebreak \spt\ in its spinor space, namely the orbits of \suf\ and \sps-\inv\ spinors.
 A similar scenario occurs in five dimensions.
 Referring to the invaluable work of Bryant \cite{bry}, we find that in the present case there are also just two non-trivial orbits for non-zero spinors in $\Dsm$, namely those with stabiliser group \sut\ and those with stabiliser \rt.
 We denote these orbits by \orbsut\ and \orbrt.
 In the following, we will solve the \kses\ for the former case.
\section{An Hermitian basis of gamma matrices}
In the analysis of the \kses\ it will prove very useful to work in an Hermitian basis of gamma matrices, defined by
\bea
\Gh_{\a} &=& \frac{1}{\sqrt2} \left( \G_{\a} - i \G_{\a +2}  \right)  \\
\Gh_{\betab} &=& \frac{1}{\sqrt2} \left( \G_{\b} + i \G_{\b +2}  \right) \quad\quad \a, \b = 1,2 \; .
\eea
Holomorphic and antiholomorphic indices are raised and lowered using the standard Hermitian metric $g_{\a \betab} = \delta_{\a \betab}$ on $\bC^2$. The Clifford algebra relations in this basis are $\{\Gh_{\a},\Gh_{\betab}\}=2g_{\a \betab}$ and $\{\Gh_{\a},\Gh_{\b} \} = \{\Gh_{\ab},\Gh_{\betab}\} = 0$.

Using these conditions and the fact that $e_{12} =
\frac{1}{4}\ve_{\ab\betab}\Gh^{\ab\betab}(1)$ where
$\ve_{\bar1\bar2}=1$ is the antiholomorphic volume form, it is easy
to deduce the following simple results, which will become very
useful in analysing the Killing spinor equation:
\be
\begin{array}{ccc}
\Gh_{\a}(1) = \sqrt{2} e_{\a} & \quad & \Gh_{\a\b}(1) = 2e_{\a \b} \\
\Gh_{\a}(e_{12}) = 0 & \quad & \Gh_{\ab \betab}(e_{12}) = -  2\ve_{\ab\betab}  \\
\Gh_{\betab}(1) = 0 & \quad & \Gh_{\a \betab}(1) = - g_{\a \betab}  \\
\Gh_{\betab}(e_{12}) = \ve_{\betab \db}\Gh^{\db}(1) & \quad &
\Gh_{\a \betab}(e_{12}) = \frac{1}{4}g_{\a \betab} \ve_{\gb\db}\Gh^{\gb\db}(1)
\end{array}
\ee
From these relationships, we see that an alternative basis for the spinors $\Delta^{\bC}$ is given by
\be
\{ 1, \; \Gh^{\bar1}(1), \; \Gh^{\bar2}(1), \; \Gh^{\bar{12}}(1) \} \; .
\ee
Thus the Hermitian basis is an oscillator basis consisting of creation and annihilation operators, with the Clifford vacuum represented by the spinor $1$.

In the case of \sut-\inv\ spinors, this basis will greatly simplify the solution of the \kses.
 To this we now turn.
\section{The $SU(2)$ orbit} \label{orbsut}
In this section, we first derive a canonical form for spinors in $\Dsm$ with stability subgroup $SU(2)$.
 This enables us to explicitly compute their associated spacetime exterior forms.
 Then we substitute the canonical form into the \kses\ and derive constraints on the field strength, the spinor parameters and the geometry itself.
 We will then seek to interpret these as geometric information about five-dimensional backgrounds which admit \sut-\inv\ \kss.
 In doing so, we verify some results of \cite{5dclass}.
\subsection{A Canonical Form for \orbsut} \label{sutcan}
We begin by decomposing the complex spinor representation of \spf\ under the action of \spfr.
 It splits into a direct sum of two inequivalent irreducible complex representations of \spfr\ (see Appendix \ref{complreprs}), as
\be \nn
\D^{\bC} = \D^+ \oplus \D^- \; ,
\ee
where $\D^{\pm}$ both have complex dimension of $2$.
 Following Wang in \cite{wang}, we now decompose the \spfr\ representations into \sut\ representations as follows:
\bea
&& \D_{\mathbf2}^+ \cong \L_{\mathbf1}^0(\bC^2) \oplus \L_{\mathbf1}^2(\bC^2) \\
&& \D_{\mathbf2}^- \cong \L_{\mathbf2}^1(\bC^2) \;,
\eea
where the bold subscripts denote the complex dimension of the representation.
 We see that the one-dimensional representations are carried by the $0$-forms and the $2$-forms, which make up the space $\D^+$.
 Therefore these components transform trivially under \sut\ and so are \sut-\inv.
 Since there are no non-trivial $1$-forms which are \sut-\inv, it must be the case that \orbsut\ is spanned by the $0$-forms and $2$-forms.
 So, using the SM condition, we see that the most general \sut-\inv\ SM spinor is of the form
\be
\z = \begin{pmatrix} a \\ -ib^* \end{pmatrix} 1 + \begin{pmatrix} b \\ ia^* \end{pmatrix} e_{12} \; , \quad \hbox{for }a,b,\in\bC\;.
\ee
Expanding this into real and imaginary parts shows that $\z$ is a linear combination with \textit{real} coefficients, of the following four linearly independent \sut-\inv\ SM spinors:
\begin{center}
\begin{tabular}{ccc}
    $\s_1 = \begin{pmatrix} 1 \\  0 \end{pmatrix} 1 + \begin{pmatrix} 0 \\ i \end{pmatrix} e_{12} \;, $
&   \qquad
&       $\s_2 = \begin{pmatrix} i \\  0 \end{pmatrix} 1 + \begin{pmatrix} 0 \\ 1 \end{pmatrix} e_{12} \;, $ \\
    $\s_3 = \begin{pmatrix} 0 \\ -i \end{pmatrix} 1 + \begin{pmatrix} 1 \\ 0 \end{pmatrix} e_{12} \;, $
&   \qquad
&   $\s_4 = \begin{pmatrix} 0 \\ -1 \end{pmatrix} 1 + \begin{pmatrix} i \\ 0 \end{pmatrix} e_{12} \;, $ \;.
\end{tabular}
\end{center}
Thus, $\{\s_1,\s_2,\s_3,\s_4\}$ forms a basis for the \sut-\inv\ SM spinors, \orbsut.

The basis spinors are related to each other by \spfr\ transformations:
\be
\s_1 = -\G_{13}\s_2 = -\G_{12}\s_3 = \G_{23}\s_4 \;.
\ee
This means that they all lie in the same orbit of \spfr\ in \orbsut. Consequently, we can use $Spin(1,4)$ gauge transformations and re-scaling to bring any \sut-\inv\ SM spinor into the form of any one of $\s_1,\s_2,\s_3,\s_4$. We choose our canonical form for \orbsut\ to be:
\be\boxed{
\zsut = \s_1 = \begin{pmatrix} 1 \\  0 \end{pmatrix} 1 + \begin{pmatrix} 0 \\ i \end{pmatrix} e_{12}
} \;.
\ee
Using this simple representative, it is straightforward to construct the associated \linebreak \sut-\inv\ \st\ exterior forms.
\subsection{Spacetime forms Associated to $\z^{SU(2)}$}
Let $\zsut$ be as above. Applying the results of section \ref{sutip}, all of the independent spacetime forms associated to the orbit \orbsut\ can be computed. Using the notation $\a^{kl}_{A_1\cdots A_p}=\a_{A_1\cdots A_p}(\z^k,\z^l)$, the following non-vanishing components can be found:

\beann
\a^{12} &=& -\a^{21} = -i \;,\quad\quad\quad
\a^{12}_0 = -\a^{21}_0 = 1 \;, \\
\a^{11}_{ab} &=& -(\a^{22}_{ab})^* = -\a^{11}_{(a+2)(b+2)} = (\a^{22}_{(a+2)(b+2)})^* = -\e_{ab} \;,\\
\a^{12}_{ab} &=& \a^{21}_{ab} = -\a^{12}_{(a+2)(b+2)} = -\a^{21}_{(a+2)(b+2)} = i\e_{ab} \;,\\
\a^{11}_{a(b+2)} &=& -(\a^{22}_{a(b+2)})^* = -i\e_{ab} \;,\quad\quad\quad
\a^{12}_{a(b+2)} = \a^{21}_{a(b+2)} = -\d_{ab} \;, \\
\eeann
where $a,b=1,2$. Hence we have the following spacetime forms associated to $\z^{SU(2)}$:
\begin{itemize}
\item \textbf{Zero-form}:
\be
\vp_{SU(2)}^{12} = -\vp_{SU(2)}^{21} = -i  \\
\ee
\item \textbf{One-form}:
\be \label{sut1form}
\k_{SU(2)}^{12}  = - \k_{SU(2)}^{21} = e^0  \\
\ee
\item \textbf{Two-forms}:
\bea
\o_{SU(2)}^{11}  &=& \!\!\! -(\o_{SU(2)}^{22})^* = -\left(e^1 \we e^2 - e^3 \we e^4 \right)
 - i \left(e^1 \we e^4 - e^2 \we e^3 \right)\\
\o_{SU(2)}^{12}  &=& \o^{21}_{SU(2)} = - \left(e^1 \we e^3 + e^2 \we e^4 \right)
\eea
\end{itemize}
We see that the scalar is pure imaginary and the one-form is real.

Actually, the two-forms provide the manifold with an \sut-structure.
 To see this, we first define an Hermitian basis for $\Vc$ by
\bea
&& \hat{e}^1 = \frac{1}{\sqrt2}(e_1+ie_3) \;,\qquad \hat{e}^2 = \frac{1}{\sqrt2}(e_2+ie_4)  \\
&& \hat{e}^{\bar1} = \frac{1}{\sqrt2}(e_1-ie_3) \;,\qquad \hat{e}^{\bar2} = \frac{1}{\sqrt2}(e_2-ie_4) \;.
\eea
Now, let $\hat\o$ denote the K\"ahler form associated to the complex structure $J$ and Hermitian metric $g$.
 Then,
\be
\hat\o=\frac{1}{2}g_{ik}J^k{}_j e^i\we e^j = -i\d_{\a\betab}\hat{e}^\a \we \hat{e}^{\betab}
= -\left( e^1\we e^3 + e^2\we e^4 \right) = \o_{SU(2)}^{12} \;.
\ee
Therefore the two-form $\o_{SU(2)}^{12}$ is in fact the K\"ahler form on $\Vc$.

Next, the holomorphic volume-form associated to the Hermitian basis is given by
\be
\e = \hat{e}^1\we\hat{e}^2 =\left(e^1 \we e^2 - e^3 \we e^4 \right) + i \left(e^1 \we e^4 - e^2 \we e^3 \right)
 = - \o_{SU(2)}^{11} \;.
\ee
Thus, $\o_{SU(2)}^{11}$ specifies the holomorphic form.
 This serves to show that the spinor $\zsut$ completely determines an \sut-structure $(\hat\o,\e)$ on $\Vc$.

Furthermore, we can obtain three linearly independent real two-forms $\Z^1$, $\Z^2$ and $\Z^3$ as follows.
 Define
\bea
\o^{11}_{SU(2)} =  \Z^1+i\Z^2  \qquad \hbox{and} \qquad
\o^{12}_{SU(2)} =  \Z^3   \;. \eea It can then be seen that the
forms \beann
\Z^1 &=& -\left(e^1 \we e^2 - e^3 \we e^4 \right) \\
\Z^2 &=& -\left(e^1 \we e^4 - e^2 \we e^3 \right) \\
\Z^3 &=& -\left(e^1 \we e^3 + e^2 \we e^4 \right)
\eeann satisfy the algebra of the imaginary unit quaternions, \be
\nn (\Z^I)_c{}^e(\Z^J)_e{}^d = -\d^{IJ}\d_c{}^d +
\ve^{IJK}(\Z^K)_c{}^d \; , \quad I,J,K = 1,2,3  \;\;\hbox{and}\;\;
c,d,e=1,2,3,4 \; . \ee
\section{Killing Spinor Equation Analysis} \label{sutkses}
We will now turn to the solution of the \kses\ for spinors which lie in the orbit \orbsut.
 The process is entirely analogous to the $D=11$ case.

We begin by expanding the two-form field strength into electric and magnetic parts:
\be \nn
F_{(2)} = \frac{1}{2!} F_{AB}e^{AB} = G_a e^{0a} + \frac{1}{2}F_{ab}e^{ab} \;.
\ee
Let $\xi\in\Dsm$.
 Putting the field strength expansion into the \kses, then separating the derivative terms and the gamma matrix terms into their respective time and spatial components results in the following pair of equations for each component $\xi^k$:
\bea \nn
0  &=&  \partial_0 \xi^k - \frac{1}{2}\O_{0,0I}\G_0\G^I\xi^k + \frac{1}{4}\O_{0,IJ}\G^{IJ}\xi^k
+ \frac{i}{\sqrt3} \left( \frac{1}{4}\G_0\G^{IJ}F_{IJ} -  \G^I G_I  \right) \xi^k \\ \nn
%
0 &=& \partial_K  \xi^k - \frac{1}{2}\O_{K,0I}\G_0\G^I\xi^k + \frac{1}{4}\O_{K,IJ}\G^{IJ}\xi^k \\ \nn
& & + \frac{i}{\sqrt3}\left( \frac{1}{4}\G_K{}^{IJ}F_{IJ} - \G^I F_{KI} + \frac{1}{2} \G_{K}{}^{I}\G_0 G_I
+ \G_0 G_{K} \right) \xi^k  \; ,
\eea
where $I,J,K=1,2,3,4$ are the spatial indices.

Now we can use the properties
\be
\{\G_0,\G^a\}=0 \;,\;\; \G_01= i \;,\;\; \G_0e_a=-ie_a \;,\;\; \G_0e_{12}=ie_{12} \;,
\ee
to commute $\G_0$ through each term in which it appears, so that it hits the spinor and produces a factor of $\pm i$. In this way, the equations can be expressed entirely in terms of the four-dimensional spatial gamma matrices as
\bea \nn
0 &=& \partial_0 \xi^k + \frac{i}{2}\O_{0,0I}\G^I\xi^k + \frac{1}{4}\O_{0,IJ}\G^{IJ}\xi^k
+ \frac{i}{\sqrt3} \left( \frac{i}{4}\G^{IJ}F_{IJ} - \G^I G_I \right)\xi^k  \label{5d:timecpt} \\  \nn
%
0 &=& \partial_K \xi^k + \frac{i}{2}\O_{K,0I}\G^I\xi^k + \frac{1}{4}\O_{K,IJ}\G^{IJ}\xi^k \\ \nn
& & + \frac{i}{\sqrt3}\left( \frac{1}{4}\G_K{}^{IJ}F_{IJ} - \G^I F_{KI} + \frac{i}{2} \G_{K}{}^{I} G_I
- i G_{K} \right) \xi^k  \label{5d:spatialcpt}  \; .
\eea

The next step is to expand each gamma matrix term and each tensor with respect to the Hermitian basis.
 Then, for each equation, collect together the terms proportional to the basis elements $1,\Gh^{\ab}(1)$ and $\Gh^{\bar1\bar2}(1)$.
 This will result in an expression of the form
\be
0 = C_0(1) + (C_1)_{\ab}\G^{\ab}(1) + C_2 \G^{\bar1\bar2}(1) \;,
\ee
for both the time and spatial components of the \kses.
 Since this basis expansion is equated with zero, the coefficient of each basis element must vanish separately, which gives rise to a set of four conditions associated to each component of the \kses.
 These are the constraints which we shall seek to interpret.

We must also consider the equations for $\xi^2$, but as we will see in each case, they give rise to an equivalent set of conditions. Therefore only one set needs to be investigated.
\subsection{The $\n=1$ Killing Spinor Equations for \orbsut}
We will now solve the Killing spinor equations for an \sut-\inv\ SM spinor $\z\in$\orbsut.
 The first step is to bring the spinor to canonical form, using \spf\ transformations to line $\z$ up with $\zsut$.
 However, now we are considering differential equations, we must treat the parameters in the spinors as \st\ functions rather than constants.
 Therefore, we can bring $\z\in$\orbsut\ into the form
\be
\z=r(x)\zsut= \begin{pmatrix} r(x) \\  0 \end{pmatrix} 1 + \begin{pmatrix} 0 \\ ir(x) \end{pmatrix} e_{12}\;,
\ee
where $r(x)$ is a \textit{real} function, as explained in section \ref{sutcan}.

Consider the first component $\z^1=r(x)1$.
 Substituting this spinor into the equations from section (\ref{sutkses}) and expanding into the Hermitian basis gives the following independent constraints:
\paragraph{Time component:}
\bea
%
%
\label{1sut:t1}
0 &=&  \partial_0 r(x) + r(x)\left( \frac{1}{2}\O_{0,\a}{}^{\a} - \frac{1}{2\sqrt3} F_{\a}{}^{\a}  \right) \\
\label{1sut:t2}
0 &=& r(x)\left( \frac{i}{2}\O_{0,0\gb} - \frac{i}{\sqrt3}G_{\gb}  \right)  \\
\label{1sut:t3}
0 &=& r(x)\left( \O_{0,\bar1\bar2} - \frac{1}{\sqrt3} F_{\bar1\bar2}  \right)
\eea
\paragraph{Holomorphic spatial component:}
\bea
\label{1sut:h1}
0 &=&   \partial_{\a} r(x) + r(x) \left( \frac{1}{2}\O_{\a,\g}{}^{\g} + \frac{\sqrt3}{2} G_{\a}  \right)  \\
\label{1sut:h2}
0 &=& r(x) \left(   \frac{i}{2}\O_{\a,0\gb} + \frac{i}{2\sqrt3}F_{\d}{}^{\d}g_{\a\gb}
-\frac{\sqrt3i}{2} F_{\a\gb}  \right) \\
\label{1sut:h3}
0 &=& r(x) \left(  \O_{\a,\bar1\bar2} + \frac{1}{\sqrt3} G^{\b} \ve_{\b\a}    \right)
\eea
\paragraph{Antiholomorphic spatial component:}
\bea
\label{1sut:a1}
0 &=& \partial_{\ab}r(x) + r(x) \left( \frac{1}{2}\O_{\ab,\g}{}^{\g} + \frac{1}{2\sqrt3}G_{\ab}  \right) \\
\label{1sut:a2}
0 &=& r(x) \left(  \frac{i}{2} \O_{\ab,0,\gb} - \frac{i}{2\sqrt3} F_{\ab\gb}   \right)  \\
\label{1sut:a3}
0 &=& r(x) \O_{\ab,\bar1\bar2}
\eea
We find an equivalent set of conditions for $\z^2$.
 Let us proceed to solve these constraints, to express the field strength in terms of the spin-connection and the spacetime function $r$.
\subsection{Solving the Killing Spinor Equations} \label{1sut:solution}
Some immediate relationships arising from equations (\ref{1sut:t2}), (\ref{1sut:t3}), (\ref{1sut:h3}) and (\ref{1sut:h3}) are, respectively,
\bea
G_{\gb}&=&\frac{\sqrt3}{2}\O_{0,0\gb} \label{1sut:G1}\\
F_{\gb_1\gb_2}&=&\sqrt3\O_{0,\gb_1\gb_2} \label{1sut:201} \\
G_{\gb}&=&-\sqrt3\e_{\gb}{}^{\a}\O_{\a,\bar1\bar2} \label{1sut:G2}\\
\O_{\ab,\betab\gb} &=& 0 \label{1sut:conn1} \;.
\eea
The expressions for $G_{\gb}$ impose that
\be \label{1sut:conn2}
\frac{1}{2}\O_{0,0\gb}=-\e_{\gb}{}^{\a}\O_{\a,\bar1\bar2} \; .
\ee
Now, adding (\ref{1sut:t1}) to its complex conjugate implies
\be \label{1sut:funct1}
\partial_0 r(x) = 0 \;,
\ee
so that the \ks\ is time-independent.
 Then, (\ref{1sut:t1}) gives
\be \label{1sut:tr1}
F_{\a}{}^{\a}=\sqrt3\O_{0,\a}{}^{\a} \;.
\ee
From (\ref{1sut:a2}) we obtain
\be \label{1sut:202}
F_{\gb_1\gb_2}=\sqrt3\O_{\gb_1,0\gb_2} \; ,
\ee
which, in conjunction with (\ref{1sut:201}) gives the following condition on the connection:
\be \label{1sut:conn3}
\O_{0,\betab\gb}=\O_{\betab,0\gb} \;,
\ee
and hence
\be \label{1sut:conn3a}
\O_{\betab,0\gb} = - \O_{\gb,0\betab} \;.
\ee
Next, we can trace (\ref{1sut:h2}) to find
\be \label{1sut:tr2}
F_{\a}{}^{\a}=\sqrt3\O_{\a,0}{}^{\a} \;,
\ee
which, together with (\ref{1sut:tr1}) gives
\be \label{1sut:conn4}
\O_{0,\a}{}^{\a}=\O_{\a,0}{}^{\a} \;.
\ee
Now, (\ref{1sut:h2}) gives
\be \label{1sut:11}
F_{\b\gb}=\frac{1}{\sqrt3}\left( \O_{\b,0\gb} + g_{\b\gb}\O_{0,\d}{}^{\d} \right) \;.
\ee
This equation and its complex conjugate implies that
\be \label{1sut:conn4a}
\O_{\b,0\gb}=-\O_{\gb,0\b} \;.
\ee
Turning to the remaining two equations, we can add the complex conjugate of (\ref{1sut:h1}) to (\ref{1sut:a1}) to find
\be \label{1sut:G3}
G_{\gb}=-\sqrt3\partial_{\gb} \log r \;,
\ee
or equivalently,
\be \label{1sut:conn5}
0 = 2\partial_{\gb} \log r + \O_{0,0\gb} \;.
\ee
Substituting back into (\ref{1sut:h1}) tells us that
\be \label{1sut:G4}
G_{\ab}=\sqrt3\O_{\ab,\g}{}^{\g} \;,
\ee
from which we find a final connection constraint:
\be \label{1sut:conn6}
\O_{\ab,\g}{}^{\g}=\frac{1}{2}\O_{0,0\ab} \;.
\ee

This exhausts the constraints which may be derived from the \kses\ for one SM spinor belonging to the orbit \orbsut.
\subsection{Summary}
The $\n=1$ \kses\ ensure that every component of the field strength is determined in terms of the geometry, and we can write
\bea \label{1sut:F}
F_{(2)}&=&\frac{\sqrt3}{2}\left( \O_{0,0\a} e^0\we e^\a + \O_{0,0\ab} e^0\we e^{\ab} \right)
+ \frac{\sqrt3}{2}\left( \O_{0,\a\b} e^\a\we e^\b + \O_{0,\ab\betab} e^{\ab}\we e^{\betab} \right) \nn \\
&+& \sqrt3\left( g_{\a\betab}\O_{0,\d}{}^{\d} \right)
+ \frac{1}{\sqrt3}\left( \O_{\a,0\betab} - \frac{1}{2}g_{\a\betab}\O_{0,\d}{}^{\d} \right) \;,
\eea
where the $(1,1)$ piece has been separated into its trace plus a traceless $(1,1)$-form. This expresses the decomposition of the field strength into irreducible representations of \sut.
 We can also interpret this expression in terms of the intrinsic torsion modules of an $\so(1,4)$ manifold which possesses an \sut-structure, using the results from Appendix \ref{GrayHerv}.

As well as the constraints on the field strength, there are a number of conditions relating certain components of the connection to one another.
 As we will see in the next section, there is a choice of frame in which these conditions hold. This leads us to some important geometric properties of the solution.
\subsection{The Geometry of the Spacetime}
Let us first consider the one-form defined by $\hat{\k}=r^2\k_{SU(2)}^{12} = r^2 e^0$, where $r(x)$ is the real \st\ function which parametrises $\z$. To see if $\hat{\k}$ is associated to a Killing vector field, we must verify that each independent component of the equation $2\nabla_{(A}\hat{\k}_{B)}=0$ is satisfied, where $\nabla_A\hat{\k}_B=\partial_A\hat{\k}_B -\O_{A,}{}^{C}{}_{B}\hat{\k}_C$. We find that
\bea
2\nabla_{(0}\hat{\k}_{0)} &=& \partial_0 r^2 \label{sutkv1} \\
2\nabla_{(0}\hat{\k}_{\a)} &=& \partial_\a r^2 + r^2 \O_{0,0\a} \label{sutkv2} \\
2\nabla_{(\a}\hat{\k}_{\b)} &=& r^2\left( \O_{\a,0\b} + \O_{\b,0\a} \right) \label{sutkv3} \\
2\nabla_{(\a}\hat{\k}_{\betab)} &=& r^2 \left( \O_{\a,0\betab} + \O_{\betab,0\a} \right) \label{sutkv4}
\eea
These equations all vanish due to constraints (\ref{1sut:funct1}), (\ref{1sut:conn5}), (\ref{1sut:conn3a}) and (\ref{1sut:conn4a}), respectively. Furthermore, $\hat{\k}^2=\d_{AB}\hat{\k}^A\hat{\k}^B=-r^4$, so that $\hat{\k}$ is indeed associated to a timelike Killing vector field, $k$.

Given a timelike Killing vector field, we can choose adapted coordinates so that $k=\partial_t$ and the metric can be written in the form
\be
ds^2=-r^4(dt+\a)^2+ds_4^2 \;,
\ee
where $ds_4^2$ is the metric of the four-dimensional space $B$ which is orthogonal to the orbits of $k$, and $\a$ is a one-form which is independent of $t$. An obvious choice of frame is to take $e^0=r^2(dt+\a)$ and let $\{e^J\}_{J=1}^4$ be a frame for $ds^2_4$, so that
\be \label{1sut:frame}
ds^2=-(e^0)^2 + (e^1)^2 + (e^2)^2 + (e^3)^2 + (e^4)^2 \;.
\ee

Now, the existence of the timelike Killing vector $k$ ensures that the frame does not depend on $t$. This means that for the torsion-free condition of the Levi-Civita connection to be satisfied,
\benn
\hbox{i.e.}\quad de^A+\O_{B,}{}^{A}{}_{C}e^B\we e^C=0 \;,
\eenn
we require that $\O_{0,JK}=\O_{J,0K}$. This is consistent with the constraints on the connection given by (\ref{1sut:conn3}), (\ref{1sut:conn3a}), (\ref{1sut:conn4}) and (\ref{1sut:conn4a}).
 It remains to interpret conditions (\ref{1sut:conn1}), (\ref{1sut:conn2}), (\ref{1sut:conn5}) and (\ref{1sut:conn6}).

Firstly, consider equation (\ref{1sut:conn1}), which says that the $(3,0)+(0,3)$ part of the connection vanishes.
 This is the condition which implies that the complex structure on $B$ is integrable, so that $B$ is complex (cf. section \ref{11d:2su5}).
 Therefore, since $B$ also possesses an \sut-structure, it is in fact a \textit{special Hermitian} manifold \cite{cabrera}.

Condition (\ref{1sut:conn2}) tells us that
\be
\frac{1}{2}\O_{0,0\gb} = \O_{\a,}{}^\a{}_{\gb} \;,
\ee
which expresses $\O_{0,0\gb}$ in terms of the geometry of $B$.
 In terms of intrinsic torsion modules (see Appendix \ref{GrayHerv}), we have
\be \label{1sut:w4}
(\rm y_4)_{\ab} = 2(\rm w_4)_{\ab} \;.
\ee
In other words, this is a relationship between the $\so(1,4)$- and $\su(2)$-structures.
 Furthermore, condition (\ref{1sut:conn6}) says that
\be
(\rm y_4)_{\ab} = 2(\rm w_5)_{\ab} \;,
\ee
so that in fact,
\be \label{w4=w5}
\rm{w}_4=\rm{w}_5 \;.
\ee

Let us investigate how this condition is related to the Grey-Hervella classification of almost Hermitian manifolds.
 Defining $(W_5)_k=\e^{ij}\na_{[i}\e_{jk]}$ we find that
\be
(W_5)_\a = \O_{k,}{}^k{}_\a - \e^{\b_1\b_2}\O_{\b_1,}{}^\g{}_{\b_2}\e_{\g\a} \;.
\ee
Now, using (\ref{1sut:w4}) and the fact that $(\rm{w}_5)_\a=\O_{\a,\gb}{}^{\gb}$, equation (\ref{1sut:conn5}) implies that $W_5=2(\rm{w_4})=2(\rm{w_5})$, and therefore
\be \label{1sut:w5}
W_5 = - d( \log r)  \;.
\ee
This is a geometric condition on the four-dimensional base-space $B$, which says that the Gray-Hervella class $W_5$ is exact.

In summary, the existence of an \sut-\inv\ \ks\ provides the following conditions on the spacetime:
\begin{itemize}
\item The field strength is completely determined by the Levi-Civita spin-connection, as in (\ref{1sut:F}).
\item There is a timelike Killing vector field $k$ associated to the spacetime, which enables us to choose adapted coordinates in which we can write the metric in the form (\ref{1sut:frame}).
\item The four-dimensional spatial manifold which is transverse to the orbits of $k$ is special Hermitian with an \sut-structure specified by the tensors $(g,\hat\o,\e)$ or equivalently, the connection constraints found from the \kses.
\item The \kses\ impose the geometric constraint that the Gray-Hervella class $W_5$ is exact.
\end{itemize}
We note the similarity between these results and the case of \suf-\inv\ \kss\ in an eleven-dimensional background.
\subsection{The $\n=2$ Killing Spinor Equations for \orbsut}
Now, let us investigate the solution of the \kses\ for two spinors belonging to the orbit \orbsut.
 This configuration is chosen so that their common stability subgroup is non-trivial.
 We assume that the first \ks\ is of the form $\z_1=r(x)\zsut$, so that all the constraints derived above may utilised in the equations for the second spinor.
 However, since we have used our $Spin(1,4)$ gauge freedom to transform $\z_1$ into canonical form, we are restricted to using \sut\ transformations to simplify $\z_2$.
 However, since the second spinor is also \sut-\inv, it must be of the generic \orbsut\ form, as described in section (\ref{sutcan}):
\be
\z_2 = \begin{pmatrix} a(x) \\ -ib^*(x) \end{pmatrix} 1 + \begin{pmatrix} b(x) \\ ia^*(x) \end{pmatrix} e_{12} \; ,
\ee
where $a$ and $b$ are \textit{complex} functions of the \st. Note that if we expand these functions into real and imaginary parts as $a=a_0+ia_1$ and $b=b_0+ib_1$, then we have
\be
\z_2 = a_0\s_1 + a_1\s_2 + b_0\s_3 + b_1\s_4 \;,
\ee
where $\{\s_1,\s_2,\s_3,\s_4\}$ is the basis of \orbsut\ found previously.
%
%
%
\subsubsection{Killing Spinor Equations for $\z_2$}
We obtain the \kses\ for $\z_2$ from the equations found in section (\ref{sutkses}). On substituting the $\n=1$ constraints into these equations, we find that three of them vanish identically, while the remaining six give the following conditions:
%
\bea
\label{2sut:time}
 \partial_0 a(x) \;=&0&=\;  \partial_0 b(x)  \\
\label{2sut:a}
\partial_{\a}  a(x) -  a(x)\O_{\a,\g}{}^{\g} \;=&0&=\; \partial_{\ab} a(x) +  a(x)\O_{\ab,\g}{}^{\g} \\
\label{2sut:b}
\partial_{\a}  b(x) -  b(x) \O_{\a,\g}{}^{\g} \;=&0&=\; \partial_{\ab} b(x) +  b(x) \O_{\ab,\g}{}^{\g}
\eea
Thus, there are no further constraints on either the geometry or the field strength, so the presence of a second \sut-\inv\ spinor whose parameters satisfy the above equations, will not alter the \sg\ background.
 In other words, given the first \sut-\inv\ \ks, the \kses\ are trivially solved for this second \ks.
 The constraints on the functions $a$ and $b$ are the only conditions which must be satisfied in order for the background to possess two \kss\ instead of just one.
\subsubsection{Constraints on $\z_2$}
As we have seen, there are no further constraints on the five-dimensional background.
 However, the conditions which arise on the complex parameter functions $a$ and $b$ give us important information about the form of $\z_2$.

Firstly, taking real and imaginary parts of equation (\ref{2sut:time}), we see that
\be
\label{2sut:t}
\partial_0 a_0 = \partial_0 a_1 = \partial_0 b_0 = \partial_0 b_1 = 0 \;,
\ee
so that $\z_2$ is \textit{time-independent}.

Now, equations (\ref{2sut:a}) and (\ref{2sut:b}) provide identical constraints on the functions $a$ and $b$.
 Let us first consider $a$.
 Adding and subtracting equations (\ref{2sut:a}) and their complex conjugates gives
\bea
\label{2sut:a1}
0 &=& \partial_{\ab} \log \{|a|^2\} + 2\O_{\ab,\g}{}^{\g}  \\
\label{2sut:a2}
\partial_{\ab} \log a^* &=& \partial_{\ab} \log a \; .
\eea
Now, equations (\ref{2sut:a1}) and (\ref{1sut:conn5}) along with their complex conjugates imply that
\be
\partial_i \log |a|^2 = \partial_i \log r^2 \;,
\ee
where $i=1,2,3,4$ labels the coordinate indices on the special Hermitian manifold $B$.
 Integrating this, we see that
\be
|a|^2=r^2 \;,
\ee
where we have set the integration constant to zero, since \kss\ are determined up to an overall scale \cite{jguggp}.
 This means that the real and imaginary parts of $a$ can be written as follows:
\be
a_0 = |r|sin \th_a \;, \qquad a_1 = |r|cos \th_a \;,
\ee
for some angle $\th_a\in[0,2\pi]$.
 Moreover, since we have identical constraints on $b$, there exists another parameter $\th_b\in[0,2\pi]$ such that
\be
b_0 = |r|sin \th_b \;, \qquad b_1 = |r|cos \th_b \;.
\ee

Now, equation (\ref{2sut:a2}) and its complex conjugate together imply that
\be
\partial_i \log a = \partial_i \log a^* \;.
\ee
Integrating and taking real and imaginary parts, we see that the constant of integration must be zero and also $cos\th_a=0$.
 This means that $\th_a=\frac{\pi}{2}$ or $\frac{3\pi}{2}$, which in turn implies that $a=a_0=\pm r$.
 Similarly, we also find that $b=b_0=\pm r$.
 Therefore, since $\z_2$ is determined up to an overall constant, there are two independent possibilities,
\be \label{2sut:z2}
\boxed{
\z_2 = |r|\left( \s_1 \pm \s_3 \right) \;.
}
\ee

In conclusion, if a $D=5$ \sg\ background possesses one \sut-\inv\ \ks\ of the form $\z_1$, then it admits a second \sut-\inv\ spinor of the form (\ref{2sut:z2}) without any additional constraints on the spacetime.
\section{One \sut-\inv\ Killing Spinor Implies Four}
Given one \sut-\inv\ \ks, the \kses\ for any other linearly independent spinor of \orbsut\ reduce to a set of constraints purely on the spacetime functions which parametrise the spinor, see equations (\ref{2sut:time}), (\ref{2sut:a}) and (\ref{2sut:b}).
 These are the only additional constraints, and there are no extra conditions on the field strength or the geometry.
 This means that the \sut-structure determined by the constraints on the Levi-Civita connection given in section \ref{1sut:solution} or equivalently by the tensors $(g,\hat\o,\e)$, is identical to that of the $\n=1$ case, and is reduced no further.
 Moreover, since the spinors lie in the same orbit of \spf, they give rise to equivalent spacetime forms.
 In particular, they yield the same timelike Killing vector, so that the spacetime metric may be written in the form (\ref{1sut:frame}) with no further conditions on the Hermitian base manifold $B$.

These considerations lead us to conclude that backgrounds admitting any number of \sut-\inv\ Killing spinors are equivalent.
 However, as was shown in section \ref{sutcan}, \orbsut\ is four dimensional, therefore there can only ever be a  maximum of four linearly independent \sut-\inv\ \kss\ in the background.
 From the above discussion we observe that the presence of four such spinors change neither the fields nor the \sut-structure which arise from the $\n=1$ case.
 In other words, if there exists one \sut-\inv\ \ks\ in the background, then the \kses\ are reduced in such a way that there are generically four linearly independent \kss\ of \orbsut.
 These have common stabiliser \sut\ and induce an \sut-structure on the four-dimensional Hermitian manifold $B$ which lies orthogonal to the orbits of the timelike Killing vector associated to the orbit \orbsut.
 To summarise, the existence of one \sut-\inv\ \ks\ implies the existence of four.

We know that the other stabilisers of spinors in $\Dc$ are \rt and $\{\id\}$, therefore the only way that a $D=5$ background will possess an \sut-structure is if its \kss\ all lie in the orbit \orbsut.
 Now, since there are a maximum of eight supersymmetries in $N=1$, $D=5$ \sg, and $dim_{\bR}(\mathcal{O}_{SU(2)})=4$, we can conclude that backgrounds possessing an \sut-structure are \textit{half-supersymmetric}.
 This is in agreement with the results of \cite{5dclass}.
%
%
%
%
%
%
%
\chapter{Killing Spinor Equation Analysis for Two $SU(4)$-invariant Spinors} \label{app:sufr}
In section \ref{11d:2su4} we consider $D=11$ backgrounds in which there are two \sufr-\inv\ \kss\ of the form
\bea
&&\eta_1 = f(x)(1+e_{12345})
\\
&&\eta_2 = \sqrt{2}g(x)(e_5+e_{1234}) \;,
\eea
where $f$ and $g$ are real functions of the spacetime.
 The former is \suf-\inv\ and the latter is \sufr-\inv.
 In this section, we give the solution of the Killing spinor equations for the $SU(4)$-invariant spinor.

To begin with, we substitute the expressions for the field strength which arise in the $\n=1$ case (see section \ref{11d:n=1}) into the \kses\ for $\eta_2$.
 Then we use the resulting equations to determine further conditions on the components of the field strength and spin-connection.
 In particular, we will find many more constraints on the geometry, resulting in an \sufr-structure on the ten-dimensional spatial manifold $B$.

Throughout this calculation, we utilise the torsion-free condition $\O_{i,0j}=\O_{0,ij}$ which arises from the \kses\ for $\eta_1$ and also equation (\ref{11d:conn2}), which expresses $\O_{0,0i}$ in terms of the geometry of the space $B$.
\section{Field Strength Substitution}
Let us then consider the expressions for the field strength which arise from the $\n=1$ \kses\ for $\eta_1$, and substitute them into the \kses\ for $\eta_2$.
 First, we analyse the constraints arising from the component $\mathcal{D}_0\eta_2=0$, see equations (\ref{11d:n=22t1})-(\ref{11d:n=22t5}).
 Using equations (\ref{11d:F1}) and (\ref{11d:G1}), we see that (\ref{11d:n=22t1}) and its complex conjugate
imply
\be
\O_{0,0\bar5}=\O_{0,05} \;,
\la{soloone}
\ee
and
\be
-2 \O_{0,05}+ (\O_{5,5\bar5}-\O_{\bar5,5\bar5})
+(\O_{5,\b}{}^\b-\O_{\bar5,\b}{}^\b)=0~.
\la{solotwo}
\ee
Next, using (\ref{11d:F2}) and (\ref{11d:G(0,3)}), condition (\ref{11d:n=22t2}) implies that
\be
4 \O_{0, \bar \a 5}+i \O_{\b_1,\b_2\b_3} \e^{\b_1 \b_2\b_3}{}_{\bar\a}=0~.
\la{solothree}
\ee
Equation (\ref{11d:n=22t3}) and its complex conjugate give the constraints
\be
\partial_0\log g=0
\la{solofour}
\ee
and
\be
\O_{0,\b}{}^\b-\O_{0,5\bar 5}+{i\over12} F_\a{}^\a{}_\b{}^\b
-{i\over 6} F_\a{}^\a{}_{5\bar5}=0~.
\la{soloo}
\ee
Using (\ref{11d:F2}), the latter can be rewritten as
\be
\O_{0,5\bar5}=\O_{0,\g}{}^\g~.
\la{solofive}
\ee

Using (\ref{11d:F(1,3)}) to eliminate the field strength components in (\ref{11d:n=22t4}), we find
\be
i \O_{5, \bar\b_1\bar\b_2}+ i \O_{[\bar\b_1, \bar\b_2] \bar5}
-{1\over2} \O_{0, \g_1\g_2} \e^{\g_1\g_2}{}_{\bar\b_1\bar\b_2}=0~.
\la{solosix}
\ee
Similarly, using (\ref{11d:G2}) and (\ref{11d:F(1,3)}), we get from (\ref{11d:n=22t5}) the condition
\be
2\O_{\bar 5, \bar
5\bar\a}+2\O_{5,\bar\a\bar5}-5\O_{0,0\bar\a}-2\O_{\b,\bar\a}{}^{\b}=0~.
\la{soloseven}
\ee

Now, we analyse the constraints which arise from the component $\mathcal{D}_{\ab}\eta_2=0$, see equations (\ref{11d:n=22a1})-(\ref{11d:n=22a10}).
 Equations (\ref{11d:F2}) and (\ref{11d:G(0,3)}) show that (\ref{11d:n=22a1}) gives \linebreak (\ref{solothree}), so that (\ref{11d:n=22a1}) is not independent.
 Next, substituting  (\ref{11d:G2}) and (\ref{11d:F(1,3)}) into equation (\ref{11d:n=22a2}), we find that
\bea
\O_{\bar\a, \bar\b5}-\O_{\bar\a, \bar\b\bar5} +{1\over3}
\O_{5,\bar\a\bar\b}-\O_{\bar 5,\bar\a\bar\b} -{2\over3}
\O_{[\bar\a,\bar\b]\bar5} -{i\over3} \O_{0,\g_1\g_2}
\e^{\g_1\g_2}{}_{\bar\a\bar\b}=0~.
\la{solaone}
\eea
Using conditions (\ref{11d:F(1,3)}) and (\ref{11d:G2}) to eliminate the field strength components from (\ref{11d:n=22a3}), we get
\bea
\partial_{\bar\a}\log g+{1\over2} \O_{\bar\a,\g}{}^\g-{1\over2}\O_{\bar\a,5\bar5}-{1\over6} \O_{\b,\bar\a}{}^\b
+{1\over6}\O_{5,\bar\a\bar5}+{2\over3}\O_{\bar5, \bar5\bar\a}-
{1\over6}\O_{0,0\bar\a}=0~.
\la{solatwo}
\eea
Similarly using (\ref{11d:F(1,3)}) and (\ref{11d:G2}), equation (\ref{11d:n=22a4}) gives
\bea
\O_{\bar\a,\g_1\g_2}\e^{\g_1\g_2}{}_{\bar\b_1\bar\b_2}
+ \left( {1\over3}\O_{5,5\d}-{1\over3}\O_{\bar\b,\d}{}^{\bar\b}+{1\over3}\O_{\bar5,\d 5}
+{1\over6}\O_{0,0\d} \right) \e^\d{}_{\bar\a\bar\b_1\bar\b_2}=0~.
\la{solathree}
\eea
Next, using (\ref{11d:F(1,3)}) and (\ref{11d:G2}) we see that condition (\ref{11d:n=22a5}) implies (\ref{solosix}).
 Also, using conditions (\ref{11d:tr2}) and (\ref{11d:F2}), we cannot eliminate all the components of the field strength from (\ref{11d:n=22a6}), and we obtain
\bea
i \O_{\bar\a, 0\g}+{1\over2}
F_{\bar\a\g5\bar5}-\frac{i}{3}\O_{0,\b}{}^\b g_{\bar\a\g}-
\frac{2i}{3} \O_{0,5\bar5} g_{\bar\a\g}=0~.
\la{solafour}
\eea
Substituting (\ref{11d:G2}) and (\ref{11d:F2}) into (\ref{11d:n=22a7}) gives
\bea
{1\over4} F_{\bar\a\bar5\g_1\g_2}-{1\over8}
\left( \O_{\bar\a,\bar\b_1\bar\b_2}
+\O_{[\bar\a,\bar\b_1\bar\b_2]} \right) \e^{\bar\b_1\bar\b_2}{}_{\g_1\g_2}
+{i\over3} g_{\bar\a[\g_1} \O_{\bar5, 0\g_2]}=0~.
\la{solafive}
\eea

Condition (\ref{11d:n=22a8}) relates the partial derivative of $g$ to the geometry of spacetime.
 Using (\ref{11d:G2}), we find that
\bea
\partial_{\bar\a}\log g-{1\over2} \O_{\bar\a,\g}{}^\g
-{1\over2} \O_{\g,\bar\a}{}^\g
+{1\over2} \O_{\bar\a,5\bar5}
+{1\over2}\O_{5,\bar\a\bar5}-{1\over2}\O_{0,0\bar\a}=0~.
\la{solasix}
\eea
Eliminating the field strength components from (\ref{11d:n=22a9}) using (\ref{11d:G2}) and (\ref{11d:F1}), we find that
\bea
\!\!\!\!\!\!\!\!{1\over3} (-{1\over2}\O_{0,05}-{1\over2}\O_{0,0\bar5}-\O_{5,\g}{}^\g
+\O_{\bar5,\g}{}^\g-\O_{5,5\bar5}+\O_{\bar5,5\bar5} ) g_{\bar\a\b}
+\O_{\bar\a,\b\bar5}+\O_{\b,\bar\a\bar5}=0~.
\la{solaseven}
\eea
Similarly, using (\ref{11d:F(1,3)}), we get from (\ref{11d:n=22a10})  that
\bea
4i\O_{\bar\a,0\bar5}-\O_{\g_1,\g_2\g_3} \e^{\g_1\g_2\g_3}{}_{\bar\a}=0~.
\la{solaeight}
\eea

We now turn to investigate the conditions that arise from the component $\mathcal{D}_{\bar5}\eta_2=0$, see equations (\ref{11d:n=2251})-(\ref{11d:n=2259}).
 Firstly, using conditions (\ref{11d:tr2}) and (\ref{11d:F2}), equation (\ref{11d:n=2251}) gives
\be
2\O_{0,5\bar5}+\O_{0,\g}{}^\g=0\;,
\ee
and together with equation (\ref{solofive}) implies that
\be \O_{0,5\bar5}=\O_{0,\g}{}^\g=0~.
\ee

Next, we use (\ref{11d:G2}) and (\ref{11d:F(1,3)}) to write (\ref{11d:n=2252}) as
\bea
\O_{\bar5,\bar\b5}-{2\over3} \O_{5,\bar\b\bar5}+{1\over3}
\O_{\bar5,\bar5\bar\b} -{1\over3}
\O_{\a,\bar\b}{}^\a-{5\over6}\O_{0,0\bar\b}=0~.
\la{solfone}
\eea
Using (\ref{11d:G2}), we can express the partial derivative of $g$ in (\ref{11d:n=2253}), as
\bea
\partial_{\bar5} \log g-\O_{\bar5,5\bar5}-{1\over2}\O_{0,0\bar5}=0~.
\la{solftwo}
\eea

Using (\ref{11d:G2}) and (\ref{11d:F(1,3)}) to eliminate the field strength components from (\ref{11d:n=2254}),
we find that
\bea
{1\over2} \O_{5,\g_1\g_2}\e^{\g_1\g_2}{}_{\bar\b_1\bar\b_2}
+{5\over6}\O_{\g_1,\g_25} \e^{\g_1\g_2}{}_{\bar\b_1\bar\b_2}
+{1\over3} \O_{\bar5, \g_1\g_2} \e^{\g_1\g_2}{}_{\bar\b_1\bar\b_2}
-{4i\over3} \O_{0,\bar\b_1\bar\b_2}=0~.
\la{solfthree}
\eea
We will see that (\ref{solfthree}) is not an independent condition.
 Similarly, using  (\ref{11d:F(1,3)}) we  find that condition (\ref{11d:n=2255}) implies (\ref{solaeight}), and so it is not independent.
 Furthermore, using (\ref{11d:G(0,3)}) and (\ref{11d:F2}) it is straightforward to see that (\ref{11d:n=2256}) implies (\ref{solothree}), and therefore it is also not an independent constraint.

Next, using (\ref{11d:G(0,3)}) we find that the condition (\ref{11d:n=2257}) implies
\bea
\O_{5, \b_1\b_2}+\O_{[\b_1,\b_2]5}=0~.
\la{solffour}
\eea
Also, one can show using (\ref{11d:F1}) and (\ref{11d:G(0,3)}) that (\ref{11d:n=2258}) yields
\bea
\partial_{\bar5}\log g-{2\over3} (\O_{\bar5,\g}{}^\g-\O_{5,\g}{}^\g)
+{1\over3} \O_{\bar 5, 5\bar5}+{2\over3} \O_{5, 5\bar5}
-{1\over6} \O_{0,0\bar5}+{1\over3} \O_{0,05}=0~,
\la{solffive}
\eea
and similarly using (\ref{11d:F1}), that (\ref{11d:n=2259}) gives
\bea
\O_{0,0\a}+2\O_{\a,5\bar5}+2\O_{\a,\g}{}^\g+2 \O_{\bar5,\a\bar5}=0~.
\la{solfsix}
\eea
This concludes the procedure of substituting the $\n=1$ conditions for the field strength components into the constraints which arise from the \kse\ for $\eta_2$.
\section{Geometric Conditions}
The conditions that have been derived which involve the spin-connection can be interpreted as restrictions on the geometry of spacetime.
 Here, we analyse these constraints to determine the independent components of the connection.
 We will utilise equation (\ref{11d:conn2}), which expresses $\O_{0,0i}$ in terms of the geometry of the ten-dimensional spatial manifold $B$.
 In particular, the connection component $\O_{0,05}$ can be expressed in terms of the geometry of $B$.
 Consequently, (\ref{soloone}) and (\ref{solotwo}) can be written as
\be
(\O_{\bar\b}{}^{\bar\b}{}_5-\O_{\b}{}^{\b}{}_{\bar5})- (\O_{5,\b}{}^\b+\O_{\bar5,\b}{}^\b)
- (\O_{5,5\bar5}+\O_{\bar5,5\bar5})=0
\la{anonea}
\ee
and
\be
-(\O_{\bar\b}{}^{\bar\b}{}_5+\O_{\b}{}^{\b}{}_{\bar5})
+ 2 (\O_{5,\b}{}^\b-\O_{\bar5,\b}{}^\b)
+2 (\O_{5,5\bar5}-\O_{\bar5,5\bar5})=0~,
\la{anoneb}
\ee
respectively.
 Alternatively, we can use (\ref{11d:daf}) to show that (\ref{soloone}) implies the constraint
\be
\partial_{5}f-\partial_{\bar 5} f=0~.
\la{anone}
\ee

Condition (\ref{solothree}) can be solved in terms of $\O_{0, \bar\a 5}$ to find
\be \O_{0, \bar\a 5}=
-{i\over 4} \O_{\b_1,\b_2\b_3}
\e^{\b_1\b_2\b_3}{}_{\bar\a}~,
\la{antwo}
\ee
which together with (\ref{solaeight}) implies that
\be \O_{0,\bar\alpha
5}=\O_{0,\bar\alpha \bar 5}~.
\la{anthree}
\ee
Equation (\ref{solofour}) implies that $g$ is independent of the frame time direction.

Next, condition (\ref{solosix}) can be solved to yield the relationship
\be
\O_{0,\b_1\b_2}={i\over4} (\O_{5,\bar\g_1\bar\g_2}+
\O_{\bar\g_1,\bar\g_2 \bar 5})
\e^{\bar\g_1\bar\g_2}{}_{\b_1\b_2}~.
\la{anfour}
\ee
Using equation (\ref{11d:conn2}), we see that condition (\ref{soloseven}) restricts the geometry of the space $B$.
 Substituting (\ref{anfour}) into (\ref{solaone}), we find that
\be
\O_{\bar\a,\bar\b 5}-\O_{\bar\a,\bar\b
\bar5}+\O_{5,\bar\a\bar\b} -\O_{\bar5,\bar\a\bar\b}=0~.
\la{anfive}
\ee
In particular, this gives
\be \O_{(\bar\a,\bar\b)
5}-\O_{(\bar\a,\bar\b) \bar5}=0
\la{ansix}
\ee and
\be
\O_{[\bar\a,\bar\b] 5}+\O_{5,\bar\a\bar\b}=0~,
\la{anseven}
\ee
where we have used (\ref{solffour}) in the final step.
 We will examine condition (\ref{solathree}) later.

Constraint (\ref{solafour}) determines the flux $F_{\bar\a\g5\bar 5}$ in terms of the connection.
 Similarly, (\ref{solafive}) expresses $F_{\bar\a\bar 5\g_1\g_2}$ in terms of the connection, and by taking its trace and comparing with (\ref{11d:F(1,3)}), we find that
\be
F_{\g\bar 5\d}{}^\d=\O_{0,\g\bar 5}=0~.
\la{annine}
\ee
Using condition (\ref{antwo}), this in turn implies that the totally anti-symmetric part of the connection vanishes,
\be
\O_{[\b_1,\b_2\b_3]}=0 \;.
\ee

Next, by adding (\ref{solatwo}) and (\ref{solasix}), and using (\ref{soloseven}), we obtain
\be
\partial_{\bar\a}g=\partial_{\bar\a}f~.
\la{anten}
\ee
The difference between (\ref{solatwo}) and (\ref{solasix}), again using (\ref{soloseven}), is
\be
\O_{\bar\a,\b}{}^\b-\O_{\bar\a, 5\bar5}+\O_{\bar5,\bar5\bar\a}-{1\over2} \O_{0,0\bar\a}=0~.
\la{antenb}
\ee
Condition (\ref{solaseven}) together with (\ref{solotwo}) implies that
\be
\O_{(\bar\a,\b),\bar 5}=\frac{1}{2}g_{\bar \a\b}\O_{0,0\bar 5}~.
\la{aneleven}
\ee
Also, by taking (\ref{soloone}) into account we find
\be
\O_{(\bar\a,\b)5}=\O_{(\bar\a,\b)\bar 5}~.
\la{antwelve}
\ee
Taking the trace of the preceding two relations yields
\bea
&&\O_{\bar\g,}{}^{\bar\g}{}_{\bar 5}+\O_{\g,}{}^\g{}_{\bar 5}=4\O_{0,0\bar 5}~,\la{anthirteena}\\
&&\O_{\bar\g,}{}^{\bar\g}{}_{ 5}+\O_{\g,}{}^\g{}_{5}=\O_{\bar\g,}{}^{\bar\g}{}_{\bar 5}+\O_{\g,}{}^\g{}_{\bar 5}~.
\la{anthirteenb}
\eea

Now, by adding (\ref{solftwo}) and (\ref{solffive}), we find that
\be
\partial_{\bar 5}\log g+\frac{1}{3}(\O_{5,\g}{}^\g-\O_{\bar 5\g}{}^\g)+\frac{1}{3}(\O_{5,5\bar 5}-\O_{\bar 5,5\bar
5})-{1\over6} \O_{0,05}=0~.
\la{anfourteen}
\ee
Together with its complex conjugate, this equation gives
\be
(\partial_{\bar 5}-\partial_5)g=0~.
\la{anfifteen}
\ee
Instead, if we subtract (\ref{solffive}) from (\ref{solftwo}), we get
\be
-(\O_{\bar 5,\g}{}^\g-\O_{5,\g}{}^\g)+2\O_{\bar 5,5\bar
5}+\O_{5,5\bar 5}+\O_{0,05}=0~,
\la{ansixteen}
\ee
where we have also utilised (\ref{soloone}).
 Now, the latter equation and its complex conjugate imply that
\be
\O_{5,5\bar 5}+\O_{\bar 5,5\bar 5}=0~,
\la{anseventeen}
\ee
and
\be
-(\O_{\bar 5,\g}{}^\g-\O_{5,\g}{}^\g)+\O_{\bar5,5\bar5}+\O_{0,05}=0~.
\la{anseventeenb}
\ee
Comparing this condition with (\ref{solotwo}), we see that
\be
\O_{0,05}=\O_{5,5\bar 5}
\la{anseventeenc}
\ee
and
\be
\O_{5,\g}{}^\g=\O_{\bar 5,\g}{}^\g~.
\ee

Substituting (\ref{anseventeenc}) into (\ref{solftwo}), we get
\be
\partial_{\bar 5}g=\partial_{\bar 5}f~,
\la{antenc}
\ee
where we have also used (\ref{11d:daf}).
 Therefore, equation (\ref{anten}) and (\ref{antenc}) imply that
\be
\boxed{ f=g } ~.
\ee
Equation (\ref{solfthree}) is not independent.
 This can be seen using (\ref{solffour}) and by comparing with (\ref{anfour}).

Let us now turn to investigate (\ref{solathree}).
 This can be written as
\be
\O_{\bar\a, \g_1\g_2}+ \left( {1\over3} \O_{5,5[\g_1}-{1\over3} \O_{\bar\b,[\g_1}{}^{\bar\b}-
{1\over 3} \O_{\bar 5,5[\g_1} +{1\over6} \O_{0,0[\g_1} \right) g_{\g_2]\bar\a}=0~.
\la{solathreeb}
\ee
Taking its trace, we get
\be
\O_{\bar \b,\a}{}^{\bar\b}+ \O_{5,5\a}+\O_{\bar 5,\a
5}+{1\over2} \O_{0,0\a}=0~.
\la{solathreetr}
\ee

It remains to investigate conditions (\ref{solathreetr}), (\ref{solfsix}), (\ref{antenb}), (\ref{solfone}) and (\ref{soloseven}).
 Adding (\ref{solfsix}) and (\ref{antenb}) we find that
\be
2\O_{\a, 5\bar5}+\O_{\bar 5,\a \bar 5}+\O_{5,5\a}=0 \;,
\la{twobi}
\ee
and by subtracting we obtain
\be
2\O_{\a, \b}{}^\b+ \O_{0,0\a}+\O_{\bar 5, \a\bar 5}-\O_{5,5\a}=0~.
\la{twostar}
\ee
Now, using (\ref{solathreetr}) and (\ref{twostar}) from the $\a$-component of (\ref{11d:conn2}), we can eliminate $\O_{\bar \b,\a}{}^{\bar\b}$ and $\O_{\a, \b}{}^\b$ to give
\be
\O_{5,5\a}+\O_{\bar 5, \a\bar5}-2\O_{\a, 5\bar5}=0~.
\la{sixa}
\ee
Comparing this with (\ref{twobi}), we get
\be
\O_{\a, 5\bar5}=0~,~~~~~~\O_{5,5\a}+\O_{\bar 5,\a\bar5}=0~.
\la{sixb}
\ee
Eliminating $\O_{\bar \b,\a}{}^{\bar\b}$ using (\ref{solathreetr}) from  (\ref{solfone}) and (\ref{soloseven}), we get
\be
\O_{5,\a\bar5}-{1\over3}\O_{\bar 5,\a5}+{2\over3} \O_{5,5\a}-{2\over3}\O_{0,0\a}=0
\ee
and
\be
\O_{5,5\a}+\O_{\bar 5,\a 5}-\O_{0,0\a}=0~.
\la{sixc}
\ee
Using the latter, the former becomes
\be
\O_{5,\a\bar5}+\O_{5,5\a}-\O_{0,0\a}=0~.
\la{sevena}
\ee
Now, subtracting (\ref{sixc}) from (\ref{sevena}), we find that
\be
\O_{5,\a\bar5}=\O_{\bar5,\a5}~.
\la{sevenb}
\ee
Substituting (\ref{sevena}) into (\ref{solathreetr}) and using (\ref{sevenb}), we get
\be
\O_{\bar\b,\a}{}^{\bar\b}=-{3\over2} \O_{0,0\a}~.
\la{sevenc}
\ee
Also, substituting (\ref{sevena}) and (\ref{sevenc}) into (\ref{solathreeb}) yields
\be
\O_{\bar\a, \b_1\b_2}+\O_{0,0[\b_1} g_{\b_2]\bar\a}=0~.
\ee
Furthermore, condition (\ref{twostar}) implies that
\be
2\O_{\a,\b}{}^\b+\O_{\bar5,\a\bar5}+\O_{\bar5,\a5}=0~.
\ee
This exhausts the analysis of the connection constraints, and we have determined the independent components.
 The final results are summarised in equations (\ref{summary00})-(\ref{summary5}), and it can be shown that these constraints provide an \sufr-structure on the spatial manifold $B$.
%
%
%
%
        \bibliographystyle{alpha}
    \end{document}